# OPTIMIZATION-BASED DESIGN AND ANALYSIS OF TAILOR-MADE IONIC LIQUIDS

By

AMIRHOSSEIN MEHRKESH

B.S. Isfahan University of Technology, 2006

M.S. University of Isfahan, 2009

A thesis submitted to the

Faculty of the Graduate School of the

University of Colorado in partial fulfillment

of the requirements for the degree of

Doctor of Philosophy

Civil Engineering

2015

This thesis for the Doctor of Philosophy degree by

Amirhossein Mehrkesh

has been approved for the

Civil Engineering Program

by

Arunprakash T. Karunanithi, Advisor

Kannan Premnath, Chair

Azadeh Bolhari

Indrani Pal

Fernando Rosario-Ortiz

21 November 2015

Mehrkesh, Amirhossein (PhD, Civil Engineering)

Optimization-based Design and Analysis of Tailor-made Ionic Liquids

Thesis directed by Associate Professor Arunprakash T. Karunanithi

## ABSTRACT

Solvents comprise two thirds of all industrial emissions. Traditional organic solvents easily reach the atmosphere as they have high vapor pressure and are linked to a host of negative environmental effects including climate change, urban air-quality and human illness. Room temperature ionic liquids (RTIL), on the other hand, have low vapor pressure and are not flammable or explosive, thereby resulting in fewer environmental burdens and health hazards. However, their life cycle environmental impacts as well as freshwater ecotoxicity implications are poorly understood. RTILs are molten salts that exist as liquids at relatively low temperatures and have unique properties. Ionic liquids consist of a large organic cation and charge-delocalized inorganic or organic anion of smaller size and asymmetric shape. The organic cation can undergo unlimited structural variations through modification of the alkyl groups attached to the side chain of the base cation skeleton and most of these structural variations are feasible, from chemical synthesis point of view, due to the easy nature of preparation of their components. Functionally, ionic liquids can be tuned to impart specific desired properties by switching anions/cations or by incorporating functionalities into the cations/anions. It is estimated that theoretically more than a trillion ionic liquid structures can be formed. Due to their tunable nature, these molten salts have the potential to be used as solvents for variety of applications.

This dissertation presents a computer aided IL design (CAILD) methodology with an aim to design optimal task-specific ionic liquid structures for different applications. We utilize group-contribution based ionic liquid property prediction models within a mathematical programming framework to reverse engineer functional ionic liquid structures. The CAILD model is then utilized to design optimal ionic liquids for solar energy storage, as a solvent for aromatic-aliphatic separation, and as an absorbent for carbon capture process. Using the developed CAILD model, we were able to computationally design new ionic liquid structures with physical and solvent properties that are potentially superior to commonly used ILs. The accuracy of the developed model was back tested and verified using available experimental data of common ILs.  However, we would like to note that the computational design results from this dissertation needs to be experimentally validated.

This dissertation also developed ecotoxicity characterization factors for few common ILs. The developed characterization factors (CFs), can be used in future studies to perform holistic (cradle-to-grave) life cycle assessments on processes using ILs to understand their environmental and ecological impacts.

The form of and content of this abstract are approved.  I recommend its publication.

Approved: Arunprakash Karunanithi

# DEDICATION

This dissertation is dedicated to my brilliant and outrageously loving and supporting wife, Anna and to my always encouraging, ever faithful mother, Soraya, and to the memory of my late father, Eskandar, who taught me how to live a peaceful and happy life, the person who will be missed forever.

# ACKNOWLEDMENTS

I am grateful to have had Dr. Arunprakash Karunanithi as my advisor. Without his knowledge, guidance, support, and enthusiasm towards this research, I would not have been able to complete this dissertation. He has taught me to be optimistic, persistent and confident in the work I am doing. He provided me all the tools needed to accomplish this research from financial support, to computer software to grants for attending conferences.

I also would like to thank the faculty, staff and my friends (fellow graduate students in our research group) whom I interacted with during my graduate program at the University of Colorado Denver.

I also want to extend my acknowledgment to Dr. Azadeh Bolhari, Dr. Mike Tang and Mr. Eric Ziegler who helped me in editing and proofreading. I also would like to thank all my committee members for their constructive suggestions and feedback.

## TABLE OF CONTENTS

**Chapter**

## LIST OF TABLES

**Table**

## LIST OF FIGURES

**Figure**

# Chapter 1: Introduction

Ionics liquids (ILs) are an emerging new class of chemicals that show tremendous promise in creating customized designer compounds (solvents, electrolytes, energy storage media …), which can be used for new applications or to replace current materials that lack flexibility or don't meet ecological safety concerns.[1] Ionic liquids are normally comprised of a large organic cation with positive charge and a charge-delocalized organic or inorganic anion of smaller size (can be monoatomic such as $Cl^-$) and asymmetrical shape.[2] The molecules possess a strong positive and negative charge which lends to its name as an ionic liquid. First ionic liquid – triethylammonium nitrate – was discovered more than a century ago.[3]

Compared to the case of naturally occurring ionic salts (e.g. $Na^+Cl^-$), the larger size of cations and anions in ionic liquids will result in distribution of a small charge (+1 or -1) over a much larger surface area. This fact, along with the asymmetric nature of cations and anions, explain the lower melting points of ionic liquids. Ionic liquids (ILs) are salts that normally melt at 100°C or less.[1,4,5] A schematic of an ionic liquid, 1-Butyl-3-methylimidazolium tetrafluoroborate, [Bmim] $BF_4^-$ is shown in Figure 1-1. The cation head group, imidazolium, with positive charge (+1) is shown in green, side chain groups attached to the cation-base, butyl and methyl are shown in yellow and anion with negative charge (-1) is shown in brown.

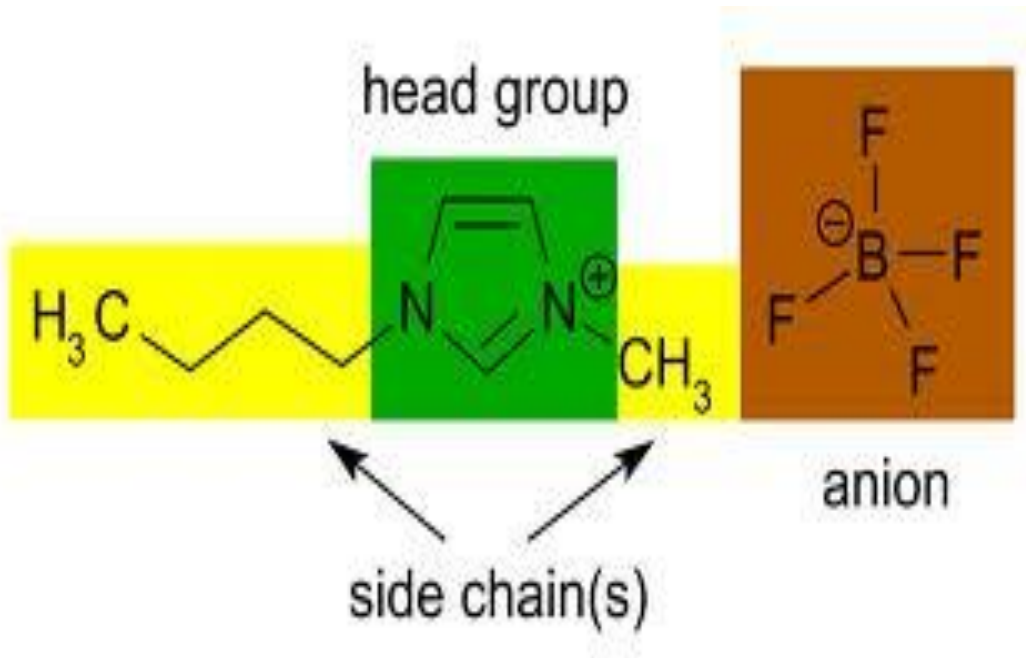

**Figure 1-1:** A schematic of an ionic liquid, $[Bmim]^{+}\ BF_4^{-}$

Interest in ionic liquids has continued to build in the academia and industry due to their interesting tunable properties and potential to provide environmentally friendly alternative to volatile organic compounds (VOCs) currently used in chemical/industrial processes. The properties of ionic liquids (pure physical properties such as viscosity and mixture properties such as solvency power) vary enormously as a function of their molecular structure, i.e. the type of cation-base, anion and number/type of side chain alkyl/functional groups present in the structure. ILs also offer a wide window of liquid state making them attractive as liquid solvents since they normally have high boiling points and very low vapor pressures. ILs are rarely flammable or explosive, thereby presenting fewer environmental burdens and health hazards.

During the past few years, considerable effort has been devoted to identifying and understanding ILs that have superior properties. The desirable properties of ionic liquids include:

- Negligible vapor pressure
- Ability to dissolve organic, inorganic, and polymeric materials
- High thermal stability (i.e. they do not decompose over a large temperature range)
- Versatile and customizable for task specific applications
- Nonvolatile and rarely flammable or explosive
- Strong regeneration properties that allow for their reuse and recycle
- Room temperature ionic liquids (RTILs) can be fluid at temperatures as low as -96°C
- Liquid phase temperature range from -96°C to 300°C; thermally stable up to 200°C
- Moderate to high electrical conductivity

These properties of ionic liquids (ILs) make them attractive as potential alternatives to current chemical compounds. Ionic liquids can be tailor-made for different applications (task-specific ionic liquids) by varying the building blocks of ionic liquids (i.e. cations, anions or groups attached to the cation-base). Thus Ionic liquids present a fantastic opportunity to step away from the status quo of utilizing volatile organic compounds, which are caustic to the environment, as solvents in current chemical processes. As the continued push towards environmentally conscious decisions at all level of industry continues, ionic liquids have the necessary properties and customizability to deliver better alternatives with reduced environmental impacts.

Ionic liquids can easily be removed from water. The ability to separate solvents from water is a critical property as industrial water pollution is increasingly becoming a major issue. Tolerance towards ecologically unsound industrial practices are diminishing on a global scale as the anthropogenic effects on the planet becomes more evident every year.

Ionic liquids have the potential to reduce the overall cradle to grave environmental/ecological impacts of current processes by offsetting upstream pollutant release from energy use and by-products which are manufactured through the VOC creation process. Ionic liquids have the potential of lower unusable by-products to designed product ratio. Their regenerative properties further their green profile as their potential for reuse and recycle exceeds those of currently used VOCs.

As regulatory oversight of emission of chemicals released into water and the atmosphere continues to tighten, finding alternatives will reduce the economic burden on industry. Industry has a strong concern towards negative externalities that result from their economic activity. In addition, adoption of environmentally friendly technologies can lead to greater acceptance from general public and can reduce the opportunity of public outcry or protest. It is far superior to develop alternatives now, than to wait for a forced decision.

Design of alternate ionic liquids can be made easy if we know how different structural groups present in them will influence the properties of interest. For example, if we are interested in an ionic liquid with high solvency power towards a specific chemical, we need to know which cations or anions contribute to higher values of the solvency power towards that compound. After identifying the best cations and anions, addition of functional/alkyl groups to the side chain of the cation-base can further help us fine tune the desirable properties (e.g. relatively lower values of melting point or low viscosity). In the next section, we discuss how computer models can help us expedite the process of selecting optimal ionic liquids for different applications. A computer-based ionic liquid design model can reduce the enormous number of experiments needed to find the optimal candidate thereby saving time and money.

## 1.1 Computer-aided ionic liquid design (CAILD)

The vast number of combinations of ionic liquids (ILs) is what provides their versatility and customization properties (estimated to be as many as $10^{14}$ ionic liquids feasible). Ionic liquids are still considered as a new generation of chemicals, which are garnering attention from academia and industry. Therefore, there is limited information on the properties of less common ionic liquids in chemical libraries and databases. Without the necessary information, random synthesis of ionic liquids and testing of their properties is costly and time consuming. Computer-aided molecular design (CAMD) is a promising approach that has been used for molecular systems to design compounds (e.g. solvents) for a variety of applications.[6-12] CAMD method integrates property prediction models and optimization algorithms to reverse engineer molecular structures with unique properties of interest. Due to the fact that ionic liquids are made of replaceable building blocks (structural groups), we believe that a similar approach is even more relevant for designing tailor-made ionic liquids.[13]

Use of computer-aided models to design optimal compounds, reduces the costs and allows engineers to model a multitude of potential candidates for a specific application. We propose that CAMD approach can be adapted specifically towards ionic liquid design [Computer-aided Ionic Liquid Design (CAILD)] where we take into account cations, anions and functional groups attached to cation core.[13-16] In this dissertation, we successfully show that the CAILD model is capable of creating ionic liquids with optimal desired properties for different applications (e.g. an ionic liquid with high thermal storage capacity ($\eta = \rho C_p$) can be a good candidate for a solar thermal storage process).

Based on the above discussion, it is clear that in order to find optimal ionic liquids for different applications using a computer-aided design framework, we need to know how different structural groups in an ionic liquid will contribute towards the properties of interest. For example, let's consider a situation where we want to design a good solvent to remove toluene from a multicomponent mixture. If we know that imidazolium cation-base usually results in higher values of solvency power towards toluene, compared to the other cation-bases, then an ionic liquid with imidazolium cation should always be chosen as the optimal ionic liquid unless it violates other physical properties or process constraints (such as the selected ionic liquid has an unacceptably high melting point or viscosity values).

In order for us to be able to design an optimal ionic liquid for an application of interest, we need to make sure that first it is a theoretically feasible chemical structure (chemical feasibility constraints) and secondly the designed ionic liquid meets other process criteria necessary for it to be used in large industrial scales. Therefore, the optimization framework needs an objective function (the value of the property of interest) that needs to be minimized or maximized along with a set of constraints that should be satisfied to guaranty that the designed ionic liquid is a feasible candidate. A good example of the type of constraints needed in a CAILD model relates to the design of an ionic liquid which is liquid at room temperature. Here a constraint of $T_m$<25 ˚C should be enforced within the optimization model.

Based on the above discussion, it is clear that we need models capable of predicting different properties of ionic liquids based on the type and number of structural groups present in them.[17-24] These models commonly referred to as group contribution (GC) models have been -to some extent- developed for ionic liquids. The GC models can be used within a CAILD framework to enable prediction of physical properties of the ionic liquids during the design process. Without comprehensive group contribution models capable of predicting different properties of ionic liquids, it is not possible to utilize the power of CAILD models to their full extent. In other words, CAILD models are most useful when they have the capability of exploring *all* possible combinations of ILs towards finding the optimal candidate for a given application and this is not possible unless we have group contribution models covering all cations, anions and side chain groups.

Therefore, when group contribution models for certain properties or contribution parameters for certain structural groups are not available one needs to develop these from scratch. This is where unavailability of experimental data on the properties of interest could be problematic since accurate group contribution models cannot be developed without sufficient amount of experimental data.

It would be important to point out that for certain cases first order group contribution models for of ionic liquids are not able to predict certain properties accurately. The first order group contribution models simply consider one value for the contribution of a particular group irrespective of where that group is located within the ionic liquid (e.g. they do not distinguish between a $CH_2$ group directly attached to the aromatic carbon and a $CH_2$ group attached to other aliphatic side chain groups). In situations where group contribution models do not work properly, other approaches such as computational chemistry based correlative models or Quantitative Structure Property Models can be utilized to predict the physical properties of ionic liquids.

In this research, we utilized COSMO-RS (Conductor like Screening MOdel for Real Solvents), a quantum chemistry-based equilibrium thermodynamics model with the purpose of predicting the chemical potentials of compounds in the liquid phase, to predict pure properties (e.g. melting point or viscosity) or mixture properties (e.g. activity coefficients and solubility) of ionic liquids. When an optimal ionic liquid is designed for a specific application using the CAILD model predictions based on COSMO-RS model can serve to validate the results and show us whether the design solution is suitable for use in large industrial scales.

Further, computational chemistry models can provide a strong foundation of information to build libraries of data to draw upon for future research and development. Modeling reduces the overall cost by eliminating ionic liquids whose properties do not meet the desired application.

## 1.2 Current applications

During the past few years, ILs have been studied for variety of applications. Some of them are listed below:[2, 5, 25-30]

- Battery Technologies
- Advanced fuel cell concepts
- Dye sensitized solar cells (DSSC)
- Thermo-electrical cells
- Supercapacitors
- Hydrogen generation through water splitting
- Carbon ($CO_2$) capture
- Nuclear fuel processing
- Solar (thermal) energy storage

BASF's commercial investigation of ILs reveals that the compounds have strong potential as solvents that can provide efficiency improvements in a several applications including:

- in chemical reactions and separation processes
- as hydraulic fluid and lubricant
- as polymer additives (antistatic)
- in metal deposition processes
- in dissolving and processing cellulose
- as electrolytes in electronic devices

## 1.3 Environmental impacts of ionic liquids

The non-volatile nature of ionic liquids greatly limits the impact on air quality by reducing or completely eliminating their direct emissions to the atmosphere. For this reason ionic liquids are often considered as inherently green/environmentally benign solvents with the potential to completely replace traditional volatile organic solvents in several applications. However toxicological studies have shown that some ionic liquids are very toxic towards freshwater organisms or human cell lines,[31-34] but due to their immense variety, ionic liquids can be designed/tuned to be environmentally benign.[35] In order to conclude that ionic liquids (ILs) are benign alternatives to molecular solvents, their environmental impacts need to be analyzed in a holistic manner. Life cycle assessment (LCA), which is a technique for assessing the environmental aspects associated with different steps in the production of a product, can be performed on ionic liquids like any other chemical compound to evaluate their true greenness. It is worth mentioning that, even though life cycle analysis of ionic liquids could be very beneficial for the informed selection of these compounds, it is quite challenging. The challenge arises from the fact that: ionic liquids are not yet produced in the large commercial scales. Consequently, no primary data is available on material/energy consumption and direct environmental discharges during their production. On the other hand there is very little data on the environmental fate, transport and toxicity of ionic liquids in the literature.

As environmental impact studies continue to reveal the negative impacts that VOC compounds have on the environment, chemist and engineers are striving to develop alternatives that reduce the overall ecological impact of current chemical processes. Although the ionic liquid field is developing rapidly it is important to consider the environmental, ecological, and human health impacts at the design stage for their successful use and long-term acceptance. Currently, there is very little understanding of the environmental impacts of producing ionic liquids as well as their impacts on fresh water ecotoxicity once they are released to the environment.

## 1.4 Ionic liquids safety

The low volatility/negligible vapor pressure of ionic liquids eliminates an important pathway for their release into the environment. The diversity of the ionic liquids' variants available makes the process of selecting the ones that meet the defined safety requirements easier. A study shows that ultrasound waves can convert a solution of imidazolium-based ionic liquids with acetic acid and hydrogen peroxide ($H_2O_2$) to less harmful compounds.[36] Despite the fact that ionic liquids mostly have negligible vapor pressure, few of them have shown combustible properties and therefore should be handled carefully.[37] A brief exposure of some ionic liquids (~5 seconds) to a flame torch can ignite them.

# Chapter 2: Forward Problem, Prediction of Melting Point and Viscosity of ILs

In this chapter, we present two empirical correlations to predict the melting point and viscosity of ILs in a way that does not require experimental input or complex simulations, but rely on inputs from simple calculations based on standard quantum chemistry (QC). To develop these correlations, we used data related to size, shape, and electrostatic properties of cations and anions that constitutes the ionic liquids.

## 2.1 Introduction

As it can be interpreted from their name, ionic liquids are composed of ions, a cation and an anion, but their properties can significantly vary from their relatives, salts, in two main ways. First, the properties of salts can be mostly attributed to their ionic nature since strong ionic bonds hold the particles together. Ionic salts are mostly made of small monoatomic ions, which are in the close vicinity of each other in their crystal network. Since the lattice energy of a crystalline compound is proportionally related to the inverse of the distance between the two components, the ionic bonds of salts are very strong, which contributes to properties such as very high melting point and high viscosity. On the other hand, ionic liquids are made of larger multiatomic cations and anions that result in weaker ionic bonds compared to that of salts. This explains the considerably lower melting point (many of them are in liquid state at room temperature) and viscosity of ionic liquids.  Secondly, contrary to salts, ionic liquids do not occur naturally in the environment and must be artificially synthetized.

The multiatomic nature of cations and anions in ionic liquids presents a great opportunity for researchers to fine tune ILs' properties and tailor them for different applications. In ionic liquids, mainly cations and occasionally anions are composed of several alkyl side chain groups ($CH_2$, $CH_3$…) and functional groups (OH, $NH_2$, COOH...). A vast number of different ionic liquids (an estimated number of $10^{14}$ ionic liquids)[1] can be potentially synthesized through distinct combinations of different cation-bases, alkyl groups, functional groups (attached to cations or anions), and anions. Careful evaluation of experimental data from literature on the physical and thermodynamic properties of ILs shows that substituting one type of functional group or anion with a different type can drastically alter the property of interest, such as its solvency power towards a specific compound. Such behavior and trends can be seen in all different categories of ionic liquids.

Despite the fact that the ionic bonds in ILs are relatively weak, their properties can still be attributed to their ionic nature as even a weak ionic bond is still much stronger than other types of intermolecular forces. Studies show that after ionic forces, hydrogen bonds between ionic particles (cations and anions) are the most important contributor to physical properties of ILs. Even though there are, potentially millions of different ILs that are possible, to date only a few hundred of them have been actually synthetized. It is not humanly possible to synthesize every feasible ionic liquid; therefore, we need to customize and intelligently design them before synthesizing for task-specific applications. Computer-aided optimization frameworks can help us design optimal ionic liquids suitable for a wide range of applications from an extraction solvent to thermal energy storage.[38] Ionic Liquids are generally salts that are liquid below 100°C. Therefore, not all ILs are in the liquid state at room temperature.

When it comes to the melting point of ILs, those with significantly lower melting points or "room temperature ionic liquids [RTILs]" ($T_m$<25°C), are of great interest to researchers seeking new application for ILs. The reason for the desirability of low melting point ionic liquids is the fact that ILs are being considered as separation solvents for selective dissolution of gaseous (e.g. $CO_2$), liquid (e.g. toluene), and solid (e.g. cellulose) solutes. They are also widely considered as liquid solvents to promote chemical reactions.[39,40] From a practical view point, for an IL to be used as an industrial solvent it needs to be transported (pumped) across multiple unit operations and therefore it must be in the liquid phase.

Another significant barrier towards commercialization of IL based applications is their high viscosity that occurs due to their ionic nature (existence of strong ionic bonds) making them difficult to transport. It is necessary to have powerful pumping equipment and efficient process equipment to handle viscous fluids. Therefore, looking for and customizing ionic liquids that have relatively low viscosity and melting point will greatly aid in commercialization of ionic liquids.

Studies related to the ionic materials show that strong ionic bonds are mainly responsible for holding charged particles together. Crystal lattices of ionic materials (e.g. ionic liquids) are made of cations and anions held together by electrostatic attraction. The ionic force between charged particles is directly proportional to the charge of each particle and inversely to the distance between the two ions. The larger the cations and anions are, the weaker the ionic bonds between them would be. This is due to the fact that by increasing the distance between two ions the electrostatic attraction, which holds them together, will be reduced.

## 2.2 Methods

Widely available information related to melting point and viscosity of salts explicitly shows that there is a meaningful relationship between the magnitude of the two discussed properties and the lattice energy of ionic bonds. Generally, in an ionic compound, size (volume and area), shape (sphericity), molecular weight, and dielectric constant of ions play an important role in determining the strength of the ionic bond of the compound. Normally, the larger size of the positive and negative ions (cations and anions) results in longer distances between the ions in the crystal making the ionic bond weaker. On the other hand, the shape of the ions is also important as they are better packed together when they are more symmetrical in shape. It has been suggested that asymmetry of ions in an ionic compound, most likely, will decrease the melting point since ions are more loosely connected and can be separated from each other more easily (by applying lower amount of energy). When we compared the size of cations and anions of a variety of ionic liquids with their melting points and viscosities, we were able to observe that in the case of ionic liquids the relationship is much more complex. We came across ionic liquids, which violated the above discussed trends where certain ILs with relatively larger cations and anions did not necessarily have lower melting point or viscosity compared to smaller ILs. One reason for this is the fact that ionic bonds are not the only intermolecular forces responsible for holding the particles together and other types of forces such as hydrogen bonding and polar-polar forces also come into play. Therefore we developed new correlations to predict melting point and viscosity of ionic liquids using information related to the size, shape and electrostatic properties of their ions. In order to account for deviations related to the above discussed trends, in addition to the three descriptors, we included several other quantum chemical descriptors to refine the correlations with an aim to cover wide variety of ionic liquids. Another issue was that for many ionic liquids, there were multiple experimental values reported for the two physical properties. This inconsistency was especially observed in the case of melting point, primarily due to the fact that the process of synthesis and existence of impurities affects this property. We avoided considering ILs that had inconsistent experimental values during the development of the correlation. Ionic liquids selected for this study were all 1:1 (one cation and one anion) with delocalized charges. These types of ILs are normally able to avoid crystallization and form glasses compounds far below room temperature.[41]

Currently, the most widely used approach to predict the melting point of ILs is quantitative structure-property relationship (QSPR) methods, mostly combined with artificial neural networks (ANNs).[41] In this approach, there is a reasonably good correlation between actual and predicted melting points within a standard deviation of less than 10°C.[41] The limited availability of experimental data on physical properties of ILs is the main drawback of constructing good QSPR models. In recent years, simulations with molecular dynamics (MD) have evolved to study the behavior of ILs. The quality of these simulations strongly depends on the employed force fields. Several groups have tuned them specifically for ILs, while others have modified previously existing ones. For example, Alavi and Thompson have used MD simulations to predict the melting temperature of $[C_2MIm]^+PF_6^-$. The demanding simulation indicated a melting point that was approximately 43°C too high.[41] Maginn used a similar model for the two polymorphs of $[C_4MIm]^+Cl^-$ and obtained a $T_{fus}$ that was between 20 and 55°C too high.[41] The drawbacks of all MD simulations are the high load of computational calculations and need to know the crystal structure.

In order to collect quantum chemistry data of cations and anions, we used TURBOMOLE, which is a powerful, general-purpose Quantum Chemistry (QC) program, which can be used for ab-initio electronic structure calculations.[41] This software allows accurate prediction of cluster structures, conformational energies, excited states, and dipoles that can be used in a broad variety of applications. When a chemical compound, in our case a cation or anion, is simulated using TURBOMOLE it can be exported as a Cosmo file, which can be later used in COSMOtherm software. COSMOtherm is a universal tool, which combines quantum chemistry (QC) and thermodynamics to calculate properties of liquids.[41] This tool is able to calculate the chemical potential of different molecules (in pure or mixed forms) at different temperatures. In contrast to other available methods, COSMOtherm is able to predict thermodynamic properties of compounds as a function of concentration and temperature by equations, which are thermodynamically consistent.[41]

Previously, computational methods such as BP86/SV(P) optimization approaches were carried out within the TURBOMOLE program package through the resolution of identity (RI) approximation. The imported/created geometries were then used for further optimization with the TZVP basis set. Next, when program converged a file with .cosmo format was exported, which later was used in COSMOtherm software for further calculations. At the next step, the computational chemistry data associated with cations and anions in the selected ionic liquids required for the development of aforementioned empirical correlations were collected. The data obtained from COSMOtherm software was later used to develop the empirical correlations for the prediction of melting point and viscosity of ILs. In the COSMOtherm software BP_TZVP_C30_1401 dataset was chosen for all of the calculations.

## 2.3 Results and discussion

Molar mass of cation ($MW_C$), molar mass of anion ($MW_A$), volume of anion $[A^\circ]^3$ ($Vol_A$), volume of cation $[A^\circ]^3$ ($Vol_C$), Area of anion $[A^\circ]^2$ ($Area_A$), Area of cation $[A^\circ]^2$ ($Area_C$), dielectric energy of cation ($Di_C$), dielectric energy of anion ($Di_A$), symmetrical value of ions (σ), density $[\frac{kg}{m^3}]$ (ρ), radii of cation $[A^\circ]$ ($R_C$) and radii of anion $[A^\circ]$ ($R_A$), and the average distance between one cation and one anion in the network $[A^\circ]$ ($R_t$) were used as descriptors to develop empirical correlations to predict melting point and viscosity of ILs. The accurate measurement of volume, area and ionic radii of ions is only possible through imaging approaches such as X-ray diffraction. In the case of unavailability of X-ray data correlative or approximation approaches (e.g. van der Waals model) can be used to estimate the characteristics related to size and shape of particles. This data is very sparse for ILs and since we are interested in the development of a universal correlation covering a wide range of IL structures we have to rely on predictive data. Typical cations and anions in ILs, do not usually have spherical shapes, so it is necessary to estimate their ionic radii through correlations for further use. To develop a predictive approach to estimate approximate values for radii of the cations and anions, we selected ions for which experimental ionic radii data were available. Next, the radii of cations and anions were estimated using the van der Waals model in which all cations and anions were assumed to have spherical shape. Volume of corresponding ions were estimated from COSMOtherm software and were converted to the radii through eqn. (2-1).

$$R_{Calc} = \frac{3}{4} \times \frac{1}{\pi} \times Vol^{1/3} \qquad (2\text{-}1)$$

A linear correlation between the experimental radii of cations and anions (for which X-ray values were available) and the corresponding values of their van der Waals radii is displayed in Figure 2-1. As it can be seen from Figure 2-1, the actual and model predicted values for the radii of ions are perfectly correlated through a linear relationship with $R^2$=0.98622 and the corresponding equation shown in eqn. (2-2).

$$R_{Pred} = 0.48601 \times R_{Calc} - 0.01725 \qquad (2\text{-}2)$$

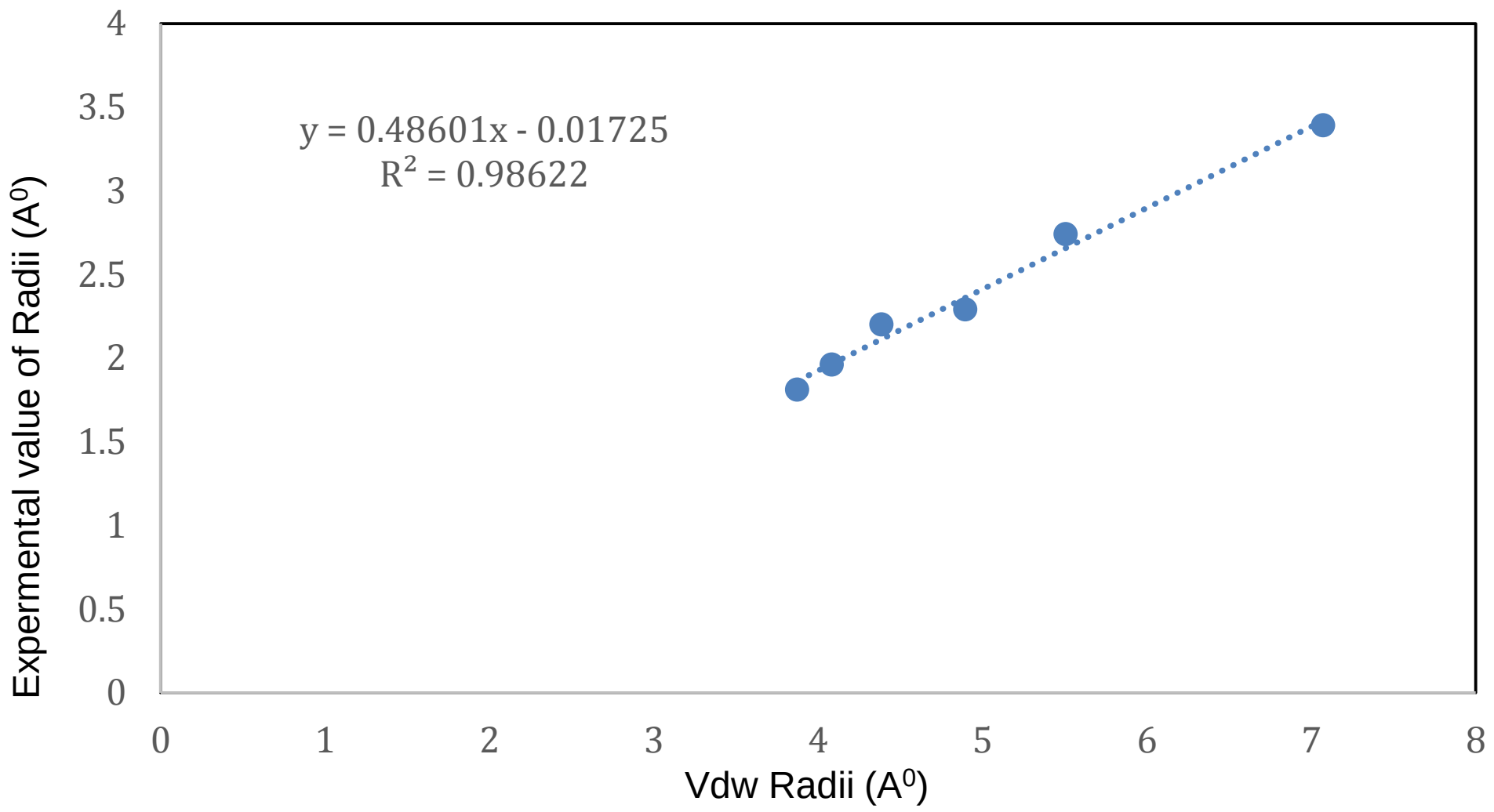

**Figure 2-1:** A correlation between van der Waals (VdW) and experimental Radii

Although, there were only few experimental data available for use in this correlation, since it covers a large range of VdW ionic radii from 3.8 A˚ to 7.5 A˚, we used this correlation to predict the unknown values of ionic radii for the rest of cations and anions considered in this study. The distance between anions and cations in the studied ILs, Rt, were estimated as the sum of ionic radii of cations and anions present in the crystal network of ILs. The symmetrical value of ionic liquids, σ, were calculated using the sphericity of cations and anions. Sphericity is a measure of how spherical an object is and can be calculated using the formula shown in eqn. (2-3).

$$Sph = \frac{\pi^{\frac{1}{3}}(6V_p)^{\frac{1}{3}}}{A_p} \quad (2\text{-}3)$$

where, $V_p$ and $A_p$ are the volume and surface area of the particle, respectively.

Further, the symmetrical value of ionic liquids were calculated through the following equation, eqn. (2-4).

$$\sigma = \sqrt{Sph_C * Sph_A} \quad (2\text{-}4)$$

where, $Sph_C$ and $Sph_A$ are the sphericity of cations and anions, respectively.

### 2.3.1 Melting point

Experimental data on the melting point of several ILs covering different categories (different type of cation head groups, and anions), were gathered from the literature.[42-59] A multivariate correlation with several inputs based on quantum chemistry parameters gathered from COSMOtherm, and the output parameter, experimental values of $T_m$, were performed utilizing Eureqa software, a powerful data analysis tool developed by Nutonian, Inc.

In the case of predicting the melting points of ionic liquids, 37 points of data on the actual melting point values of ILs were used, out of which 17 data points (i.e. 45% of the data points) were chosen solely for validation set, thereby not participating in the training process and are depicted as the points in green color in Figure 2-2.

The experiential data of the melting point of ILs and the multivariate trend line, representing the empirical correlation developed to predict the melting point, are shown in Figure 2-2. As it can be seen the trend line is capable of predicting the melting point of the selected ILs.

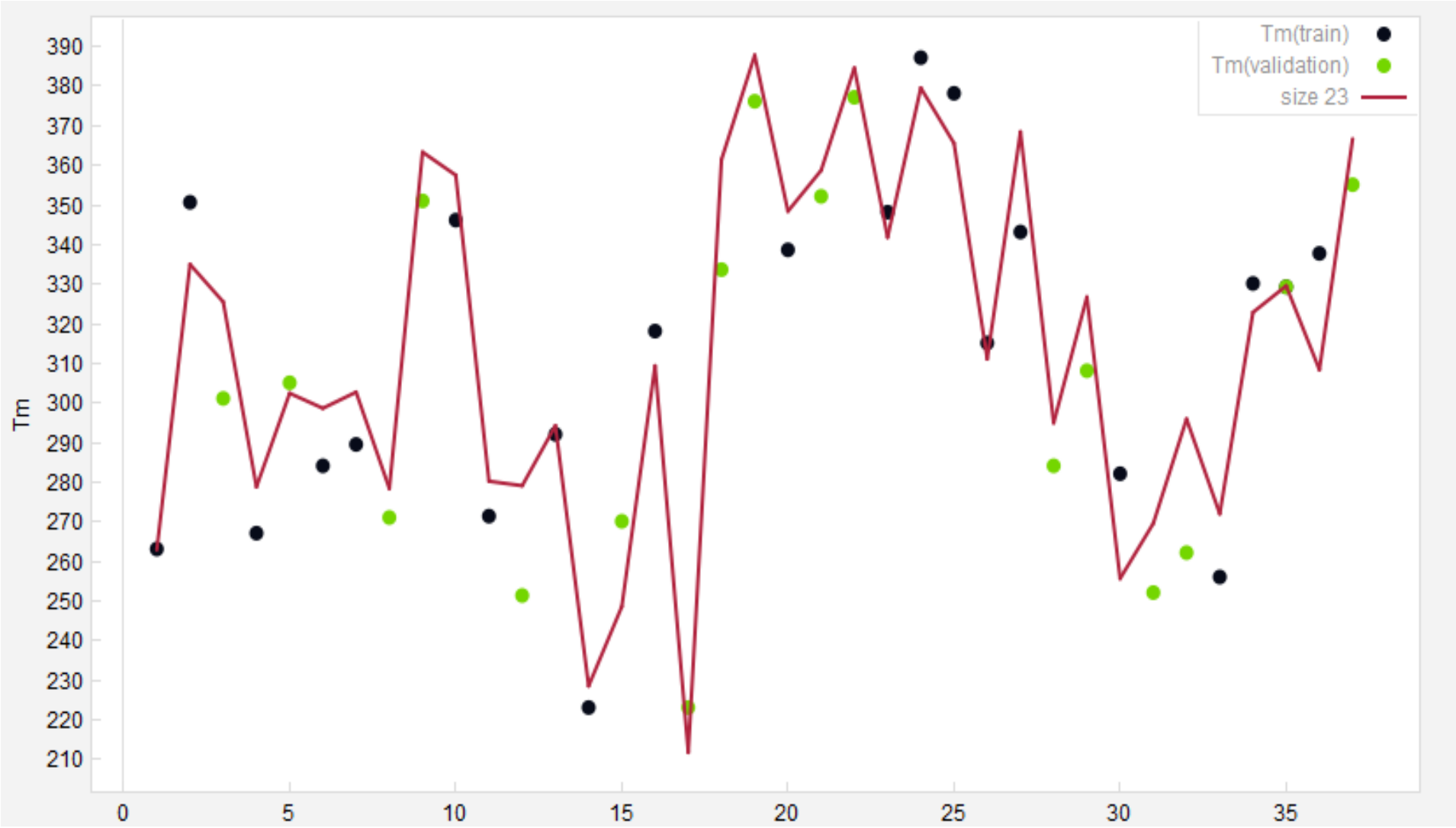

**Figure 2-2:** Actual vs. model predicted melting points for training and test data sets

At the next level, a comparison between the experimental (observed) values of melting points and their corresponding values, predicted by the correlative model, are shown in Figure 2-3.

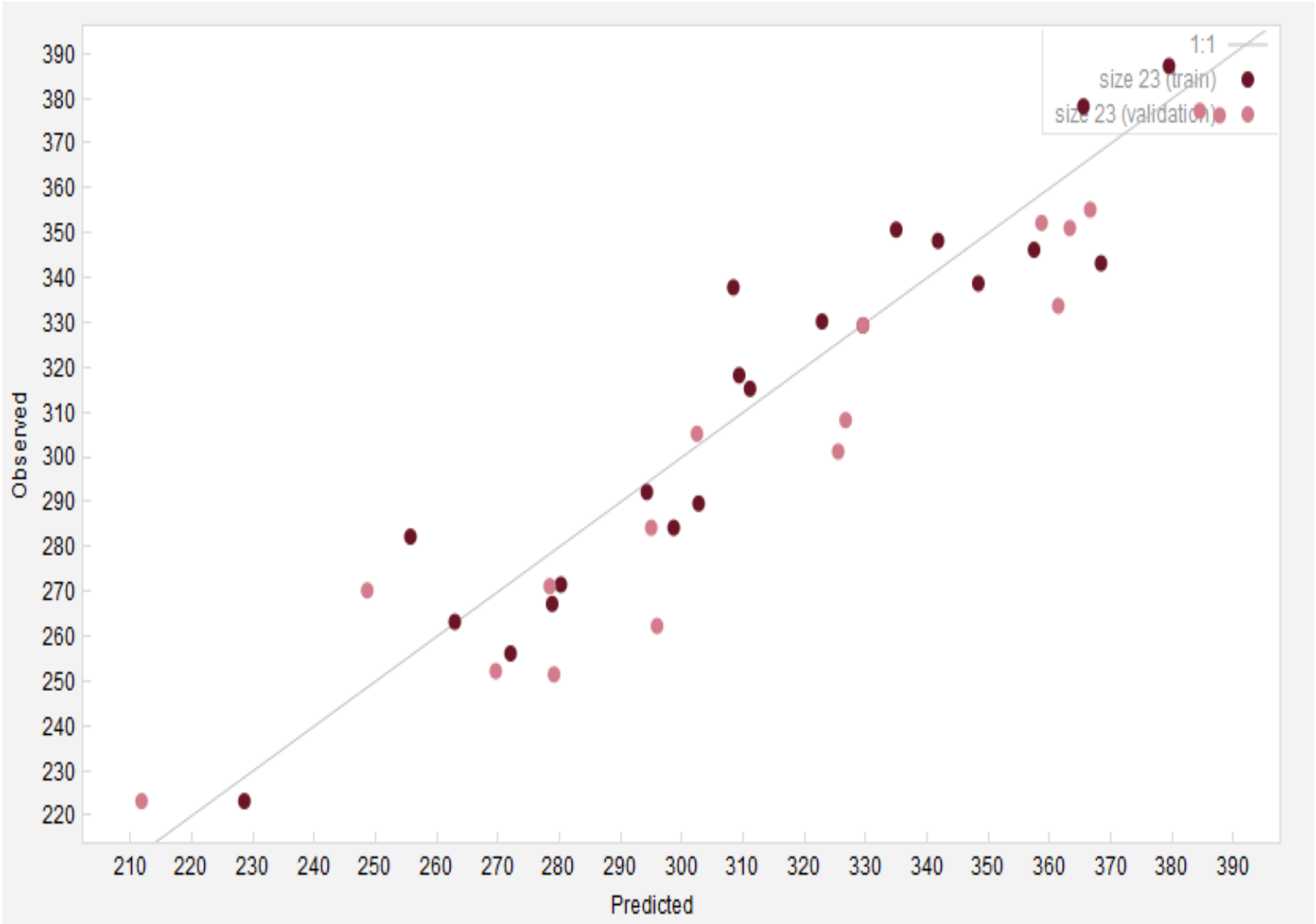

**Figure 2-3:** Goodness of the model for predicting melting points of the selected ionic liquids

The best mutivariative correlation (possessing highest $R^2$) developed by the software was used to predict the melting point of ILs as shown in eqn. (2-5).

$$T_m = a\,(R_A) + b\,(\rho) + c\,(R_C)(\sigma) - d - e(MW_C) - f(Di_A) \qquad (2\text{-}5)$$

a=11.38, b=0.05413, c=196.6, d=434.3, e=0.649, f=1661

Table 2-1 lists the characteristics related to the selected predictive empirical correlation.

**Table 2-1:** Model characteristics in prediction of melting points (calculated on validation data

| Parameter | Value |
|---|---|
| $R^2$ (Goodness of Fit) | 0.8534 |
| Correlation Coefficient | 0.9588 |
| Maximum Error | 33.7208 |

| Parameter | Value |
|---|---|
| Mean Absolute Error | 14.8818 |
| Maximum Relative Error | 11.9865 |
| Mean Relative Error (%) | 4.6378 |

### 2.3.2 Viscosity

In the next step, to develop an empirical correlation for predicting viscosity of ILs, experimental data on the viscosity of several different ILs at different temperatures, were gathered from the literature.[60-75] Once more, a multivariate correlation with several quantum chemistry descriptors along with temperature, as inputs parameters, and Ln (viscosity), as the output parameter, was developed using Eureqa software.

In the case of predicting the viscosity, we used 78 points of data on the actual viscosity of ILs at different temperatures, out of which 23 data points were chosen to be in our validation set, thereby not participating in the training process and are depicted as the points with green coloring in Figure 2-4.

The experiential data for the viscosity of ILs and the multivariate trend line, representing the empirical correlation developed to predict the viscosity, are shown in Figure 2-4. As it can be seen the trend line is capable of predicting the viscosity of selected ILs.

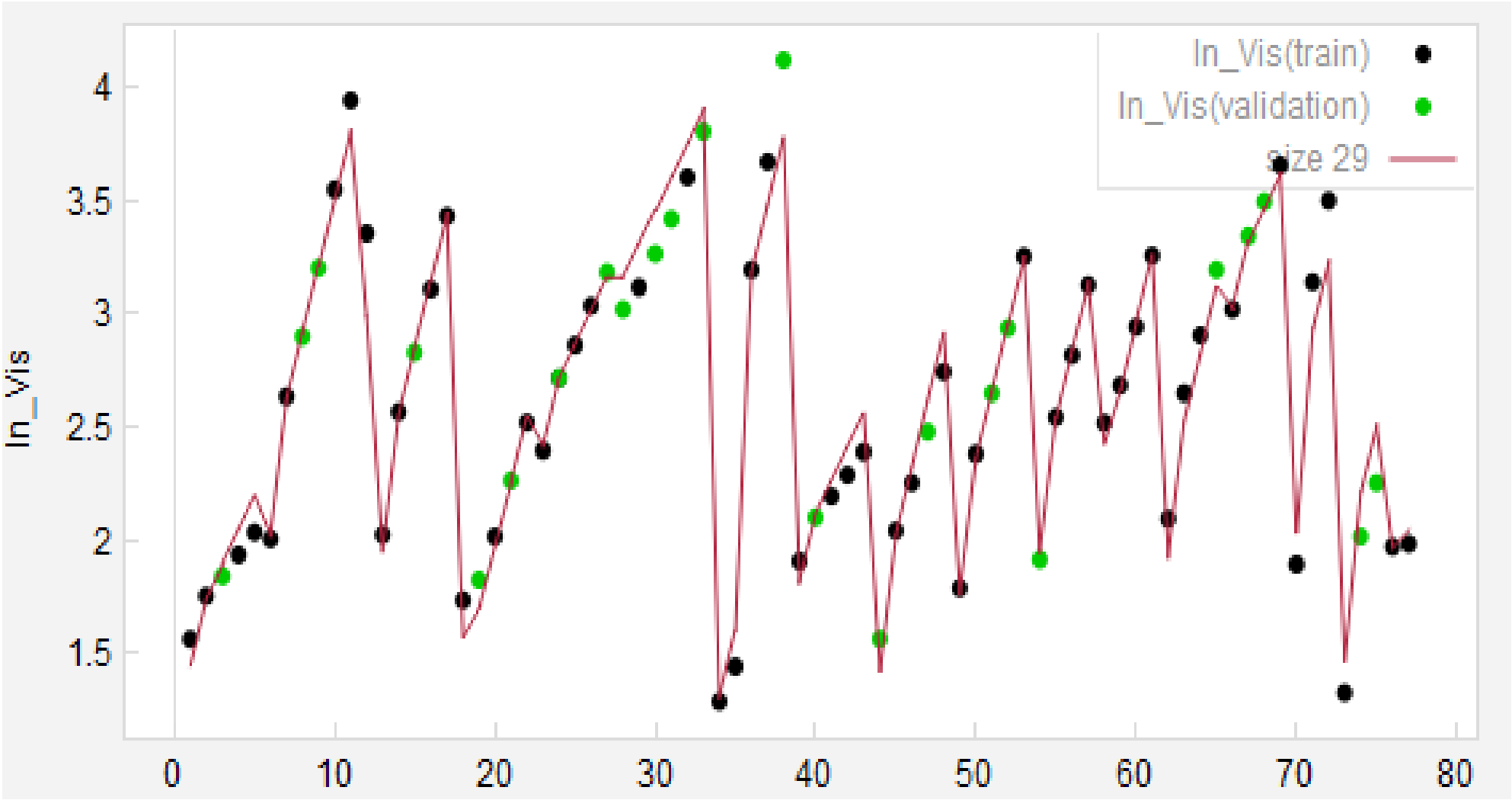

**Figure 2-4:** Actual vs. model predicted viscosities for training and test data sets

Once more, a comparison between the experimental and model predicted values of viscosity are depicted in Figure 2-5.

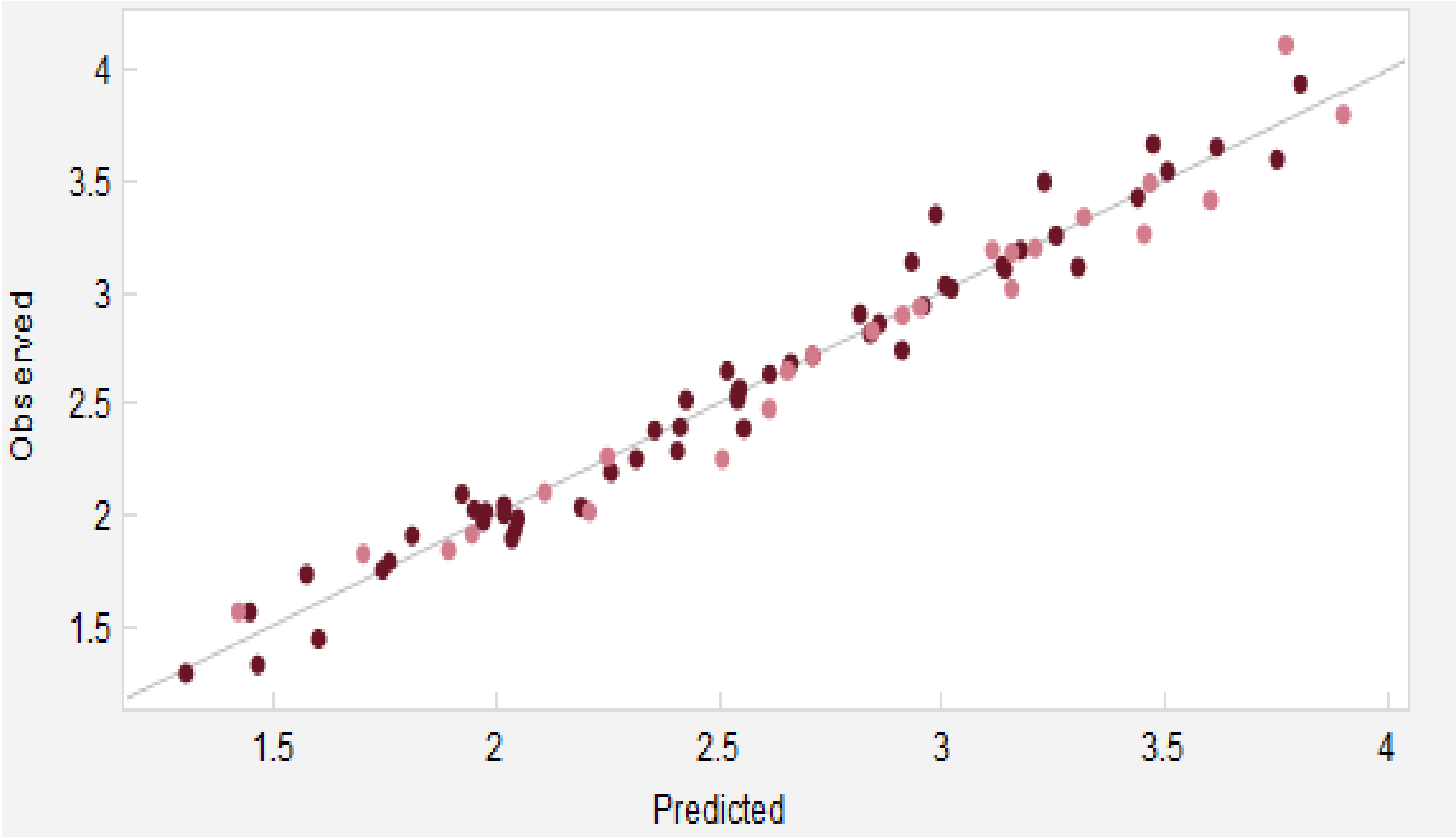

**Figure 2-5:** Goodness of the model for predicting viscosities of the selected ionic liquids

The best correlation (possessing highest $R^2$) to predict the viscosity of ILs is shown in eqn. (2-6).

$$\ln(\text{vis}) = a\,(\sigma) + b\,(R_t) + c\,(\text{Vol}_A) - d - e\,(\text{Area}_A) - f\,(T) - g\,(\text{Di}_A) - h\,(\text{Di}_C) \qquad (2\text{-}6)$$

a=16.513, b=2.2179, c=0.00892, d=15.0073, e=0.02686, f=0.02975, g=15.8297, h=48.1367

Table 2-2 lists the characteristics related to the empirical model used to predict the viscosity of ILs as a function of QC parameters and temperature.

**Table 2-2:** Model characteristics in prediction of viscosity (calculated on validation data)

| **Parameter** | **Value** |
|---|---|
| $R^2$ (Goodness of Fit) | 0.9633 |
| Correlation Coefficient | 0.9823 |
| Maximum Error | 0.3430 |
| Mean Absolute Error | 0.08895 |
| Maximum Relative Error (%) | 11.3247 |
| Mean Relative Error (%) | 3.366 |

## Chapter 3: Reverse Problem; Computer-aided Design of Ionic Liquids

In this chapter a general computer-aided IL design (CAILD) framework along with 4 case studies used to evaluate the ability of the model to select optimal ionic liquids for different applications, are presented.

### 3.1 Introduction

There exists a large library of anions and cations.[76,77] Similar to organic compounds, where the atoms carbon, hydrogen and oxygen can be combined to form thousands of alternative molecular structures ionic liquids can be formed through any combination of cations, anions, and alkyl groups attached to the cation core leading to several structural possibilities (estimated to be as many as $10^{14}$ combinations).[78,79] This is due to the fact that ILs are composed of organic cations and these organic compounds can have unlimited structural variations due to the easy nature of preparation of many components.[80] Moreover, synthesis of a wide range of ionic liquids is relatively straightforward. This presents a great opportunity to engineer ionic liquids that have specific properties. Task specific ionic liquids can be designed for a particular application by controlling the physicochemical properties by judicious selection/modification of the cation, the anion, and/or the alkyl chains attached to the cation. This also presents an unusual challenge, where synthesizing, screening, and testing the limitless possibilities becomes an impossible task.[81]

This is where *in silico* methods could act as a valuable tool for discovering new ionic liquids with tailored properties. Up until now, the majority of ionic liquid computational studies are based on *ab initio* methods such as molecular dynamics, and quantum chemical calculations.[82-85] These methods are extremely important and offer useful insights as they are able to predict properties without performing costly experiments. However, as in the case of experimental studies, one needs to perform several individual simulations which again is impractical due to the long simulation times required for statistical averaging. Both molecular dynamic simulations and experimentation are necessary steps in the selection of task specific ionic liquids. These are important steps to be applied at the final stages of ionic liquid selection. The missing piece is a method for fast exploration, design and identification of a subset of promising candidates, from the millions of ionic liquid alternatives that are available.

Computer-aided molecular design (CAMD), is a promising approach that has been widely applied for molecular systems to design organic solvents for a variety of applications.[86-92] It integrates property prediction models and optimization algorithms to reverse engineer molecular structures with unique properties. We believe that this approach is even more needed for the design of ionic liquids due to the numerous ionic combinations that are possible.

## 3.2 Computer-aided ionic liquid design (CAILD)

In this study we present an overarching framework that aims to identify ionic liquids that exhibit certain desirable behavior. Here, the identity of the compound (in this case an ionic liquid) is not known *a priori*, but we can specify the properties that the compound (i.e. ionic liquid) needs to have.[86] This approach termed as computer-aided ionic liquid design (CAILD) can be defined as "given a set of ionic liquid functional groups (i.e. base Cations: $Im^+$, $Py^+$, $NH_4^+$ ….; anions: $Cl^-$, $Tf_2N^-$, $BF_4^-$….; side chain groups: $CH_3$, =O, -S-, OH….) and a specified set of target properties (e.g. melting point, electrical conductivity, viscosity, solubility…) we can find an ionic liquid structure that matches these properties".

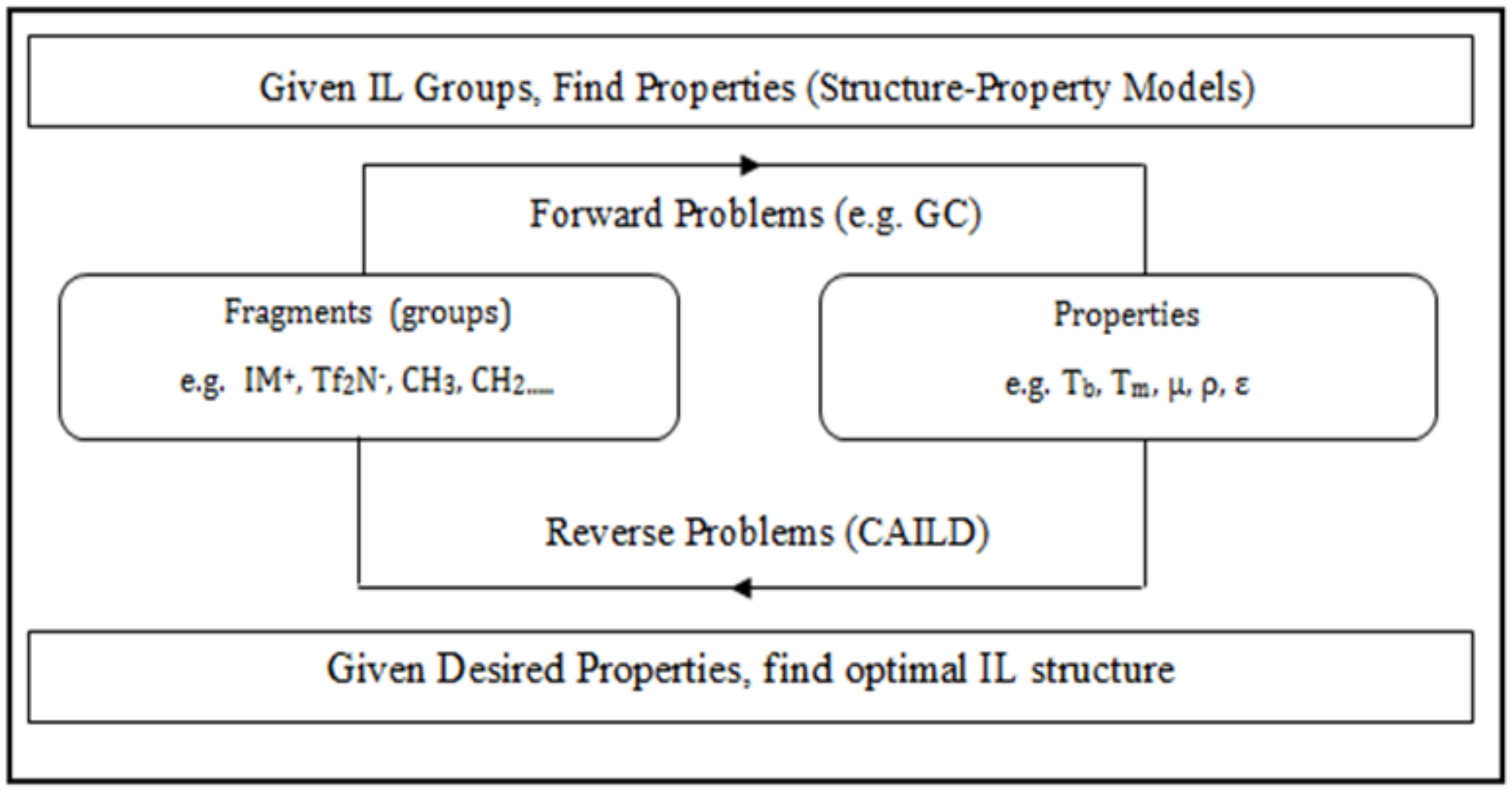

**Figure 3-1:** CAILD framework

Intuitively, CAILD can be thought of as a reverse problem of structure (or group) based property prediction as shown in Figure 3-1. In the forward problem (property prediction) we know the ionic liquid structure and are interested in its properties. In the reverse problem (CAILD) we know the target property values (or ranges) and are interested in feasible ionic liquid structures. To implement CAILD, we need: 1) a framework to fragment ionic liquids into groups; 2) combination and feasibility rules to identify chemically feasible ionic liquids; 3) structure (or group) based models for property prediction; and 4) an optimization framework to search through millions of available alternatives. Mathematical programming approaches provide a useful mechanism to solve such CAILD problems. Different methods to solve such problems include 'generate and test type approaches' (e.g. Harper *et al.*)[93], or 'optimization based approaches' (e.g. Sahinidis *et al.*).[94] The solution to the underlying Mixed Integer Non-Linear Programming (MINLP) model results in the optimal molecular structure for a given application. The objective function is usually an important property related to the design problem, while the constraints relate to structural feasibility, pure component properties, solution (mixture) properties and equilibrium relationships. A conceptual representation of this approach is shown in Figure 3-2.

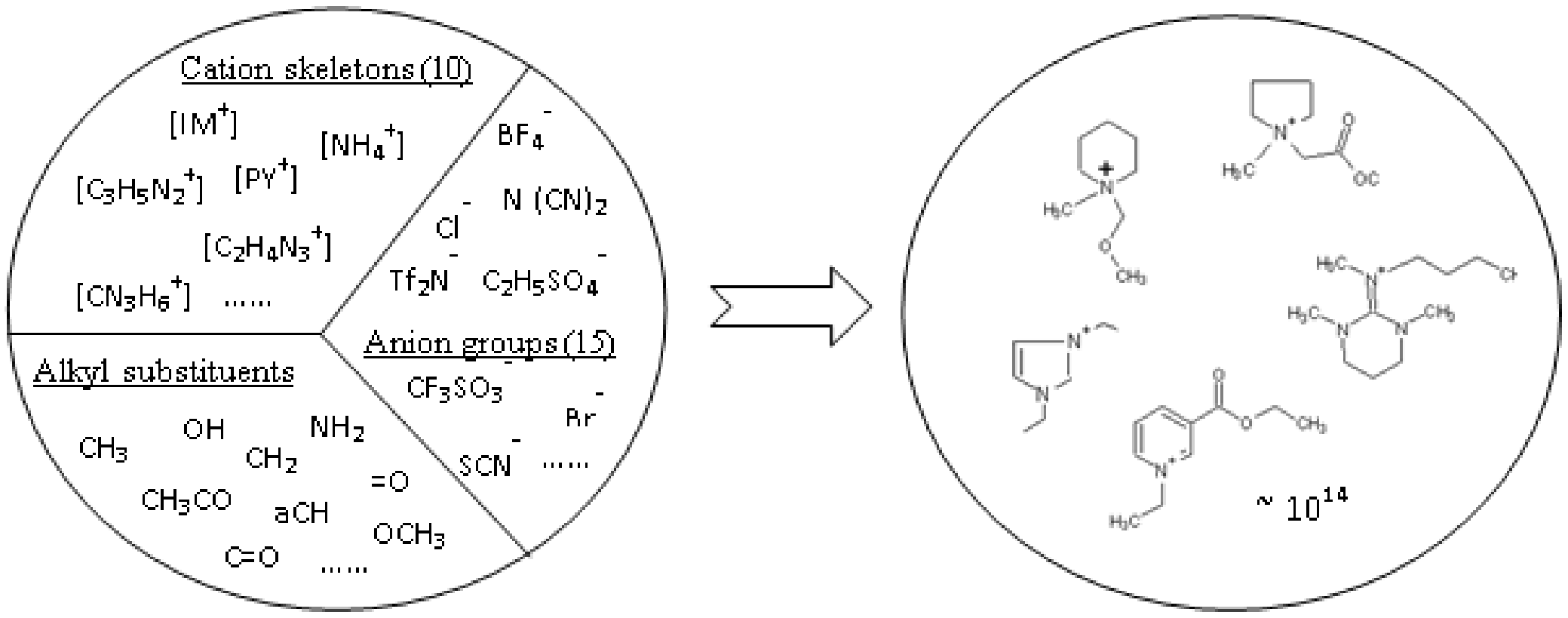

**Figure 3-2:** Conceptual description of CAILD

Section 3.2.1 will focus on describing the general mathematical framework of the proposed approach; section 3.2.2 deals with structural constraints, providing an in-depth mathematical treatment of feasibility, complexity and bonding rules required to design chemically feasible ionic liquids; section 3.2.3 deals with physical property constraints where group contribution based structure-property models are discussed for predicting ionic liquid physical (pure component) properties; section 3.2.4 focuses on solution property constraints that utilize the functional group concept based models, such as UNIFAC, for calculation of solution (mixture) properties through activity coefficients; section 3.2.5 focuses on operations research methods for solving the proposed optimization model.

### 3.2.1 Mathematical framework

The generic mathematical formulation of the CAILD model as an optimization problem is shown in eqns. (3-1) to (3-11). This formulation takes the form of a mixed integer non-linear programming (MINLP) model.

$$f_{obj} = \max\ f(c, a, y, ng, x) \tag{3-1}$$

$$h_1(c, a, y, ng) = 0 \tag{3-2}$$

$$h_2(c, a, y, ng) \leq 0 \tag{3-3}$$

$$g_2(c, a, y, ng) \leq 0 \tag{3-4}$$

$$d_1(c, a, y, ng, x) = 0 \tag{3-5}$$

$$d_2(c, a, y, ng, x) \leq 0 \tag{3-6}$$

$$c \in \mathbb{R}^m \tag{3-7}$$

$$a \in \mathbb{R}^n \tag{3-8}$$

$$y \in \mathbb{R}^u \tag{3-9}$$

$$ng \in \mathbb{R}^q \tag{3-10}$$

$$x \in \mathbb{R}^r \tag{3-11}$$

where $h_1$ is a set of structural feasibility and complexity equality constraints, $h_2$ is a set of structural feasibility and complexity inequality constraints, $g_2$ is a set of pure component physical property inequality constraints, $d_1$ is a set of equality design constraints, $d_2$ is a set of solution (mixture) property inequality design constraints, $c$ is a $m$-dimensional vector of binary variables denoting cation base groups, $a$ is a $n$-dimensional vector of binary variables denoting anion, $y$ is a $u$-dimensional vector of binary variables denoting the alky side chains, $ng$ is a $q$-dimensional vector of integer variables representing number of groups in the alkyl side chains, and $x$ is a $r$-dimensional vector of continuous variables representing compositions, flow rates etc.

### 3.2.2 Ionic liquid structural constraints

The designed ionic liquids need to satisfy certain rules to ensure chemical feasibility. These rules, termed as structural constraints, include feasibility rules such as the octet rule, the bonding rule and complexity rules. Similar rules have been previously developed for molecular compounds.[95,96] Eqns. (3-12) to (3-24) represent a comprehensive set of constraints that were developed to ensure design of ionic liquid candidates that are chemically feasible.

$$\sum_{i \in c} c_i = 1 \tag{3-12}$$

$$\sum_{j\in a} a_j = 1 \quad (3\text{-}13)$$

$$\sum_{l=1}^{6} y_l = \sum_{i\in c} c_i \upsilon_{ci} \quad (3\text{-}14)$$

$$\sum_{i\in c}(2-\upsilon_{ci})c_i + \sum_{l=1}^{6}\sum_{k\in G}(2-\upsilon_{Gkl})y_l ng_{kl} = 2 \quad (3\text{-}15)$$

$$\sum_{k\in G} y_l ng_{kl}\,(2-\upsilon_{Gkl}) = 1 \quad (3\text{-}16)$$

$$\sum_{l=1}^{6}\sum_{k\in G} y_l ng_{kl} \le n_G^U \quad (3\text{-}17)$$

$$\sum_{k\in G} y_l ng_{kl} \le n_{Gl}^U \quad (3\text{-}18)$$

$$\sum_{k\in G^*} y_l ng_{kl} \le t_1 \quad (3\text{-}19)$$

$$\sum_{k\in G^*} y_l ng_{kl} \ge t_2 \quad (3\text{-}20)$$

$$\sum_{k\in G^*} y_l ng_{kl} = t_3 \quad (3\text{-}21)$$

$$\sum_{l=1}^{6}\sum_{k\in G^{**}} y_l ng_{kl} \le t_4 \quad (3\text{-}22)$$

$$\sum_{l=1}^{6}\sum_{k\in G^{**}} y_l ng_{kl} \ge t_5 \quad (3\text{-}23)$$

$$\sum_{l=1}^{6}\sum_{k\in G^{**}} y_l ng_{kl} = t_6 \quad (3\text{-}24)$$

where $c_i$ is a vector of binary variables representing the cations and $a_i$ is a vector of binary variables representing the anions. $y_l$ is a vector of binary variables representing the alkyl chains *l*. $ng_{kl}$ is a vector of integer variables representing the number of groups of type *k* in the alkyl side chain *l*. $\upsilon_{ci}$ , $\upsilon_{Gkl}$ are vectors of group valencies of the cations and alkyl groups, respectively. G is the set of all alkyl groups available for the cation side chains. eqns. (3-12) and (3-13) ensure a maximum of one cation base and one anion respectively for each IL candidate. eqn. (3-14) fixes the number of alkyl side chains attached to the cation based on available free valence of the cation base. *Modified octet rule*: The implementation of the modified octet rule, eqn. (3-15), ensures that any designed cation is structurally feasible and that each valence in all structural groups of the cation is satisfied with a covalent bond. Note that this formulation has already accounted for the positive charge associated with the cation. eqn. (3-16) ensures that the octet rule is implemented for each side chain *l* to ensure that the valences in the individual chains are satisfied with a covalent bond. *Cation size*: The size of the cation is controlled by introducing an upper bound on maximum number of groups ($n_G^U$) that are allowed in the cation, eqn. (3-17). *Alky chain size*: The size of the alkyl chains are controlled by introducing an upper bound on the maximum number of groups $n_{Gl}^U$ that can be present in each alkyl side chain, eqn. (3-18). Eqns. (3-19), (3-20) and (3-21) can be utilized to place restrictions on number of occurrences ($t_1$, $t_2$ and $t_3$) of a certain group, $G^*$, in each side chain *l*. In other words, eqn. (3-19) can be used to make sure that a certain main group such as aldehyde or alcohol not being present more than a certain number of times in each side chain of the cation and eqn. (3-20) can be applied when we want a certain group to be present at least $t_2$ times and eqn. (3-21) can be used when an exact number of occurrence of a certain group is desired e.g. when we want to have exactly one aldehyde group in a certain side chain in the cation. Eqns. (3-22), (3-23) and (3-24) can be utilized to place restrictions on number of occurrences ($t_4$, $t_5$ and $t_6$) of a certain group $G^{**}$, in the cation, which can be calculated as summation of number of occurrences of the particular group in all the side chains in the cation. The purpose of eqns. (3-22) to (3-24) is exactly similar to that of eqns.

(3-19) through (3-21) with the difference of placing restrictions on the number of occurrences of a particular group in whole cation (summation of all of the side chains) instead of only one side chain.

The cation related structural feasibility constraint, eqn. (3-14), is explained using the generic cation dialkylimidazolium (shown in Figure 3-3a) as an example. According to the proposed formulation the Valence for this cation is 2 (i.e.$\upsilon_{ci} = 2$), as there are 2 alkyl side chains ($R_1$ and $R_2$) that are allowed. Similarly, the Valence of a trialkylimidazolium (shown in Figure 3-3b) is 3. For dialkylimidazolium, the right hand side of constraint 3, eqn. (3-14), will translate into $\sum_{i \in c} c_i \upsilon_{ci} = (1)(2) = 2$, which will fix the left hand side of constraint 3 as $\sum_{l=1}^{6} y_l = 2 \;\; i.e. [y_1 = 1]$ & $[y_3 = 1]$. Therefore the vector $y$, will take the following values [1 0 1 0 0 0] with the two ones representing the presence of 2 alky side chains at positions 1 and 3. Next, with the use of few feasible and infeasible examples shown in Figure 3-4, we explain how the whole set of feasibility constraints, eqns. (3-12) to (3-16), work. Figure 3-4a shows a feasible ionic liquid, 1,3-diethylimidazolium tetrafluoroborate. The cation Valence and alkyl group valences related to this ionic liquid are listed in Table 3-1.

**Table 3-1:** Cation and alkyl side chain groups valences

| **Cation groups (c)** | $\boldsymbol{i}$ | $\boldsymbol{\upsilon_{ci}}$ |
|---|---|---|
| 1-alkyl-3-alkyl-Im | 1 | 2 |
| **Alkyl groups *(k)*** | $\boldsymbol{l}$ | $\boldsymbol{\upsilon_{Gkl}}$ |
| $CH_3$ | 1 | 1 |
| $CH_2$ | 1 | 2 |
| $CH_3$ | 3 | 1 |
| $CH_2$ | 3 | 2 |

The values related to the vectors $y_l$ and $ng_{kl}$ for this ionic liquid are listed in Table 3-2.

**Table 3-2:** Values of $\boldsymbol{y_l}$ and $\boldsymbol{ng_{kl}}$ for 1,3-diethylimidazolium tetrafluoroborate

| $l$ | $y_l$ | $k$ | $ng_{kl}$ |
|---|---|---|---|
| 1 | 1 | $CH_3$ | 1 |
| | | $CH_2$ | 1 |
| 3 | 1 | $CH_3$ | 1 |
| | | $CH_2$ | 1 |

In this case there is only one cation base (eqn. (3-12) is satisfied) and one anion (eqn. (3-13) is satisfied). The number of side chains are 2 which equates to the valence of cation (eqn. (3-14) is satisfied). Now, left hand side (LHS) of eqn. (3-15) translates into (2-2)(1)+[(2-2)(1)(1)+(2-1)(1)(1)]+[(2-2)(1)(1)+(2-1)(1)(1)]=2 which is equal to right hand side (RHS) of the equation (eqn. (3-15) is satisfied). For both of the side chain positions (*l*) 1 and3, LHS of eqn. (3-16) translates into [(1)(1)(2-1)+(1)(1)(2-2)]=1 which is equal to RHS of the equation (eqn. (3-16) is satisfied). Eqns. (3-17) through (3-24) are only used to control the cation size and place restrictions on the type and number of occurrences of select groups. As such these are not feasibility constraints but user specified structural design constraints. Figure 3-4b shows an infeasible imidazolium-based ionic liquid. In this case there is only one cation base (eqn. (3-12) is satisfied) and one anion (eqn. (3-13) is satisfied). The number of side chains are 2, which equates to the valance of cation (eqn. (3-14) is satisfied). The LHS of constraint 4, eqn. (3-15), translates to (2-2)(1)+[(2-2)(1)(1)+(2-2)(1)(1)]+[(2-1)(1)(1)+(2-1)(1)(1)]=2,

which equates to the RHS of the equation and hence constraint 4, related to the whole cation, is satisfied. However, constraint 5, eqn. (3-16), related to the side chains are violated as follows: for side chain $l$=1 containing 2 ethyl groups, eqn. (3-16), translates to, [(1)(1)(2-2)+(1)(1)(2-2)]$\neq$ 1 , and for side chain $l$=3 containing 2 methyl groups eqn. (3-16) translates to, [(1)(1)(2-1)+(1)(1)(2-1)]$\neq$ 1. Therefore, the structure shown in Figure 3-4b is infeasible. For the structure shown in Figure 3-4c, eqns. (3-12), (3-13) and (3-14) are satisfied. The LHS of constraint 4, eqn. (3-15), translates to (2-2)(1)+[(2-2)(1)(2)+(2-1)(1)(1)]+[(2-1)(1)(2)]=3, which does not equate to the RHS of the equation (i.e. 2) and hence constraint 4, related to the whole cation, is violated. Constraint 5, eqn. (3-16), related to the side chains translates to the following: For side chain position 1 (i.e. $l$=1), LHS of eqn. (3-16) is [(1)(2)(2-2)+(1)(1)(2-1)]=1, which equates to the RHS of the equation (i.e. 1), but for side chain position 3 (i.e. $l$=3) the same equation translates to [(1)(2)(2-1)]=2 which does not equate to the RHS of the equation (i.e. 1) and hence constraint 5, related to side chain 3, is violated. Therefore, the structure shown in Figure 3-4c violates two of the feasibility constraints and hence is infeasible.

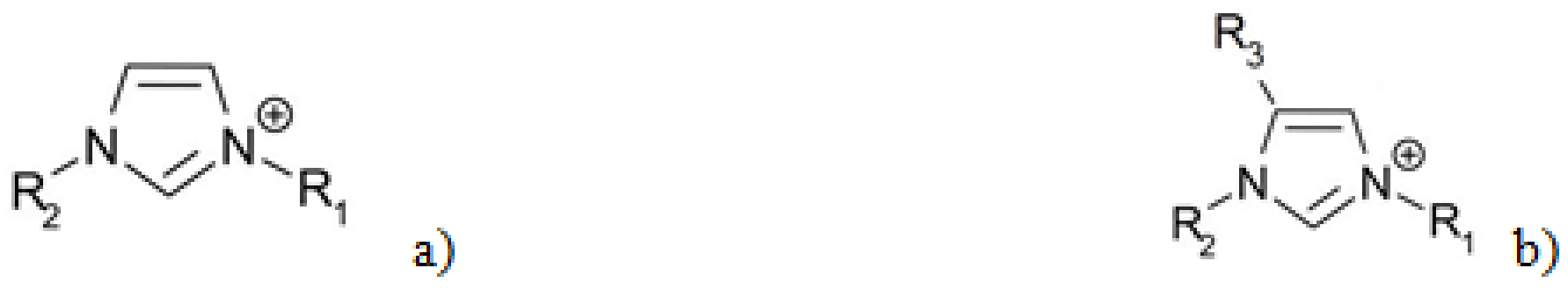

**Figure 3-3:** A general Scheme of two feasible cation structures

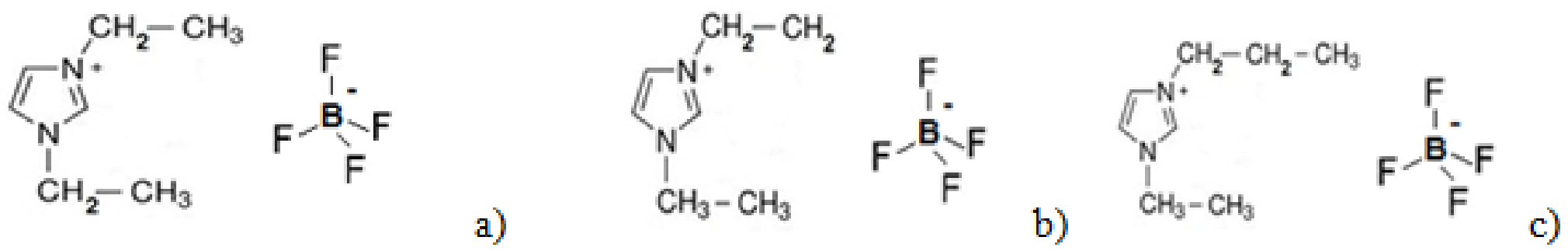

**Figure 3-4:** Examples of feasible and non-feasible Ionic liquids

### 3.2.3 Physical property constraints

Ionic liquid structures play a key role in determining their unique physical properties. Physical property constraints utilize structure-property models which provide insights into the relationship between molecular structures and their properties. The particular type of structure-property relationships suited for CAILD are group contribution (GC) models. As discussed before, GC models for physical properties are based on functional group additive principle. The ionic liquid is fragmented into characteristic groups and the property of interest is predicted as an additive function of the number of occurrence of a given group times its contribution to the pure component property. The contribution parameters of different groups are derived by correlating experimental data to a group additive expression. These models exist for several IL physical properties such as viscosity[97], density[98-101], melting point[102], electrical conductivity[103], thermal conductivity[103], heat capacities[104], solubility parameter[105] and toxicity.[106]

### 3.2.4 Solution property constraints

Thermodynamic properties of non-ideal solutions are important for evaluating intermolecular interactions between multiple components (both ionic and molecular) present in a mixture. These thermodynamic properties are essential to evaluate the potential of ionic liquids as solvents for reaction (solid solubility and liquid miscibility) and the separation of fluid mixtures (liquid-liquid extraction and gas-liquid absorption). An essential requirement is the ability to predict excess Gibbs free energy (activity coefficients) of systems involving ionic liquids which enable prediction of equilibrium concentrations. These constraints are not only a function of binary/integer structure variables but also relate to the compositions of the various components of the mixture. The proposed CAILD framework requires models for the prediction of activity coefficients that are based on 'solution of groups' concept. The basic hypothesis of the solution of groups' concept is that interactions between molecules can be approximated as interactions between functional groups. To illustrate this concept, the different interactions amongst the groups of a mixture of an ionic liquid, [Mim] $Tf_2N$, and $CH_3OH$ is shown in Figure 3-5.

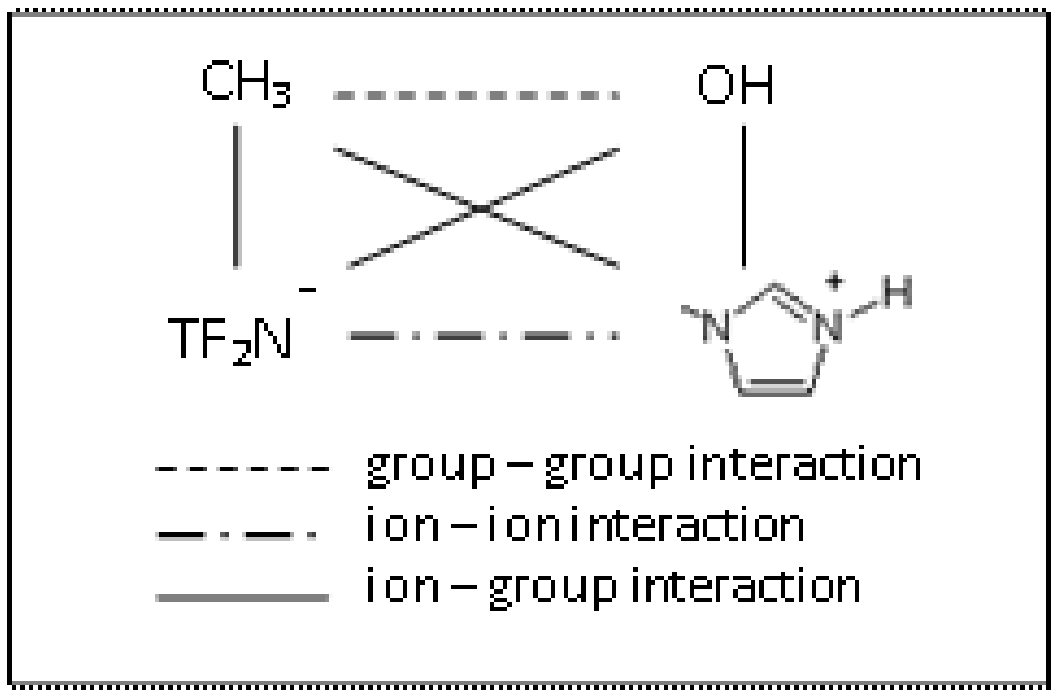

**Figure 3-5:** Different types of group interactions involved in solutions containing ionic

The number of distinct cation head groups, anion groups and alkyl groups are much less in comparison to the number of distinct ionic liquids that can be generated from them. Therefore, a relatively small number of group interaction parameters are required to represent all possible ionic liquids. UNIFAC[107] (UNIversal quasi-chemical Functional-group Activity Coefficients) is a widely used group contribution model to predict phase equilibrium in non-electrolyte systems. The UNIFAC model combines the concept of functional groups with a model for the activity coefficient based on UNIQUAC (Universal Quasi Chemical). The activity coefficient has a combinatorial contribution (due to differences in size and shape of molecules) and a residual contribution (due to energetic interactions).

$$\ln \gamma_i = \ln \gamma_i^C + \ln \gamma_i^R \tag{3-25}$$

The group volume ($R$) and surface area ($Q$) parameters of the combinatorial part are calculated as summation of group parameters (volume $R_k$ and surface area $Q_k$) while binary group interaction parameters ($a_{mn}$ and $a_{nm}$) are required for the calculation of the residual component. The UNIFAC approach was originally used for non-electrolyte systems, however in recent studies several research groups have utilized this approach for ionic liquids by careful representation of ionic groups and/or incorporating assumptions that factor the ionic nature of the groups. In order to apply the UNIFAC model, in its current form, to ionic liquids Wang *et al.*[108] and Lei *et al.*[109] treated ionic groups as a single non-dissociate neutral entity. For example, the ionic liquid [Bmim] $BF_4^-$, was decomposed into two $CH_3$ groups, three $CH_2$ groups and one [Im] $BF_4^-$ group. Using the above representation Lei *et al.*[109] have added 12 new ionic groups (e.g. [Im] $PF_6^-$) to the existing UNIFAC table. Ionic liquid groups are included in the modified UNIFAC (Dortmund) model.[110] Most recently, Roughton *et al.*[105] have characterized the ionic groups in the same way as proposed in our proposed CAILD formulation; i.e. as separate cation base, anion and alkyl groups. The underlying assumption is that the ionic groups can be treated separately and the interactions between the ionic groups can be assumed to be zero due to the strong interaction and weak dissociation between ion pairs.[105] More detailed treatment of UNIFAC approach for ionic liquids can be found in Roughton *et al.*[105], and Wang *et al.*[108], Lei *et al*.[109]

***Liquid-Liquid equilibrium***

Designing industrial scale liquid-liquid separation systems using ionic liquid requires modeling equilibrium relationships. In a non-ideal liquid mixture, species which have limited mutual solubility in the given liquid phase exhibit positive deviations from Raoult's Law. The quantitative measure of non-ideality is the liquid activity coefficient $\gamma$, which is a function of composition and temperature. If we identify the two liquid phases as $l_1$' and \`$l_2$', their respective mole fractions in the two phases are related by the equilibrium condition as follows:

$$\gamma^{l_{1,\mathrm{i}}} x_{1,i} = \gamma^{l_{2,\mathrm{i}}} x_{2,\mathrm{i}} \qquad (3\text{-}26)$$

Where, $\gamma^{l_{1,\mathrm{i}}}$ and $\gamma^{l_{2,\mathrm{i}}}$ are activity coefficients of component *i* in the liquid phases 1 and 2 respectively, and $x_{1,i}$ and $x_{2,\mathrm{i}}$ are mole fractions of component *i* in the two phases.

***Solid-Liquid equilibrium***

Estimation of equilibrium saturation concentrations of solid-liquid systems is essential to model processes that involve solute dissolution, and crystallization. The liquid phase activity coefficient predictions discussed previously and the pure component properties of solute ($\Delta_{fus}H, T_m$), can be utilized for these calculations.

$$\ln x_1^{sat} - \frac{\Delta_{fus}H}{T_m}\left(1 - \frac{T_m}{T}\right) + \ln \gamma_1^{sat} = 0 \quad (3\text{-}27)$$

where $\Delta_{fus}H$ , $T_m$ and $T$ represent enthalpy of fusion (J/mol), melting point (K) and temperature (K), respectively. $\gamma_1^{sat}$, represents activity coefficient of solute at saturation and $x_1^{sat}$ is the solubility of solute.

### 3.2.5 Solution of the underlying MINLP

The presented CAILD model is a non-convex, mixed integer non-linear programming (MINLP) problem, involving large number of integer and binary variables. Consideration of mixture properties through the UNIFAC model results in non-linearity and most of the binary design variables (structural) participate in the non-linear terms. Combinatorial complexity is an inherent issue in CAMD – MINLP models due to the nature of the search space. The most direct approach for solving the underlying MINLP model is complete enumeration. Generate and test methods fall under this category.[93] Solution to the MINLP model can also be achieved through mathematical programming using deterministic (e.g. branch and bound[111], branch and reduce[112]) and stochastic optimization methods (e.g. simulated annealing[113], genetic algorithms[114], and tabu search[90]). Approaches that combine features from both domains, such as decomposition methods, have also been previously developed.[94,115] Achenie *et al.*[86] provide a detailed description of various solution techniques in the context of molecular design problems. In this section we focus on solving the CAILD framework utilizing two different methods: the decomposition methodology (includes generate and test algorithms) and genetic algorithm based optimization. Our purpose is to demonstrate that different types of solution approaches can be used towards a solution of the proposed CAILD formulation. Main details about the two approaches are provided below while in depth analysis can be found elsewhere.[115,116]

***Decomposition Method***

In this approach the CAILD-MINLP model is decomposed into an ordered set of subproblems where each subproblem requires only the solution of a subset of constraints from the original set. As each subproblem is being solved large numbers of infeasible candidates are eliminated leading to a final smaller subproblem. The first subproblem usually consists of the structural constraints and it equates to enumeration. The second subproblem consists of pure component (physical) property constraints while the third subproblem consist of mixture property constraints. These three subproblems taken together equate to generate and test methods. The ionic liquid candidates that pass through all of the above subproblems are the only ones that will be considered in the final optimization subproblem that involves the objective function, equilibrium relationships and process models (if considered in the design problem). Most often, the solution to the final subproblem can be achieved by solving a set of non-linear programming (NLP) problems.

### *Genetic Algorithm*

Genetic algorithm (GA) is a method that can be used to solve optimization problems based on the natural selection process that mimics biological evolution. It can be applied to solve problems that are not well suited for standard optimization algorithms, including problems in which the objective function is discontinuous, non-differentiable, stochastic, or highly nonlinear.[116] Unlike traditional search and optimization methods, GAs perform a guided stochastic search where improved solutions are achieved by sampling areas of the search space that have a higher probability for good solutions.[116] The optimization process starts with a collection of chromosomes (candidates). The fitter candidates are selected as 'parents' and allowed to exchange or alter their genetic information, through crossover and mutation operations, with an aim to create more fitter off springs. At every iteration new populations of off springs are created to replace the existing population. This process of evolution is repeated for a pre-determined number of generations or until the solution is found.[116] In GA, the selection of fitter parents for next generation is based on their fitness values as determined by a fitness function. The fitness function is usually very closely related to the original objective function of the search problem (in all of the case studies presented in this study, the fitness function was identical to the objective function). The GA solution of CAILD model was implemented in the MATLAB environment with most parameters taking the default values. Specifically, for all the case study problems, the population size was fixed at 20 and the initial population was generated randomly. The crossover fraction was fixed at 0.8 while the mutation probability was fixed at 0.2. We allowed two candidates with the best fitness values (elite candidates) in the current generation to automatically survive to the next generation.

## 3.3 Proof of concept examples

In this section several case studies have been presented to illustrate the usefulness of the proposed approach. Table 3-3 lists the entire set of groups and their respective valences, from which the basis sets for the four case studies were derived. Note that this basis set covers only a small set of cations, anions and functional groups for which group contribution parameters are currently available for the properties of interest. However, the design approach itself is universal in nature and upon availability of group contribution models and parameters can be easily extended (for example, to all groups in Appendix A) to cover the entire spectrum of possible ionic liquids. The maximum value allowed for the number of groups were fixed at 6 for each side chain and 12 for the whole cation.

**Table 3-3:** The basis set used for ionic liquid design

| Cations | Valence | Anions | Groups | Valence |
|---|---|---|---|---|
| Im | 2 | $BF_4^-$ | $CH_3$ | 1 |
| Mim | 1 | $PF_6^-$ | $CH_2$ | 2 |
| Py | 2 | $Tf_2N^-$ | | |
| Mpy | 1 | $Cl^-$ | | |

### 3.3.1 Electrolytes

This case study demonstrates the design of an ionic liquid that has high electrical conductivity. Electrical conductivity measures the ability of a material to conduct electric current. It is an important property for the development of electrochemical devices such as high energy batteries. Other design requirements include the following: the electrolyte (i.e. ionic liquid) needs to be a room temperature ionic liquid (RTIL); and it should have reasonably low viscosity.

The electrical conductivity of ionic liquids can be estimated using a Vogel-Tamman-Fulcher (VTF) type equation shown in eqn. (3-28).

$$\ln \lambda = \ln A_\lambda + \frac{B_\lambda}{(T - T_{0\lambda})} \qquad (3\text{-}28)$$

where $A_\lambda$, and $B_\lambda$, are adjustable parameters that can be obtained through group contribution expressions, eqn. (3-29) and eqn. (3-30), as proposed by Gardas *et al.*[103], and $T_{0k}$ has the value of 165K for all considered IL types.

$$A_\lambda = \sum_{i=1}^{k} n_i \, a_\lambda \qquad (3\text{-}29)$$

$$B_\lambda = \sum_{i=1}^{k} n_i \, b_\lambda \qquad (3\text{-}30)$$

where $n_i$ is the number of groups of type $i$ and $k$ is the total number of groups considered. Table 3-4 shows the group contribution parameters used.

**Table 3-4:** Group contributions for parameters $A_\lambda$ and $B_\lambda$

| **Species** | **$a_\lambda$** | **$b_\lambda$ (K)** |
|---|---|---|
| Im | 77.8 | -501.5 |
| Mim | 77.9 | -537.6 |
| Py | 69.6 | -544.9 |
| Mpy | 69.7 | -581.0 |
| $BF_4^-$ | 85.8 | -129.4 |
| $PF_6^-$ | 117.3 | -278.6 |
| $Tf_2N^-$ | 10.1 | -46.4 |
| $CH_3$ | 0.1 | -36.1 |
| $CH_2$ | 0.1 | -36.1 |

The viscosity of ionic liquids is calculated using an Orrick-Erbar-type approach.[117] In this method, viscosity $(\eta, \text{in cP})$ can be predicated as a function of density $(\rho, \text{in g cm}^{-3})$ molecular weight$(M)$, temperature $(T)$ and parameters A and B through the use of eqn. (3-31).

$$ln \frac{\eta}{\rho M} = A + \frac{B}{T} \tag{3-31}$$

We employ the group contribution technique proposed by Gardas *et al*.[103] to estimate the parameters A and B as follows

$$A = \sum_{i=1}^{k} n_i \, A_{v,i} \tag{3-32}$$

$$B = \sum_{i=1}^{k} n_i \, B_{v,i} \tag{3-33}$$

where $n_i$ is the number of occurrences of group $i$ (cation, anion and functional groups) and $A_{v,i}$ and $B_{v,i}$ are the contributions of group $i$ to the parameters A and B respectively. The ionic liquid densities are estimated using the below formula.

$$\rho = \frac{M}{NV(a+bT+cP)} \tag{3-34}$$

where $\rho$ the density in kg $m^{-3}$, $M$ is the molecular weight in kg $mol^{-1}$, $N$ is the Avogadro number, $V$ is the molar volume in $\dot{A}^3$, $T$ is the temperature in K and $P$ is the pressure in MPa. Based on the data provided in Gardas *et al.*[103] we developed group contribution parameters for molar volume with expressions similar to eqns. (3-32) and (3-33). The values of coefficients a, b and c are $0.8005 \pm 0.0002$ , $6.652 \times 10^{-4} \pm 0.007 \times 10^{-4}\ K^{-1}$ and $-5.919 \times 10^{-4} \pm 0.024 \times 10^{-4}\ MPa^{-1}$ . Table 3-5 shows the group contribution parameters used in this model.

**Table 3-5:** Group contributions for parameters A, B and V

| **Species** | **ΔV** | **$A_v$** | **$B_v$** |
|---|---|---|---|
| Im | 84 | 8.04 | 1257.1 |
| Mim | 119 | 7.3 | 1507.1 |
| Py | 111 | 7.61 | 1453.6 |
| Mpy | 146 | 6.87 | 1703.6 |
| $BF_4^-$ | 73 | -18.08 | 1192.4 |
| $PF_6^-$ | 109 | -20.49 | 2099.8 |
| $Tf_2N^-$ | 248 | -17.39 | 510.0 |
| $Cl^-$ | 47 | -27.63 | 5457.7 |
| $CH_3$ | 35 | -0.74 | 250.0 |
| $CH_2$ | 28 | -0.63 | 250.4 |

The melting point of ionic liquids is estimated by a group contribution approach using eqn. (3-35) as proposed by Aguirre *et al.*[118]

$$.T_m = \frac{\sum_i n_i T_{m,i}}{a+\sigma_c+c\tau_c} \qquad (3\text{-}35)$$

where $n_i$ is the number of occurrences of group *i* (cation, anion and functional groups) and $T_{m,i}$ is the contribution of group *i* to the melting point, *a* and *c* are constants with values of 0.1 and 0.012 respectively. $\tau_c$ , which is related to the cation flexibility, is estimated using eqn. (3-36) and $\sigma_c$ is a cation symmetry parameter having values shown in Table 3-6.

$$\tau_c = \sum_k (n(CH_2)_k - 1) \qquad (3\text{-}36)$$

**Table 3-6:** $\boldsymbol{\sigma_c}$ values for different cations

| **Type of cation** | $\boldsymbol{\sigma_c}$ **value** |
|---|---|
| $\mathrm{R = \acute{R}\ (Im, Pyr, Pip)}$ | 0.265 |
| $\mathrm{R \neq \acute{R}\ (Im, Pyr, Pip)}$ | 0.317 |
| $\mathrm{R\ or\ \acute{R}\ dimethylamino\ in (C(4)Py)}$ | 0.265 |

The contribution, $T_{m,i}$, of different groups are listed in Table 3-7.

**Table 3-7:** Group Contributions for Ionic Liquids' Melting Point

| **Group** | $\boldsymbol{T_{m,i}}$ |
|---|---|
| Im | 107.99 |
| Mim | 109.88 |
| Py | 117.212 |
| Mpy | 119.102 |

| **Group** | **$T_{m,i}$** |
|---|---|
| $BF_4^-$ | -0.479 |
| $PF_6^-$ | 16.746 |
| $Tf_2N^-$ | -0.966 |
| $Cl^-$ | 35.852 |
| $CH_3$ | -1.463 |
| $CH_2$ | -1.463 |

The CAILD design problem expressed in mathematical form is shown in eqns. (3-37) to (3-40).

*Objective function*

$$f_{obj} = \max(\lambda) \quad (3\text{-}37)$$

*Constraints*

Ionic liquid Structural Feasibility (3-38)

$$\eta < 65\ cP \quad (3\text{-}39)$$

$$T_m < 298.15\ K \quad (3\text{-}40)$$

***Results:*** The design statistics for this problem are summarized in Table 3-8. A total of 138 feasible IL structures were enumerated in subproblem 1. Out of these, 26 ILs satisfied the physical property constraints (viscosity and melting point) in subproblem 2. There were no mixture properties considered (subproblem 3) and the solution to final subproblem (subproblem 4) resulted in the optimal ionic liquid structure (1-methylimidazolium bis-trifluoromethylsulfonyl imide) with the highest electrical conductivity (shown in Figure 3-6). The properties of the designed ionic liquid are listed in Table 3-9.

**Table 3-8:** Decomposition approach: Subproblem Results

| | |
|---|---|
| **Subproblem 1:** Number of ionic liquids (ILs) generated, 138 | |
| **Subproblem 2:** Number of ILs satisfying pure component properties, 26 | |
| **Subproblem 3:** No mixture properties | |
| **Subproblem 4:** Optimal IL, $[Mim]^+[Tf_2N]^-$ | |

The same design problem, eqns. (3-37) to (3-40) was solved using the genetic algorithm toolkit in MATLAB and the program picked the exact same structure (shown in Figure 3-6) as the optimal solution.

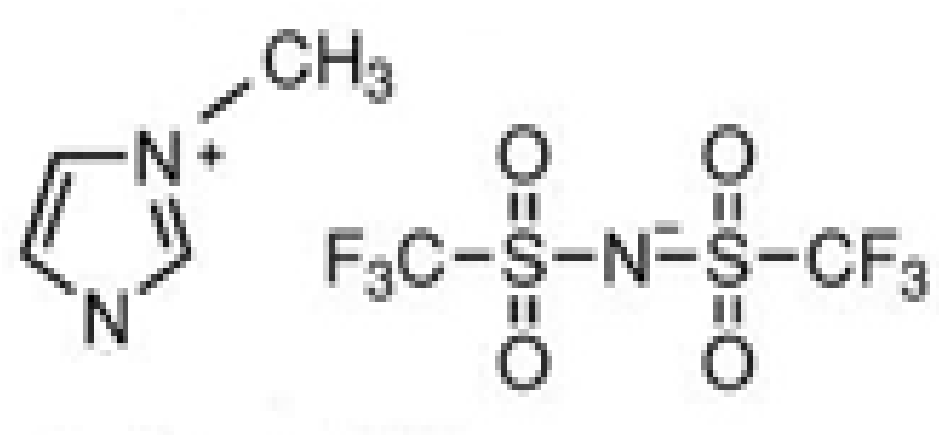

**Figure 3-6:** 1-methylimidazolium $[Tf_2N]^-$

**Table 3-9:** Design Results of the optimal IL, 1-methylimidazolium $[Tf_2N]^-$

| Properties | Value |
|---|---|
| Melting point (K) | 270.15 |
| Viscosity (cP) | 21.293 |
| Electrical Conductivity (S/m$^{-1}$) @ 25$^o$C | 1.0956 |

***Analysis*:** In this section we focus on the validation of design results through careful consideration and analysis of available experimental data. Table 3-10 lists the available experimental electrical conductivity, λ, (S/m) at 25 °C for 7 different ionic liquids that are based on the cation, anion and side groups considered in this case study (Im, Py, $PF_6$, $BF_4$, $Tf_2N$, $CH_3$ and $CH_2$). Unfortunately, we could not find the electrical conductivity data for the designed IL (1-methylimidazolium bis (trifluoromethylsulfonyl) imide). Therefore, we perform a qualitative IL structure-property trend analysis to validate the design results.

The electrical conductivity values have the following trend: $\lambda_{[C_4mim]\ Tf_2N^-} > \lambda_{[C_4mim]\ BF_4^-} > \lambda_{[C_4mim]\ PF_6^-}$. Since, all of the above ionic liquids have the same cation ($C_4$mim) but different anions ($Tf_2N^-$, $PF_6^-$ and $BF_4^-$) we can infer that electrical conductivities of ionic liquids with $Tf_2N^-$ anions are greater than those with $PF_6$ and $BF_4$ anions (for same cation and alkyl groups). The design result is consistent with this observation as the optimal structure has $Tf_2N$ anion. Similarly, by comparing the electrical conductivities of ionic liquids having the same anion ($Tf_2N^-$) we can see that [$C_2$mim] $Tf_2N^-$ > [$C_4$mim] $Tf_2N^-$ > [$C_6$mim] $Tf_2N^-$. Therefore, we can conclude that increasing the number of alkyl groups on the cation side chain decreases the electrical conductivity. The design result (Figure 3-6) is also consistent with this observation as there is only one methyl group (minimum needed to satisfy the cation Valence) present in the cation side chain. Next, we compare the λ of [$C_4$mim] $Tf_2N^-$ and [$C_4$Py] $Tf_2N^-$ which have the same anion and different cations. We found experimental electrical conductivity data for [$C_4$Py] $Tf_2N^-$ but there was no data available for [$C_4$mPy] $Tf_2N^-$. Since we already know that addition of alkyl groups to the cation base will decrease the electrical conductivity, we can infer that λ [$C_4$mPy] $Tf_2N^-$ < 0.33 S/m (i.e. λ [$C_4$Py] $Tf_2N^-$) which in turn is much less than λ [$C_4$mim] $Tf_2N^-$ (0.406 S/m). Therefore, we can conclude that electrical conductivity of ionic liquids with imidazolium based cations are greater than Pyridinium-based cations (for same anion and alkyl groups). The design result is consistent with this observation as the optimal structure selected had an imidazolium cation. Overall, the model results are in full agreement with the observed trends from experiments, thereby validating the proposed approach.

**Table 3-10:** Experimentally Measured Electrical Conductivities of Ionic Liquids

| Ionic liquid | T (°C) | λ (S/m) | Ref |
|---|---|---|---|
| 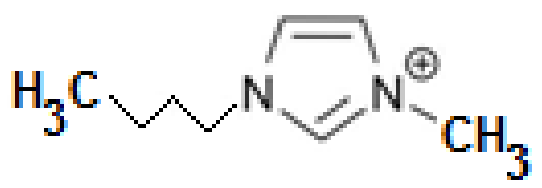 $PF_6$ | 25 | 0.146 | [119] |
| 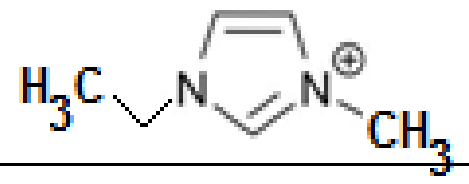 $Tf_2N$ | 25 | 0.912 | [119] |
| 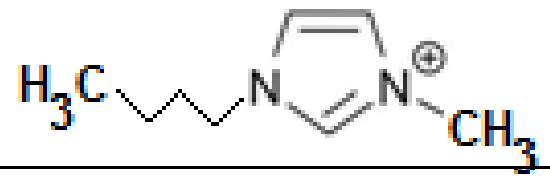 $Tf_2N$ | 25 | 0.406 | [119] |
| 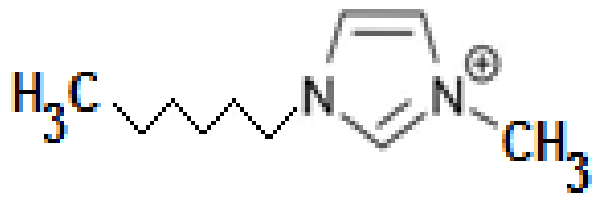 $Tf_2N$ | 25 | 0.218 | [119] |
| 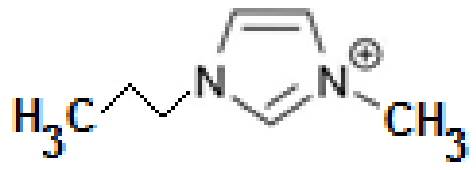 $BF_4$ | 25 | 0.59 | [120] |
| 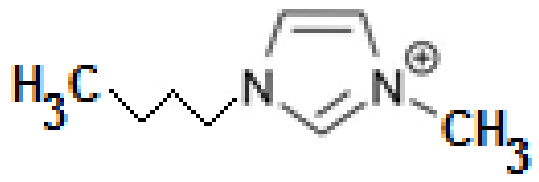 $BF_4$ | 25 | 0.35 | [120] |
| 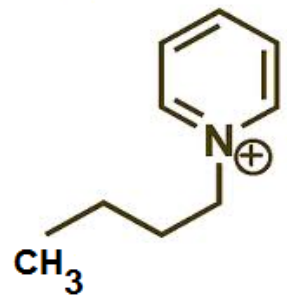 $Tf_2N$ | 25 | 0.33 | [121] |

### 3.3.2 Heat transfer fluids

Ionic liquids show great promise as heat transfer fluids and heat storage medium. High thermal conductivity is an important property for such applications. Thermal conductivity measures the ability of a material to conduct heat. Thermal conductivity of ionic liquids is weakly dependent on temperature and could be fitted with the following linear correlation.

$$k = A_k - B_k T \tag{3-41}$$

where, $k$ and T are the thermal conductivity in $Wm^{-1}K^{-1}$ and temperature in K respectively. We utilize a method[96] that employs group contribution approach to estimate the parameters $A_k$ and $B_k$.

$$A_k = \sum_{i=1}^{k} n_i \, a_k \tag{3-42}$$

$$B_k = \sum_{i=1}^{k} n_i \, b_k \tag{3-43}$$

Table 3-11 shows the group contribution parameters that were used. Pyridinium and methyl pyridinium have not been considered in this part since their group contributions were not found.[103]

**Table 3-11:** Group contributions for parameters $A_k$ and $B_k$

| **Species** | **$a_k$** | **$b_k$ ($K^{-1}$)** |
|---|---|---|
| Im | 0.1272 | 0.000000104 |
| Mim | 0.1314 | 0.00000787 |
| $BF_4^-$ | 0.0874 | 0.00008828 |
| $PF_6^-$ | 0.0173 | 0.000009088 |
| $Tf_2N^-$ | 0.0039 | 0.00002325 |
| $Cl^-$ | 0.0166 | 0.00001 |
| $CH_3$ | 0.0042 | 0.000007768 |
| $CH_2$ | 0.0010 | 0.000002586 |

Melting point and viscosity of ionic liquids were calculated through the same methods proposed in case study 1. The CAILD design problem expressed as an optimization model is shown in eqns. (3-44) to (3-47).

*<u>Objective function</u>*

$$f_{obj} = \max\,(k) \tag{3-44}$$

*<u>Constraints</u>*

Ionic liquid Structural Feasibility (3-45)

$$\eta \;<\; 65\; cP \tag{3-46}$$

$$T_m < 298.15\; k \tag{3-47}$$

***Results:*** Decomposition approach: The design statistics for this problem are summarized in Table 3-12. A total of 92 feasible IL structures were enumerated in subproblem 1. Out of these, 15 ILs satisfied the physical property constraints (viscosity and melting point) in subproblem 2. There were no mixture properties considered (subproblem 3) and the solution to final subproblem (subproblem 4) resulted in the optimal ionic liquid structure (1-ethyl-3-methylimidazolium tetrafluoroborate) with the highest thermal conductivity (shown in Figure 3-7). The properties of the designed ionic liquid are listed in Table 3-13.

**Table 3-12:** Decomposition approach: Subproblem Results

| |
|---|
| **Subproblem 1:** Number of ionic liquids (ILs) generated, 92 |
| **Subproblem 2:** Number of ILs satisfying pure component properties, 15 |
| **Subproblem 3:** No mixture properties |
| **Subproblem 4:** Optimal IL, 1-Ethyl-3-methylimidazolium tetrafluoroborate |

The same design problem, eqns. (3-44) to (3-47) was solved using the genetic algorithm toolkit in MATLAB and the program picked the exact same structure (shown in Figure 3-7) as the optimal solution.

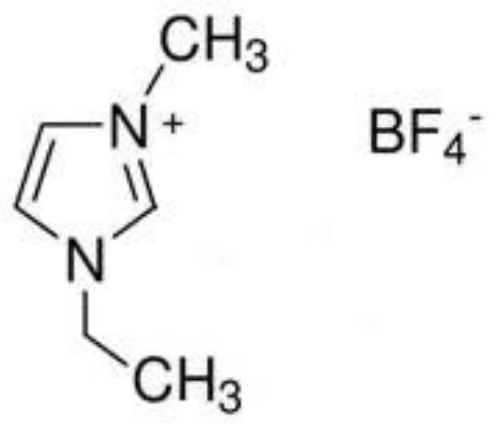

**Figure 3-7:** 1-ethyl-3-methylimidazolium $[BF_4]^-$

**Table 3-13:** Design Results of the Optimal IL, 1-ethyl-3-methylimidazolium $[BF_4]^-$

| **Properties** | **Value** |
|---|---|
| Melting point (K) | 291.71 |
| Viscosity (cP) | 60.75 |
| Thermal Conductivity ($Wm^{-1}K^{-1}$) @ 25$^{o}$C | 0.193 |

***Analysis*:** Table 3-14 shows experimental thermal conductivity, $k$ [$Wm^{-1}K^{-1}$] for 10 different ionic liquids that are based on the cation, anion and side groups considered in this case study (Im, $PF_6^-$, $BF_4^-$, $Tf_2N^-$, $CH_3$ and $CH_2$). The designed ionic liquid (Figure 3-7) is same as the IL with the highest thermal conductivity value in Table 3-14 ([$C_2$mim] $BF_4^-$). This partially validates the results. However, for a more holistic assessment we perform a qualitative ionic liquid structure-property trend analysis to determine whether the designed results are consistent with observed data. By comparing the thermal conductivity values (Table 3-14), we note $k_{[C_4mim]\,BF_4} > k_{[C_4mim]\,PF_6} > k_{[C_4mim]\,Tf_2N}$. Since all of these ionic liquids have the same cation ($C_4$mim) but different anions ($Tf_2N^-$, $PF_6^-$ and $BF_4^-$), we can infer that thermal conductivities of ionic liquids with $BF_4^-$ anion are greater than those with $PF_6^-$ and $Tf_2N^-$ anions. The design result is consistent with this observation as the optimal structure has $BF_4^-$ anion. The only base cation considered in this design problem is imidazolium. By comparing the $k$ values of different ionic liquids with the same base cation and same anion, but different side groups (i.e. [$C_4$mim] $PF_6^-$ vs [$C_6$mim] $PF_6^-$ vs [$C_8$mim] $PF_6^-$, and [$C_2$mim] $Tf_2N^-$ vs [$C_4$mim] $Tf_2N^-$ vs [$C_6$mim] $Tf_2N^-$ vs [$C_8$mim] $Tf_2N^-$ vs [$C_{10}$mim] $Tf_2N^-$, we can see that the contribution of alkyl side chain groups are not as high as that of anion, and there is no uniform trend that is observed in relation to varying number of alkyl side chain groups. Therefore, the optimal numbers of alkyl side chain groups relate to other requirements such as the IL needing to be a liquid (i.e. $T_m$< 25 ° C for RTILs) and have relatively low viscosity. Overall we can conclude that the design results are consistent with the observed experimental structure-property trends of thermal conductivity data.

**Table 3-14:** Experimental thermal conductivity data

| Ionic liquid | T (K) | k ($Wm^{-1}K^{-1}$) | Ref |
|---|---|---|---|
| 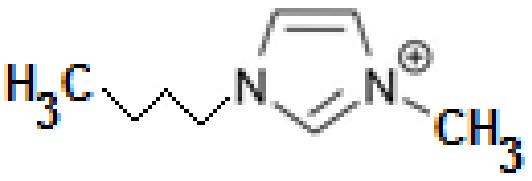 $PF_6$ | 315 | 0.145 | [122] |
| 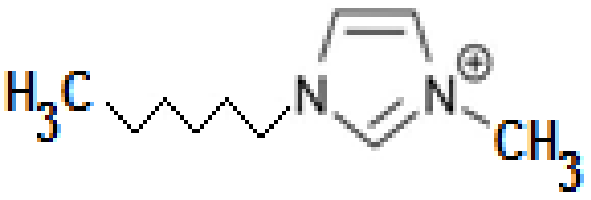 $PF_6$ | 315 | 0.146 | [122] |
| 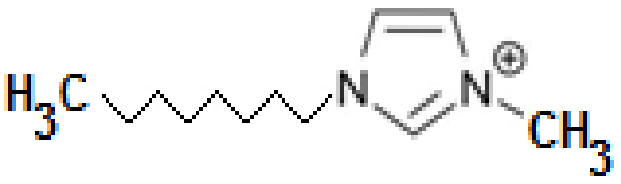 $PF_6$ | 315 | 0.145 | [122] |
| 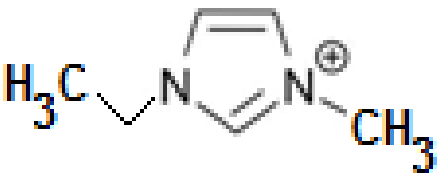 $BF_4$ | 315 | 0.1968 | [123] |
| 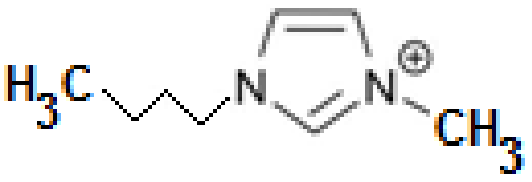 $BF_4$ | 315 | 0.1847 | [123] |
| 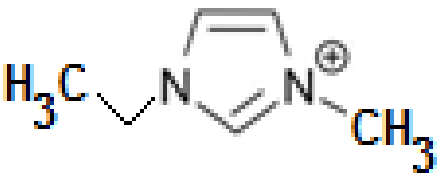 $Tf_2N$ | 315 | 0.1294 | [124] |
| 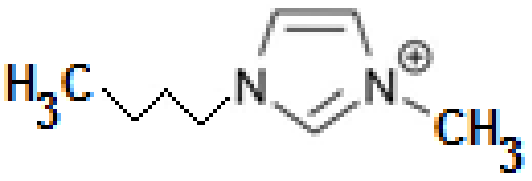 $Tf_2N$ | 315 | 0.1264 | [124] |
| 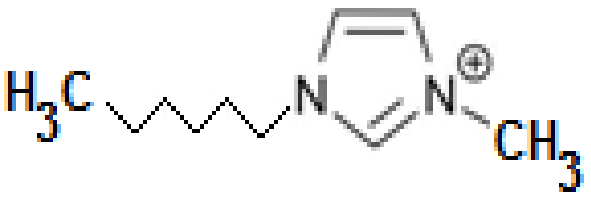 $Tf_2N$ | 315 | 0.1263 | [124] |
| 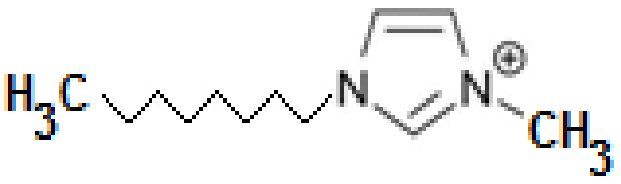 $Tf_2N$ | 315 | 0.12715 | [124] |
| 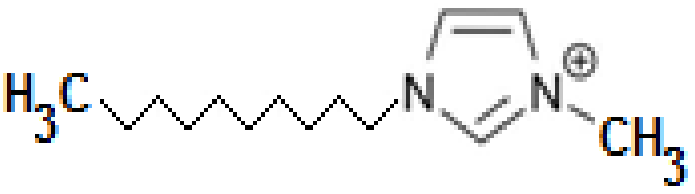 $Tf_2N$ | 315 | 0.1299 | [124] |

### 3.3.3 Toluene-heptane separation

A common use of solvents in industrial processes is as a separating agent to isolate two liquid components. This case study relates to the design of optimal ionic liquid to separate toluene (aromatic) and n-heptane (aliphatic). Sulfolane ($C_4H_8O_2S$) is a molecular solvent that is commonly used for this purpose. The design objective is to find an ionic liquid that can improve performance in comparison to sulfolane. One key requirement is to select an ionic liquid with as low viscosity as possible since, viscous solvents are not ideal from the stand point of industrial equipment design. The other requirement is to ensure that the designed solvent is a room temperature ionic liquid (RTIL) as the process requires a liquid solvent. A constraint on melting point, eqn. (3-56), is necessary to ensure design of RTILs only. Melting point and viscosity of ionic liquids were calculated through the same methods proposed in case study 1. A good separation solvent should have a high value for selectivity, eqn. (3-48), and solvent power, eqn. (3-49), and low value for solvent loss, eqn. (3-50).

Selectivity: $\beta = \frac{\gamma_{B,S}^{\infty}}{\gamma_{A,s}^{\infty}}$ (3-48)

Solvent power: $SP = \frac{1}{\gamma_{A,s}^{\infty}}$ (3-49)

Solvent loss: $SL = \frac{1}{\gamma_{S,B}^{\infty}}$ (3-50)

The three properties are a function of infinite dilution activity coefficients of the n-heptane/toluene/IL solution. The activity coefficients are calculated using the UNIFAC model (discussed in section 2.4) and the interaction parameters for ionic liquids were taken from Roughton *et al.*[105] Another important consideration is that, the addition of IL to the binary liquid mixture should result in the creation of two liquid phases. The appearance of new phases in a multi-component system can be checked through the implementation of necessary and sufficient conditions for phase stability. These conditions for a ternary system, are shown in eqns. (3-51) and (3-52) were derived by Bernard *et al.* (1967). The activity coefficients were again calculated using the UNIFAC method as discussed before.

$$(1 - x_2)\left(\frac{1}{x_2} + \frac{\partial ln\gamma_2}{\partial x_2}\right) - x_3\frac{\partial ln\gamma_2}{\partial x_3} < 0 \qquad (3\text{-}51)$$

$$\left(\frac{1}{x_2} + \frac{\partial ln\gamma_2}{\partial x_2}\right) + \left(\frac{1}{x_3} + \frac{\partial ln\gamma_3}{\partial x_3}\right) - \frac{\partial ln\gamma_2}{\partial x_3}.\frac{\partial ln\gamma_3}{\partial x_2} < 0 \qquad (3\text{-}52)$$

The CAILD design problem expressed as an optimization model is shown in eqns. (3-53) to (3-58).

It is worth mentioning that currently cost data is not available for ionic liquids as they are for the most part not commercially produced and it is also difficult to utilize cost information within a computer-aided molecular design framework. Hence cost was not considered for minimization.

*Objective function*

$f_{obj}$ = max (β) (3-53)

*Constraints:*

Ionic liquid Structural Feasibility (3-54)

$$\eta < 65\, cP \tag{3-55}$$

$$T_m < 298.15\, k \tag{3-56}$$

$$SL < 0.0065 \tag{3-57}$$

$$SP > 0.3719 \tag{3-58}$$

***Results*****:** The design statistics for this problem are summarized in Table 3-15. A total of 185 feasible IL structures were enumerated in subproblem 1 (structural constraints). Out of these, 27 ILs satisfied the physical property constraints (viscosity and melting point) in subproblem 2. Out of these, 1 ionic liquid satisfied the mixture property constraints, eqns. (3-57) and (3-58). The optimal ionic liquid structure (1-methylpyridinium tetrafluoroborate) with the highest selectivity is shown in Figure 3-8. The properties of the designed ionic liquid are listed in Table 3-16. Finally, we verified whether the designed ionic liquid created two phases when added to a hypothetical binary mixture consisting of 70% n-heptane (aliphatic) and 30% toluene (aromatic). This was accomplished by solving eqns. (3-51) and (3-52) for a range of ternary compositions (keeping n-heptane to toluene ratio constant). We identified that a phase split occurs at solvent composition range of 0.4 to 0.9.

**Table 3-15:** Decomposition Approach: Subproblem Results

| |
|---|
| **Subproblem 1:** Number of ionic liquids (ILs) generated, 185 |
| **Subproblem 2:** Number of ILs satisfying pure component properties, 27 |
| **Subproblem 3:** Number of ILs satisfying mixture properties, 1 |
| **Subproblem 4:** Optimal candidate, $[Mpy]^{+}[BF_4]^{-}$ |

The structure of the optimal ionic liquid that satisfies all the constraints and has the maximum selectivity (β) is shown in Figure 3-8. The properties of the designed ionic liquid are listed in Table 3-16.

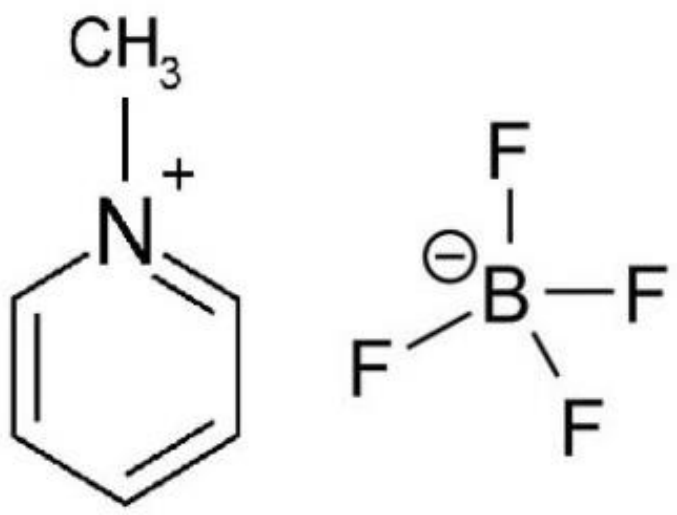

**Figure 3-8:** 1-methylpyridinium $[BF_4]^-$

**Table 3-16:** Design Results of the Optimal IL, 1-methylpyridinium $[BF_4]^-$

| Properties | 1-methylpyridinium $[BF_4]^-$ $[BF_4]$ | Sulfolane |
|---|---|---|
| Melting point (K) | 294.8 | 300.65 |
| Viscosity (cP) | 55.084 | 10.07 |
| SL | 0.006471 | 0.0065 ($t_1$) |
| SP | 0.67193 | 0.3719 ($t_2$) |
| $\beta$ | 87.262 | 6.8023 ($t_3$) |

***Analysis:*** Table 3-17 shows experimental selectivity values for the separation of aromatics from an aromatic/aliphatic mixture using different ionic liquids.[125] By comparing the selectivity values for separation of benzene from benzene/heptane mixture we see that $\beta_{[hmim][BF_4]} > \beta_{[hmim][PF_6]}$. Since both of these ionic liquids have the same cation (hmim) but different anions ($PF_6^-$ and $BF_4^-$), we can infer that selectivity values of ionic liquid with $BF_4^-$ anions are greater than those with $PF_6^-$ anions. By comparing the selectivity values of [bmim] $Tf_2N^-$ and [bmim] $PF_6^-$, (same cation and different anions), for separation of toluene/heptane mixture we can infer that selectivity values with $PF_6^-$ anion are greater than selectivity values with $Tf_2N^-$ anion. Therefore we can conclude that among the anions used in the design problem (*$Tf_2N^-$, $PF_6^-$ and $BF_4^-$*), ionic liquids having $BF_4$ anion should have the highest selectivity towards aromatic compounds. The design result is consistent with this observation as the optimal ionic liquid has $BF_4^-$ anion. Similarly, by comparing the selectivity values for separation of toluene/heptane mixture we can see that $\beta_{\text{[mmim] Tf2N}} > \beta_{\text{[emim] Tf2N}} > \beta_{\text{[bmim] Tf2N}}$. This shows that increasing number of alkyl groups on the cation side chain decreases selectivity. The design result (Figure 3-8) is consistent with this observation also as there is only one methyl group (minimum needed to satisfy the cation Valence) present in the cation side chain. With respect to cation, several studies have reported that Pyridinium-based cations have higher selectivity than imidazolium-base cations for aliphatic/aromatic separation. This is also consistent with our design results as the optimal IL had Pyridinium-based cation. This trend analysis qualitatively validates the CAILD methodology as well as the group interaction parameters (e.g. UNIFAC parameters provided in Roughton *et al.*) used in the model.

**Table 3-17:** Experimentally Measured Selectivity values for Aromatic/Aliphatic Separations

| Solvent | Separation | T (°C) | β (Selectivity) |
|---|---|---|---|
| 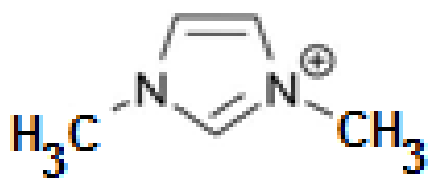 $Tf_2N$ | Toluene/heptane | 40 | 29.8 |
| 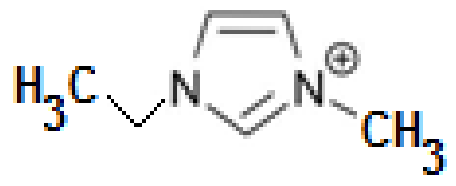 $Tf_2N$ | Toluene/heptane | 40 | 22.2 |
| 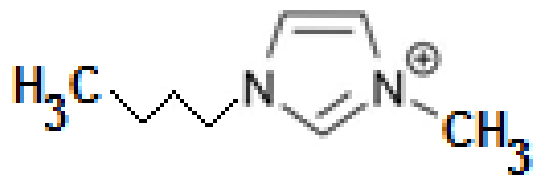 $Tf_2N$ | Toluene/heptane | 40 | 16.7 |
| 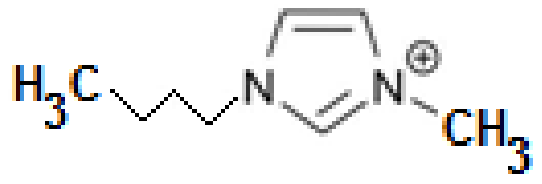 $PF_6$ | Toluene/heptane | 40 | 21.3 |
| 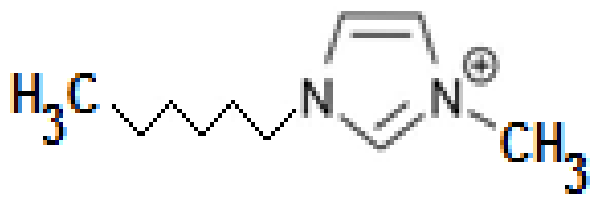 $PF_6$ | Benzene/heptane | 25 | 8.20 |
| 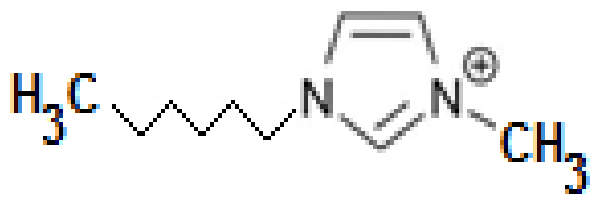 $BF_4$ | Benzene/heptane | 25 | 8.40 |
| 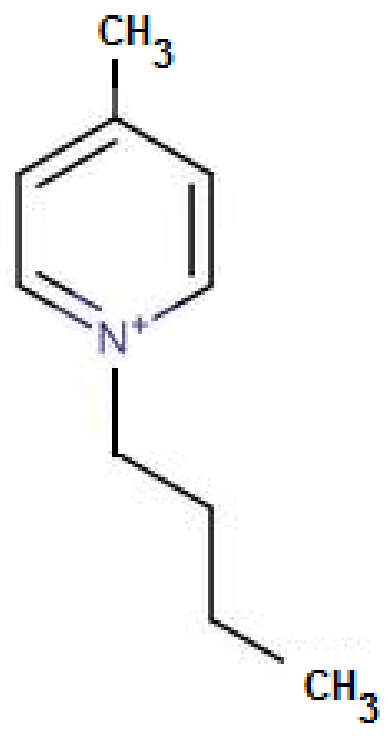 $BF_4$ | Toluene/heptane | 40 | 32.8 |

These qualitative trends from experiments are consistent with our CAILD results. The optimal design has a $BF_4^-$ anion and has minimal number of alkyl side groups (note that the selected methylpyridinium base cation needs at least one $CH_3$ group; i.e. the minimum possible number of groups to satisfy the valence requirement).

### 3.3.4 Naphthalene solubility

In this case study we consider the design of an ionic liquid solvent for the dissolution of organic compound naphthalene. A molecular solvent having high solubility for naphthalene and commonly used for its dissolution is chloroform. The measured solubility of naphthalene in chloroform is 0.473 mole fraction.[126] Our objective is to find an ionic liquid that has higher solubility for naphthalene than chloroform. In addition, the ionic liquid needs to be an RTIL, and have a reasonably low viscosity. The melting point and viscosity are calculated using the same models described in the case study 1. The expression shown in eqn. (3-61) is invoked to ensure solid-liquid phase equilibrium conditions, in order to determine the saturation concentration of solute (i.e. solubility).[115] The CAILD design problem expressed as an optimization model is shown in eqns. (3-59) to (3-63).

*Objective function*

$$f_{obj} = \max\ (x_1^{sat}) \qquad \text{`} \qquad (3\text{-}59)$$

*Constraints*

Ionic liquid Structural Feasibility (3-60)

$$\ln x_1 - \frac{\Delta_{fus}H}{T_m}\left(1 - \frac{T_m}{T}\right) + \ln \gamma_1^{sat} = 0 \quad (3\text{-}61)$$

$$\eta < 65\ cP \quad (3\text{-}62)$$

$$T_m < 298.15\ k \quad (3\text{-}63)$$

***Results*:** The design statistics for this problem are summarized in Table 3-18. A total of 185 feasible IL structures were enumerated in subproblem 1. Out of these, 27 ILs satisfied the physical property constraints (viscosity and melting point) in subproblem 2. The optimal ionic liquid structure (1-Butyl-3-ethylimidazolium [Tf2N]) with the highest solubility is shown in Figure 3-9. The properties of the designed ionic liquid are listed in Table 3-19.

**Table 3-18:** Decomposition Approach: Subproblem Results

| |
|---|
| **Subproblem 1:** Number of ionic liquids (ILs) generated, 185 |
| **Subproblem 2:** Number of ILs satisfying pure component properties, 27 |
| **Subproblem 3:** Number of ILs satisfying mixture properties, 27 |
| **Subproblem 4:** Optimal candidate, 1-Butyl-3-ethylimidazolium $[Tf_2N]^-$ |

The structure of the optimal ionic liquid that satisfies the constraints is shown in Figure 3-9. The optimal properties of the designed ionic liquid are shown in Table 3-19.

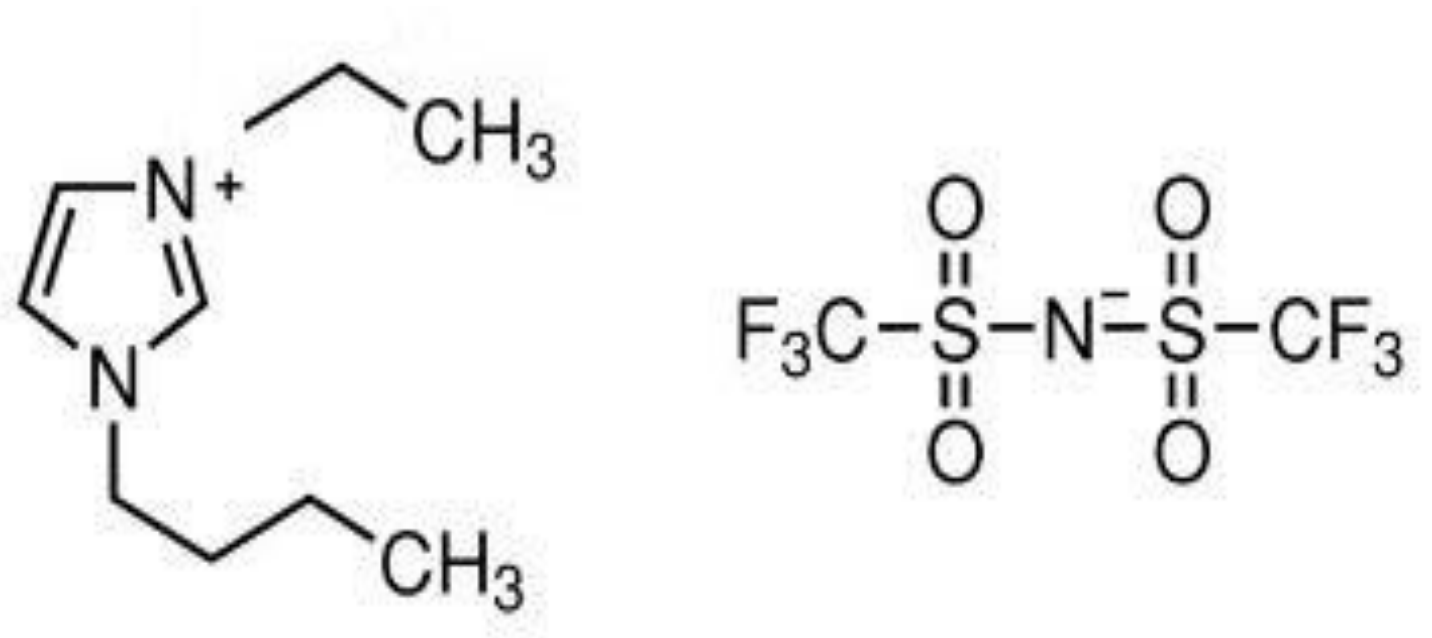

**Figure 3-9:** 1-butyl-3-ethylimidazolium [Tf2N]$^-$

**Table 3-19:** Physical properties of 1-butyl-3-ethylimidazolium $[Tf_2N]^-$

| Properties | 1-Butyl-3-ethyl imidazolium $[Tf_2N]$ | Chloroform |
|---|---|---|
| Melting point (K) | 222.78 | 209.65 |
| Viscosity (cP) | 55.59 | 0.542 |
| Naphthalene solubility @ 25 $^oC$ | 0.5069 | 0.473 |

An important component of green chemistry relates to the solvent medium in which synthetic transformations are carried out.[127] Traditional volatile organic solvents - which act as common reaction media for several chemical processes - are linked to a host of negative environmental and health effects including climate change, urban air-quality and human illness. Jessop[128] states that one of the major challenges in the search for environmentally benign solvents is to ensure availability of green solvents as replacements for non-green solvents of any kind. He uses the Kamlet-Taft plots to show that current list of green solvents populate only a small region of the entire spectrum of solvents needed for various applications and argues that large unpopulated areas of this diagram mean that future process chemists and engineers need solvents having certain desirable properties and are green. Ionic liquids offer great potential to satisfy this need.

This study presents an overarching framework that can be utilized to design optimal ionic liquids for a given application through the theoretical/computational consideration of all possible combinations. Currently, the few ionic liquid structure-property models that are available can be applied to a small subset of all available ionic liquid types. However, for this method to be fully effective, we need group contribution models and parameters that span the entire spectrum of ionic liquids. It is indeed possible to overcome this challenge as one needs property values of only few representative compounds in each class of ionic liquid (for example, covering the groups shown in Appendix A) to regress the contributions of the various groups. We propose that future research should focus on experimental property measurements and data collection of ionic liquids that cover a diverse set of cations, anions and functional groups. The second challenge is the lack of ionic liquid structure-property models (i.e. solution to the forward problem) for various thermo-physical properties of interest. There, is a great need to develop structure-property models of pure-component physical properties and thermodynamic solution (mixture) properties for a comprehensive set of ionic groups. The third challenge would relate to the accuracy of the group contribution models. However, as discussed earlier, since the primary aim of the CAILD method would be to narrow down to a small set of ionic liquids from the millions of available alternatives, reasonably accurate predictive models are sufficient. The final IL can be selected by *ab initio* computational chemistry calculation or experimental verifications of these small set of designed compounds.

Progress towards designing ionic liquids through proposed CAILD, framework, will not only contribute towards our understanding of the relationship between cation-anion structures and ionic liquid properties, but will also provide a mechanism to engineer new environmentally benign ionic liquids for critical applications.

# Chapter 4: Application 1: Design of Ionic Liquids for Thermal Energy Storage

In this chapter, we present a computer-aided framework to design task-specific ionic liquids (ILs) for solar energy storage, using structure-property models and optimization methods. Thermal energy storage density (capacity) was used as a measure of the ability of an IL to store thermal (solar) energy.

## 4.1 Introduction

Advancements in solar trough and solar tower technologies have enabled concentration of thermal energy to the extent that it can be used to drive traditional steam cycles providing an alternative to fossil fuel use.[129] Therefore, harvesting solar energy using arrays of parabolic trough collectors will enable generation of electricity at a large scale. In solar power plants, thermal energy storage (TES) is necessary to extend production periods of low or no sunlight. An important component of TES systems is thermal fluid which is needed to transfer and store heat for relatively short periods. A parabolic trough system typically consists of a series of collectors that are big mirror-like reflectors used to concentrate solar energy. When the solar radiation is received by these collectors the reflected light is concentrated at the center of the collector. A heat transfer fluid (HTF) passed through tubes present at the center of the collector absorbs the accumulated heat. The collected thermal energy is then transferred from the HTF to a storage medium or is stored in a reservoir using the heat capacity of the HTF itself.[130] The storage media can then release the thermal energy when needed for further conversion to electrical energy. Thermal energy storage (TES) therefore makes solar energy a more reliable and economic alternative source of energy.

Fluids that have high potential to store heat energy such as thermal oil (e.g. VP-1$^{TM}$), or nitrate salts (e.g. HITEC-XL$^{TM}$) are suited for thermal energy storage applications. However, nitrate salts have melting points (freezing point) above 200°C while mineral oils have upper temperature limit of 300°C[130] thus limiting their use to a narrow temperature range and thereby reducing the overall efficiency of the process. Ionic liquids (and salts) have properties that are ideal for thermal storage applications. These attractive properties include high heat capacity, high decomposition temperature and relatively high density at operating conditions. Ionic liquids (ILs) are a new generation of materials that that have a wide range of applications.[1] Similar to salts ILs are composed of ions but have much lower melting points. Several ILs are in liquid state at room temperature (commonly referred as room temperature ionic liquids [RTILs]). ILs consist of an organic cation (a cation base with alkyl side chain) and a charge-delocalized inorganic or organic anion.[131] They usually possess good thermal stability (i.e. high decomposition temperature) making them appropriate for processes operating at high temperatures. Ionic liquids can be customized through appropriate selection of cations, anions and alkyl side chains. Therefore, ILs can be tuned to impart specific functionalities for a given application by changing cation/anions/side chain groups.

In this study, we focus on the computational design of optimal ionic liquids with high thermal storage density for solar energy storage applications. The key requirements of a thermal storage media include high thermal storage capacity ($\rho . Cp\ [\frac{MJ}{m^3 K}]$), high thermal stability[129], and a wide liquid range. Therefore the properties of ionic liquid that need to be optimized for thermal storage applications include: density, heat capacity, thermal decomposition temperature and melting point. Heat capacity measurements of diverse ionic liquids reveal wide ranging values.[132] The melting point of ionic liquids can be easily adjusted by the choice of cation, anion or the groups attached to the cation side chains. Thermal stability of ionic liquids has been previously studied and it has been reported that many of them have high thermal decomposition temperatures (~400 $^{0}$C). In order for us to find the optimal ionic liquid structure having desired melting point and decomposition temperature as well as maximum thermal storage capacity, thousands of different ionic liquids need to be tested. Since this is not feasible experimentally, a computer-aided approach is suggested in this study.

Computer-aided molecular design (CAMD) is a promising technique that has been widely used to design compounds for different applications.[133-139] Gani and co-workers initially conceptualized this method for screening solvents based on UNIFAC group contribution approach.[133] CAMD usually integrates structure based property prediction models (e.g. group contribution models) and optimization algorithms to design molecular compounds with desired properties. More recently, this approach has been extended to the design of ionic liquids.[140,141,142,143,144] A comprehensive framework for computer-aided ionic liquid design (CAILD) was recently published by our group.[142] Key to the successful development and use of CAILD methods is the availability of predictive models for the properties of interest. In this work we present a new CAILD model to design novel ionic liquids as thermal fluids for solar energy storage applications. This CAILD model utilizes existing group contribution methods to predict physical and thermal properties of ionic liquids. By considering a structurally diverse set of building blocks we are able to demonstrate that new and novel structural variants of ionic liquids can be tailored specifically for this application.

## 4.2 Formulation of the design problem

In this section, we focus on presenting a computer-aided ionic liquid design (CAILD) model to find (design) optimal ionic liquid structures with high thermal storage capacity, reasonably low melting point and high decomposition temperature. In this method a variety of cation head groups, cation side chain groups (including certain functional groups), and anions were selected as ionic liquid building blocks. Typical CAILD approach requires solution to the forward problem (i.e. property prediction) as well as the reverse problem (i.e. structure generation). In a mathematical programming based CAILD approach the physical properties of ionic liquids are estimated using structure based predictive models such as group contribution (GC) models (forward problem) and optimal ionic liquid structures are generated by solving a mixed-integer non-linear programming (MINLP) formulation of the design problem (reverse problem). This study utilizes GC methods from literature[145,146,147,148a,148b] to predict the physical properties density, heat capacity, melting point and decomposition temperature. The CAILD framework proposed in Karunanithi and Mehrkesh[142] was utilized to formulate the thermal storage fluid (TSF) design problem as an MINLP model. Structural constraints, eqns. (4-1) to (4-5), were included to design feasible ionic liquid structures. These constraints are a sub-set of a comprehensive set of structural constraints presented in Karunanithi and Mehrkesh.[142]

Linear physical property constraints based on GC predication, eqns. (4-11) to (4-15), were integrated with the structural feasibility constraints. The objective of the design problem was to identify the optimal ionic liquid structure that has the highest thermal storage capacity. Therefore, the objective function was formulated to maximize the product of density and heat capacity of the ionic liquid. The solar thermal storage process is typically carried out at temperatures of around 300℃. Therefore, the thermal storage fluid -in this case the designed ionic liquid- should be operable at temperatures slightly above 300℃. To ensure that the ionic liquid does not decompose during the process, we enforce a constraint on thermal decomposition temperature to be above 400℃, eqn. (4-24). The temperature of ionic liquid after energy exchange should be higher than its melting point. To ensure this, a constraint on melting point to be above 140℃ is imposed, eqn. (4-25). The temperature window between melting point and decomposition temperature is the range at which the process can operate. The basis set (the structural building blocks) considered for this problem included 5 cation head groups, 9 anions and 5 side chain groups (alkyl and functional groups) which are shown in Table 4-1. This selection was based on groups for which group contribution parameters were available for all the properties of interest.

**Table 4-1:** Ionic liquid building blocks (groups) considered for thermal fluid design

| Cation | Anions | Groups |
|---|---|---|
| Imidazolium 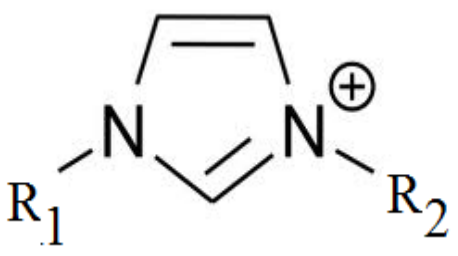 | Tetrafluoroborate $[BF_4]^-$ | Methylene ($CH_2$) |

| Cation | Anions | Groups |
|---|---|---|
| Pyridinium 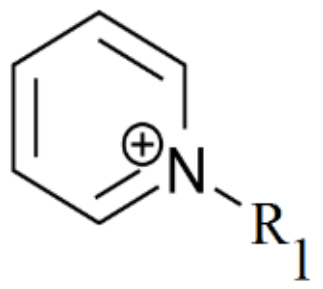 | Hexafluorophosphate $[PF_6]^-$ | Methyl ($CH_3$) |
| Ammonium 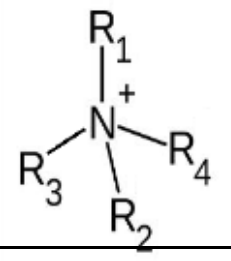 | bis(trifluoromethylsulfonyl) imide 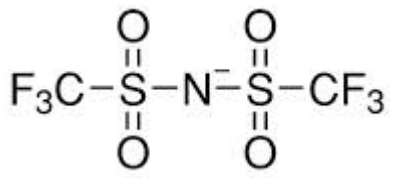 | Benzyl 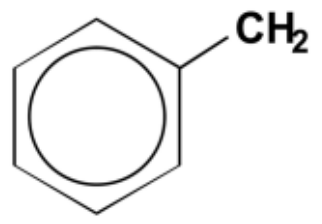 |
| Phosphonium 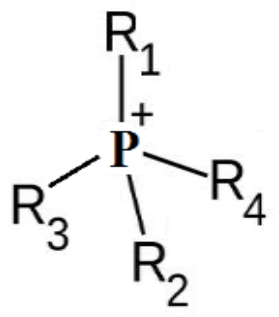 | Chloride $Cl^-$ | Methoxy 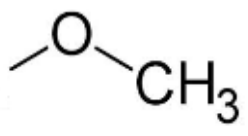 |
| Pyrrolidinium 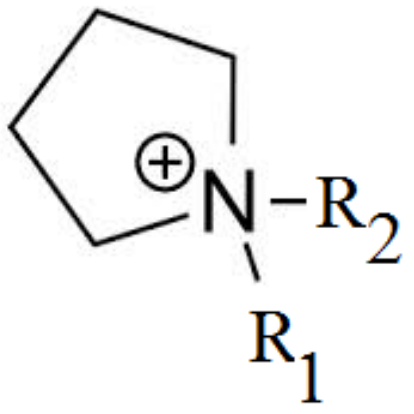 | Bromide $Br^-$ | Hydroxyl (-OH) |
| | Trifluoromethanesulfonate 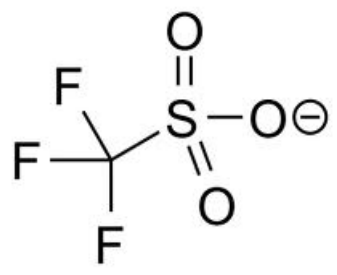 | |
| | Benzoate 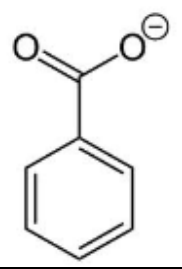 | |
| | Nitrate 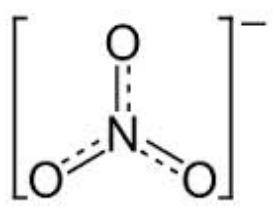 | |

| Cation | Anions | Groups |
|---|---|---|
| | Acetate 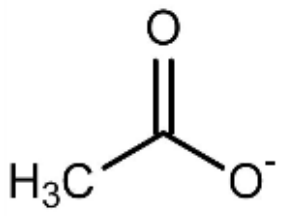 | |

***Objective function***

$$f_{obj} = \max (\rho . C_P)$$

***Constraints***

1) Structural (feasibility) Constraints

2) $T_m < 140$ ℃

3) $T_d > 400$ ℃

### 4.2.1 Ionic liquid structural constraints

A key requirement in computational design of ionic liquids is the ability to guarantee solutions that are theoretically feasible chemical structures. Further, one would have to account for practical considerations - such as limits on the size or presence/absence of certain groups etc. - that will lead to design candidates that are workable from a synthesis view point. In order to incorporate these considerations the proposed method requires that the designed compound satisfies certain rules, broadly termed as structural constraints, that includes chemical feasibility rules such as octet rule, bonding rule and complexity rules as well as rules that restrict/determine the size and the constituents of possible solutions. More details about these structural constraints can be found in Karunanithi and Mehrkesh.[142] The specific structural constraints from Karunanithi and Mehrkesh[142] that are invoked for this problem are discussed below.

The first two constraints, eqns. (4-1) and (4-2), ensures the selection of only one cation and one anion from the basis set.

$$\sum_{i \in C} c_i = 1 \tag{4-1}$$

$$\sum_{j \in A} a_i = 1 \tag{4-2}$$

Where $C$ is a set of all cation head groups considered (i.e. imidazolium, pyridinium, ammonium, phosphonium and pyrrolidinium); and $A$ is a set of all anions considered. The next set of feasibility constraints deal with valence requirements of cations which enables us to add appropriate side chain groups. Per our definition, anion is considered as a single group and does not require constraints related to addition of groups and therefore valence constraints are relevant only for the cation part. The three equations below make sure that octet rule for the cation as a whole as well as for each side chain (branch) is not violated.

$$\sum_{l=1}^{n^*} y_l = \sum_{i \in c} c_i v_{ci} \tag{4-3}$$

$$\sum_{i \in C} (2 - v_{ci}) c_i + \sum_{l=1}^{n^*} \sum_{k \in G} (2 - v_{Gkl}) y_l n g_{kl} = 2 \tag{4-4}$$

$$\sum_{k \in G} y_l n g_{kl} \, (2 - v_{Gkl}) = 1 \tag{4-5}$$

Where we fixed $n^*$ to be 2, 1, 4, 4, and 2 for imidazolium, pyridinium, ammonium, phosphonium and pyrrolidinium respectively. This was based on the fact that for imidazolium, pyridinium and pyrrolidinium the side chains are commonly connected to only the nitrogen atoms in the cation ring. However, note that if we want to broaden the design problem we can allow side chains to be attached to the carbon atoms in the cation rings by adjusting these constraints. More specific details and definitions about these constraints can be found in Karunanithi and Mehrkesh.[142] The next set of constraints deal with restricting the size of cation as well as putting limits on the presence/absence of certain groups. Limits on the total number of groups that can occur on each side chain as well as the whole cation was fixed using eqns. (4-6) and (4-7). These two constraints make sure that we do not design a very large cation, which cannot be practically synthesized. Limits on the number of functional groups with valence 1 (i.e. OH, $OCH_3$, and benzyl) that can occur in the whole cation was fixed using eqn. (4-8).

$$\sum_{l=1}^{n^*} \sum_{k \in G} y_l n g_{kl} \leq (n^* \times 16) \tag{4-6}$$

$$\sum_{k \in G} y_l n g_{kl} \leq 16 \tag{4-7}$$

$$\sum_{l=1}^{n^*} \sum_{k \in G_1} y_l n g_{kl} \leq t_1 \tag{4-8}$$

Where, $G$ is a set of all groups considered (i.e. $CH_2$, $CH_3$, OH, $OCH_3$, and benzyl) and $G_1$ is a subset of $G$ which consist of functional groups OH, $OCH_3$, and benzyl. Eqn. (4-8) is invoked three times for each $k \in G_1$ while $t_1$ for each of these three equations was assigned to be 2,1,2,2,2 for imidazolium, pyridinium, ammonium, phosphonium and pyrrolidinium head groups respectively.

### 4.2.2 Ionic liquid property prediction

This section describes in detail the physical properties of ionic liquids that are relevant for thermal storage applications and the structure based property prediction models and correlations that are utilized to predict these properties within the proposed computer-aided ionic liquid design (CAILD) framework.

***Thermal storage density***

Thermal storage density ($\eta\ [\frac{MJ}{m^3K}]$) of an ionic liquid is defined as a product of its heat capacity and density.

$$\eta = \rho . Cp \qquad (4\text{-}9)$$

Since both density and heat capacity are a function of temperature, thermal storage densities of ionic liquids vary with temperature. $\eta$ is the most critical design parameter for thermal fluids as higher value of $\eta$ will result in lower volume of thermal fluid requirement.

***Heat Capacity***

The heat capacities of ionic liquids were predicted using the approach developed by Valderrama *et al.*[147] In this method the authors used the concept of mass connectivity index (MCI) to build a structure based predictive model. Molecular connectivity was first introduced by Randic[147], and has been used then by several authors for property prediction. As suggested by Valderrama *et al.*[147], the MCI concept can be used to quantify the extent of branching in ionic liquids thereby enabling us to predict IL heat capacities better. This index considers the mass of structural groups as well as the type of connections between them as following:

$$\lambda = \sum \left(\frac{1}{\sqrt{m_i m_j}}\right)_i \qquad (4\text{-}10)$$

where, $m_i$ and $m_j$ are the mass of neighboring groups i and j in a molecule. In this expression, the sequence of groups is important, as i and j are two distinct groups and hence the connection $m_i m_j$ is different from $m_j m_i$. Valderrama *et al.*[147] showed that the MCI approach is capable of predicting the heat capacity as a function of temperature for a variety of ionic liquids with an acceptable level of accuracy. To predict the ionic liquid heat capacity, a reference value of $C_{P0}$ at reference condition $T_0$ is used as follows

$$C_P = C_{P0} + \lambda[p(T - T_0) + q(T - T_0)^2] \qquad (4\text{-}11)$$

Here "p" and "q" are constants specified for each ionic liquid and have been correlated from experimental data of ionic liquid heat capacities. The overall $C_{p,}$ as a temperature dependent variable, can be estimated as a function of molar mass of cation and anion, molar volume of ionic liquid and mass connectivity index as follows

$$C_P = a + bV_m + c\lambda + d\eta + \lambda[e(T - T_0) + f(T^2 - T_0^2)] \qquad (4\text{-}12)$$

where, $T_0$ is the reference temperature, 298.15 K, a = 15.80, b = 1.663, c = 28.01, d = -7.350,

e = 0.2530, f = -1.372×10$^{-5}$,$V_m$ is the molar volume ($cm^3$/mol), λ is the connectivity index (MCI), and $\eta = \frac{MW^{cat}}{MW^{ani}}$ of ionic liquid.

***Density***

Density of ionic liquids were predicted using the approach presented by Valderrama *et al.*[145] In this approach the density of each ionic liquid at a given temperature was estimated as a secondary property using its critical properties ($T_c$ and $V_c$) and normal boiling point ($T_b$) as primary properties through the following expression, eqn. (4-13):

$$\rho = \frac{A}{B} + \left(\frac{2}{7}\right)\left\{\frac{A \ln B}{B}\right\}\left(\frac{T - T_b}{T_c - T_b}\right) \tag{4-13}$$

where, $A = a + \frac{bM}{V_c}$, $B = (\frac{c}{V_c} + \frac{d}{M})V_c^{\delta}$, a= 0.3411, b= 2.0443, c=0.5386, d = 0.0393, δ= 1.0476.

Critical properties and normal boiling point of the ionic liquids were estimated using the group contribution method proposed by Valderrama *et al.*[146]

***Melting Point***

An ionic liquid with low melting point would be desirable since at all times the operating temperature should be kept above the melting point of the thermal storage fluid to avoid solid formation in the system. Alternately, when the melting point is too high a high process temperature needs to be maintained which will decrease the system efficiency by reducing the rate of heat transfer (sensible heat). An appropriate thermal storage fluid should possess a melting point lower than 140 °C. To calculate the melting point of ionic liquids a group contribution method proposed by Lazzus[148b] was used. In this approach two different sets of contribution for melting point were used: 1) the contribution of cation head group (cation base) and alkyl groups/functional groups attached to the side chains of the cation head groups; 2) the contribution of groups associated with the anion. Cation head groups are tabulated as whole (e.g. imidazolium or pyridinium) but side chain groups and anions are split into smaller structural fragments. The melting point of any given ionic liquid can be calculated by the summation of contribution of cations, anions and side chain groups as following:

$$T_m(k) = 288.7 + \sum n_i \Delta tc_i + \sum n_j \Delta ta_j \qquad (4\text{-}14)$$

where $n_i$ is the number of occurrence of group i, $\Delta tc_i$ is the contribution of cation to the melting point and $\Delta ta_j$ is the contribution of anion to the melting point of the given ionic liquid.

***Thermal Decomposition Temperature***

The final property of interest is thermal decomposition temperature. This property is an estimate of the highest temperature at which the ionic liquid will remain in the associated state (non-decomposed). It is important for thermal storage as it will determine the maximum applicable temperature at which the thermal fluid can be utilized. To predict the decomposition temperature of ionic liquids a group contribution method proposed by Lazzus[148a] was utilized.

This method is similar to the melting point prediction described above.

$$T_d(k) = 663.85 + \sum n_i \Delta tc_i + \sum n_j \Delta ta_j \quad (4\text{-}15)$$

where $n_i$ is the number of occurrence of group i, $\Delta tc_i$ is the contribution of cation to the decomposition temperature and $\Delta ta_j$ is the contribution of anion to any given ionic liquid.

### 4.2.3 CAILD model solution

The complete CAILD-MINLP model is shown below:

***Objective function***

$$\text{f}_{\text{obj}} = \text{max}\,(\rho . C_P)$$

***Constraints***

$$\sum_{i \in C} c_i = 1 \quad (4\text{-}16)$$

$$\sum_{j \in A} a_i = 1 \quad (4\text{-}17)$$

$$\sum_{l=1}^{n^*} y_l = \sum_{i \in c} c_i v_{ci} \quad (4\text{-}18)$$

$$\sum_{i \in c} (2 - v_{ci}) c_i + \sum_{l=1}^{n^*} \sum_{k \in G} (2 - v_{Gkl}) y_l ng_{kl} = 2 \quad (4\text{-}19)$$

$$\sum_{k \in G} y_l ng_{kl} (2 - v_{Gkl}) = 1 \quad (4\text{-}20)$$

$$\sum_{l=1}^{n^*} \sum_{k \in G} y_l n g_{kl} \leq (n^* \times 16) \tag{4-21}$$

$$\sum_{k \in G} y_l n g_{kl} \leq 16 \tag{4-22}$$

$$\sum_{k \in G_1} y_l n g_{kl} \leq t_1 \tag{4-23}$$

$$T_d(k) = 663.85 + \sum n_i \Delta tc_i + \sum n_j \Delta ta_j > \ 400\ ℃ \tag{4-24}$$

$$T_m(k) = 288.7 + \sum n_i \Delta tc_i + \sum n_j \Delta ta_j < \ 140\ ℃ \tag{4-25}$$

$$\rho = \frac{A}{B} + \ (\frac{2}{7})\{\frac{A\ ln\ B}{B}\}(\frac{T - T_b}{T_c - T_b}) \tag{4-26}$$

$$A = a + \frac{bM}{V_c} \tag{4-27}$$

$$B = (\frac{c}{V_c} + \frac{d}{M})V_c^{\delta} \tag{4-28}$$

$$a = 0.3411, b = 2.0443, c = 0.5386, d = 0.0393, \delta = \ 1.0476$$

$$V_c(cm^3/mol) = 6.75 + \sum n_i \Delta V_{ci} \tag{4-29}$$

$$T_b(K) = 198.2 + \sum_i n_i \Delta T_b \tag{4-30}$$

$$T_C(K) = {T_b}\Big/{[A + B \sum_i n_i \Delta T_c - (\sum_i n_i \Delta T_c)^2]} \tag{4-31}$$

$$A = 0.5703, B = 1.012$$

$$C_P = a + bV_m + c\lambda + d\eta + \lambda[e(T - T_0) + f(T^2 - T_0^2)] \tag{4-32}$$

$T_0 = 298.15\text{ K}, a = 15.80, b = 1.663, c = 28.01, d = -7.350, e = 0.2530, f = -1.372 \times 10^{-5}$

$$V_m = \sum_i V_{m,atoms} \qquad (4\text{-}33)$$

$$\lambda = \sum \left(\frac{1}{\sqrt{m_i m_j}}\right)_i \qquad (4\text{-}34)$$

$$\eta = \frac{MW^{cat}}{MW^{ani}} \qquad (4\text{-}35)$$

The above computer-aided ionic liquid design (CAILD) model is a Mixed Integer Non-Linear Programming (MINLP) formulation. MINLP models combine combinatorial aspects with nonlinearities and are more difficult to solve than mixed integer programming (MIP) and non-linear programming (NLP) problems. The most precise approach to solve this MINLP model would be to fully enumerate each possible combination within the entire search space. The number of possible combinations increase exponentially with number of groups considered resulting in combinatorial explosion requiring large computational times. Other multi-level approaches have been proposed to address this issue and avoid complete enumeration.[149,150] Several deterministic optimization based methods have been employed to solve CAMD-MINLP models.[137,138,151] Different stochastic methods have also been used to solve CAMD problems.[152,153,154] Property clustering technique used within a reverse problem formulation is yet another approach that has been successfully used to solve CAMD problems.[155,156] In this work a genetic algorithm (GA) based solution approach is utilized to solve this optimization model. GA is a stochastic approach that mimics nature's process of biological evolution (principle of natural selection) and has been previously used to successfully solve complex optimization (minimization) problems of different formulations.

GA can be an attractive alternate solution approach for problems that are not well suited for standard optimization algorithms such as problems with discontinuous, stochastic, or highly nonlinear objective functions.[157] The optimization process usually starts with a random collection of initial candidates from which the fitter candidates are selected as “parents” and can exchange their genetic information, through crossover and mutation operations. At each iteration, a new population of fitter candidate structures is generated to replace the existing population and this process is repeated for a pre-specified number of iterations or until the pre-defined value of tolerance for objective function is met.[157] The GA solution of CAILD-MINLP problem was implemented in the MATLAB environment with most parameters set at their default values. A large population size which was generated randomly by the program was used to start the search process. The crossover fraction was fixed at 0.8, whereas a uniform mutation probability with rate value of 0.01 was used.

## 4.3 Results and discussion

The optimal ionic liquid consists of a hydroxyl-functionalized imidazolium cation and tetrafluoroborate anion whose structure is shown in Figure 4-1.

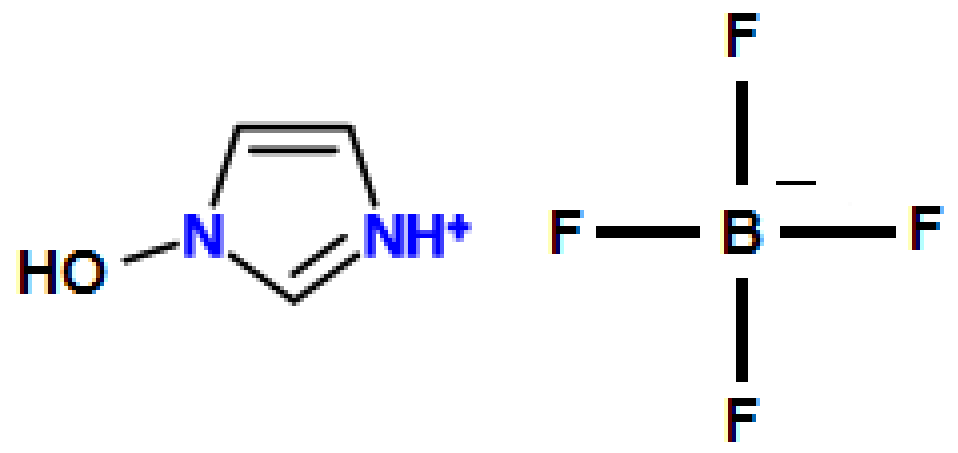

**Figure 4-1:** A schematic of the structure of optimal IL with highest thermal storage capacity

In Table 4-2, we list the physicochemical properties of the optimal IL along with two common commercial products (VP-1™ a heat transfer oil and Hitec XL™ a molten salt) used as thermal storage medium.[130] The designed ionic liquid has a higher thermal storage capacity than VP-1™ oil and has about the same value of Hitec molten salt. However, the estimated viscosity of the designed ionic liquid is an order of magnitude lower than that of the molten salt making it a better alternative. The lower viscosity is extremely important as the overall commercial feasibility of the process depends on pumping energy and low viscous TES medium would result in less operating costs. Overall, the design results suggest that the optimal ionic liquid has better thermal storage properties than existing commercial products.

**Table 4-2:** Thermo-physical properties of VP-1™, Hitec XL™ and [3-hydroxy-Imidazolium]$^{+}$[$BF_4$]$^{-}$

| **Properties (@ T)** | **VP-1™** | **Hitec XL™** | **[3-hydroxy-Imidazolium]$^{+}$ [$BF_4$]$^{-}$** |
|---|---|---|---|
| Melting Point (℃) | 13 | 120 | 129.8 |
| Decomposition Temp (℃) | 400 | 500 | 578.1 |
| Density (kg/$m^3$) | 815 (300℃) | 1992 (300℃) | 1477 |
| Heat Capacity $C_p$ (J/Kg K) | 2319 (300℃) | 1447 (300℃) | 1947 (300℃) |
| Storage density η (MJ/$m^3$ K) | 1.9 (300℃) | 2.9 (300℃) | 2.88 (300℃) |
| Viscosity (cP) | 0.2 (300℃) | 6.27 (300℃) | 0.88 (300℃) |

We performed a comprehensive set of analysis to study in more detail the relationship between ionic liquid structural components and their thermal storage properties. First, we considered each category (head group) of cation separately and identified the optimal (highest η) cation structure within each cation type. Then we fixed the five optimal cation structures and studied the effect of different anions considered in this study in terms of their influence on the TES properties (Table 4-3). It is clear that among all the anions considered, $[BF_4]^-$ always resulted in the highest thermal storage capacity irrespective of the cation type. This finding is also consistent with our optimal IL candidate, which too had $[BF_4]^-$ anion. Further, from these tables it is also clear that imidazolium cation always resulted in the highest thermal storage capacity, which is consistent with our optimal IL candidate that had an imidazolium cation

**Table 4-3:** Effect of anion variation on the thermal storage properties of ionic liquids

| **Cation** | 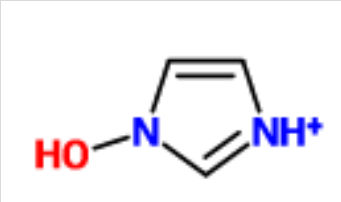 | | | | |
|---|---|---|---|---|---|
| **Anion** | **$BF_4^-$** | **$PF_6^-$** | **$Tf_2N^-$** | **$Cl^-$** | **$Br^-$** |
| Density ($kg/m^3$) | 1477.5 | 1548.9 | 1697.7 | 1415.6 | 1788.3 |
| $C_p$ (J/kg K) | 1947.36 | 1640.3 | 1527.9 | 1843.4 | 1453.4 |
| $\eta=\rho.C_p$ ($MJ/m^3$ K) | 2.88 | 2.54 | 2.59 | 2.61 | 2.6 |
| $T_m$ (K) | 402.9 | 422.1 | 350.0 | 438.1 | 436.3 |
| $T_{app}$ (K) | 851.3 | 857.3 | 882.0 | 721.8 | 757.9 |
| **Cation** | 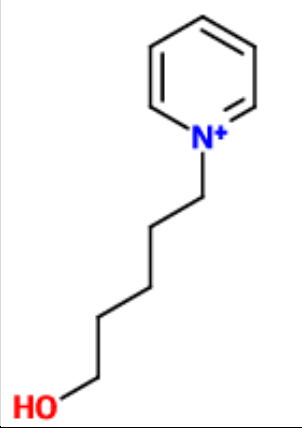 | | | | |

| **Anion** | **$BF_4^-$** | **$PF_6^-$** | **$Tf_2N^-$** | **$Cl^-$** | **$Br^-$** |
|---|---|---|---|---|---|
| Density (kg/m$^3$) | 1181.5 | 1260.5 | 1414.7 | 1101.4 | 1286.7 |
| $C_p$ (J/kg K) | 2256.9 | 1982.2 | 1798.5 | 2226.5 | 1934.5 |
| $\rho.C_p$ (MJ/m$^3$ K) | 2.67 | 2.50 | 2.54 | 2.45 | 2.49 |
| $T_m$ (K) | 409.7 | 428.9 | 356.7 | 444.9 | 443.05 |
| $T_{app}$ (K) | 765.6 | 771.5 | 796.3 | 636.1 | 672.2 |
| **Cation** | 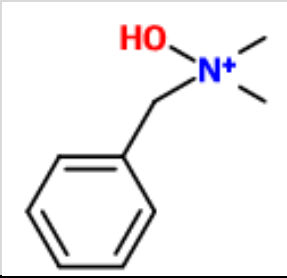 | | | | |
| **Anion** | **$BF_4^-$** | **$PF_6^-$** | **$Tf_2N^-$** | **$Cl^-$** | **$Br^-$** |
| Density (kg/m$^3$) | 1227.6 | 1309.4 | 1474.4 | 1156.5 | 1363.1 |
| Cp (J/kg K) | 2198.1 | 1919.5 | 1743.6 | 2155.8 | 1852.8 |
| $\rho$.Cp (MJ/m$^3$ K) | 2.70 | 2.51 | 2.57 | 2.49 | 2.52 |
| Tm (K) | 386.4 | 405.7 | 333.5 | 421.6 | 419.8 |
| Tapp (K) | 828.6 | 834.6 | 859.3 | 699.2 | 735.3 |
| **Cation** | 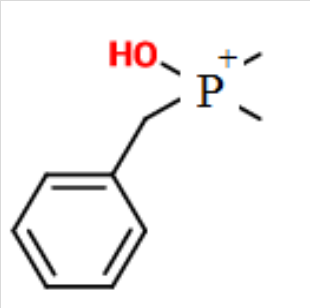 | | | | |
| **Anion** | **$BF_4^-$** | **$PF_6^-$** | **$Tf_2N^-$** | **$Cl^-$** | **$Br^-$** |
| Density (kg/m$^3$) | 1241.7 | 1317.4 | 1470.0 | 1174.0 | 1367.4 |
| Cp (J/kg K) | 2036.7 | 1812.1 | 1675.7 | 1960.7 | 1724.6 |
| $\rho$.Cp (MJ/m$^3$ K) | 2.53 | 2.39 | 2.46 | 2.3 | 2.36 |
| Tm (K) | 378.0 | 397.2 | 325.1 | 413.2 | 411.3 |
| Tapp (K) | 850.8 | 856.8 | 881.5 | 721.3 | 757.4 |
| **Cation** | 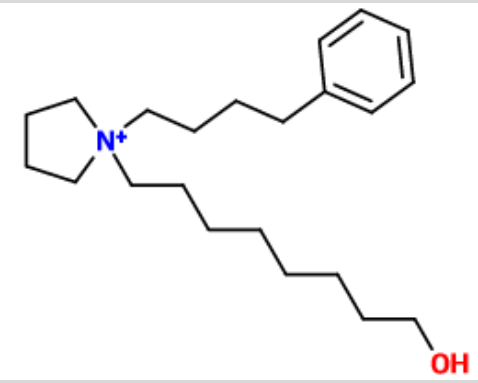 | | | | |
| **Anion** | **$BF_4^-$** | **$PF_6^-$** | **$Tf_2N^-$** | **$Cl^-$** | **$Br^-$** |
| Density (kg/m$^3$) | 1022.8 | 1075.9 | 1190.8 | 965.9 | 1063.5 |
| Cp (J/kg K) | 2538.3 | 2340.9 | 2127.8 | 2502.6 | 2337.3 |
| $\rho$.Cp (MJ/m$^3$ K) | 2.59 | 2.52 | 2.53 | 2.42 | 2.49 |
| Tm (K) | 410.4 | 429.6 | 357.4 | 445.6 | 443.7 |
| Tapp (K) | 691.6 | 697.6 | 722.3 | 562.2 | 598.3 |

Next, in order to study the effect of methylene groups present in the cation side chain on the TES properties of ILs we fixed the optimal anion as $[BF_4]^-$. For each category of cation, a variant of the optimal structure (in terms of functional groups) was identified and the number of methylene group was varied from 0 to 4. The results are shown in Table 4-4. It is clear from Table 4-4 that addition of methylene group decreases the thermal storage capacity, melting point and decomposition temperature consistently across all categories. This shows the tradeoff that needs to be made, as higher thermal storage capacity and decomposition temperature are preferred while lower melting point is required. Therefore, in general, absence of $CH_2$ group will favor higher $\eta$ but we would sometimes require the presence of $CH_2$ groups (e.g. 11 $CH_2$ groups in the case of pyrrolidinium) to meet the tight melting point constraint. These findings are also consistent with the designed IL, which did not have any methylene groups.

**Table 4-4:** Effect of number of $CH_2$ groups on $\eta$, $T_m$ and $T_{app}$ [ILs with $BF_4^-$ anion]

| **Number of $CH_2$** | | | | | | | | | | | |
|---|---|---|---|---|---|---|---|---|---|---|---|
| **0** | | | | **2** | | | | **4** | | | |
| **Cation** | $\eta$ (MJ/m$^3$ K) | $T_m$ (K) | $T_{app}$ (K) | **Cation** | $\eta$ | $T_m$ | $T_{app}$ | **Cation** | $\eta$ | $T_m$ | $T_{app}$ |
| Imidazolium 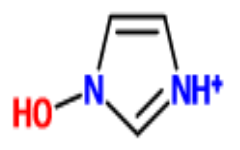 | 2.88 | 402.9 | 851.3 | 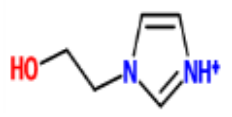 | 2.77 | 395.4 | 843.2 | 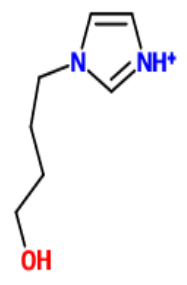 | 2.71 | 387.9 | 835.1 |

| Number of $CH_2$ | | | | | | | | | | | |
|---|---|---|---|---|---|---|---|---|---|---|---|
| **0** | | | | **2** | | | | **4** | | | |
| **Cation** | η ($MJ/m^3$ K) | $T_m$ (K) | $T_{app}$ (K) | **Cation** | η | $T_m$ | $T_{app}$ | **Cation** | η | $T_m$ | $T_{app}$ |
| Pyridinium 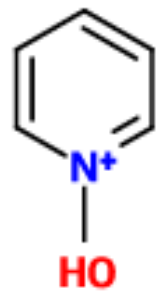 | 2.80 | 428.5 | 785.8 | 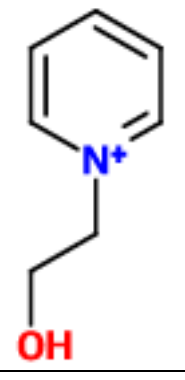 | 2.73 | 420.9 | 777.7 | 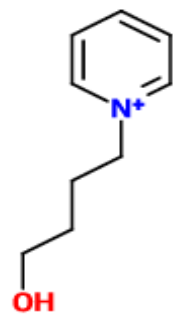 | 2.68 | 413.4 | 769.6 |
| Ammonium 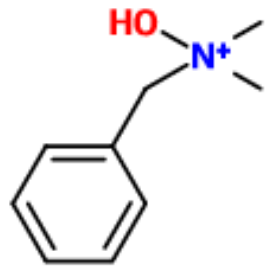 | 2.70 | 386.4 | 828.6 | 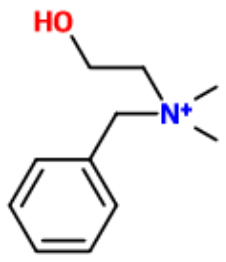 | 2.66 | 378.9 | 820.5 | 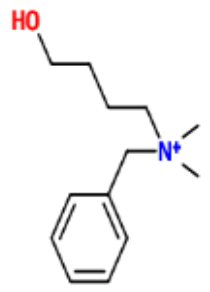 | 2.64 | 371.4 | 812.4 |
| Phosphonium 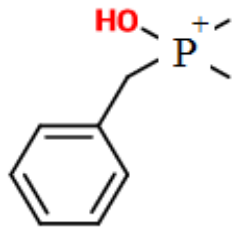 | 2.53 | 378.0 | 850.8 | 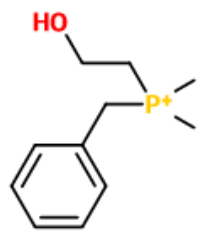 | 2.52 | 370.4 | 842.7 | 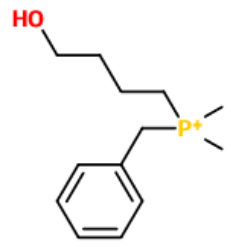 | 2.50 | 362.9 | 834.6 |
| Pyrrolidinium 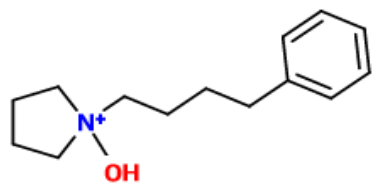 | 2.67 | 440.4 | 724.0 | 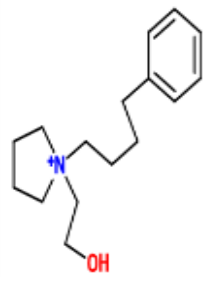 | 2.65 | 432.9 | 715.9 | 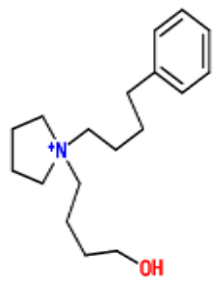 | 2.63 | 425.4 | 707.8 |

Next, to study the influence of the functional group (FG=hydroxyl or benzyl or ether or methyl) present in the cation side chain on the TES properties of ILs we fixed the optimal anion as $[BF_4]^-$. For each category of cation, a variant of the optimal structure was identified and the functional group (position marked as FG in Table 4-5) was varied. From Table 4-5 we see that cations functionalized with benzyl group always had the highest thermal storage capacity followed by hydroxyl-functionalized cations. However, we also see that benzyl-functionalized cations have significantly lower thermal decomposition temperature in comparison to hydroxyl-functionalized cations thereby mostly violating the decomposition temperature requirements (constraints). Further, benzyl-functionalized cations usually had a slightly higher melting point than hydroxyl- functionalized cations. These findings are consistent with our optimal IL, which consisted of a hydroxyl-functionalized imidazolium cation.

**Table 4-5:** The effect of variation of functional groups (FG) connected to the cation head group on η, $T_m$ and $T_{app}$

| | **Functional group (FG)** | | | | | | | | | | | |
|---|---|---|---|---|---|---|---|---|---|---|---|---|
| | **$CH_3$** | | | **Benzyl** | | | **$OCH_3$** | | | **OH** | | |
| **Cation** | η | $T_m$ | $T_{app}$ | η | $T_m$ | $T_{app}$ | η | $T_m$ | $T_{app}$ | η | $T_m$ | $T_{app}$ |
| Imidazolium 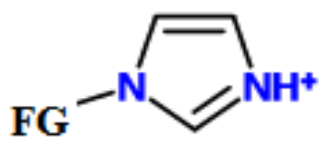 | 2.76 | 413.5 | 661.6 | 2.90 | 411.8 | 613.1 | 2.79 | 403.0 | 534.3 | 2.88 | 402.9 | 851.3 |
| Pyridinium 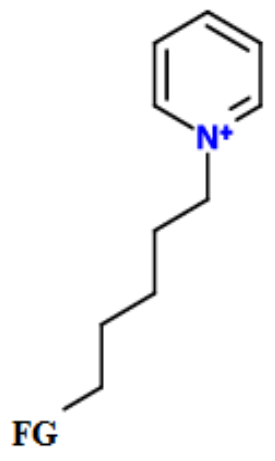 | 2.61 | 420.3 | 575.9 | 2.74 | 418.6 | 527.4 | 2.63 | 409.8 | 448.6 | 2.67 | 409.7 | 765.6 |

| | Functional group (FG) | | | | | | | | | | | |
|---|---|---|---|---|---|---|---|---|---|---|---|---|
| | **$CH_3$** | | | **Benzyl** | | | **$OCH_3$** | | | **OH** | | |
| **Cation** | η | $T_m$ | $T_{app}$ | η | $T_m$ | $T_{app}$ | η | $T_m$ | $T_{app}$ | η | $T_m$ | $T_{app}$ |
| Ammonium 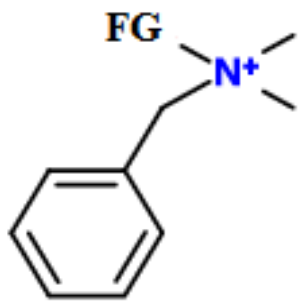  | 2.63 | 397.1 | 638.9 | 2.77 | 395.4 | 590.5 | 2.65 | 386.6 | 511.7 | 2.70 | 386.4 | 828.6 |
| Phosphonium 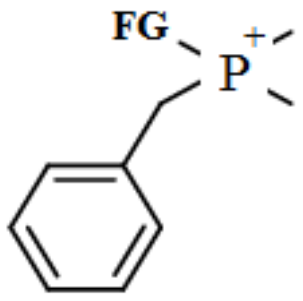  | 2.48 | 388.6 | 661.1 | 2.64 | 386.9 | 612.6 | 2.50 | 378.1 | 533.8 | 2.53 | 378.0 | 850.8 |
| Pyrrolidinium 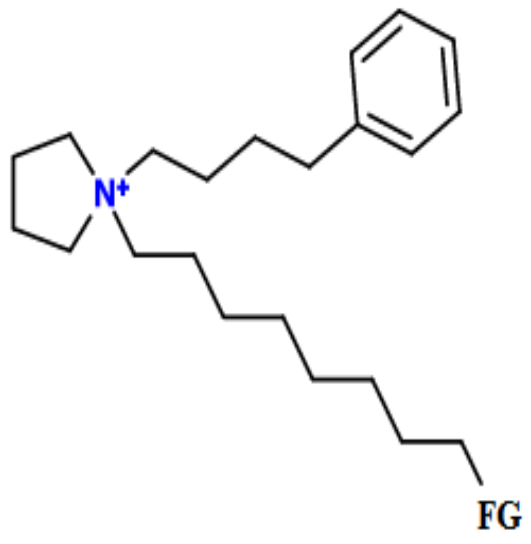  | 2.57 | 421.0 | 501.9 | 2.62 | 419.3 | 453.50 | 2.58 | 410.5 | 374.7 | 2.60 | 410.4 | 691.6 |

Since the hydroxyl-functionalized optimal IL is, a novel and non-traditional IL there were no experimental data available for validation purposes. Therefore, we relied on QSPR predictions of the physical properties (ρ, $C_p$) based on quantum chemical calculation inputs implemented within COSMOtherm software to cross check our results. The optimal ionic liquid (3-hydroxy-imidazolium $[BF_4]^-$) was simulated in Turbomole, which was linked to COSMOtherm software. In the first step to evaluate the accuracy of QSPR correlation in COSMOtherm we compared experimental heat capacity ($C_p$) and density ($\rho$) data of nine common imidazolium based ionic liquids with corresponding prediction from COSMOtherm QSPR models. The comparison results are tabulated in the Table 4-6. As it can be seen from Table 4-6 the QSPR/COSMOtherm predictions were very close to the experimental data with an average relative error of 1.49%. Next, Table 4-7 shows the comparison of the physical properties of the optimal ionic liquid (3-hydroxy-imidazolium $[BF_4]^-$) predicted by QSPR/COSMOtherm approach and the group contribution models used in this study. As it can be seen from Table 4-7, the predictions are close which further validates our design results.

**Table 4-6:** Comparison of COSMO-predicted $C_p$ and $\rho$ of selected ILs with the corresponding exp. data

| | $BF_4^-$ | | | | $PF_6^-$ | | | | $Tf_2N^-$ | | | |
|---|---|---|---|---|---|---|---|---|---|---|---|---|
| | $C_p$ | | $\rho$ | | $C_p$ | | $\rho$ | | $C_p$ | | $\rho$ | |
| | Exp. | COS. | Exp. | COS. | Exp. | COS. | Exp. | COS. | Exp. | COS. | Exp. | COS. |
| [emim] | 308.1 | 313.29 | 1285.5 | 1300.65 | --- | 350.43 | 1480.5 | 1496.59 | 505 | 490.17 | 1517 | 1545.94 |
| [bmim] | 364.8 | 368.91 | 1201.6 | 1205.37 | 397.6 | 406.05 | 1367.7 | 1376.93 | 564 | 545.79 | 1436.1 | 1451.68 |
| [hmim] | 426.1 | 424.21 | 1145.5 | 1139.33 | 419.25 | 431.35 | 1292.7 | 1291.14 | 623 | 601.08 | 1372.0 | 1377.78 |
| **Avg. Relative error (%)** | | | | | 1.49 % | | | | | | | |
| **Max Relative error (%)** | | | | | 3.52% | | | | | | | |

**Table 4-7:** Comparison between COSMO-predicted and GC-predicted values for $C_p$ and $\rho$, and $\eta$ of the optimal IL

| | Method | | |
|---|---|---|---|
| **Properties** | **COSMO-RS** | **Group Contribution** | **Abs Relative Difference (%)** |
| Cp ($J.mol^{-1}.K^{-1}$) | 240.98 | 231.42 | 3.97 |
| $\rho$ ($kg/m^3$) | 1600.34 | 1624.1 | 1.48 |
| $\eta$ ($MJ/m^3$ K) | 2.23 | 2.174 | 2.54 |

# Chapter 5: Application 2: Design of Ionic Liquids for Aromatic-Aliphatic Separation

In this chapter, a computer-aided design framework with the capability of predicting ionic liquid selective dissolution power towards aromatic compounds is presented. COSMO-RS method was first evaluated for its accuracy for the prediction of the solubility of liquid solutes in ionic liquids and then these prediction were utilized within an optimization framework to fit the binary interaction and group parameters for UNIFAC model.. Next, the fitted parameters along with the UNIFAC model was integrated within the CAILD framework to design optimal ionic liquids for this separation problem.

## 5.1 Introduction

In order for a chemical compound to be considered as a green substitute of a common solvent, it must be able to address industries' concerns as well as any adverse impact these solvents might have on the environment.[158] A liquid-liquid extraction process is normally used when other methods of separation (e.g. distillation) are not feasible. Aromatic/Aliphatic separation is one of the well-known industrial processes in which a chemical (solvent) is utilized to withdraw/remove aromatic compounds from a multi-component mixture of aromatic and non-aromatic compounds. Current industrial plants, which use organic solvents such as furfural or sulfolane as the solvent in the extraction process, show several adverse environmental impacts.[159-162] Thus, substitution of commonly used solvents with their ionic liquid counterparts can help address some of these deficiencies. Several studies have shown that when ILs are added to a binary mixture of one aromatic and one non-aromatic compound, they mostly follow a type II ternary system, meaning that they are technically able to process feed streams containing aromatic compounds at different compositions.[163] On the other hand, ionic liquids are known to have the potential to act as solvents in separation processes where a low concentration of a solute is available, and thereby, can be economically beneficial when compared to their conventional organic counterparts.[164] Ionic liquids have also shown a lower tendency towards dissolution in hydrocarbon compounds, leading to lower solvent losses during the regeneration step.[165] In order for an ionic liquid (IL) to be considered as a solvent for an aromatic/aliphatic separation process, it needs to show high values for selective extraction capacities towards aromatic compounds.[166]

As discussed in our previous work,[166] a large inventory of anions, cations, and functional groups exist or can be synthetized.[167,168] ILs can be formed through different combinations of cations, anions, and alky side chain groups (attached to the cation-base) leading to a vast number of feasible ionic liquid structures (estimated to be as many as $10^{14}$ combinations).[169,170] This presents a very good opportunity to tailor ILs for specific applications. Task-specific ILs can potentially be designed for different applications. The design process is usually accomplished through controlling the properties of ILs by informed selection or modification of the cation, the anion, and/or the alkyl chains connected to the cation-base. While this can be very exciting to potentially screen a large number of feasible ionic liquids, it can also present an unusual challenge, since synthesizing, screening, and testing the limitless number of ionic liquids can become a daunting task.[171]

As mentioned before, the ILs can be tuned to possess the properties of interest. Similarly to what has been done before for molecular compounds, we can tailor ILs to increase their desired properties (such as solvency power or thermal storage capacity), while decreasing their undesired features (e.g. high viscosity or high toxicity) at the same time.

Although ILs have many interesting properties, they have not been commercialized yet. One of the reason for this could be that ILs possess, high viscosities and relatively high melting points. These drawbacks must be addressed in order to make ILs more appealing to industrial stakeholders.

Computer-aided molecular design (CAMD) has been a promising approach for years now and has been widely used to design organic (molecular) solvents for different applications.[172-178] A CAMD framework integrates models used for property prediction with optimization algorithms to design molecular structures with desired properties. In our previous work[166], we successfully showed that a similar approach can be applied to the case of designing new ILs through the numerous possible combinations of cations, anions, and side chain groups.

## 5.2 Computer-aided ionic liquid design (CAILD)

For this study, the CAILD framework that we developed[166] is used again to identify ionic liquids (ILs) that possess desirable properties for aromatic/aliphatic separation. In the CAILD approach a set of IL structural groups (i.e., cations bases such as: Imidazolium ($Im^+$), Pyridinium ($Py^+$), ammonium ($NH_4^+$), anions such as: $Tf_2N^-$, $BF_4^-$ and side chain groups such as: CH2, $CH_3$, OH etc.) are considered as building blocks and are combined together in different ways to design ILs which have specific desired properties of interest (e.g., melting point, viscosity, solubility ….).

To design an optimal candidate (ILs) as an extraction solvent for toluene/n-heptane separation process, a comprehensive approach consisting of five stages was used. The presented CAILD model in this study is a nonconvex, Mixed Integer Non-linear Programming (MINLP) model, involving a large number of integer and binary variables. Consideration of mixture properties through the UNIFAC model results in nonlinearity and most of the binary design variables (structural) participate in the nonlinear terms. Combinatorial complexity is an inherent issue in CAMD–MINLP models due to the nature of the search space. In this study, we focus on solving the CAILD framework utilizing GA-based optimization.

***Genetic algorithm (GA)***

GA is a stochastic method used to solve optimization problems based on the natural selection process mimicking the evolution phenomenon which occur in biological systems. GA can be used to solve problems that are not well suited for standard optimization algorithms.[179] In the presented case study, the fitness function of the GA is identical to the objective function.

### 5.2.1 Forward problem

In order to evaluate the intermolecular interactions between different components present in a mixture (both ionic and molecular), the thermodynamic properties of non-ideal solutions needs to be calculated. An essential requirement to predict the equilibrium conditions of a system involving ionic liquids is to calculate the activity coefficients of compounds in that system. These activity coefficients can be used to predict the equilibrium conditions of the system and are not only a function of binary (group-group) interactions, but also relate to the compositions of the chemical components in the mixture. Therefore, the proposed CAILD framework requires a model for predicting activity coefficient that is based on the concept of solution of groups. The basic assumption related to the prediction of activity coefficients of compounds in a mixture using solution of group approach is that the interactions between different molecules can be approximated as the summation of interactions between the groups present in those molecules. The total number of cation head groups, anions, and alkyl side chain groups are much less than the number of distinct ILs which can be generated from their combinations. Therefore, a relatively small number of group interaction parameters are required (are enough) to represent all feasible ILs. UNIFAC[180] model is a widely used group contribution (GC) method to predict the phase equilibrium conditions in nonelectrolyte systems. The UNIFAC method makes use of the concept of functional groups (their contribution and the number of their occurrences in each compound) to predict the activity coefficients. The activity coefficient calculated using the UNIFAC model is the summation of two terms: a combinatorial term that considers the differences in size and shape of groups, and a residual term that accounts for the energetic interactions of different groups.

$$\ln \gamma_i = ln\, \gamma_i^C + ln\gamma_i^R$$

The volume ($R$) and surface area ($Q$) parameters related to the combinatorial part of the activity coefficients of each compound are calculated as the summation of group parameters ($R_k$ for volume and $Q_k$ for surface area), while binary interaction parameters ($a_{mn}$ and $a_{nm}$) between different functional groups present in the components of the system are required to calculate the residual part. The UNIFAC model was originally developed for nonelectrolyte systems, however, recently several groups have utilized this method for systems containing ILs by careful representation of ionic groups in the system. In order to apply the UNIFAC method to systems with ILs, Wang *et al.*[181] and Lei *et al.*[182] considered ionic liquids as single non-dissociated neutral entities. ILs as separate functional groups are included in the modified UNIFAC.[183] Recently, Roughton *et al.*[184] characterized ionic groups in the same way as proposed in our developed CAILD framework; i.e. as separate cation-bases, anions, and side chain groups. The assumption here is that the cation and anion can be treated separately and the interaction between them is assumed to be equal to zero.[184]

Due to paucity of experimental data on activity coefficients of chemical compounds dissolved in wide variety of ILs, fitting the relevant UNIFAC parameters for systems involving ILs is not possible. For the time being, the main limitation on the use of computer-aided models to design optimal ionic liquids for multi-component separation processes can be attributed to the lack of enough experiential data on activity coefficients. Even though millions of different ILs can be theoretically feasible, due to the gaps in the experimental data and non-availability of property prediction models, a vast majority of ionic liquids cannot be considered within a computer-aided design framework. In order to overcome this limitation, we use a quantum chemistry (QC) approach to simulate the chemical compounds and ionic liquids (based on energy minimization), to predict their activity coefficients in the mixture. COMSO-RS (COnductor like Screening MOdel for Real Solvents), is a quantum chemistry-based equilibrium thermodynamics model that can predict the chemical potentials of compounds in the liquid phase. COSMO-RS is able to calculate the distribution of charge density (σ) on the surface of molecules to predict the chemical potential (μ) of each compound in the solution. The chemical potentials calculated through COSMO-RS approach are the basis for the calculation of other thermodynamic equilibrium properties such as activity coefficients, solubility, and free energy of solvation. The advantage of this method is that it was developed as a general prediction method that does not require any system specific adjustments. Since COSMO-RS uses charge density (σ), it does not require functional group parameters, meaning that we do not require experimental data to predict activity coefficients. COSMO-RS has shown to be a promising tool for a solvent screening task in which the most powerful solvent for a specific liquid-liquid extraction process can be selected. During the past few years, many researchers have used COMSO-RS to calculate the solubility and/or activity coefficients of different chemical compounds in mixtures containing ILs. In order to study the ability of COSMO-RS method to accurately predict activity coefficients of chemical compounds in ILs, a comparison was made between experimental values of $\gamma^{\infty}$ and COSMO-RS predictions. These results, along with the absolute error (%) and the temperature at which the activity coefficients were measured/estimated, are tabulated in Table 5-1. Results in Table 5-1 show that in most cases, COSMO-RS has been able to predict the activity coefficients (i.e. solubility) of chemical compounds in their mixtures with ILs to a reasonable degree of accuracy.

**Table 5-1:** Experimental infinite dilution activity coefficients ($\gamma^{\infty}$) vs. Cosmo-predicted values

| IL | T (k) | $\gamma^{\infty,exp}_{benzene}$ | $\gamma^{\infty,cosmo}_{benzene}$ | Err. (%) | T(k) | $\gamma^{\infty,exp}_{hexane}$ | $\gamma^{\infty,cosmo}_{hexane}$ | Err. (%) | T(k) |
|---|---|---|---|---|---|---|---|---|---|
| **P66614 Cl** | 308.15<br>318.15<br>328.15 | 0.408<br>0.403<br>0.401 | 0.432<br>0.430<br>0.429 | 5.84<br>6.78<br>7.23 | 308.15<br>318.15<br>328.15 | 0.766<br>0.753<br>0.746 | 0.695<br>0.688<br>0.681 | 9.23<br>8.65<br>8.75 | 308.15<br>318.15<br>328.15 |
| **Bmim $Tf_2N$** | 303.15<br>313.15<br>323.15 | 0.881<br>0.892<br>0.903 | 1.180<br>1.138<br>1.099 | 33.98<br>27.63<br>21.75 | 303.15<br>313.15<br>323.15 | 14.2<br>13.5<br>12.7 | 14.294<br>12.445<br>10.911 | 0.66<br>7.81<br>14.09 | 303.15<br>313.15<br>323.15 |
| **BmPyr $Tf_2N$** | 303.15<br>313.15<br>323.15<br>333.15 | 0.84<br>0.86<br>0.88<br>0.89 | 1.121<br>1.081<br>1.045<br>1.011 | 33.46<br>25.73<br>18.70<br>13.55 | 303.15<br>313.15<br>323.15<br>333.15 | 13.8<br>13.3<br>12.1<br>11 | 10.027<br>8.821<br>7.814<br>6.968 | 27.34<br>33.68<br>35.42<br>36.65 | 303.15<br>313.15<br>323.15<br>333.15 |
| **Hmim $Tf_2N$** | 301.65<br>312.25<br>333.25<br>343.75 | 0.779<br>0.776<br>0.777<br>0.769 | 0.977<br>0.945<br>0.888<br>0.863 | 25.37<br>21.74<br>14.31<br>12.22 | 298.15<br>313.15<br>333.15 | 8.330<br>7.710<br>6.580 | 8.766<br>7.340<br>5.911 | 5.23<br>4.79<br>10.16 | 298.15<br>313.15<br>333.15 |
| **BmPy TfO** | 298.15<br>318.15<br>338.15<br>358.15 | 1.41<br>1.46<br>1.5<br>1.54 | 1.697<br>1.593<br>1.497<br>1.410 | 20.37<br>9.11<br>0.20<br>8.43 | | | | | |
| | **Average error (%)** | **16.41** | | | | | | | |

Based on our results, we suggest that we can use the activity coefficients predicted by COSMO-RS method as surrogates for missing experimental data. These predicted activity coefficients were used within an optimization framework (minimizing the relative error) to fit both group parameters (R and Q) and the binary interaction parameters ($a_{ij}$ and $a_{ji}$) of the UNIFAC model. The advantage of this approach is that these fitted parameters cover a wider range of binary coefficients related to multitude of cations, anion, and functional groups compared to what would have been possible using experimental data. A list of cations, anions, and side chain groups considered in this study are listed in Appendix B. The UNIFAC parameters (group and binary interaction parameters) fitted through the discussed procedure (listed in Appendix C) was used in a computer-aided design (CAILD) framework to explore a wide variety of ionic liquids (~29,000) with an aim to design the most optimal solvent for the liquid-liquid extraction process. These UNIFAC parameters for imidazolium and pyridinium cation head groups (which appeared in all the 8 optimal ionic liquids), along with all considered anions and functional groups are listed in Appendices C-1 and C-2 respectively.

The accuracy of the fitted UNIFAC model was tested using available experimental data on the solubility of toluene and n-heptane in different ionic liquids and the results along with the relative errors are tabulated in Table 5-2.

**Table 5-2:** Experimental solubility data vs. CAILD predicted data

| Ionic Liquid | Solute | T(K) | $x_{Ext,exp}^{solute}$ | $x_{Ext,calc}^{solute}$ | Error (%) | Ref |
|---|---|---|---|---|---|---|
| 1-hexylpyridinium $BF_4^-$ | Toluene | 298.15 | 0.421 ± 0.03 | 0.49458 | 17.48 | 185 |
| 1-hexylpyridinium $BF_4^-$ | Toluene | 318.15 | 0.441 ± 0.03 | 0.518175 | 17.500 | 185 |
| 1-hexylpyridinium $BF_4^-$ | Toluene | 338.15 | 0.453 ± 0.03 | 0.542791 | 19.82 | 185 |
| 1-hexyl-3-methylimidazolium $Tf_2N^-$ | Toluene | 298.15 | 0.487 ± 0 | 0.35888241 | 26.31 | 186 |
| 1-ethyl-3-methylpyridinium $Tf_2N^-$ | Toluene | 293.15 | 0.295 ± 0.012 | 0.3237049 | 9.73 | 187 |
| 1-ethyl-3-methylpyridinium $Tf_2N^-$ | Toluene | 303.15 | 0.298 ± 0.012 | 0.3269968 | 9.76 | 187 |
| 1,3-dimethylimidazolium $Tf_2N^-$ | Toluene | 313.2 | 0.5891 | 0.521006 | 11.56 | 188 |
| 1-pentyl-3-methylimidazolium $Tf_2N^-$ | Toluene | 298.15 | 0.577 ± 0.013 | 0.2745038 | 52.42 | 189 |
| 1-octyl-3-methylimidazolium $PF_6^-$ | Heptane | 313.2 | 0.0095 ± 0.0014 | 0.0080512 | 15.25 | 190 |
| 1,3-dimethylimidazolium $Tf_2N^-$ | Heptane | 313.2 | 0.0178 ± 0.002 | 0.0123246 | 30.76 | 188 |
| 1-hexyl-3-methylimidazolium $Tf_2N^-$ | Heptane | 313.2 | 0.1375 ± 0.0061 | 0.154986 | 12.72 | 188 |
| 1-ethyl-3-methylpyridinium $Tf_2N^-$ | Heptane | 298.15 | 0.042 ± 0.014 | 0.0336166 | 19.96 | 191 |

### 5.2.2 Reverse problem

Naser and Fournier[192] used a strategy for the computer-aided design of molecular compounds in which the features related to the process model were also included (as the optimization constraints). This framework was used to find an optimal solvent (i.e. an ionic liquid) for the extraction process. One of the advantage of this approach is that it enables us to predict the compositions of the three components (toluene, n-heptane and IL) in the raffinate and extract phases as well. In order to illustrate the use of the CAILD method using the UNIFAC parameters developed in the Forward problem, we solve a case study for an aromatic/aliphatic separation process.

**Case Study: Toluene/n-heptane separation.** Recovering toluene, an aromatic compound, from a toluene/n-heptane mixture is desired. Due to the fact that boiling points of toluene and n-heptane are in the close vicinity of each other, the distillation process is not a feasible method for this separation and instead a solvent extraction process (liquid-liquid extraction) is commonly used. The feed was assumed to be a mixture of toluene (A) and heptane (B) with mole fraction of toluene, $x_A^F$, and heptane, $x_B^F$, being 0.27 (24 wt% ) and 0.73 (76 wt%), respectively with the molar flow rate of F=1000 mole/hr. This data is based on a real case scenario adopted from an industrial plant producing lubricating base-oil through the removal of petroleum-based aromatics from the lube-oil cut. The solvent was assumed to be a pure ionic liquid that is to be designed and pure furfural for comparison (common solvent for aromatic/aliphatic separation) with the molar flow rate of 2450 mole/hr.

A schematic of the single stage extraction compartment is shown in Figure 5-1.

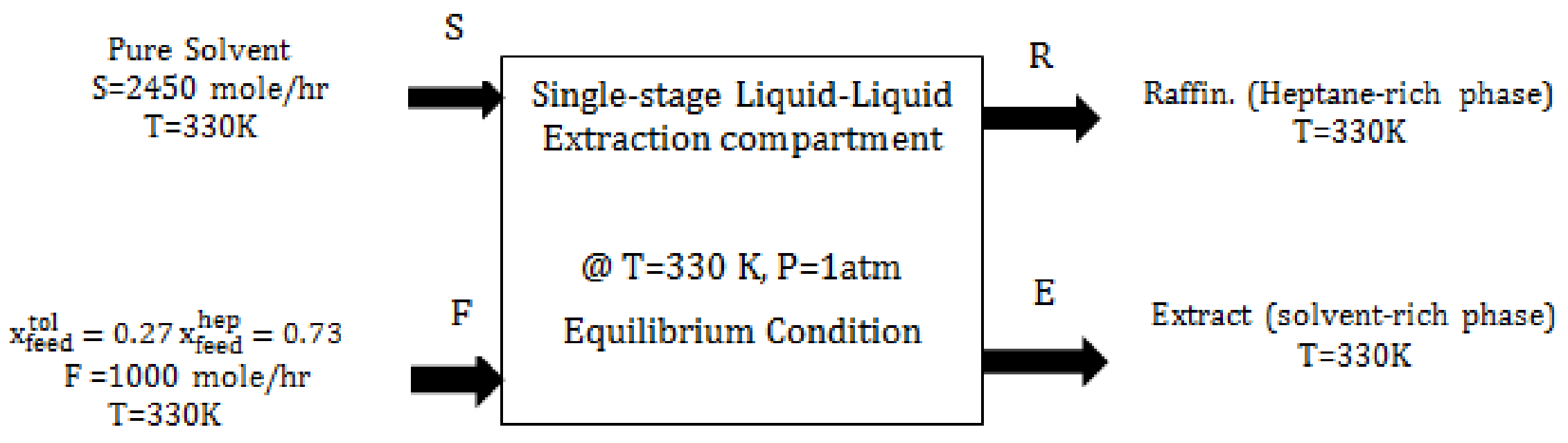

**Figure 5-1:** A schematic of the single stage extraction process

***Detailed Formulation of the Problem***

In this optimization framework, we seek to minimize the extraction of n-heptane from the feed stream while extracting as much toluene as possible since our goal is to maximize the concentration of n-heptane in the raffinate phase. In order to find a proper objective function, a performance index (PI) was used through the division of the yield of extraction of toluene by yield of extraction of heptane squared (since the lower amount of the yield of extraction of n-heptane is more important). These two yield of extractions were defined as the fractions of toluene and n-heptane in the feed stream which was separated from it during the solvent extraction process.

$$\text{Objective function} = \text{Max}\ (PI = \frac{Yield\ of\ extraction\ of\ toluene}{(Yield\ of\ extraction\ of\ heptane)^2} \qquad (5\text{-}1)$$

***Feasibility constraints***

$$\sum_{i \in c} c_i = 1 \qquad (5\text{-}2)$$

$$\sum_{j\in A} a_i = 1 \quad (5\text{-}3)$$

$$\sum_{l=1}^{6} y_l = \sum_{i\in C} c_i v_{ci} \quad (5\text{-}4)$$

$$\sum_{i\in c}(2-v_{ci})c_i + \sum_{l=1}^{6}\sum_{k\in G}(2-v_{Gkl})y_l ng_{kl} = 2 \quad (5\text{-}5)$$

$$\sum_{k\in G} y_l ng_{kl}\,(2-v_{Gkl}) = 1 \quad (5\text{-}6)$$

$$\sum_{k\in CH_2} y_l ng_{kl} \leq 15 \quad (5\text{-}7)$$

***Solution property constraints***

$$\mathrm{SL}(solvent\ loss) = \frac{1}{\gamma_{S,B}^{\infty}} \leq 0.0582 \quad (5\text{-}8)$$

$$\beta(selectivity) = \frac{\gamma_{B,S}^{\infty}}{\gamma_{A,S}^{\infty}} \geq 10.5918 \quad (5\text{-}9)$$

$$\mathrm{SP}(solvent\ power) = \frac{1}{\gamma_{A,S}^{\infty}} \geq 0.4539 \quad (5\text{-}10)$$

$$\gamma_A^E \times x_A^E - \gamma_A^R \times x_A^R = 0 \quad (5\text{-}11)$$

$$\gamma_B^E \times x_B^E - \gamma_B^R \times x_B^R = 0 \quad (5\text{-}12)$$

$$\gamma_S^E \times x_S^E - \gamma_S^R \times x_S^R = 0 \quad (5\text{-}13)$$

$$x_A^E \times E + x_A^R \times R = F \times x_A^F + S \times x_A^S \quad (5\text{-}14)$$

$$x_B^E \times E + x_B^R \times R = F \times x_B^F + S \times x_B^S \quad (5\text{-}15)$$

$$x_S^E E + x_S^R \times R = F \times x_S^F + S \times x_S^S \quad (5\text{-}16)$$

$$x_A^E + x_B^E + x_S^E = 1 \quad (5\text{-}17)$$

$$x_A^R + x_B^R + x_S^R = 1 \quad (5\text{-}18)$$

As discussed before, the optimal ILs designed through our CAILD framework must satisfy certain rules to ensure the chemical feasibility of the proposed compounds. These rules also termed as structural constraints, include feasibility rules (e.g. the octet rule), the bonding rule, as well as other complexity rules. Eqns. (5-2) to (5–6) represent a set of constraints developed to ensure that the designed ILs are chemically feasible which are adopted from our previous work. [166] Eqn. (5-2) ensures that only one cation from the vector $c_i$, representing the entire inventory of cations, is chosen. In the same way, eqn. (5-3) ensures that only one anion from the vector $a_i$, is chosen. $y_l$ is a vector of binary variables (0,1) dealing with the existence (1) or non-existence (0) of vacant valences on the cation-base. For example $y_{l=1} = 1$ means that there is an open/vacant position on the atom number 1 (location 1) of the cation-base to which a side chain group can be attached. $ng_{kl}$ is a vector of integer variables representing the number of groups of type *k* in the alkyl side chain *l*.

$\upsilon_{ci}$ is the Valence of the selected cation (i.e. the number of open positions a cation base has for side chains) and $\upsilon_{Gkl}$ is the Valence of the functional or alkyl side chain groups (e.g. $CH_2$ has a Valence of 2, so it can be connected to two other groups; $CH_3$ has a Valence of 1 and can be a terminal group attached to the end of a chain or to the cation-base directly).

Eq. 4 fixes the number of alkyl side chains attached to the vacant valences of the cation base. The implementation of the modified octet rule (Eq. 5) ensures that any designed cation is structurally feasible, meaning that each Valence in all structural groups of the cation is filled with a covalent bond. Eq. 6 ensures that the octet rule is implemented for each side chain (*l*) in the sense that the valences of each individual chains are satisfied with a covalent bond. Eq. 7 restricts the size of the side chain groups connected to the cation base and in this case has an upper bound of 15 for the number of $CH_2$ groups present in each side chain.

Eqns. (5-8) to (5-10) deal with the properties of interest of the designed solvent to show how well the optimal candidate can satisfy our needs. As aforementioned we need to design a solvent for toluene/n-heptane separation which has the solvent properties better than those of furfural. In order to design the ionic liquid, three solvent properties were considered: solvent loss, selectivity and solvent power. These solvent properties were calculated for a system consisting of toluene/n-heptane and furfural (as the extraction solvent) using the infinite dilution activity coefficients of different binary mixtures of the compounds as shown in eqns. (5-8) to (5-10). Since the designed IL (optimal candidate) needs to be a better solvent than furfural it must have a lower solvent loss and higher solvent power (towards toluene) and selectivity (higher tendency towards dissolution of toluene rather than n-heptane).

In this problem, the final/equilibrium compositions (mole fractions) of the three components in the raffinate ($x_A^R$ , $x_B^R$ , $x_S^R$ ) and extract ($x_A^E$ , $x_B^E$ ,$x_S^E$ ) phases as well as the molar flow rate of the raffinate (R) and extract phases (E) were unknown. In the problem, the feed and solvent flow rates and the compositions of the three compounds in these two input streams are given in moles/hr and mole fraction, respectively. Eqns. (5-11) to (5-13) ensure equilibrium conditions and hence the concentration of components in the two phases (raffinate and extract) do not change. Eqns. (5-14) to (5-16) deals with mass balance of each component (toluene, n-heptane and IL) around the system. Eqns. (5-17) and (5-18) are necessary to ensure that the summation of the mole fractions of the three components equal to unity in both t phases (raffinate and extract), showing that no normalizations are required. It is worth mentioning that due to systematic errors and uncertainties that arose from the models used to predict the solubility and activity coefficients of components it was impossible to satisfy the equivalences of eqns. (5-11) to (5-18) with 100% accuracy. A ($\pm$5%) tolerance was applied to these constraints to enable the optimization program to converge.

## 5.3 Results

The CAILD program was run several times to ensure that it converged to a global optima rather than a local optima. A high number for the population size input parameter in the GA ensured convergence of the program to the same result for every run. We used the optimization model to identify 8 ionic liquids (ILs) with optimal solvent properties. The design results related to the solubility of toluene and n-heptane in optimal ionic liquids along with the molar flow rates of raffinate and extract phases for each case are tabulated in Table 5-3. The results related to the same extraction process using furfural as a solvent was also calculated and listed on the last row of Table 5-3 for comparison purposes. All of the calculations were made at T=330 K which is a typical temperature for an aromatic/aliphatic separation (liquid- liquid extraction) process.

**Table 5-3:** CAILD results for the optimal ILs and furfural at T=330 K

| | Raffinate | | Extract | | | |
|---|---|---|---|---|---|---|
| **Compound** | **No. of moles** | **x (mole fraction)** | **No. of moles** | **x (mole fraction)** | **E (mole/hr)** | **R (mole/hr)** |
| Toluene<br>n-heptane<br>**1-(ethoxyethyl)pyridinium $AlCl_4$** | 117.69357<br>701.24674<br>4.59405 | 0.14291<br>0.85151<br>0.00558 | 150.80643<br>31.25326<br>2453.4059 | 0.05722<br>0.01186<br>0.94092 | 2635.465 | 823.534 |
| Toluene<br>n-heptane<br>**1-(benzyl)-3-methylpyridinium $AlCl_4$** | 93.01245<br>686.16365<br>5.34647 | 0.11856<br>0.87463<br>0.00681 | 178.08755<br>41.63635<br>2449.6535 | 0.06672<br>0.01560<br>0.92549 | 2669.377 | 784.522 |
| Toluene<br>n-heptane<br>**1-propyl-3-methylpyridinium $AlCl_4$** | 107.35496<br>690.78479<br>12.95947 | 0.13236<br>0.85166<br>0.01598 | 159.14504<br>42.61521<br>2439.0405 | 0.06026<br>0.01655<br>0.92360 | 2640.801 | 811.099 |
| Toluene<br>n-heptane<br>**1-ethyl-3-methylpyridinium $AlCl_4$** | 106.12357<br>686.83145<br>16.01235 | 0.13118<br>0.84902<br>0.01979 | 163.37643<br>42.56855<br>2440.9876 | 0.06278<br>0.01608<br>0.92219 | 2646.932 | 808.967 |
| Toluene<br>n-heptane<br>**1-(3-ethoxypropyl)pyridinium $AlCl_4$** | 108.87295<br>689.43320<br>11.07183 | 0.13451<br>0.85181<br>0.01368 | 160.02705<br>43.06680<br>2436.4282 | 0.06063<br>0.01632<br>0.92416 | 2639.522 | 809.377 |
| Toluene<br>n-heptane<br>**1-benzyl-3-methyl-imidazolium $AlCl_4$** | 92.64559<br>678.73373<br>8.54488 | 0.11879<br>0.87026<br>0.01096 | 177.7544<br>51.7663<br>2442.955 | 0.06651<br>0.01954<br>0.91412 | 2672.476 | 779.924 |
| Toluene<br>n-heptane<br>**1-(methybenzyl)-3-propylimidazolium $PF_6$** | 99.420<br>674.886<br>6.828 | 0.12728<br>0.86398<br>0.00874 | 174.079<br>58.3139<br>2448.172 | 0.06494<br>0.02095<br>0.91330 | 2680.56 | 781.13 |
| Toluene<br>n-heptane<br>**1-methylbenzyl-3-methoxymethylbenzyl methylsulfate** | 123.469<br>683.141<br>0.0751 | 0.15306<br>0.84685<br>0.000092 | 145.331<br>49.259<br>2447.925 | 0.05645<br>0.01864<br>0.92636 | 2642.51 | 806.68 |
| Toluene<br>n-heptane<br>**Furfural** | 55.262<br>488.500<br>72.951 | 0.08961<br>0.79210<br>0.11829 | 214.738<br>241.499<br>2377.050 | 0.07579<br>0.08478<br>0.83897 | 2833.29 | 616.71 |

As discussed before, it is critical for the designed ILs to have relatively low viscosities and melting points to enable its use as a solvent for extraction in an industrial setting. The initial selection of anions considered (as part of the basis set) in this study was based on the fact that they contribute towards lower values of melting points making it unlikely that any IL designed will be in solid state. The viscosities and melting points of the 8 optimal ILs, listed in Table 5-3, were calculated post design using the correlations developed in chapter 2 and presented in the previous study.[193] These predictions along with solvency power calculated through the CAILD model are listed in Table 5-4.

**Table 5-4:** Physical properties and solvency power of optimal ILs and furfural

| **Compound** | **Vis_Solvent (cP)** | **Tm_Solvent (K)** | **Yield of Ext_hep (%)** | **Yield of Ext_tol (%)** | **PI** |
|---|---|---|---|---|---|
| toluene<br>n-heptane<br>**1-(ethoxyethyl)pyridinium $AlCl_4$** | 15.6798 | 262.9530 | 4.2813 | 55.8542 | 304.727 |
| toluene<br>n-heptane<br>**1-(benzyl)-3-methylpyridinium $AlCl_4$** | 20.6084 | 271.5854 | 5.7036 | 65.9584 | 202.754 |
| toluene<br>n-heptane<br>**1-propyl-3-methylpyridinium $AlCl_4$** | 11.3945 | 258.1737 | 5.8377 | 58.9426 | 172.960 |
| toluene<br>n-heptane<br>**1-ethyl-3-methylpyridinium $AlCl_4$** | 11.3151 | 262.9873 | 5.8313 | 60.5098 | 177.948 |
| toluene<br>n-heptane<br>**1-(3-ethoxypropyl)pyridinium $AlCl_4$** | 19.7611 | 275.2263 | 5.8996 | 59.2693 | 170.290 |
| toluene<br>n-heptane<br>**1-benzyl-3-methyl-imidazolium $AlCl_4$** | 21.3088 | 270.2397 | 7.0913 | 65.8350 | 130.920 |
| toluene<br>n-heptane<br>**1-benzyl-3-propylimidazolium $PF_6$** | 52.9785 | 312.9393 | 7.9882 | 64.4739 | 101.038 |
| toluene<br>n-heptane<br>**1-methylbenzyl-3-ethoxymethylbenzyl-imidazolium methylsulfate** | 105.0221 | 304.6925 | 6.7478 | 53.8262 | 118.215 |
| toluene<br>n-heptane<br>**furfural** | 1.0630 | 236.0000 | 33.0822 | 79.5326 | 7.2670 |

The chemical structure of the 8 optimal ionic liquids which were designed through the CAILD model are listed in Table 5-5.

**Table 5-5:** A schematic of the structure of optimal ILs

| Ionic Liquid (IL) | Structure |
|---|---|
| 1-(ethoxyethyl)pyridinium Tetrachloroaluminate | 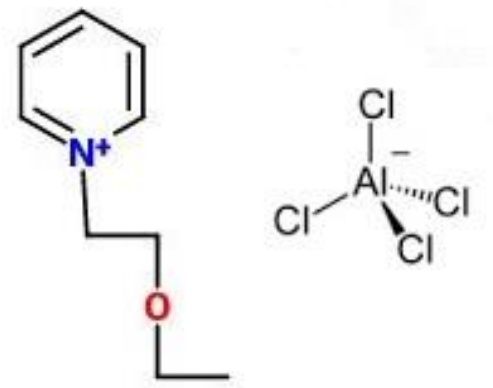 |
| 1-(benzyl)-3-methylpyridinium Tetrachloroaluminate | 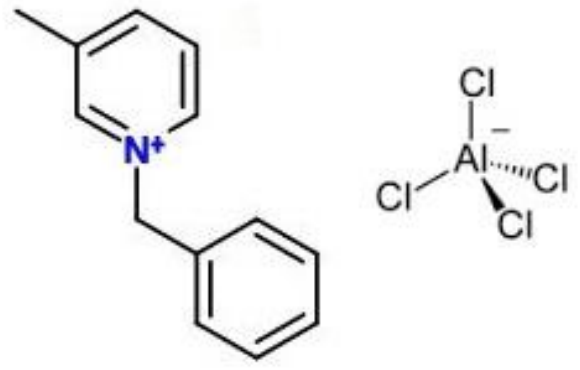 |
| 1-propyl-3-methylpyridinium Tetrachloroaluminate | 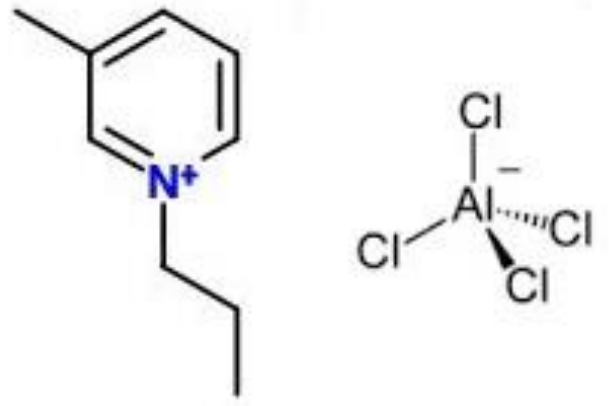 |
| 1-ethyl-3-methylpyridinium Tetrachloroaluminate | 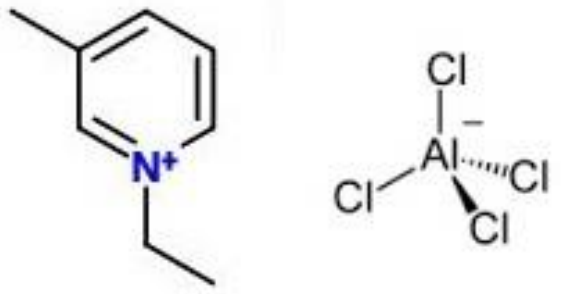 |
| 1-(3-ethoxypropyl)pyridinium Tetrachloroaluminate | 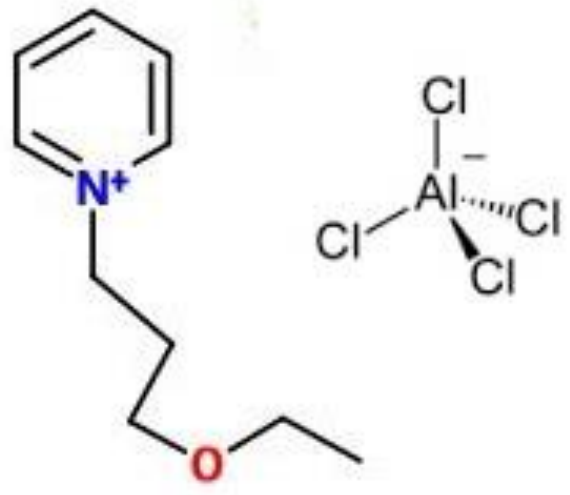 |
| 1-benzyl-3-methyl-imidazolium Tetrachloroaluminate | 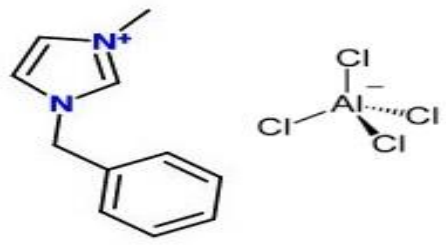 |

| Ionic Liquid (IL) | Structure |
|---|---|
| 1-benzyl-3-propyl-imidazolium Hexafluorophosphate | 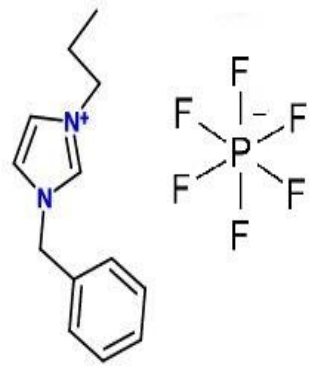 |
| 1-methylbenzyl-3-methoxymethylbenzyl-imidazolium Methyl sulfate | 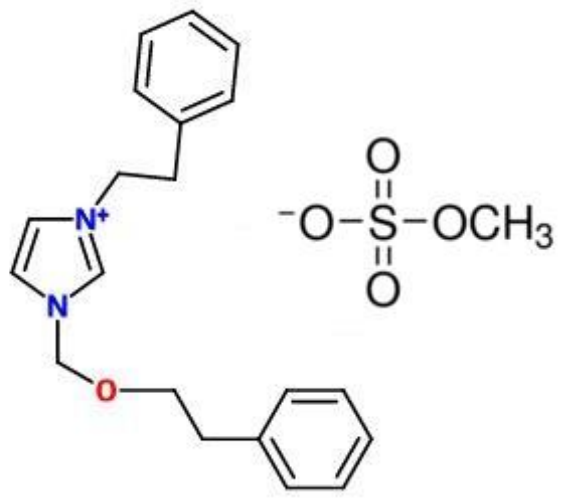 |

# Chapter 6: Application 3: Design of Ionic Liquids for $CO_2$ Capture

## 6.1 Introduction

The rapidly growing challenge of global warming (GW), which is believed to be caused by human activates resulting in the release of greenhouse gases (e.g. $CO_2$, $CH_4$) into the atmosphere, needs to be dealt with in a timely manner. During the past decade, researchers working in this area are divided into two distinct groups; one group has been focusing on the substitution of fossil fuels with their renewable counterparts (e.g. solar energy, wind energy, biofuels….) to decrease/eliminate the direct release of greenhouse gases to the atmosphere and the other group has been working on carbon (mainly $CO_2$) capture, storage, and utilization (CCSU).

Absorption, adsorption, and membrane separation are among the chemical processes being considered for $CO_2$ capture process.[194] Among the aforementioned processes, absorption (chemical or physical) with a re-generable solvent, appears to be one of the most promising methodologies for $CO_2$ capture.[195] Monoethanolamine (MEA), a polar solvent, has shown the ability to be a good solvent for $CO_2$ capture.[196] Even though amine-based solvents normally possess relatively high solvency power towards $CO_2$, their relatively high vapor pressure makes the solvent regeneration step inefficient and environmentally impactful.[197] In order to deal with this issue, use of alternative solvents such as ionic liquids, has attracted a considerable amount of attention in the past few years.[198] Ionic liquids, have the potential to be tailored to possess the desired properties of organic solvents (e.g. having high solvency power towards $CO_2$) without possessing their undesired properties.[199]

Studies indicate that based on the type of anion selected, ionic liquids (ILs) can be utilized to promote physical or chemical absorption of $CO_2$. The physical absorption, which is merely based on the dissolution of $CO_2$ in the liquid phase (IL-rich phase), is only feasible when a great abundance (i.e. a high partial pressure) of $CO_2$ is available in the gaseous mixture. The pre-combustion capture is the process which deals with the removal of $CO_2$ from the fuel streams (mainly containing $CO_2$ and $H_2$) before they undergo the chemical combustion. The higher concentration of $CO_2$ in the feed stream (compared to flue gas), make the physical absorption using an ionic liquid feasible.

In chapter 3, we showed that a large inventory of anions, cations, and functional groups either exist or can be synthetized.[200-202] A large number of ILs can potentially be formed through the combination of these distinct building blocks (estimated to be as many as $10^{14}$ feasible ILs).[203,204]

## 6.2 Forward problem

Activity coefficients of compounds in a multicomponent mixture can be used to predict the equilibrium condition of a system. Similar to the approach used in the previous chapter on the design of an IL for aromatic/aliphatic separation, UNIFAC model can be utilized to predict the activity coefficients of the components based on the concept of group contribution. Due to limited availability of CO2 solubility data in ionic liquids we followed a predictive approach similar to the one explained in chapter 5.

COSMO-RS, a promising tool for solvent screening for different applications, was used to predict the activity coefficients and solubility of CO2 in selected ILs. Henry's law constant, *H*, is a partitioning coefficient between vapor (gas) and liquid phases which measures the tendency of a compound to remain in the gas phase versus being dissolved in the liquid phase. The Henry's law constant can be defined as the partial pressure of a compound (at a given temperature) in the gas phase divided by its concentration in the liquid phase when the equilibrium condition between the two phases has been reached. A higher value of Henry's constant implies that the compound of interest has higher volatility thereby less likely to stay in the liquid phase (i.e. lower concentration in the liquid phase). Henry's constant is a good index to evaluate whether a chemical (a molecular compound or an ionic liquid) is a good solvent for $CO_2$ capture. In order for an ionic liquid to be a good solvent for a carbon capture process the Henry's constant of IL-$CO_2$ system should be as low as possible. Ionic liquids generally have very low vapor pressures and hence the Henry's constant of $CO_2$ in ILs can be defined as following

$$\mathrm{H} = p_{CO_2}^{vap} \times \gamma_{CO_2} \qquad (6\text{-}1)$$

$\gamma_{CO_2}$ is the activity coefficient of CO2 in the liquid (IL-rich) phase and is inversely proportional to its solubility in the IL and $p_{CO_2}^{vap}$ is the vapor pressure of $CO_2$ at the given temperature. In this study, the experimental values of the vapor pressure of $CO_2$ were used. The activity coefficients of $CO_2$ in different ionic liquids were calculated using COSMO-RS model[213] implemented in COSMOtherm software.[214] In order to verify the ability of the COSMO-RS model to predict the activity coefficients (hence the solubility) of $CO_2$ in different ILs, a comparison between experimental and COSMO-predicted values of Henry's constants for different IL-$CO_2$ mixtures is presented.

All calculations in COSMOtherm were performed using BP level of density functional theory (DFT) and TZVP basis set (the recommended settings for predicting the thermophysical properties of chemical compounds in chemical engineering applications).[215] In order to find the most precise dataset, which is available to predict the Henry's constant of $CO_2$ in different ILs, we tested all different BP-TZVP basis sets available in COSMOtherm. We identified that for IL-$CO_2$ systems, BP_TZVP_C21_0108 parameterization set predicts the Henry's constants most closely to the experimental data. However, a systematic error between the predicted and experimental values was observed, therefore a linear relationship, shown in eqn. (6-2), between experimental and COSMO-predicted values of *H* (based on BP_TZVP_C21_0108 basis set) was developed to reduce the average errors of COSMO-based predictions.

$$H_{mod} = 0.7403 H_{pred} + 7.9697 \qquad (6\text{-}2)$$

Table 6-1 lists the experimental and COSMO-predicted values (both before and after modification) of Henry's constant for 18 different IL-$CO_2$ systems which were not part of the original data set used to develop the linear relationship.

**Table 6-1:** A comparison between experimental and COSMO-based values of Henry's constant

| $H_{exp}$ [bar] | $H_{Cosmo-pred}$ [bar] | $H_{Cosmo-modified}$ [bar] |
|---|---|---|
| 57.8 | 76.09 | 64.30 |
| 67.9 | 93.12 | 76.91 |
| 77.0 | 112.60 | 91.33 |
| 50.58 | 61.17 | 53.25 |
| 57.49 | 68.98 | 59.04 |
| 64.17 | 76.91 | 64.90 |
| 69.35 | 85.45 | 71.23 |
| 80.02 | 104.22 | 85.13 |
| 103.38 | 136.98 | 109.37 |
| 139.94 | 205.09 | 159.80 |
| 47.46 | 60.83 | 53.00 |
| 52.77 | 67.73 | 58.11 |
| 57.28 | 75.30 | 63.71 |
| 62.72 | 83.54 | 69.82 |
| 71.26 | 101.29 | 82.95 |
| 80.46 | 121.24 | 97.73 |
| 94.18 | 155.87 | 123.36 |
| 97.94 | 168.26 | 132.54 |
| **Average error** | **36.6%** | **12.7%** |

Eqn. (6-2) was used to modify the values of Henry's constants predicted by COSMOtherm software. The modified values of H along with the experimental values of vapor pressure of $CO_2$ were used to reverse calculate its activity coefficients in different ILs.

Based on the approach used in the previous chapter, the COSMO-predicted activity coefficients (after correcting for systematic errors) were used to develop group parameters (*R* and *Q*) and binary interaction parameters ($a_{ij}$ and $a_{ji}$) of the UNIFAC model for systems consisting of $CO_2$ and ILs. A list of cations, anions, and side chain groups used in this study are listed in Appendix D. The UNIFAC parameters developed in this study can be used in any computer-aided ionic liquid design (CAILD) framework to explore a wide variety of ionic liquids (estimated to be as many as 21,000) with an aim to find the most optimal ionic liquid for a $CO_2$ capture process.

The accuracy of the CAILD approach was tested using available experimental data on the solubility of $CO_2$ in several different ionic liquids and the results along with the relative errors are tabulated in Table 6-2.

**Table 6-2:** Experimental and UNIFAC predicted values of $CO_2$ solubility in different ILs

| Ionic Liquid (IL) | Solute | T (K) | P(Kpa) | $x^{exp}_{CO2,IL}$ | $x^{calc}_{CO2,IL}$ | Ref |
|---|---|---|---|---|---|---|
| 1-butyl-3-methylimidazolium $Tf_2N^-$ | $CO_2$ | 283.1 | 100 | 0.0373 ± 0.008 | 0.0498 | 216 |
| 1-butyl-3-methylimidazolium $PF_6^-$ | $CO_2$ | 313.3 | 103 | 0.0162 ± 0.0016 | 0.0135 | 217 |
| 1-(3-hydroxypropyl)pyridinium $Tf_2N^-$ | $CO_2$ | 303.58 | 64.3 | 0.01175 ± 0.00071 | 0.0102 | 218 |
| 1-(3-hydroxypropyl)pyridinium $Tf_2N^-$ | $CO_2$ | 323.47 | 69.9 | 0.00578 ± 0.00035 | 0.00635 | 218 |
| 1-(3-hydroxypropyl)pyridinium $Tf_2N^-$ | $CO_2$ | 333.46 | 66.5 | 0.0038 ± 0.00023 | 0.00451 | 218 |
| 1-ethyl-3-methylimidazolium $Tf_2N^-$ | $CO_2$ | 283.43 | 100 | 0.03996 ± 2e-05 | 0.0412 | 219 |
| 1-ethyl-3-methylimidazolium $Tf_2N^-$ | $CO_2$ | 293.39 | 100 | 0.03203 ± 1e-05 | 0.03152 | 219 |
| 1-ethyl-3-methylimidazolium $Tf_2N^-$ | $CO_2$ | 303.39 | 100 | 0.02626 ± 1e-05 | 0.02446 | 219 |

| Ionic Liquid (IL) | Solute | T (K) | P(Kpa) | $x_{CO2,IL}^{exp}$ | $x_{CO2,IL}^{calc}$ | Ref |
|---|---|---|---|---|---|---|
| 1-butyl-3-methylimidazolium $BF_4^-$ | $CO_2$ | 283.1 | 99.9 | 0.0197 ± 0.0028 | 0.032715 | 220 |
| 1-butyl-3-methylimidazolium $PF_6^-$ | $CO_2$ | 283.15 | 49.9 | 0.0139 ± 0.006 | 0.017019 | 221 |
| 1-butyl-3-methylimidazolium $PF_6^-$ | $CO_2$ | 283.15 | 99.9 | 0.0267 ± 0.006 | 0.031079 | 221 |
| 1-ethyl-2,3-dimethylimidazolium $Tf_2N^-$ | $CO_2$ | 283.15 | 49.9 | 0.016 ± 0.004 | 0.019515 | 221 |
| 1-ethyl-2,3-dimethylimidazolium $Tf_2N^-$ | $CO_2$ | 283.15 | 99.8 | 0.0323 ± 0.004 | 0.039781 | 221 |

## 6.3 Reverse problem

In this section we propose a computer-aided IL design model which integrates a process model within the solvent design framework and can design optimal ionic liquid for $CO_2$ capture process is presented.

### 6.3.1 Case study

A feed stream consisting of pure $CO_2$ at 1 bar pressure enters a single stage absorption column where it comes into contact with pure IL. The absorption process occurs at 288 K and 1 bar. The IL-enriched phase which now has absorbed $CO_2$ will enters the desorption column in which the $CO_2$ will be separated from the liquid phase by increasing the temperature. The desorption process occurs at 318 K and 1 bar. The goal of the CAILD framework is to design an IL with high $CO_2$ solubility (i.e. high solvency power towards $CO_2$) and the lowest performance index (*PI*) possible. This means that the designed ionic liquid will strongly absorb $CO_2$ at lower temperatures and releases it effectively at higher temperatures. A schematic of the coupled absorption-desorption process of interest is depicted in Figure 6-1.

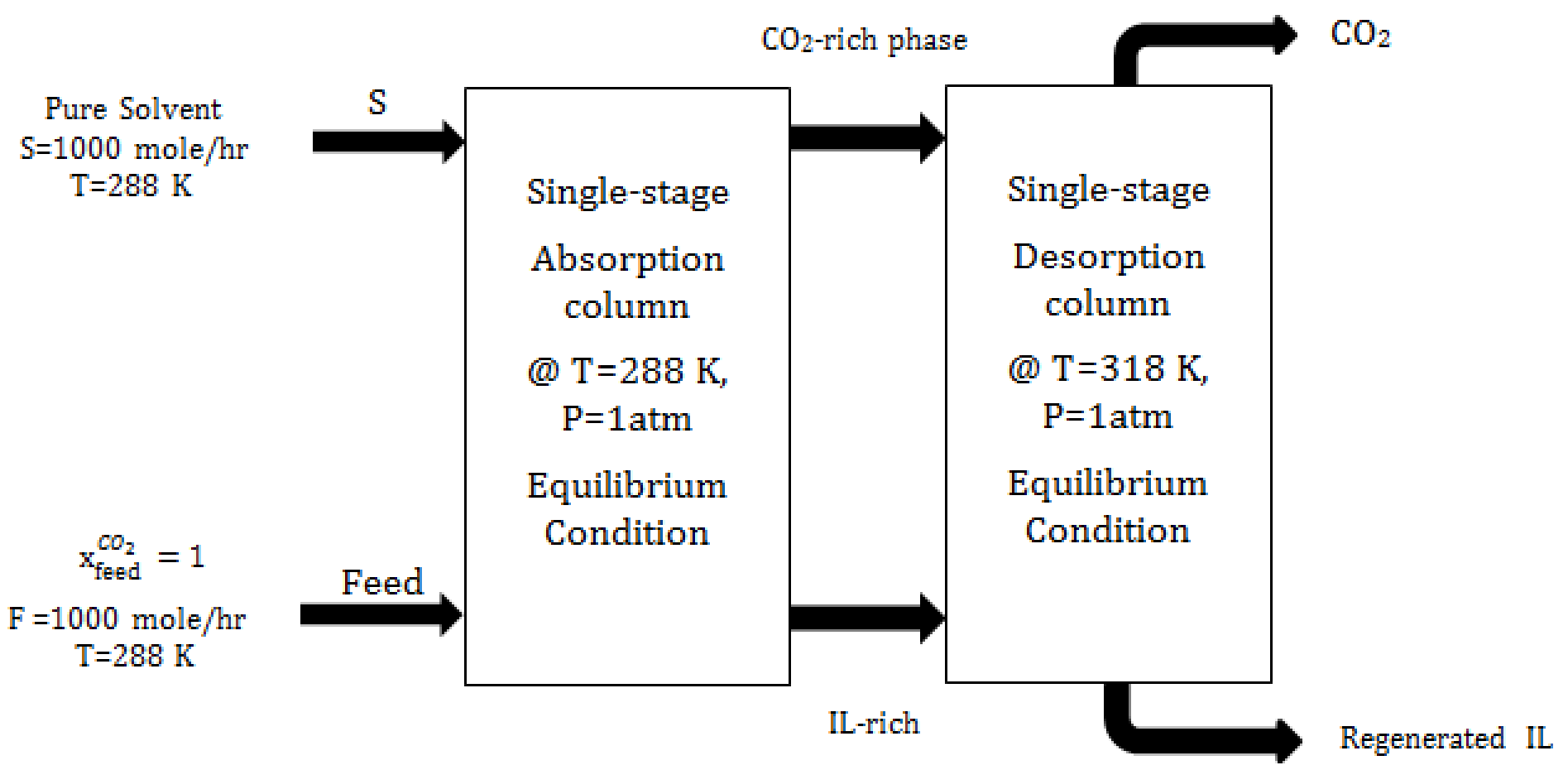

**Figure 6-1:** A schematic of single stages $CO_2$ absorption-desorption processes

***Detailed Formulation of the Problem***

In this optimization framework the objective is to maximize the solubility of $CO_2$ in the IL-phase during the absorption step while at the same time minimize the solubility of $CO_2$ in the IL at a higher temperature during the desorption step. The objective function was defined as the performance index (*PI*) as shown below:

$$\text{Objective function} = \min \left(PI = \frac{H_{288\,K}^2}{H_{318\,K}}\right) \quad (6\text{-}3)$$

***Feasibility constraints***

$$\sum_{i \in c} c_i = 1 \quad (6\text{-}4)$$

$$\sum_{j\in A} a_i = 1 \quad (6\text{-}5)$$

$$\sum_{l=1}^{6} y_l = \sum_{i\in C} c_i v_{ci} \quad (6\text{-}6)$$

$$\sum_{i\in c}(2 - v_{ci})c_i + \sum_{l=1}^{6}\sum_{k\in G}(2 - v_{Gkl})y_l ng_{kl} = 2 \quad (6\text{-}7)$$

$$\sum_{k\in G} y_l ng_{kl}\,(2 - v_{Gkl}) = 1 \quad (6\text{-}8)$$

$$\sum_{k\in CH_2} y_l ng_{kl} \leq 15 \quad (6\text{-}9)$$

***Solution property constraints***

$$\text{SP}\ (solvent\ power) = \frac{1}{\gamma_{A,S}^{\infty}} \geq 0.429 \quad (6\text{-}10)$$

$$\frac{P_{CO_2}^{partial}}{x_{CO_2-IL}^{sat}} = \gamma_{CO_2}^{sat} \times P_{CO_2}^{Vap} \quad (6\text{-}11)$$

$$x_{CO_2,absoprtion}^{IL-enriched\ phase} + x_{IL,absoprtion}^{IL-enriched\ phase} = 1 \quad (6\text{-}12)$$

$$x_{CO_2,desorption}^{IL-enriched\ phase} + x_{IL,desorption}^{IL-enriched\ phase} = 1 \quad (6\text{-}13)$$

As described in previous chapter, the optimal IL designed using the CAILD framework must satisfy specific rules to ensure the chemical feasibility of the designed compound. These rules, also named as structural constraints, include feasibility rules (e.g. octet rule), bonding rule, and complexity rules (which deals with the size of cation as well as the side chains attached to the cation-base). Eqn. (4-9) deals with the size of each side chain attached to the cation-base that is an upper bound of 15 for the number of $CH_2$ groups in this study. Eqn. (4-

10) ensures that an ionic liquid with the solvent power better than that of methanol (a conventional solvent used for physical absorption of $CO_2$) is selected as the optimal solvent.

Eqn. (4-11) guaranties that the system has reached equilibrium condition where the mole fraction of $CO_2$ in the liquid phase ($x^{sat}_{CO_2-IL}$) reaches its maximum value at that temperature. In other words, when a system consisting of $CO_2$ and an IL has reached the equilibrium condition, the composition of the two phases (IL-rich phase and $CO_2$ rich phase) will not change with time; meaning that the IL cannot dissolve more $CO_2$ at that temperature.

Since ILs mostly have negligible vapor pressures, we do not expect to detect any trace of them in the gas phase (neither in absorption nor in the desorption stages). Therefore it is safe to say that in both the stages only liquid phases consist of more than one compound (i.e. $CO_2$ and IL). Eqns. (4-12) and (4-13) ensure that the summation of the mole fraction of the two components in the liquid phase is equal to 1.

It is also worth mentioning that due to the systematic and model errors in predicting the solubility and activity coefficients of the compound in different phases, it is impossible to satisfy the equivalences of Eqns. (6-11) through (6-13) with 100% accuracy. Therefore a ($\pm$5%) tolerance was applied to the equations to enable the optimization program converge.

### 6.4 Results

The CAILD optimization program was run several times to ensure that it converged to a global optima. The optimization model (CAILD) was used to identify the top 5 optimal ionic liquids (ILs) from different categories (different cation head groups) with best solvent properties. The structure of the optimal solvents along with their structures are listed in Table 6-3.

**Table 6-3:** Name, symbol and structure of the optimal ILs

| Ionic Liquid (IL) | Symbol | Chemical Structure |
|---|---|---|
| 1-(12-phenyloxydodecyl)-3-(5-phenylpentyl)imidazolium $Tf_2N$ | $IL_1$ | 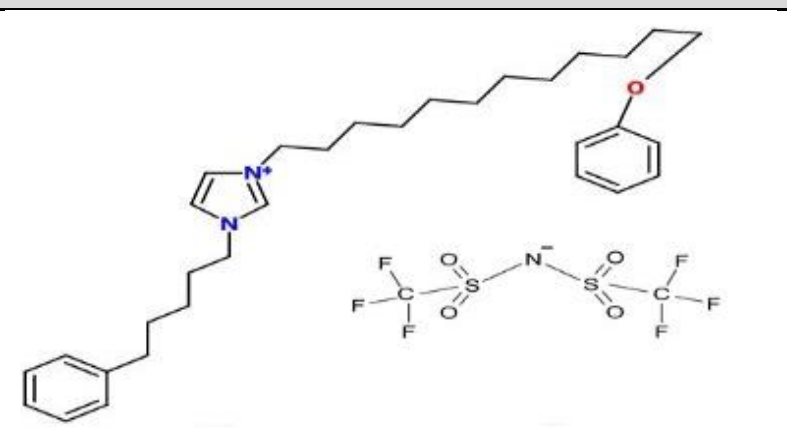 |
| 1-(3-methoxypropyl)-4-ethylpyridinium ethyl sulfate | $IL_2$ | 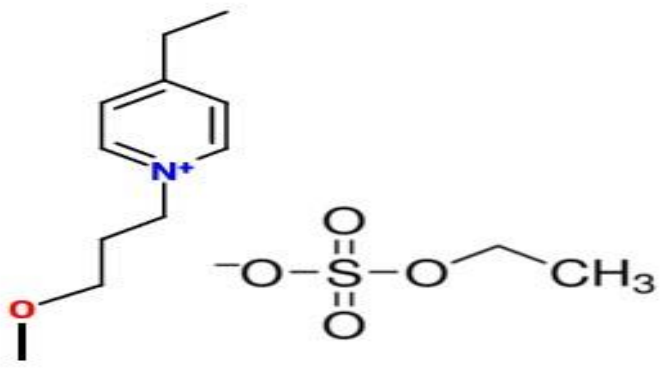 |
| 1-(15-benzyloxypentadecyl)-1-hexylpyrrolidinium $Tf_2N$ | $IL_3$ | 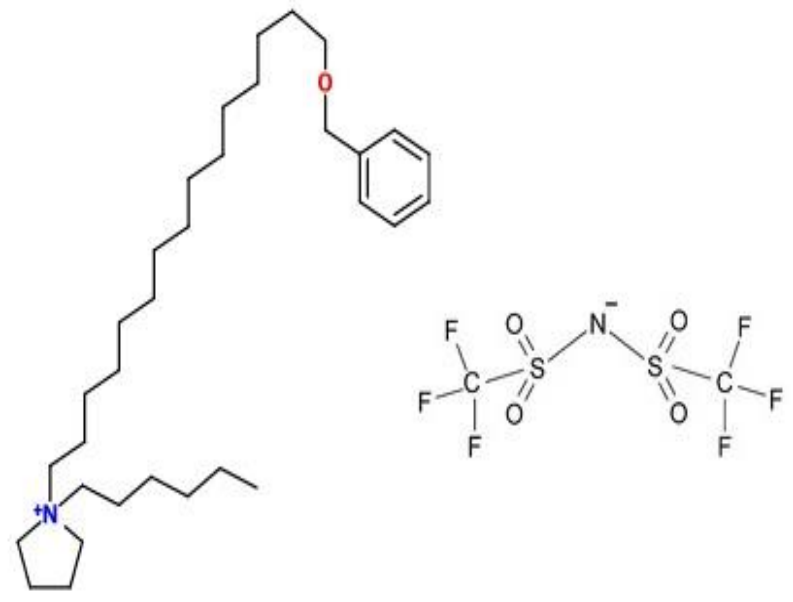 |
| Decyl-(4-pentadecylbenzyl)-methylphosphonium triflate | $IL_4$ | 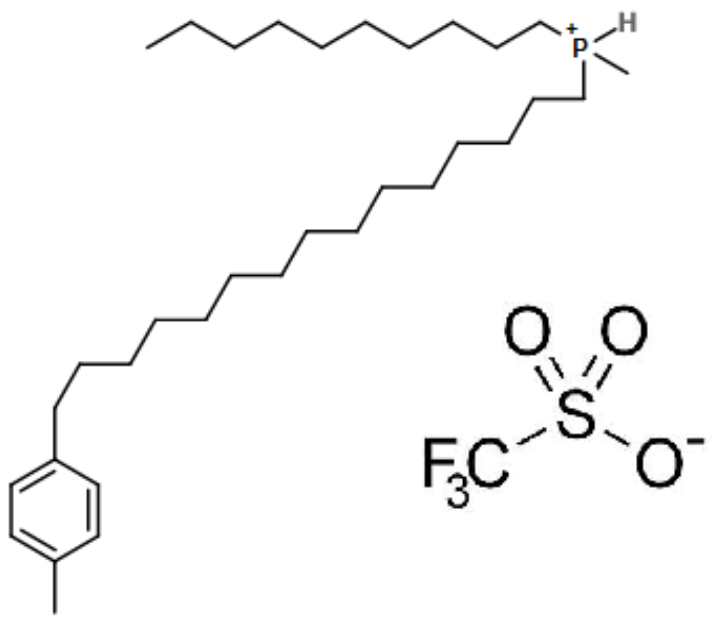 |

| Ionic Liquid (IL) | Symbol | Chemical Structure |
|---|---|---|
| 4-(8-phenoxyoctyl)-4-(phenylnonyl)morpholinium $Tf_2N$ | $IL_5$ | 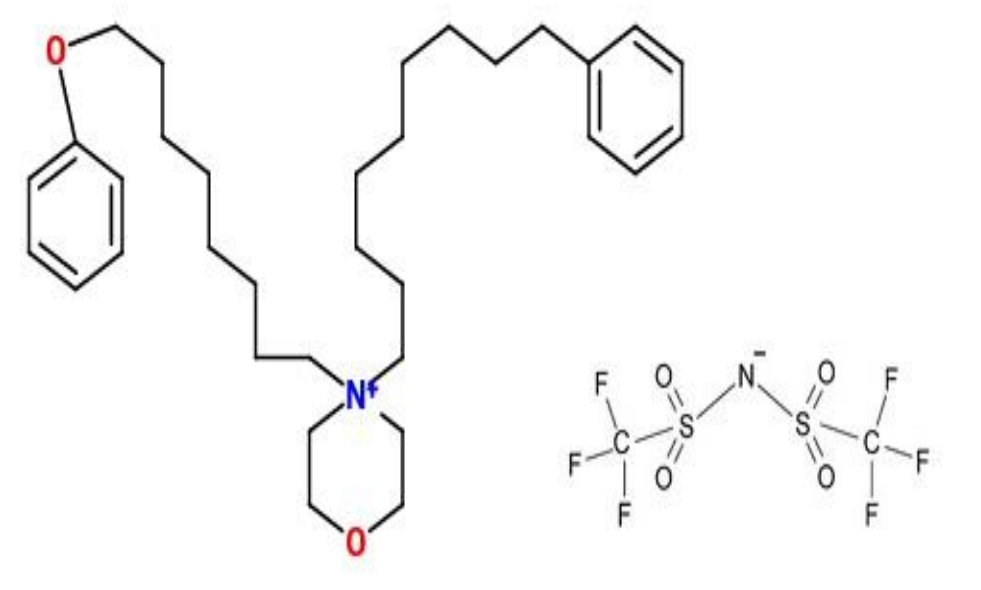 |

As discussed before, it is critical for the optimal ILs to have reasonably low viscosity and melting point to enable them to be used as the process solvent in industry. The initial selection of anions used in this study was based on this fact that they contribute to the lower values of melting points of the ILs they make. The viscosity and melting point values of the 5 optimal ILs, listed in Table 6-4, were calculated from correlations developed in chapter 2 and presented in our previous study.[222] These values, along with the predicted solubility of $CO_2$ in these ILs and the solvency power of the optimal ILs, are listed in Table 6-4.

**Table 6-4:** Pure (physical) and mixture properties of the optimal ILs

| IL | SP | $x_{IL}^{CO_2}$ | $H_{abs}$ @ 288 K [bar] | $H_{des}$ @ 318 K [bar] | PI | $T_m$ [K] | Viscosity @ 288 K [cP] |
|---|---|---|---|---|---|---|---|
| IL1 | 4.532 | 0.0706 | 9.7498 | 20.649 | 4.6034 | 201.95 | 587.91 |
| IL2 | 1.438 | 0.0309 | 10.9397 | 22.934 | 5.2183 | 312.04 | 66.10 |
| IL3 | 4.717 | 0.0798 | 9.1024 | 19.888 | 4.1660 | 228.08 | 724.92 |
| IL4 | 4.848 | 0.0843 | 7.5657 | 16.518 | 3.4653 | 230.94 | 845.34 |
| IL5 | 3.963 | 0.0677 | 11.2954 | 23.779 | 5.3654 | 224.24 | 1521.50 |

| IL | SP | $x_{IL}^{CO_2}$ | $H_{abs}$ @ 288 K [bar] | $H_{des}$ @ 318 K [bar] | PI | $T_m$ [K] | Viscosity @ 288 K [cP] |
|---|---|---|---|---|---|---|---|
| Bmim $Tf_2N$ | 1.755 | 0.0340 | 48.7733 | 90.8902 | 26.1726 | 267.1 | 81.60 |
| Bpy $BF_4$ | 0.911 | 0.0211 | 74.4796 | 136.7710 | 40.5583 | 272.1 | 313.1 |
| BmPyr triflate | 1.5522 | 0.0301 | 54.0972 | 103.0201 | 28.4071 | 277.56 | 299 |
| Methanol | 0.429 | 0.00853 | 225.0339 | 419.1725 | 155.168 | 175.55 | 0.6220 |

Bmim: 1-butyl-3-methylimidazolium

Bpy: 1-butylpyridinium

BmPyr: 1-butyl-1-methylpyrrolidinium

triflate: Trifluoromethanesulfonate

# Chapter 7: Life Cycle Environmental Implications of Ionic Liquids

## 7.1 Life Cycle Perspectives on Aquatic Ecotoxicity of Common Ionic Liquids

As it was mentioned earlier, field of ILs is developing rapidly, thus it is important to consider their environmental, ecological, and human health impacts at the design stage for their long-term acceptance. Despite the large amount of attention, ILs are still a fairly new class of materials and several of their environmental characteristics have only been studied recently. The non-volatile nature of ILs can greatly limit atmospheric pollution and for this reason, they are often promoted as green chemicals with the potential to replace volatile organic solvents.[223] However, in recent years, many studies have reported that several ILs possess relatively high level of toxicity towards freshwater organisms. While both points are valid, a realistic picture can only emerge through analysis of the life cycle ecological impacts that considers the upstream ecological impacts associated with producing them as well as the downstream ecological impacts due to their use. The current study addresses this research gap by focusing on understanding the production side and end-of-life fresh water ecotoxicity impacts related to ILs.

To date there have been very few Life Cycle Assessment (LCA) studies performed on processes that involve ILs.[224,225,260,261] These studies indicate that, from a life cycle perspective, IL based processes do not necessarily improve the environmental performance in comparison to molecular solvents based processes. One common limitation in all of these studies is that they did not consider any possible impacts associated with the direct release of ILs into the environment. Therefore, it is clear that the environmental fate as well as toxicity information related to ILs has not been incorporated in any LCA study to date. This is because toxicity-based characterization factors for ILs are not yet available in impact assessment tools such as TRACI 2.1. One of the main objective of this paper is to address this research gap by developing, freshwater ecotoxicity characterization factors for the following five common ILs: 1-butyl-3-methylimidazolium bromide ($[Bmim]^+[Br]^-$), 1-butyl 3-methylimidazolium chloride ($[Bmim]^+[Cl]^-$), 1-butyl 3-methylimidazolium tetrafluoroborate ($[Bmim]^+[BF_4]^-$), 1-butyl-3-methylimidazolium hexafluorophosphate ($[Bmim]^+[PF_6]^-$), and 1-butylpyridinium chloride ($[BPy]^+[Cl]^-$).

In the recent past, several studies have provided new insights into IL chemistry, environmental fate and toxicity. Many ILs exhibit significant solubility in water[227] and even water immiscible ILs show limited water solubility and stability.[228] This, combined with the fact that they are non-volatile, makes wastewater discharge as the most likely route through which ILs will eventually be released into the environment Once discharged, ILs will interact with aquatic ecosystems through a variety of mechanisms and can cause damage to aquatic species.[229] To evaluate the ecotoxicity of ILs, a broad range of testing models (bacteria, fungi, algae, plants, and animals) have been used.[230] Standard ecotoxicological tests show that many ILs have high toxicity towards freshwater organisms (*e.g.* algae and D. magna).[231] Wells *et al.*[231] examined four different classes of ILs, which all had relatively high ecotoxicity impacts (EC50 < 100 mg $L^{-1}$). Certain ILs, such as those with shorter side-chain 1-alkyl-3-methyl imidazolium and pyridinium, show only moderate ecotoxicity to bacteria, algae, and invertebrates, but when the side chains are longer than $C_8$ their ecotoxicity profiles become significantly worse. This is also true for phosphonium and ammonium-based

ILs, which have higher molecular weights. The environmental fate of ILs depend on several biotic and abiotic factors. Only in the past few years have the major abiotic mechanisms of ILs (*e.g.* their sorption in different type of soils) been understood.[232,233] The extent of the final toxicity impact of ILs will depend heavily on their physicochemical properties, interactions with surrounding environmental media, and chemical and biological transformations. This information is essential to understand how these compounds are transformed and how long they will persist in the environment.

On the other end of the spectrum, there is still very little understanding of the environmental and ecological impacts associated with the production of ILs. This is because ILs are emerging materials and are not yet produced in commercial scales. Consequently, no primary data is available on material/energy consumption. Further, very little data is available about the precursors needed to produce these ILs. This study integrates the existing[224,225] and newly developed life cycle inventories for the production of the above mentioned five ILs with an aim to understand their production side upstream freshwater ecotoxicity impacts.

Through an in-depth analysis, we compare the relative contributions of the different IL life cycle stages (production phase and use phase loss) towards ecotoxicity impacts and discuss their significance. We would like to point out that similar studies on other materials, such as carbon nanotubes (Eckelman et al.[248]) have revealed important information on comparing production and use phase impacts.

## 7.2 Methods

### 7.2.1 Goal, Scope, System Boundary

The main goal of this study was to understand and compare the relative freshwater ecotoxicity impacts of ILs related to their production phase and direct release to the environment during use phase. The scope of this study includes: 1) building cradle-to-gate life cycle inventory for production of five common ILs; 2) developing characterization factors for their freshwater ecotoxicity impacts; and 3) comparing the potential ecotoxicity impacts of the production and use phase release. A functional unit of 1 kg of the five ILs was considered.. The system boundary (shown in Figure E-1 in Appendix E) includes all upstream steps necessary for the production of ILs, treatment, and recovery as well as the direct release of IL to the environment during use.

### 7.2.2 Life Cycle Inventory of Ionic Liquid Production and Data Sources

Inventory of upstream steps include raw material inputs and emissions related to the production of reactants, precursors, reagents and ancillary materials as well as the extraction, conversion and delivery of energy inputs. The emissions from the construction and maintenance of chemical plants are assumed to be relatively low[225] and hence neglected. Data related to electricity, thermal energy (steam), transportation systems, and chemical production were derived either from Ecoinvent,[258] or USLCI database integrated in SimaPro software.[259] Data availability permitting, for consistency purposes we assumed that all life cycle steps involved in the production of ILs to be within the United States. Whenever non-U.S. data had to be used, every effort was made to adjust the data to corresponding US energy mix. The inventory of all materials (e.g. precursors, reagents, reactant etc.) and energy used for the production of the ILs along with the data sources are listed in Table E-2 of Appendix E. Please note that data for chemicals that were not available in standard LCI databases (ILs and some of their precursors) were assembled using mass and energy balances derived from chemical process simulations (CPS)[225,234] supplemented with theoretical calculations based on sound judgment and good engineering practice. A detailed description of the procedure adopted to generate the mass and energy balance is presented in Appendix E.

### 7.2.3 Fresh Water Ecotoxicity Impacts of Ionic Liquid Production

The life cycle impact assessment methodology, Tool for the Reduction and Assessment of Chemical and other environmental Impacts (TRACI 2.1.) developed by U.S. Environmental Protection Agency utilizes characterization factors derived from the USEtox model.[238] Using these characterization factors, emissions of organic and inorganic materials released during the upstream life cycle steps of IL production was translated into total freshwater ecotoxicity impacts.

### 7.2.4 Development of Ecotoxicity Characterization Factors for Ionic Liquids

Characterization factors are used to quantify the extent to which a given pollutant contributes to environmental impacts. In this work we utilized the USEtox model[238], which is a state-of-the-art modeling framework based on scientific consensus for characterizing ecotoxicity impacts of chemicals, to develop freshwater ecotoxicity characterization factors for the five ILs. In this approach, the aquatic ecotoxicity characterization factors are estimated as a product of fate factor (FF), exposure factor (XF) and effect factor (EF) as shown in eqn. (7-1).

$$\mathrm{CF_{ecotox}} = \mathrm{EF_{ecotox}} * \mathrm{FF} * \mathrm{XF} \quad (7\text{-}1)$$

EF relates to the inherent toxicity of the substance of interest, FF relates to the fate and transport of the substance, and XF relates to potential routes of exposure and intake of the substance. In the following section, we describe the procedure for the development of each of these factors for ILs.

***Effect Factor:*** EF reflects the relationship between the concentration of an IL and potentially affected fraction (PAF) of aquatic organisms. It is defined as the slope of the concentration response relationship up to the point when the PAF reaches 50% (eqn. (7-2)).

$$\mathrm{EF} = \frac{\mathrm{PAF}}{\mathrm{HC50}} = \frac{0.5}{\mathrm{HC50}} \quad (7\text{-}2)$$

HC50 corresponds to the geometric mean of species-specific EC50 data and EC50 represents the effective concentration of the IL of interest at which 50% of the population of a particular species experiences a response. Table 7-1 summarizes the aquatic toxicity data (EC50) of the five ILs collected from the literature. It can be observed that, there is a wide distribution of EC50 values associated with ecotoxicty of ILs towards different species. In the next step, as suggested by USEtox, the geometric mean of EC50 values for each IL was calculated. Based on the guidelines provided for the use of USEtox model acute toxicity data were converted to chronic data using acute-to-chronic ratio of two. Similarly, when no observed effect concentration (NOEC) was reported we extrapolated NOEC to EC50 by a factor of nine. Next, the geometric mean of individual EC50 data for each IL, HC50, was used to estimate the effect factor.

***Fate Factor***: Multimedia fate and transport models are used to derive the environmental fate factors. In these models the environment is represented as a number of homogeneous compartments (e.g. air, water, and soil). The intermedia transfer of chemical substances between different compartments is modeled as a set of mass balance equations.

**Table 7-1:** Toxicity values of selected Ionic Liquids

| Type | Species | Test type | Reported toxicity values (mg/L) | | | | | Geometric mean of EC50s [Chronic] (mg/L) | | | | | Ref |
|---|---|---|---|---|---|---|---|---|---|---|---|---|---|
| | | | [Bmim] [Br] | [Bmim] [Cl] | [Bmim] [BF4] | [Bmim] [PF6] | [BPy] [Cl] | [Bmim] [Br] | [Bmim] [Cl] | [Bmim] [BF4] | [Bmim] [PF6] | [BPy] [Cl] | |
| Bacteria | *Photobacterium phosphoreum* | growth inhibition (Acute) | 408.020 | --- | --- | --- | --- | 204 | --- | --- | --- | --- | 239 |
| Bacteria | *Aliivibrio fischeri* | Luminescence (Acute) | 256-2250 | 429-897 | 799-802 | 334-929 | 257-440 | 491.6 | 216.5 | 399.5 | 278.5 | 168.1 | 240,241,228 |
| Algae | *Pseudokirchneriella subcapitata* | growth of algal biomass or photosynthesis O2 evolution (Acute) | 229-5260 | 504 | 568 | 375 | 63.9 | 393.1 | 252 | 284 | 187.5 | 32 | 242-243 |
| Daphnia (Crustacean) | *Daphnia magna* | first brood number of neonates & immobilization (Acute) | NOEC = >3.2 EC50 = 8.03 | NOEC= >3.2 EC50= 6.3 | 10.7 | NOEC= >3.2 EC50= 19.9 | 20 | 10.7 | 9.5 | 5.4 | 16.9 | 10 | 229 |
| Animalia | *Physella acuta* | egestion rate & death (Acute) | NOEC = 4 LC50 = 229 | ---- | --- | LC50 = 123 | --- | 64.2 | --- | --- | 61.5 | --- | 244 |
| fish | *Danio rerio* | death & frond number (Acute) | LC50>=100 | LC50>=100 EC50=59.5 | LC50>=100 | LC50>=100 | LC50>=100 | 50 | 38.6 | 50 | 50 | 50 | 257 |
| **HC50** | | | | | | | | 105.2 | 66.9 | 74.2 | 77.0 | 40.5 | |

The fate factor, which represents the persistence of a substance (residence time in days) in a particular environmental compartment, depends on physicochemical properties, partition coefficients, and biodegradation rates of the substance of interest. Most of the IL fate and transport parameters needed as input to the model were gathered from literature, and are listed in Table 7-2. A small value of 1 x $10^{-20}$ for biodegradation rate was assigned to all ILs that were categorized as non-biodegradable but had no quantitative values.[255]

***Exposure Factor***: The environmental exposure factor for freshwater ecotoxicity is defined as the fraction of IL dissolved in freshwater and is determined using eqn. (7-3).

$$\mathrm{XF} = \frac{1}{(1+(\mathrm{Kp_{SS}.SS}+\mathrm{K_{DOC}.DOC}+\mathrm{BAF.BIOmass}).10^{-6})} \qquad (7\text{-}3)$$

Where, $Kp_{ss}$, represent the partition coefficient between water and suspended solids and $K_{DOC}$ represents the partition coefficient between water and dissolved organic carbon respectively. BAF represents the bio-concentration factor in fish. SS (15 mg $L^{-1}$), DOC (5 mg $L^{-1}$), and BIOmass (1 mg $L^{-1}$) are default values for suspended matter, dissolved organic carbon and biomass concentrations, respectively.[248] The exposure factor (XF) for each IL was also calculated based on parameters listed in Table 7-2. We would like to note that there are certain limitations in using the proposed approach to the case of ILs. The USEtox model is suited for relatively small organic and inorganic materials while ILs typically have high molecular weights. However, other studies have reported results using the USEtox model for complex materials such as carbon nanotubes.[248] In addition, even though we assume that ILs can be treated as neutral entities, the ionic nature of these compounds may influence the results. Despite these limitations, we believe that, as per current state of knowledge, the approach of using the USEtox model represents the best possible way towards ecotoxicity impact assessment of ILs.

**Table 7-2:** Environmental Properties of the studied Ionic Liquids

| Parameter | Unit | Value | | | | |
|---|---|---|---|---|---|---|
| | | $[Bmim]^+ [Br]^-$ | $[Bmim]^+ [Cl]^-$ | $[Bmim]^+ [BF_4]^-$ | $[Bmim]^+ [PF_6]^-$ | $[BPy]^+ [Cl]^-$ |
| Molecular weight | g $mol^{-1}$ | 219.12 | 174.67 | 226.03 | 284.18 | 171.67 |
| Octanol-water partition coeff. $K_{OW}$ | | 0.0033 | 0.0029 | 0.0030 | 0.0218 | 0.0020 |
| Organic carbon-water partition coeff. $K_{OC}$ | L $kg^{-1}$ | 398.11 | 398.11 | 398.11 | 398.11 | 141.61 |
| Henry's law constant at 25 °C, $(K_H)^a$ | Pa kg $mol^{-1}$ | $1\times10^{-20}$ | $1\times10^{-20}$ | $1\times10^{-20}$ | $1\times10^{-20}$ | $1\times10^{-20}$ |
| Water solubility (25 °C) | mg $L^{-1}$ | $1\times10^{6}$ | $5.5\times10^{5}$ | $1\times10^{6}$ | $1.9\times10^{4}$ | $4.32\times10^{4}$ |
| Dissolved carbon-water partition coeff. $(K_{DOC})^b$ | L $kg^{-1}$ | 18.27 | 18.27 | 18.27 | 18.27 | 6.50 |
| Suspended solids partition coeff. $(Kp_{SS})^b$ | L $kg^{-1}$ | 18.27 | 18.27 | 18.27 | 18.27 | 6.50 |
| Sediment-water partition coeff. $(Kp_{sd})^b$ | L $kg^{-1}$ | 18.27 | 18.27 | 18.27 | 18.27 | 6.50 |
| Soil-water partition coeff. $(Kp_{sl})^b$ | L $kg^{-1}$ | 18.27 | 18.27 | 18.27 | 18.27 | 6.50 |
| Biodegradation rate in air[c] | $s^{-1}$ | $1\times10^{-20}$ | $1\times10^{-20}$ | $1\times10^{-20}$ | $1\times10^{-20}$ | $1\times10^{-20}$ |
| Biodegradation rate in water[c] | $s^{-1}$ | $1\times10^{-20}$ | $3.47\times10^{-8}$ | $1.36\times10^{-8}$ | $1\times10^{-20}$ | $1\times10^{-20}$ |
| Biodegradation rate in sediment[c] | $s^{-1}$ | $1\times10^{-20}$ | $1\times10^{-20}$ | $1\times10^{-20}$ | $1\times10^{-20}$ | $1\times10^{-20}$ |
| Biodegradation rate in soil[c] | $s^{-1}$ | $1\times10^{-20}$ | $1\times10^{-20}$ | $1\times10^{-20}$ | $1\times10^{-20}$ | $1\times10^{-20}$ |
| Bioaccumulation factor in fish $(BAF_{fish})^d$ | L $kg^{-1}$ | $1.58\times10^{-4}$ | $8.55\times10^{-4}$ | $8.94\times10^{-4}$ | $5.73\times10^{-3}$ | $6.04\times10^{-4}$ |

a) The value $1\times10^{-20}$ for $K_H$ was assigned to the ionic liquids, as all ILs have very small (negligible) vapor pressures.

b) The values for soil-water partition coefficient of all ionic liquids were calculated using the formula of $K_d = K_{OC}*f_{OC}$, in which $f_{oc}$ represents the fraction of organic carbon in the soil ≅ 0.0459 and $K_{OC}$ is the organic carbon-water partition coefficient. This value of $K_d$ was assigned to all four parameters $K_{DOC}$, $Kp_{SS}$, $Kp_{sd}$, and $Kp_{sl}$

c) The small value of $1\times10^{-20}$ was used for all non-biodegradable ionic liquids which had a biodegradability rate <%1 within 28 days.

d) Bioaccumulation factor in fish ($BAF_{fish}$) of the ionic liquids were calculated using this formula: log BAF= -0.68 + 0.94 log $K_{OW}$

### 7.2.5 Fresh Water Ecotoxicity Impact of Direct Release of Ionic Liquids

Large-scale use of ILs will inevitably lead to their release into the aquatic environment through wastewater disposal or accidental leakage.[249] The ecotoxicity impact of this release needs to be considered in any type of IL assessment. At a large commercial scale it would be reasonable to expect that industrial plants using ILs will have an appropriate treatment unit to remove and recover ILs from wastewater streams. This is a likely scenario since ILs are expensive and they can also be recovered easily thereby providing significant economic incentive to recover and reuse them. Studies show that in a typical chemical plant, loss of volatile organic solvents is usually estimated to be close to 10% with the majority of the loss due to air emissions.[250] For example, monoethanolamine, a solvent used for $CO_2$ capture requires a makeup of 10% solvent with <1% emissions to water.[250] For the case of ILs, their non-volatile nature would eliminate the possibility of any air emissions while water emissions would represent the most significant source of exposure. As typical water emission of solvents is < 1% we conservatively assume a maximum of 2% loss of ILs (during the use phase) to the wastewater stream which can be further significantly reduced by management of industrial wastewater. The assumed 2% loss of ILs to wastewater is in fact a very conservative estimate as nearly complete recycle and reuse of these compounds is feasible and since ILs are lot more expensive than organic solvents an economically viable process/technology that uses ILs would involve recovery and recycle of as much IL as possible. Further, environmental regulations would limit levels of IL discharge to the environment through wastewater effluents. Several treatment methods such as oxidative,[251] thermal,[252] and photocatalytic degradation[253] have been proposed for IL removal. Other approaches such as adsorption in activated carbon[254] and use of salts[249] have also been suggested. For this study, we selected the treatment option based on salting out using aluminum based salts as proposed by Neves *et al*.[249] due to the fact that this approach makes use of inorganic salts commonly used in current waste treatment plants and also has high recovery. Neves *et al*.[249] reported a recovery efficiency of 96% to 100% for a wide variety of ILs. We assume a range of 96 to 98 percent recovery of the ILS lost to the waste water stream which results in a final release fraction of 0.04% to 0.08% of ILs into freshwater streams. The ecotoxicty impact associated with the direct release of the IL to freshwater was estimated as a product of the IL characterization factor the mass of IL used (i.e. 1 kg) and the final release fraction to the freshwater (unrecovered IL).

### 7.2.6 Uncertainty

This section deals with the treatment of uncertainty related to the ecotoxicity estimates of the two life cycle stages, production phase and release during use. Monte Carlo simulation was employed and critical uncertain parameters were varied to generate multiple samples. Upon reviewing the LCI related to the production of ILs we identified that in comparison to other emissions, metals had very high ecotoxicty characterization factors. In addition, the USEtox model developers report caution while using the characterization factors of inorganic materials (metals). Therefore, the characterization factors (CFs) for all metals were varied by $\pm 1$ order of magnitude using a uniform distribution to generate Monte Carlo inventory samples with an aim to quantify the model uncertainty related to the production of ILs at 95% confidence. To capture the uncertainties related to ecotoxicity impacts associated with release during use phase a similar Monte Carlo simulation was performed within the USEtox model. A uniform distribution with one order of magnitude variation in each direction was assigned to each model input parameter tabulated in Table 7-2 as well as to HC50 data shown in Table 7-1. A range of 0.4%–2% (0.4% representing 80% removal through water treatment and 2% representing no water treatment) of the produced ILs was assumed to be released to the freshwater compartment. The Monte Carlo simulation randomly generated 1000 input parameter sets resulting in 1000 model outputs for the freshwater ecotoxicty characterization factors. The sample statistics were used to generate the mean ecotoxicty impacts as well as uncertainty levels at 95% confidence.

## 7.3 Results and Discussion

The effect factor (EF), fate factor (FF), and exposure factor (XF) of the five ILs studied are shown in Table 7-3.

**Table 7-3:** USEtox based effect factors, fate factors, exposure factors, and characterization factors for different ionic liquids

| IL | EF [PAF $m^3$ /kg] | FF [days] | XF | CF (CTUe/kg) |
|---|---|---|---|---|
| $[Bmim]^+[Br]^-$ | 4.75 | 131.5 | 0.9996 | 624.375 |
| $[Bmim]^+[Cl]^-$ | 7.47 | 100.1 | 0.9996 | 747.448 |
| $[Bmim]^+[BF_4]^-$ | 6.73 | 122.4 | 0.9996 | 823.422 |
| $[Bmim]^+[PF_6]^-$ | 6.49 | 142.9 | 0.9996 | 927.05 |
| $[BPy]^+[Cl]^-$ | 12.34 | 143.3 | 0.9998 | 1767.97 |

The fact that XFs are very close to 1 or 100% shows that the aquatic organisms will be exposed to nearly all of the ILs once released to the freshwater bodies. The overall characterization factors for freshwater ecotoxicity of the five ILs are also shown in Table 7-3. These characterization factors along with release fraction was used to estimate ecotoxicity impacts of release during use phase. For comparison purposes, we present these IL characterization factors along with CFs of few other conventional chemical compounds (gathered from TRACI 2.1 database) in Table E-3 of Appendix E.

Cradle-to-gate LCI of the production of the five ILs, each consisting of 102 air emissions (97 for pyridinium), 121 water emissions, and 35 soil emissions was translated into freshwater ecotoxicity impacts using ecotoxicity characterization factors from TRACI 2.1 database. The error bars based on 95% confidence interval of the Monte Carlo results are reported in Figure 7-1 along with the mean value of ecotoxicty impacts for both production and release of ILs.

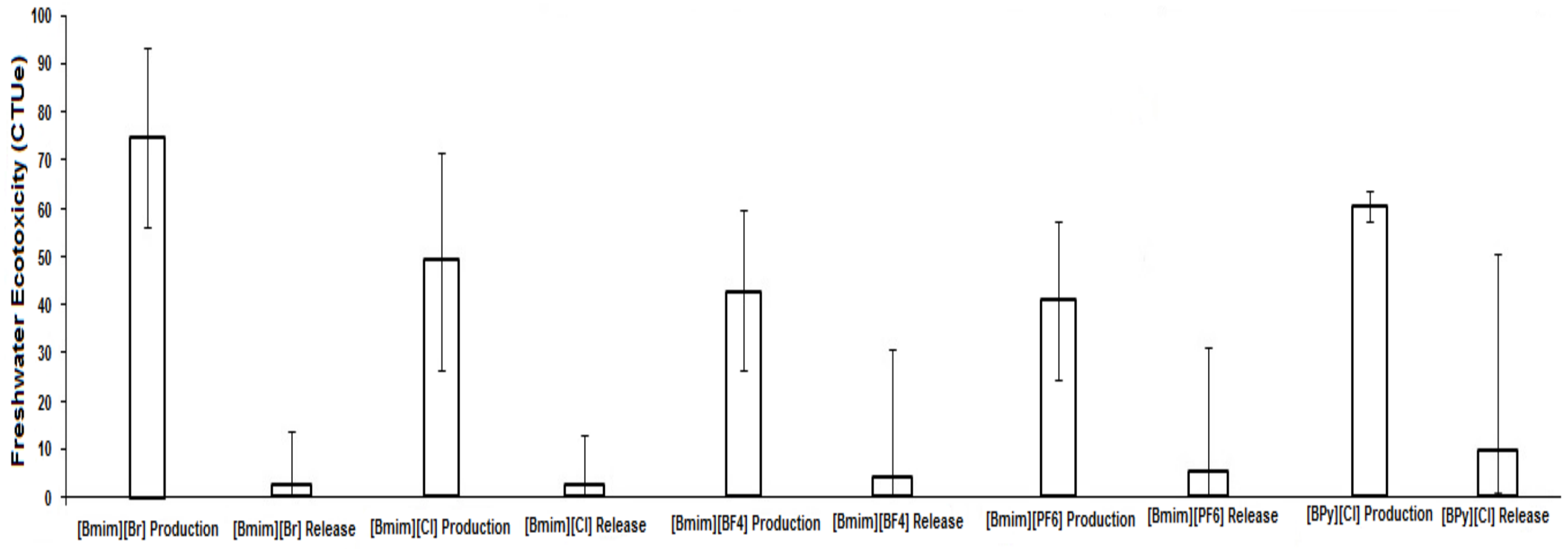

**Figure 7-1:** Ecotoxicity impacts related to production and use phase release of ILs

Results from Figure 7-1 reveal that for all studied ILs the mean freshwater ecotoxicity impacts attributable to the production phase of the ILs is nearly one order of magnitude higher than mean impacts due to their estimated release to the freshwater during use phase. This finding suggests that future research should not only focus on designing and identifying ILs with minimal toxicity but, more importantly, focus on improving the environmental profile of upstream production steps. Next, for the IL production phase, we examined relative contribution of ecotoxicity impacts due to upstream cradle-to-gate energy use versus cradle-to-gate direct release of materials and chemicals (i.e. chemical release during production of reagents, precursors, reactants etc.).  We find that majority of ecotoxicity impacts of production of ILs arise from chemical releases associated with the upstream production steps rather than energy use. Table 7-4 breaks down the contributions) of these two categories. For the five ILs considered in this study the average energy related ecotoxicity impacts was about 17% while the average chemical or material related ecotoxicity impacts was about 83%.

**Table 7-4:** Breakdown of energy and material related ecotoxicity impacts of IL

| IL | Energy Related Ecotoxicity Impacts | Material Related Ecotoxicty Impacts (%) |
|---|---|---|
| $[Bmim]^+[Br]^-$ | 7.5 | 92.5 |
| $[Bmim]^+[Cl]^-$ | 20 | 80 |
| $[Bmim]^+[BF_4]^-$ | 27 | 73 |
| $[Bmim]^+[PF_6]^-$ | 21 | 79 |
| $[BPy]^+[Cl]^-$ | 8 | 92 |

Since ILs are a combination of cations and anions, we can separate their production related ecotoxicity impacts into cation related impacts and anion related impacts. Table 7-5 lists these numbers for the IL $[Bmim]^+[Br]^-$. About 12.7% of the total ecotoxicty impacts can be attributed to cation production steps and the remaining can be attributed to anion production steps. In addition, we find that almost 86% (out of the 87.3%) of the anion production ecotoxicty impacts are due to the direct emission of the three chemicals Bromobenzene, Benzene, 1,2-dibromobenzene. As for cation production steps, metal emissions are fully responsible for the ecotoxicity impacts (12.7%). Further analysis shows that for production of $[Bmim]^+[Br]^-$ approximately 5% of the total cradle-to-gate life cycle ecotoxicity impacts are due to upstream emissions related to natural gas/electricity production for energy purposes. The remaining 95% are associated with the direct release of materials during upstream processes (including the process of natural gas extraction/processing for non-energy purposes). The above results show that we cannot decrease the production side ecotoxicity impacts of $[Bmim]^+[Br]^-$ by simply switching to alternate energy sources (such as renewables). Therefore, it is clear that researchers need to identify less impactful synthesis routes or control emissions associated with certain key precursors to further minimize freshwater ecotoxicity impacts of this IL. Note that the three chemical (Bromobenzene, Benzene, 1,2-dibromobenzene) releases that contribute the most (86%) towards $[Bmim]^+[Br]^-$ production ecotoxicity impacts are all by-products in the synthesis steps related to 1-Bromobutane. Therefore, alternate methods or routes to produce 1-Bromobutane or tight control of chemical releases during this step, has the potential to significantly decrease the total ecotoxicty impacts. Note that for $[Bmim]^+[Cl]^-$ ecotoxicity impacts due to cation and anion production are roughly equal while for $[BPy]^+[Cl]^-$ the cation related impacts are significantly higher (76%) than that of anion. As for the IL $[BPy]^+[Cl]^-$ the major portion (76.1%) of ecotoxicity impacts can be attributed to the release of just one chemical chloramine during the upstream step of production of pyridine which is a key precursor for the cation. Another interesting fact emerges when we compare the ecotoxicity impacts of $[Bmim]^+[Br]^-$ and $[Bmim]^+[Cl]^-$. Both ILs have the same cation, similar (halogen) anion, similar precursor materials and very identical synthesis routes. However, their production side ecotoxicty impacts vary a lot (48.3 CTUe and 16.4 CTUe respectively). This variation can entirely be attributed to the higher ecotoxicity impact associated with bromobenzene emission during the production of reactant 1-bromobutane in comparison to lower ecotoxicity impacts of chlorobenzene emission during the production of reactant 1-chlorobutane. These results point to the fact that overall production side ecotoxicity impacts are very sensitive to certain key chemical releases during the upstream synthesis steps. Next, we compare the sensitivity of the results to the production of one of the precursors, hydrogen chloride (HCl) or hydrochloric acid ($HCl.H_2O$), through four different routes/processes.

**Table 7-5:** Breakdown of freshwater ecotoxicity impacts of ILs associated with use phase release

| IL | Impact ($CTU^e$) | Reactants | Emissions | Impact ($CTU^e$) | % |
|---|---|---|---|---|---|
| $[Bmim]^+[Br]^-$ | 48.3 | 1-Methylimidazole | Metals | 6.12 | 12.7 |
| | | 1-Bromobutane (BuBr) | Bromobenzene<br>Benzene<br>1,2-dibromobenzene | 38.6<br>2.03<br>0.809 | 80.0<br>4.2<br>1.7 |
| $[Bmim]^+[Cl]^-$ | 16.4 | 1-Methylimidazole | Metals | 7.68 | 46.8 |
| | | 1-Chlorobutane (BuCl) | Chlorobenzene<br>Benzene<br>1,2-dichlorobenzene | 6.4<br>1.16<br>0.46 | 39<br>7.1<br>2.8 |
| $[BPy]^+[Cl]^-$ | 53.2 | Pyridine | Chloramine | 40.5 | 76.1 |
| | | 1-Chlorobutane (Bu Cl) | Chlorobenzene<br>Benzene<br>1,2-dichlorobenzene | 6.65<br>1.2<br>0.48 | 12.5<br>2.2<br>0.9 |

HCL is used in the production of 1-chlorobutane which is needed for production of both $[Bmim]^+[Cl]^-$ and $[BPy]^+[Cl]^-$. The most common industrial process for the production of HCl is the chlorination of benzene which has very high ecotoxicty impacts as seen in Table E-4 (Appendix E). In comparison, we find that three other less common alternate processes that can be used to produce HCl have very nominal impacts (See Table E-4). This analysis shows that identification and scale-up of alternate efficient processes has the potential to significantly decrease the production side ecotoxicity impacts of ILs. It is also clear that ecotoxicity impacts due to the direct release of ILs during use phase is very sensitive to the release fraction. Therefore, through efficient manufacturing processes, careful control of IL release into water, and efficient recovery and reuse we can further reduce use phase ecotoxicity impacts.

In summary, the findings of this paper show that, for the studied ILs, ecotoxicity impacts related to the emissions associated with upstream IL production steps are significantly greater than impacts due to their estimated direct release during use phase. Further analysis revealed that ecotoxicity impacts due to chemical releases during upstream production steps outweigh upstream emissions related to energy use. It is also evident that one or two key precursor chemicals involved in the upstream steps of IL production contribute disproportionately high towards overall ecotoxicity impacts. We also see that different approaches (processes) to produce key precursor chemicals can result in widely varying ecotoxicty impacts. We propose that future research should focus on developing synthesis and purification steps that reflect green chemistry and green engineering principles[256] with an emphasis on lowering the life cycle impacts of the IL production phase. Few areas of importance would include tight control of chemical release, improve reaction yields and process efficiencies, identify alternate production processes for key precursors, and chemical recovery and reuse during the synthesis steps of ILs and their precursors. The results from this paper encourage further investigation into life cycle ecotoxicity impacts of other types of ILs. Future technologies based on ILs should consider the full life cycle ecotoxicity impacts in order to assess their risks and benefits. Due to the emerging nature of IL applications, life cycle assessment (LCA) studies are crucial now, as they are most beneficial during the early stages of technology development and can help avoid unintentional shift of environmental burdens from one stage to another. Such studies will enable integration of life cycle impacts as part of design requirements and encourage chemists and engineers to develop truly "green" ILs.

## 7.4 Life cycle assessment of energetic ionic liquids

Energetic materials are used as explosives or as fuels. They release large amount of energy when they decompose. In the case of explosives all energy is released rapidly while in the case of fuels energy is released in a controlled manner. These materials derive their energy content from oxidation of the carbon backbone or from their high positive heats of formation. The general requirements for energetic materials are high energy density, thermal stability, low sensitivity to impact and low toxicity.[262] Traditional energetic materials that are commonly used in explosive formulations are HMX (1,3,5,7-tetranitro-1,3,5,7-tetraazacyclooctane), RDX (1,3,5-trinitro-1,3,5-triazacyclohexane) and TNT (2,4,6, trinitrotoluene).[263] Hydrazine derivatives are widely used as energetic fuel in rocket propulsion systems.[264,265] When discharged to the environment energetic materials will interact with biological systems. Use of energetic materials such as TNT, and RDX can leave residues which can potentially impact environmental and human receptors.[266] Monitoring studies reveal that some of these munition compounds persist at the sites where they were produced or processed.[267] Unexploded and low-order detonation residues containing TNT, RDX and HMX have been pointed out as main sources of groundwater contamination in military training ranges.[268] Indeed, munition compounds, such as RDX, have been detected in sole-source drinking water aquifers in military ranges such as Camp Edwards.[269] These chemicals have been found to be moderately to highly toxic to freshwater organisms.[266] In addition low concentrations of explosive compounds have been measured in marine sediments.[270]

With an aim to address some of the above environmental concerns researchers are exploring other *green* energetic material formulations. There is growing interest in the development of new energetic ionic salts and liquids for use as aerospace propellants and explosives.[264] As energetic materials, ionic salts offer several advantages over conventional energetic molecular compounds that include negligible volatility (ease of handling) and high density.[263] Energetic ionic salts can be prepared by combining energetic cations such as 1,2,3 triazolium with energetic anions such as nitrates, perchlorate and dinitramide. The high heats of formation of these salts are primarily due to the presence of nitrogen containing cations and anions.[263] Nitrogen-rich heterocyclic energetic salts are of particular interest.[271,275,276,277,278] A large number of ionic salts that are based on a triazole derivative have been proposed as energetic materials.[273,274,280] Triazole has a molecular formula of $C_2H_3N_3$ with a five-membered ring that contain three nitrogen atoms located at 1,2,3 or 1,2,4 positions. 1,2,4 triazole and 1,2,3 triazole have heats of formation values of 109 KJ/mol and 272 KJ/mol respectively.

As discussed previously, one of the main driving forces for the discovery and development of new energetic materials, such as ionic salts, is the mitigation of environmental and toxicological hazards associated with currently used materials. Manufacture of chemicals through environmentally friendly approaches represents a fundamental industrial challenge.

The energetic ionic salts possess lower vapor pressures and higher densities compared to non-ionic molecules.[263] Due to their negligible vapor pressure they are usually considered as “green” alternatives to volatile molecular compounds. In addition, ionic salts have tunable physical and chemical properties that enable us to tailor their structures for task specific applications such as energetic materials. Since ionic liquids and salts are of inherently less risk to human health and the environment they are considered as green chemicals. However, a more fundamental definition of green chemistry involves reducing or eliminating the use or generation of hazardous substances in the design, manufacture and application of chemical products.[281]

In order to legitimately evaluate the greenness of ionic salts as energetic materials, it is not only enough to consider the inherently benign nature of the chemical, but also need to take a holistic view that considers environment and health impacts associated with the entire life cycle of their production including direct environmental emissions during the production phase and indirect emissions associated with energy use in their production. This study presents the first comprehensive cradle-to-gate life cycle assessment that considers all stages involved in production of 1,2,3 triazolium nitrate, a triazole based energetic ionic salt, and compares it with the environmental impact associated with production of TNT on a functional unit basis. This approach will allow us to systematically investigate whether ionic salt based energetic materials provide any environmental benefits in comparison to traditional energetic materials. The chemical structure of 1,2,3 triazolium nitrate and TNT are shown in Figure 7-2.

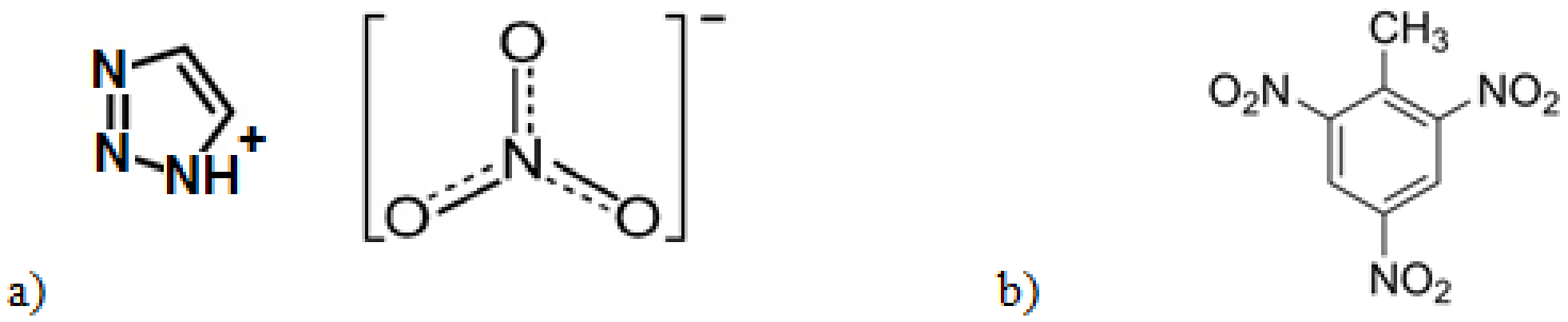

**Figure 7-2:** a) 1,2,3 triaolzium nitrate; b) TNT

There are several challenges involved in performing an LCA of ionic salts. Most of these challenges are due to the fact that ionic salts are a new class of compounds that are emerging. Ionic salts are not yet produced in large scales in commercial plants and there is no primary data available on material/energy consumption and direct environmental discharges. Process design and simulation software cannot be used to model production processes of ionic salts due to lack of comprehensive physical and thermodynamic property models for these salts and their precursors. Therefore, simulation of material and energy balances of ionic salt production processes becomes very difficult. The other important challenge in modeling environmental impacts is that, there are no emissions factors available in LCA databases such as Ecoinvent and Gabi for several precursors (reactants) that are required for IL production. Due to these limitations no LCA study has been done on ionic salts. To our knowledge, even for ionic liquids, there have been only few LCA studies that have considered them in their analysis.[282,283,284] In order to overcome the above mentioned challenges we use a theoretical approach to estimate theoretical energy requirements for reaction and separation steps involved in ionic salt production. Then we adjust the theoretical energy requirements to actual energy consumption by accounting for energy losses through the use of data from a comparable industrial process. Direct discharges of the ionic salt and its precursors to the environment during the production phase are assumed to be negligible. This energy and associated environmental loads constitute the inventory for life-cycle assessment (LCA) method.

### 7.4.1 Process and Energetic requirements for triazolium nitrate and TNT synthesis

***Synthesis of 1,2,3 triazolium nitrate***: The main reaction for synthesis of triazolium nitrate proposed by Drake *et al.*[262] and shown in eqn. (7-10) is adopted for this study. However, the emission factors (life cycle emissions) for the reactants involved are not available in standard LCI databases such as Ecoinvent. Therefore, we consider a series of upstream reactions, eqns. (7-4a) to (7-9), that constitute the *life cycle tree* for the production of ionic slat 1,2,3 triazolium nitrate. We calculate theoretical energy requirements for each of these steps that are part of the reaction tree. Major energy consumption in these batch processes would relate to the reaction and separation stages. Wherever appropriate, we make further assumptions of minimal separation energy requirements (and therefore ignore them) when the products are in two different phases (easy to separate) or the product is of high yield (no need to separate small quantities of by product). Actual industrial scenarios involving potential future scale-up are expected to be more energy intensive, as in an industrial plant the actual energy consumption is few times greater than theoretical energy requirements due to heat and energy losses. To capture this effect, we did a comprehensive review of several studies and found this factor to vary between 3 to 5 times that of theoretical energy requirement. In order to make an adjustment we selected a comparable process (synthetic production of sodium carbonate) for which industrial energy consumption data was available.[285] We calculated the theoretical energy requirement for this process and compared it with actual energy consumed and found that actual electricity consumption is 3.2 times higher than theoretical electricity requirement while actual natural gas consumption is 4.2 times higher than theoretical natural gas requirements. We also assume that for exothermic reactions electricity is used for cooling and for endothermic reactions natural gas is used for heating. We use the two correction factors in all our calculations to transform theoretical energy requirement to actual energy consumption.

$$\mathrm{Nacl} \rightarrow \mathrm{Na}^{+} + \mathrm{Cl}^{-} \tag{7-4a}$$

$$\mathrm{Na}^{+} + e^{-} \rightarrow \mathrm{Na} \tag{7-4b}$$

$$2\ \mathrm{Na}\ (s) + 2\ \mathrm{NH_3}\ (\mathrm{l}) \xrightarrow{-40\ ^{\circ}\mathrm{C}} 2\ \mathrm{NaNH_2}\ (\mathrm{s}) + \mathrm{H_2}\ (\mathrm{g}) \tag{7-5}$$

$$2\ \mathrm{NH_3}\ (\mathrm{g}) + 2\ \mathrm{O_2}\ (\mathrm{g}) \xrightarrow{25\ ^{\circ}\mathrm{C}} \mathrm{N_2O}\ (\mathrm{g}) + 3\mathrm{H_2}\ \mathrm{O}\ (\mathrm{l}) \tag{7-6}$$

$$2\ \mathrm{NaNH_2}\ (\mathrm{s}) + \mathrm{N_2O}(\mathrm{g}) \xrightarrow{25\ ^{\circ}\mathrm{C}} \mathrm{NaN_3}\ (\mathrm{s}) + \mathrm{NaOH}\ (s) + \mathrm{NH_3}\ (\mathrm{g}) \tag{7-7}$$

$$\mathrm{NaN_3}\ (\mathrm{s}) + \mathrm{Hcl}(\mathrm{l}) \xrightarrow{65\ ^{\circ}\mathrm{C}} \mathrm{HN_3}\ (\mathrm{g}) + \mathrm{Nacl}\ (s) \tag{7-8}$$

$$\mathrm{HN_3}\ (\mathrm{l}) + \mathrm{C_2H_2}(\mathrm{g}) \xrightarrow{25\ ^{\circ}\mathrm{C}} \mathrm{C_2N_3H_3} \tag{7-9}$$

$$\mathrm{C_2N_3H_3}\ (\mathrm{l}) + \mathrm{HNO_3}(\mathrm{l}) \xrightarrow{25\ ^{\circ}\mathrm{C}} 1,2,3\ \text{triazolium Nitrate}\ (s) \tag{7-10}$$

The production of sodium is based on well-known electrolysis cell process, eqns. (7-4a) and (7-4b). The potential required to oxidize $Cl^-$ ions to $Cl_2$ is -1.36 volts and the potential needed to reduce $Na^+$ ions to sodium metal is -2.71 volts. Therefore, a potential of at least 4.07 volts is required to drive this reaction.[286] In the second step, sodium (solid) and ammonia (gas) are reacted at 375 ℃ to produce sodium amide and hydrogen, eqn. (7-5).[287] Sodium amide is in liquid phase and hydrogen is in the gas phase at this temperature. Thus, we assume that the energy requirement for separation of the two phase products in a small scale batch plant is equal to the energy required for cooling sodium amide from reaction temperature to room temperature. We calculated the theoretical heat of reaction as -2.12 MJ/Kg and theoretical heat of separation as 0.634MJ/Kg. Accounting for correction factors this translates into a total cooling load requirement of 2.0448 KWh per kg ionic salt and total heating load of 2.66 MJ per kg ionic salt (equivalent to 0.0616 $m^3$ of natural gas/kg).In the next step ammonia (gas phase) and oxygen are reacted at room temperature, to produce nitrous oxide and water (Eq. 3). Nitrous oxide is in gas phase and water is in liquid phase at this temperature. Thus, we assume no significant energy requirement for separation in a small scale batch plant. We calculated the theoretical heat of reaction to be -15.47MJ/Kg. Accounting for correction factor this translates to 11.162 KWh per kg ionic salt.

In the next step, sodium amide (solid phase) and nitrous oxide are reacted at 200 $^oC$, to produce sodium azide (solid phase), sodium hydroxide (solid phase) and ammonia (gas phase). The energy requirement for separation stage is based on solid-solid separation of the two solid products and the theoretical heat of reaction was -3.295 MJ/Kg. Accounting for correction factors, the total cooling load requirement translates to 2.87 KWh/kg and total heating load requirement translates to 2.60 MJ/Kg (0.0602 $m^3$ of natural gas per kg of sodium azide). In the next step sodium azide (solid phase) is reacted with hydrochloric acid (liquid phase) at 65 $^oC$, to produce hydrazoic acid ($HN_3$) and sodium chloride salt.[288] Since one of the products is a gas and the other a solid, the two phases can be separated through a one step flash drum. Therefore we assume that for a batch plant the energy required for separation is minimal. We calculated the theoretical heat of reaction to be –2.9 MJ/kg. Accounting for correction factor, this translates to 2.096 KWh/kg ionic salt. In the next step hydrazoic acid ($HN_3$) and acetylene gas are reacted at 25 $^oC$, to produce 1,2,3 triazole ($C_2N_3H_3$).[289] The yield for this reaction is 99 %. In view of very high yield (low un-reacted materials) and presence of no important byproducts, for all practical purposes 1,2,3 triazole ($C_2N_3H_3$) can be considered pure. Thus, we assume no significant energy is required for separation. We calculated the theoretical heat of reaction as -2.275 MJ/kg. Accounting for correction factor, this translates to 3.68 KWh per kg ionic salt. In the next step 1,2,3 triazole ($C_2N_3H_3$) and nitric acid ($HNO_3$) are reacted at 25 $^oC$, to produce 1,2,3 triazolium nitrate (the energetic salt) with a yield, 98.9 %. Due to high yield of reaction (very low non-reacted materials) and no important byproducts the product can be considered as a pure component. Thus, we assume no significant energy is required for the separation part and the main energy consumption is the reaction phase. We calculated the theoretical heat of reaction as +1.73 MJ/Kg. Accounting for correction factor, this translates to 7.27 MJ/Kg of Ionic salt (0.168 $m^3$ Natural gas per kg of Ionic salt). The entire energy and material balance of the life cycle tree to produce 1 Kg of the ionic salt is shown in Figure 7-3.

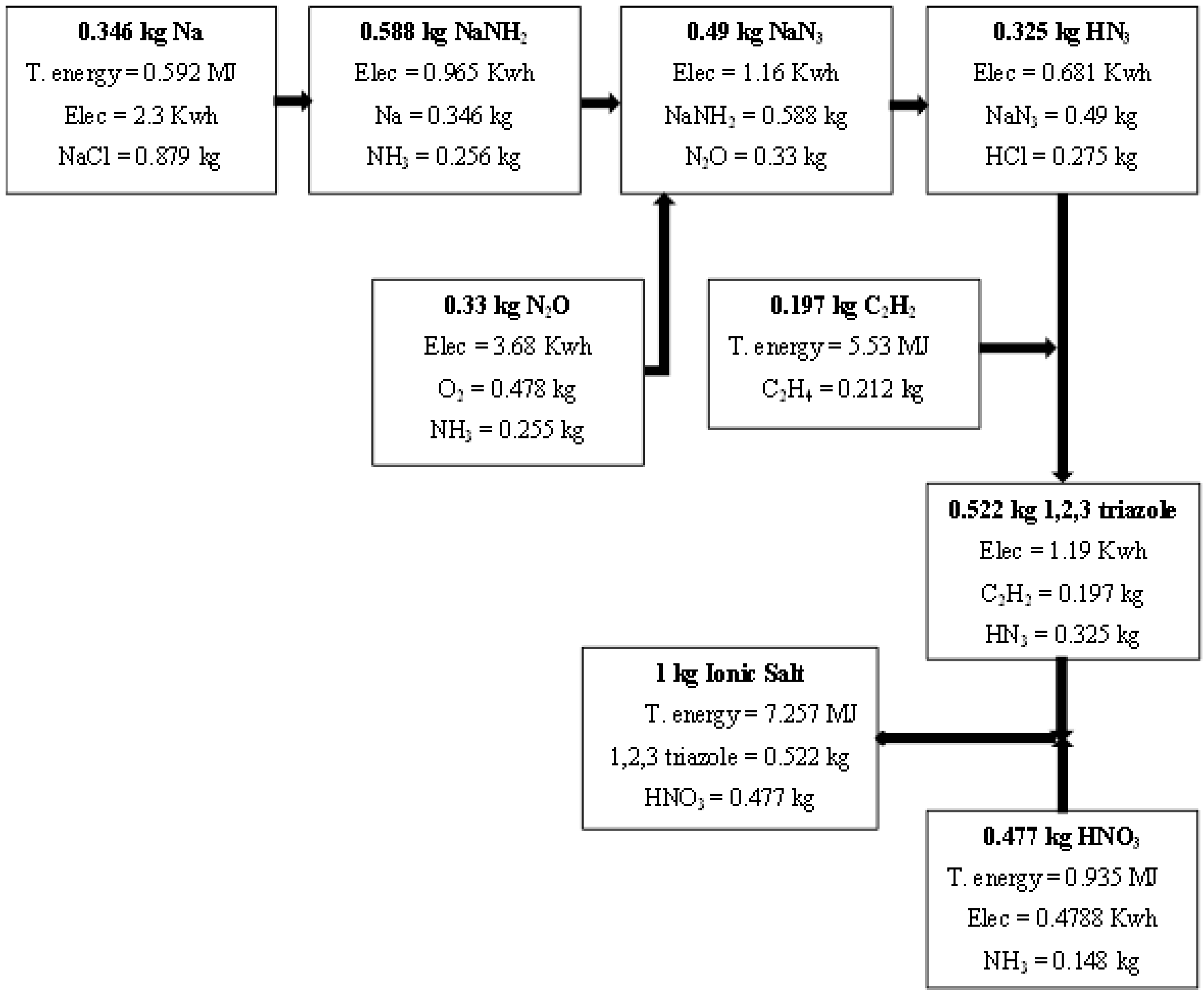

**Figure 7-3:** Material and energy flows associated with the life cycle tree for producing the ionic salt 1,2,3 triazolium nitrate

***Synthesis of TNT:*** TNT production is based on the synthesis procedure reported by Tadeusz[290] In the considered process toluene and nitric acid are reacted at 80 ℃ (both in liquid phase) to give $\alpha, \beta$ and $\gamma$ trinitrotoluene (TNT). The industrial results show that 95 % of the product is $\alpha - \mathrm{TNT}$ and the rest is distributed between $\beta - \mathrm{TNT}$ and $\gamma - \mathrm{TNT}$. We assume that the major energy requirements for TNT production relates to the reaction step and separation step (separation of $\alpha - \mathrm{TNT}$ from byproducts). The theoretical energy requirement for the reaction step equals to heat of reaction calculated as -1.768 MJ/kg. The actual energy consumed by accounting for energy losses was estimated using comparable plant data as described previously. The estimated actual electricity requirement for cooling in reaction and separation steps is 2.638 KWh per kg TNT. Energy requirement for separation purposes has been predicted to be 0.902 MJ/Kg.

### 7.4.2 Life cycle assessment (LCA) of energetic ionic salts

*Functional Unit*: Energy content is the most appropriate functional unit for this comparative study. 1,2,3 triazolium nitrate and TNT have different energy content and their energy release mechanism also differs. While energy release from TNT is based on oxidation, the ionic salt relies on heat for formation. Therefore the heat of combustion for TNT and heat of formation for ionic salt were used as measures of energy content. A reference of 1 MJ energy content was used as the basis of comparison. On a mass equivalence basis this translates to following reference flow: 1 Kg of TNT equivalent to 1.62 Kg of ionic salt.

*System Boundary:* The system boundary includes the final step of ionic salt production (reaction and separation), upstream reaction/separation steps for the precursors (as defined by the reaction tree), electricity and natural gas production, upstream processes involved in electricity and natural gas production including raw-material extraction and transportation.

*Life Cycle Inventory:* Life cycle inventory (LCI) represents the collection of data on the material and energy inputs and emissions associated with the production of the energetic ionic salt. Material and energy flows constructed in the previous section were used as inputs to the life cycle inventory (LCI). Since majority of processes are either new for which industrial scale up has not been developed or physical, chemical and thermodynamic properties of the precursors are not available and hence chemical process simulation was not possible, we used data from the approach outlined in the previous section as inputs to the inventory. Due to the challenges associated with performing an LCA of new chemicals -that were outlined earlier- we consider this simplified approach as adequate for the scope of this study. Emission factors for production of electricity, natural gas and other starting materials of the life cycle tree were obtained from the U.S. life cycle inventory database.[291] The emission factor for electricity from grid was assumed as 70% generation from bituminous coal and 30% generation from natural gas, based on 2008 U.S. grid electricity data.[292] Since contribution of other renewable energy and nuclear sources to the grid were either very small or vary significantly depending on the location we assumed all grid electricity is from coal and natural gas. Emission factors for electricity production from bituminous coal included emissions from coal mining & transport and emissions from power plant. Emission factors for electricity production from natural gas (NG) included emissions from NG extraction from ground and transport, emissions from NG processing and emissions from power plant. Emission factors for natural gas combustion include emission from NG extraction from ground, emissions from NG processing and emissions from NG combustion in an industrial boiler. Life cycle emission factors for some of the materials in the life cycle tree (sodium chloride [NaCl], ammonia [$NH_3$], oxygen [$O_2$], hydrochloric acid [HCl], ethylene [$C_2H_4$] and nitric acid [$HNO_3$]) that were available in the US LCI database were used. Life cycle emission factors of the remaining materials (sodium [Na], sodium amide [$NaNH_2$], nitrous oxide [$N_2O$], sodium azide [$NaN_3$], hydrazoic acid [$HN_3$], ethylene [$C_2H_2$], 1,2,3 triazole [$C_2N_3H_3$]) were calculated using the theoretical approach explained in section 2. These emission factors have been applied to the inputs to calculate the life cycle emissions for ionic salt and TNT production thereby completing the output side of the inventory (LCI).

*Life Cycle Impact Assessment:* The life cycle impact assessment methods describe environmental impacts based on characterization factors. These characterization factors are developed by consideration of inherent characteristics of chemicals (for example toxicity) as well as information on fate and transport and possible mode of exposure. The life cycle impact assessment (LCIA) methodology based on Tools for the Reduction and Assessment of Chemical and other Environmental Impacts (TRACI) developed by U.S. Environmental Protection Agency was used in this study. This method was considered the most appropriate since it is based on United States data and models. This study considers only mid-point impacts as end-point impact modeling brings in additional uncertainty to the results and TRACI is primarily a mid-point impact assessment method. Midpoint impact categories quantify the relevant emissions and resources from the life cycle inventory in terms of common reference substances (e.g. Kg $CO_2$eq). The impact categories considered are: 1) Global warming; 2) Acidification; 3) Eutrophication; 4) Smog formation; 5) Human heath criteria; 6) Human health cancer; 7) Human health non-cancer; and 8) Ecotoxicity. Classification and Characterization steps of LCA were applied to relate individual elementary flows in the inventory to the impact categories and to identify relevant characterization factors based on the media to which the emissions occur. Normalization was not considered in this study as normalization factors based on U.S. data were not available.

*Sensitivity Analysis*: The main source of uncertainty in this study relates to the conversion factors used for translating theoretical electricity and thermal energy requirement to actual energy consumed in an industrial plant. A sensitivity analysis is performed to study the effect of varying the conversion factors ± 30%. We examine in detail how sensitive the results are to changes in these conversion factors.

### 7.4.3 Results and discussion

This section summarizes the main findings from comparing the ionic salt with TNT. The total scores of each environmental impact category for 1,2,3 triazolium nitrate and TNT are shown in Table 7-6. The impact profiles resulting from production of 1.62 kg of ionic salt and 1 kg of TNT are shown in Figure 7-4, with ionic salt impact set at 100% and TNT displayed as a level relative to the former. Relative comparisons between the ionic salt and TNT (Figure 7-4) show that in all of the analyzed categories ionic salt had significantly higher environmental/health impact than TNT. With respect to climate change, ionic salt production has roughly 3 times higher environmental burden than TNT production. In the category of human health, ionic salt is roughly 3, 4 and 4 times more impactful than TNT for criteria, cancer and non-cancer cases respectively. The environmental burden of IL is higher by approximately 2.5, 2, 2 and 4 times that of TNT for acidification, eutrophication, smog formation and ecotoxicity respectively.

The climate change indicator, global warming air (GW), is dominated by $CO_2$ emissions during the life cycle of both IL and TNT production. $CO_2$ emissions account for 96% of total

GW for both IL and TNT with methane accounting for the remaining 4%. Sulfur dioxide (70% for IL and 58% for TNT) and nitrous oxide (27% for IL and 33% for TNT) emissions dominate acidification indicator while photochemical smog is dominated by nitrogen oxide emissions (90% for both IL and TNT). Eutrophication potential is dominated by nitrogen oxide (98.98% for IL and 96.22% for TNT) emissions. The human health criteria indicator is dominated by sulfur dioxide emissions (93% for IL and 90% TNT) while human health cancer and non-cancer indicators are entirely due to benzene emissions. The ecotoxicity indicator is also entirely due to benzene emissions.

**Table 7-6:** Impact of ionic salt and TNT (functional unit: 1 MJ energy content)

| Category | Units | Ionic salt | TNT |
|---|---|---|---|
| Global Warming Air | Kg $CO_2$ eq. | 29.4738851 | 9.07309769 |
| Acidification Air | Kg $H^+$ mole eq. | 9.489402937 | 3.752113828 |
| HH Criteria Air | Kg $PM_{10}$ eq. | 0.023712006 | 0.007939867 |
| Eutrophication Air | Kg N eq. | 0.00281911 | 0.001383166 |
| Eutrophication water | Kg N eq. | 2.89911E-05 | 5.42903E-05 |
| Smog Air | Kg $O_3$ eq. | 1.588845 | 0.777182 |
| Ecotoxicity (Fresh air) | CTU-eco | 0.111861804 | 3.3806E-06 |
| Human health (Cancer) | CTU-cancer | 2.47276E-11 | 6.34994E-12 |
| Human Health (Non-cancer) | CTU-noncancer | 6.26146E-12 | 1.60791E-12 |

The results of the cradle to gate life cycle comparison unequivocally shows that energetic ionic salts, such as 1,2,3 triazolium nitrate, have larger environmental burden than traditional energetic materials such as TNT. This disproves the commonly accepted notion that ionic liquids and ionic salts are “green”. Though ionic compounds are inherently benign due to their negligible vapor pressure, this fact alone does not make them green. A holistic analysis that includes the inherent properties of the ionic salt, emissions associated with their production, exposure and end-of-life impacts need to be considered. This study provides greater insights into the *greenness* of energetic ionic materials through a more holistic approach.

Closer examination of the results reveal that majority of the life-cycle environmental burden can be attributed to energy consumption (electricity and natural gas). This is due to the fact the emissions during energetic material production phase (reaction and separation) dominates other phases such as raw material extraction and transportation. Moreover the environmental footprint of the ionic salts is much larger than TNT due to the fact that steps involved in producing ionic salts and their precursors are much more energy intensive than the steps to produce TNT. This translates into significant increases in environment and health impacts across all categories. Since most of the emissions in this study can be attributed to electricity and natural gas consumption, using more efficient industrial plants, and/or finding alternate synthesis pathways can result in net energy savings and help offset some of the emissions. Utilization of renewable resources such as solar, wind and waste biomass to produce electricity can significantly reduce the overall environmental footprint of the production process.

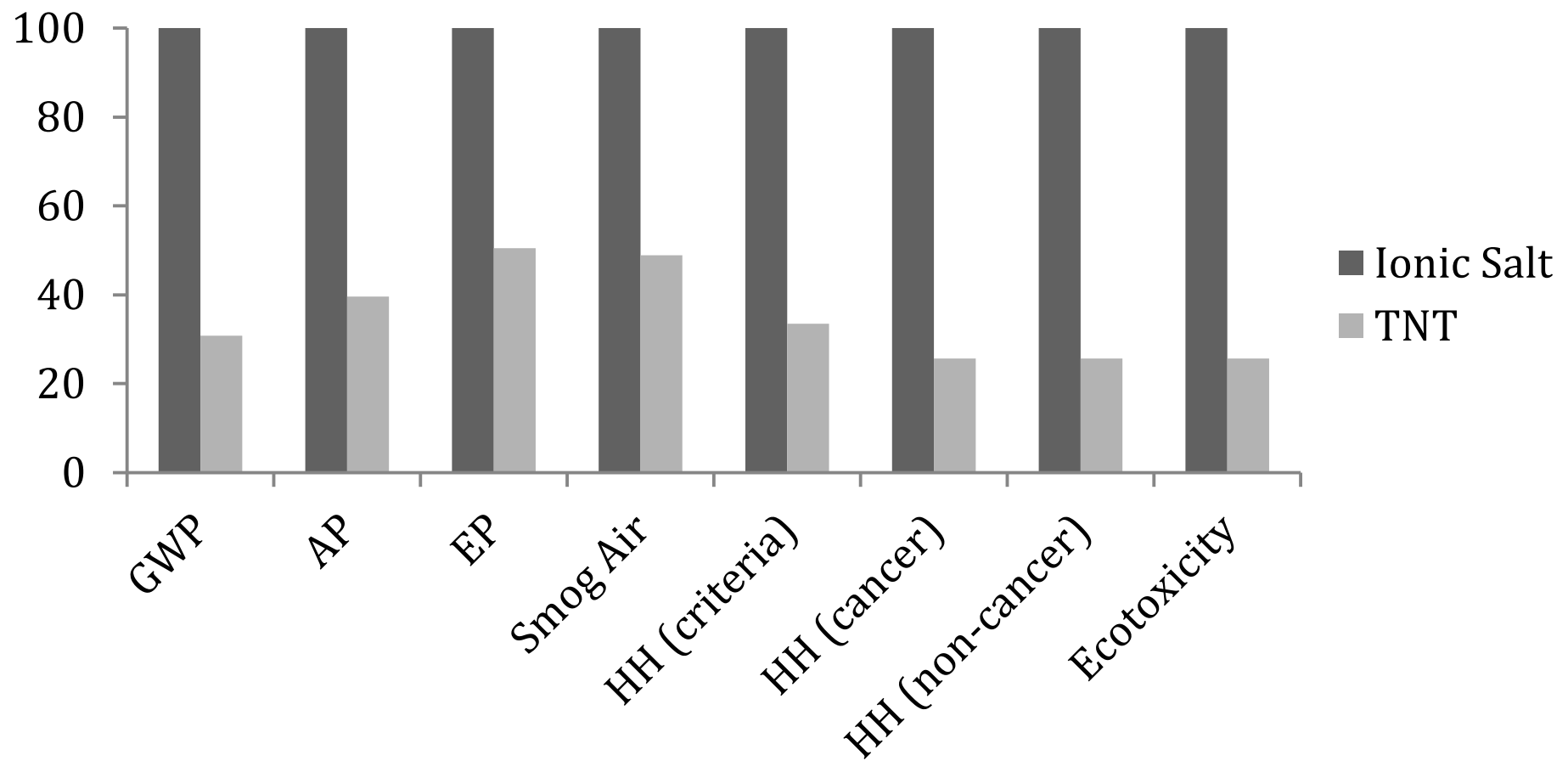

**Figure 7-4:** Comparison of scaled impacts of ionic salt and TNT (functional unit of 1MJ energy content): GWP (Global Warming Potential), AP (Acidification Potential), EP (Eutrophication Potential), HH (Human Health).

***Uncertainty***

Uncertainties related to conversion factors used for extrapolating theoretical energy calculations to actual energy consumption were addressed via sensitivity analysis. Additional sources of uncertainties are identified as follows: a) all calculations were based on the life cycle tree, eqns. (7-4a) to (7-10), and it is possible that alternate methods (reactions) exist for producing one or more of the precursors; b) the reaction yields are based on lab-scale experiments from literature; c) we assume majority of the environmental impact for ionic salt and precursor production processes (i.e. life cycle tree) can be attributed to the energy intensive reaction and separation steps which could be a source of uncertainty; d) the calculations in this study are based on the assumption of small scale batch processes for producing the ionic salt and precursors which could be another source of uncertainty if future manufacture plants are continuous.

### ***Sensitivity***

The sensitivity of the LCA results to '*theoretical energy-to-actual energy'* conversion factor is shown in Table 7-7 and Figure 7-5. The conversion factor was varied by ± 30% to study how sensitive the results are to this parameter. The error margins indicate that the results are highly robust to changes in this parameter and all conclusions that were previously arrived at are fully valid.

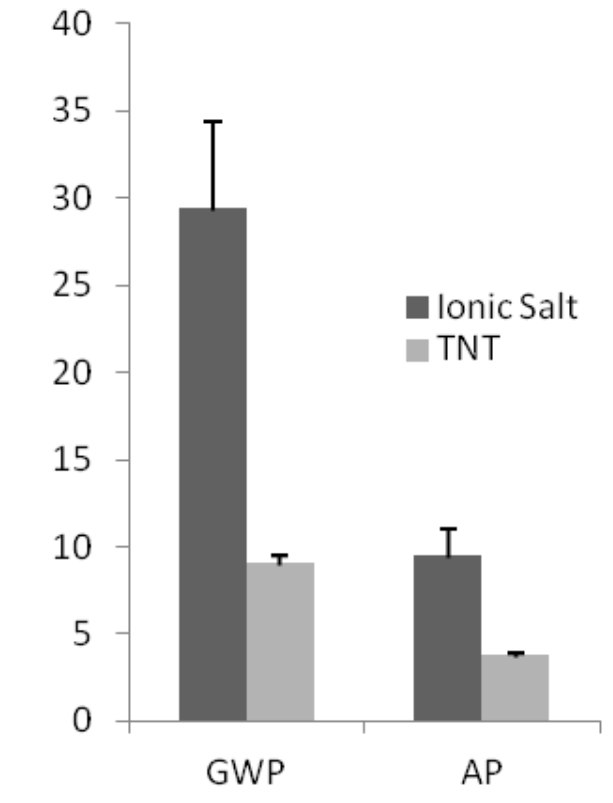

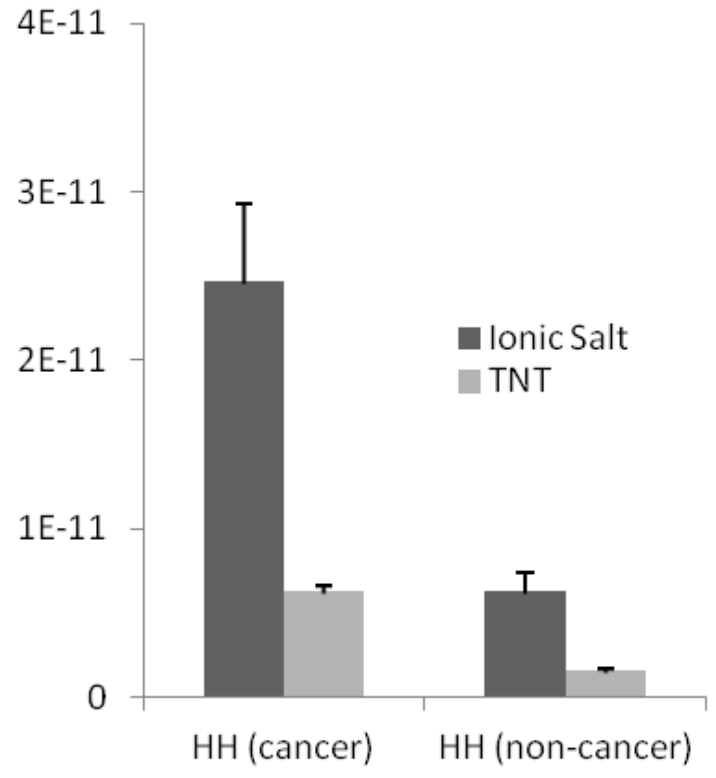

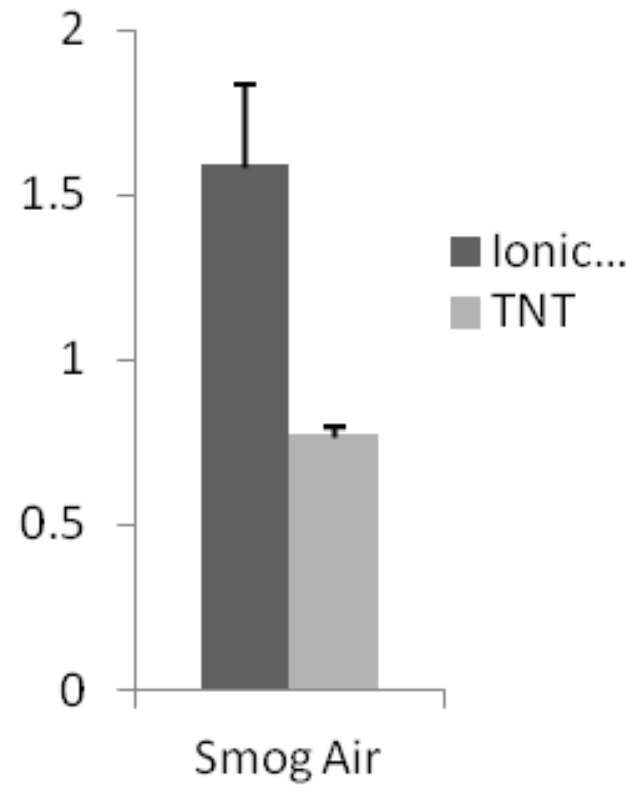

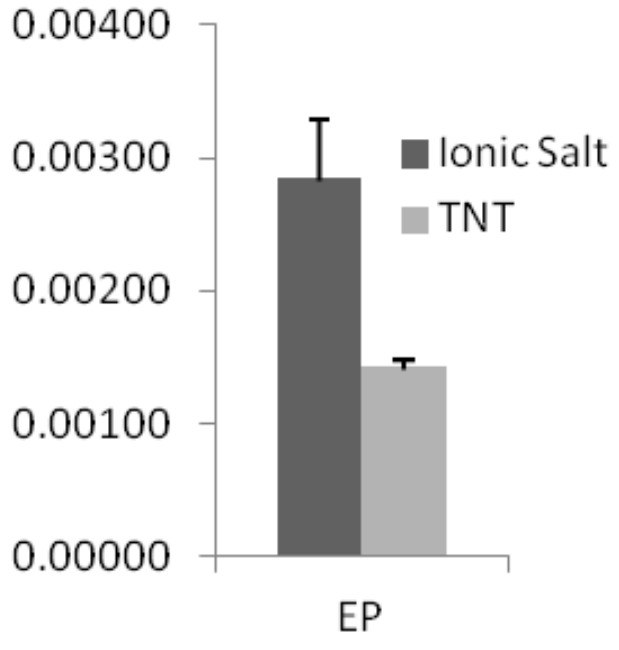

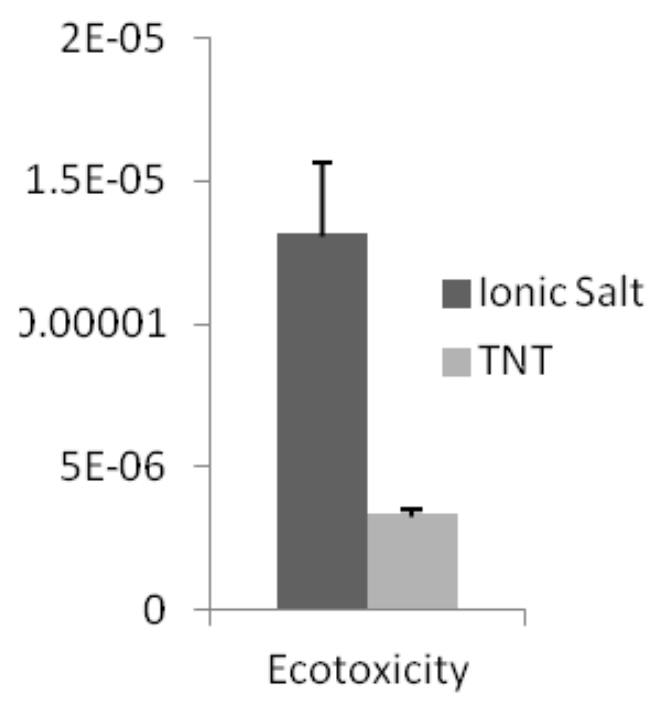

**Figure 7-5:** Sensitivity analysis of the scaled impacts of ionic salt and TNT (functional unit of 1 MJ): GWP (Global Warming Potential), AP (Acidification Potential), EP (Eutrophication Potential), HH (Human Health).

The proportional size of the error bars indicate that the impacts associated with the ionic salt are more sensitive to this parameter (conversion factor) than for TNT. This is due to the fact that there are more upstream reactions steps (shown in the life cycle tree) for ionic salt and hence the conversion factor is applied multiple times (in comparison to TNT). For bigger conversion factors the difference between ionic salt and TNT impacts will be even more than predicted. A comprehensive search of industrial data and a detailed analysis indicates that the considered conversion factors (3.2 and 4.2) fall in the lower end which implies that the LCA results presented here are conservative estimates.

**Table 7-7:** Sensitivity analysis

| Substance | GWP | AP | EP | Smog | HH (cancer) | HH (non-cancer) | Ecotox. |
|---|---|---|---|---|---|---|---|
| **Ionic Salt (IS)** | 29.4739 | 9.4894 | 0.00285 | 1.5888 | 2.473E-11 | 6.261E-12 | 1.32E-05 |
| **IS (+30%)** | 34.3751 | 11.029 | 0.00328 | 1.833 | 2.93E-11 | 7.420E-12 | 1.56E-05 |
| **IS (-30%)** | 24.5725 | 7.9499 | 0.00241 | 1.3447 | 2.015E-11 | 5.103E-12 | 1.07E-05 |
| **TNT** | 9.0731 | 3.7521 | 0.00144 | 0.7772 | 6.35E-12 | 1.608E-12 | 3.38E-06 |

| Substance | GWP | AP | EP | Smog | HH (cancer) | HH (non-cancer) | Ecotox. |
|---|---|---|---|---|---|---|---|
| **TNT (+30%)** | 9.48553 | 3.8905 | 0.00148 | 0.7988 | 6.616E-12 | 1.675E-12 | 3.52E-06 |
| **TNT (-30%)** | 8.65707 | 3.6126 | 0.00140 | 0.7554 | 6.081E-12 | 1.540E-12 | 3.24E-06 |

***Scenario analysis***

In order to investigate the influence of the use of renewable energy on ionic salt production processes, we developed two hypothetical scenarios as follows: 1) In the first scenario cooling energy (electricity) required for all materials needed during ionic salt production (final reaction/separation unit as well as all upstream reactions/separation units) comes from wind (renewable source); 2) In the second scenario wind energy is used only in the last two stages of “triazole” and “Ionic salt” production. The second scenario is more likely and meaningful due to the fact that an ionic liquid production plant is likely to purchase the primary raw materials (acetylene, hydrazoic acid and nitric acid – upstream processes) from other industries that would likely use electricity from grid. For both scenarios, natural gas is assumed to deliver the required heating energy. In both cases of comparison it is assumed that fossil energy is completely used for TNT production. The results for the two scenarios are summarized in Table 7-8 and Table 7-9 respectively. It can be concluded that under scenario one the environmental impact of ionic salt, in most categories, is lower than TNT. However, under scenario two (more realistic) the environmental impact of ionic salt is still significantly higher than TNT. It will be reasonable to assume that under a scenario where only renewable energy is used (in all stages) for both ionic salt and TNT production the magnitude of environmental impact for both cases will be lowered proportionally, however the final analysis and conclusions presented in relation to the comparison between them will still hold true (since ionic salt production processes consume significantly more life cycle energy than TNT production processes).

**Table 7-8:** Environmental Impact for scenario 1

| Category | Units | Ionic salt | TNT |
|---|---|---|---|
| Global Warming Air | Kg $CO_2$ eq. | 2.778853 | 9.073098 |
| Acidification Air | Kg $H^+$ mole eq. | 0.541855 | 3.752114 |
| HH Criteria Air | Kg $PM_{10}$ eq. | 0.000932 | 0.007940 |
| Eutrophication Air | Kg N eq. | 0.000347 | 0.001383 |
| Eutrophication water | Kg N eq. | 2.7504E-05 | 5.42903E-05 |
| Smog Air | Kg $O_3$ eq. | 0.197913 | 0.777181 |
| Ecotoxicity (Fresh air) | CTU-eco | 3.94306E-06 | 3.3806E-06 |
| Human health (Cancer) | CTU-cancer | 7.40643E-12 | 6.34994E-12 |
| Human Health (Non-cancer) | CTU-noncancer | 1.87544E-12 | 1.60791E-12 |

**Table 7-9:** Environmental Impact for scenario 2

| Category | Units | Ionic salt | TNT |
|---|---|---|---|
| Global Warming Air | Kg $CO_2$ eq. | 27.29388 | 9.0730977 |
| Acidification Air | Kg $H^+$ mole eq. | 8.758716 | 3.752114 |
| HH Criteria Air | Kg $PM_{10}$ eq. | 0.021851 | 0.007939 |
| Eutrophication Air | Kg N eq. | 0.002617 | 0.001383 |
| Eutrophication water | Kg N eq. | 2.88697E-05 | 5.42903E-05 |
| Smog Air | Kg $O_3$ eq. | 1.475256 | 0.777181 |
| Ecotoxicity (Fresh air) | CTU-eco | 1.24115E-05 | 3.3806E-06 |
| Human health (Cancer) | CTU-cancer | 2.33131E-11 | 6.34994E-12 |
| Human Health (Non-cancer) | CTU-noncancer | 5.90329E-12 | 1.60791E-12 |

The analysis presented in this study considers only the life cycle energy consumption. Environmental impacts associated with emissions were not assessed. In order to get a more complete picture it is necessary to perform a cradle-to-grave life cycle assessment that will include indirect emissions, direct material emissions (of ionic salt, precursors etc.) during production phase, as well as end-of-life (after use) impacts of the energetic materials. However, this is not currently possible as the state-of-the-art impact assessment methods, such as TRACI[293] and Eco Indicator[294], do not contain characterization factors for ionic salts and many of their precursor materials. Therefore, there is a great need for future research to focus on fate, transport, and mechanism of damage to human and ecosystem species by ionic salts, ionic liquids and their precursor materials. This will allow us to develop characterization factors for these compounds and help us investigate the exposure and end-of-life impacts of the emitted material. The results presented in this study are subject to several uncertainties.

# Chapter 8: Conclusions and Future Work

This study focused on the development of a novel computer-aided molecular design framework and environmental impact assessment methods of ionic liquids, a newer generation of materials which have attracted much attention during the past few years. Ionic liquids are shown to have the capability of replacing organic compounds used in industrial plants for a vast number of applications; from liquid-liquid extraction to thermal energy storage. One of the important components of green chemistry is attributed to the solvent medium in which a chemical reaction or an extraction process is carried out.[295] Traditional molecular solvents are proven to cause adverse environmental and human health impacts and therefore they should be replaced with greener alternatives whenever a better candidate, which can do similar tasks, is available. Jessop[296] states that one of the major challenges in the search for environmentally benign or green solvents is to ensure their availability. Jessop used the Kamlet-Taft plots to show that current list of green solvents populate only a small region of the entire spectrum of solvents needed for different applications and argues that large unpopulated areas of this diagram mean that future process chemists and engineers need solvents having specific desirable properties which are also green. Many studies have concluded that ionic liquids offer a great potential to satisfy this need by serving as the potential candidate to be used for several different applications. However, the true greenness of any chemical can be ascertained only when a holistic view of its environmental impacts is considered. Towards, this end this dissertation has made contribution in the realm of understanding and characterizing life cycle environmental performance of ionic liquids.

This dissertation also presented, an overarching framework, termed as CAILD, that can be utilized to design or tune ionic liquids through a modeling perspective. The details of the CAILD optimization approach along with several important case studies were presented. The success of ionic liquid design requires that the physical properties of these new compounds be predicted with an acceptable level of accuracy. Chapter two of this dissertation focused on contributions related to the prediction of two pure/physical properties of ionic liquids, namely melting point and viscosity, as it is a necessary step for successful application of ionic liquids. This was accomplished through a Quantitative-Structure-Property-Relationship (QSPR) approach were quantum chemistry (QC) based descriptors were used to develop physical property correlations.  The QC models which normally do not need any structural groups related parameters can be combined with QSAR/QSPR approach to predict the physical properties of chemical compounds with an acceptable precision. The used QC models rely on the energy profile and the charge density of the chemical compound of interest and can be utilized when there is no reliable group contribution approach available.

Chapters five and six of this dissertation were focused mainly on the prediction and use of solution properties, in this case activity coefficients, of ILs and solutes in a multicomponent mixture. Due to lack of data we proposed a new approach for fitting the interaction parameters based on QC and COSMO predictions. The activity coefficients estimated through the fitted UNIFAC model were used to predict the solubility of several solutes in ILs. The approach was used to find optimal solvents with desired properties (e.g. high solvency power towards the solute of interest, low melting point, etc.) for different applications. Chapter five was focused on designing an optimal IL for the extraction of aromatic compounds from an aliphatic-aromatic mixture. In chapter six a case study was presented where an optimal IL was designed to act as the solvent for a $CO_2$ capture process. Chapter 7 focused on studying the environmental impacts associated with the production and release of common ILs through development of their characterization factors. These characterization factors (CF) were used to calculate the freshwater ecotoxicty impacts associated with the direct release of the ionic liquids into the environment. In the case of freshwater ecotoxicty impacts, the impacts associated with the production of ILs far outweighed the impacts linked to the potential release into freshwater resources.

To conclude, the last section of chapter 7 described the results of comparative cradle-to-gate life cycle assessments (LCA) between two energetic materials, one being an energetic ionic liquid and the other being a convectional energetic compound. The results showed that for this specific case, when it comes to the life cycle impacts associated with the production, ILs might not be more environmentally benign than their molecular counterparts. This finding is interesting in the sense that it contradicts with the commonly accepted belief of ILs as green materials. This fact along with the high toxicity of some ILs towards aquatic organisms shows that one needs to be cautious when using the word “green” for all of the ionic liquids. This also shows that ionic liquids need to be tuned/designed specifically for a given application by considering not only their performance but also their environmental impacts before they are produced in larger commercial scales.

## 8.1 Limitations and Recommendations

For a computer-aided design framework to successfully search for optimal candidates in different cases, it needs to cover a broad range of structural groups acting as building blocks of the compound being designed through the model. That being said, the first challenge in computer-aided design of ionic liquids is that the available group contribution models of the physical properties of ILs do not span the entire spectrum of cations, anion and functional groups. Although, it is not impossible to overcome this challenge since there are only a limited number of structural groups making an enormous number of ILs, this has not yet happened. We strongly propose that future research should focus on experimental measurements and data collection of the physical properties of ionic liquids, to expand the available group contribution methods to cover a more diverse set of cations, anions and/or functional groups.

The second challenge is that for certain physical properties such as surface tension, there are no group contributions available. In addition, for certain physical properties such as viscosity, the available models either fails to predict the property of newer compounds with an acceptable range of accuracy or they are very complicated and cannot be integrated directly within a CAILD framework. In this study, we made use of quantum chemistry (QC) models combined with specific QSAR/QSPR correlations to predict viscosity and melting point of ILs. The limiting part was that even though, through QC models based correlations, we were able to predict the two physical properties of ILs with a relatively better accuracy (compared to existing group contribution methods), but it was not feasible to integrate them directly within a CAILD framework. Therefore, in certain case studies we first had to find a set of optimal and feasible IL candidates based on the objective function and other constraints and then evaluate the list of ionic liquids for their melting point and viscosity requirements using the developed correlations in chapter 2.

The third challenge relates to the accuracy of the group contribution models that we used as part of the CAILD model and the intrinsic uncertainty they carried. The group contribution (GC) models are meant to predict the physical properties of newer compounds, based on the type and number of the structural groups present in these compounds. In order to develop a good GC model for a physical property,  we need to have diverse set of chemicals with overlapping structural groups. In the next step for all of the selected chemicals, the experimental data on the physical property of interest should be either measured or collected from literature so that the data can be used to fit group contributions for different structural groups. These contributions along with the information on the type and number of groups can be utilized to predict the physical properties of newer compounds. Although this approach seems to be straightforward, the limitation resides in the uncertainty of the experimental data used to develop group contribution models. This problem can be addressed if the uncertainty of the experimental data is reported with the data, but for most ionic liquids that was not the case. In addition to that, experimental data collected from different sources normally do not have the same level of uncertainty. This means that the intrinsic uncertainty embedded in the group contribution models which are used to predict the physical properties of ILs in a CAILD model is not accurately measureable. These uncertainties will also be carried on to the design results i.e. the structure of the optimal IL. This study, was not able to capture these uncertainties as, for the most part, we were restricted to using group contribution models available in the literature for which uncertainty values were not reported.

Another challenge relates to the limitation of USEtox model which was used to develop the ecological characterization factors of common ionic liquids. The model itself is a state-of-the-art and most comprehensive model available on the fate and transport modeling of chemical compounds in the environment. Although, the USEtox model is appropriate for our calculations the liming part is that it was primarily developed for molecular compounds and the model has not been previously used for the case of ionic liquids. The large and asymmetric shape of the cations and anions in ionic liquids along with their electrical charges make the accuracy of USEtox for the case of ionic liquids questionable. Despite this limitation, other studies have used USEtox to develop characterization factors for materials such as Carbon Nanotubes (CNT) which are made of extensively large unit cells. Another limitation of using the USEtox model for the case of ionic liquids can be attributed to their potential dissociation in the water media since that would change the measured physical properties. The studies on the behavior of ionic liquids in the aquatic environments and the data on their dissociation factors are still very limited. The promising part is that some studies have suggested that ionic liquids mostly act as Lewis acids in water rather than as an ionic material such as sodium chloride. The aforementioned fact implies that ILs can be assumed to be mostly dissolved in water because of their hydrogen bonding interaction with water molecules rather than being completely disassociated. The fact that some ionic liquids are not soluble in water, despite their ionic nature, can add further credence to this hypothesis.

The results of the life cycle studies which were performed in this dissertation are limited to a few numbers of ionic liquids and should not be used to make any general conclusion or comments about ionic liquids in general. Conclusions such as ionic liquids are more impactful compared to the case of organic compounds or similar to that should be only made case by case and after careful consideration of the results obtained from the comprehensive life cycle assessments. The fact that there is no actual data on the production of ionic liquids make results of their life cycle assessment studies more limited and uncertain.

## 8.2 Contributions

Even though there is still dispute over the time when the first ionic liquid was discovered, Ethanolammonium nitrate was first reported by Gabriel and Weiner[297] in 1888 and it was as early as 1943 that the term ionic liquid was first used. Since then, thousands of scientific papers and patents have been published in the areas of synthesis and use of ionic liquids. Ionic liquids are one of the most widely studied categories of materials during the past few years. Not a single day goes by without people thinking about a new application for ionic liquids. That being said, the process of choosing the best and most optimal ionic liquid is quite challenging. Ionic liquids have a very wide range of properties so any change in their structures can change their physical and solution properties significantly. An ionic liquid which is well suited for a given application might be a bad choice for another application. Therefore the challenge of optimal IL selection can be addressed accordingly through their intelligent design.

Despite the unique characteristics and promising properties of ionic liquids, only a few hundred (compared to $10^{14}$ theoretically possible structures) ionic liquids have been synthetized and tested to this date. Therefore, people have only tried a very small subset of all feasible ionic liquids for any application of interest presenting a great platform to discover new ionic liquids.

On the other hand, property prediction through structure-property models is limited because such models have not extensively been developed. The above limitations, need to be addressed before any attempt of large scale commercialization of ILs is made.

The ultimate aim of the CAILD methodology, would be to narrow down to a smaller set of ionic liquid candidates from the millions of available alternatives. The final IL, which is most optimal for a given task can then be selected by *ab initio* computational chemistry calculations or by experimental verification of the designed compounds. The progress towards designing newer ionic liquids through the proposed CAILD model, will not only contribute towards our understanding of the relationship between cation and anion structures and the physical properties of ionic liquids, but will also provide a mechanism to engineer new ionic liquids which can also be environmentally benign for a given application.

Another important contribution of this study was to perform life cycle assessment on well-known ionic liquids. The environmental impacts associated with the production of an energetic ionic salt was compared with a conventional explosive material. The approach we used for this study was innovative and unique in the sense that certain conversion factors were exclusively developed for this study which were used to convert theoretical energy requirements of the production of the ionic salt to actual industrial data. Since there was no actual data available on the production of ionic salts, this approach made the life cycle assessment and the calculation of energy-related impacts during the production stages possible.

Finally, the freshwater ecotoxicity characterization factors related to five common ionic liquids were calculated for the first time through the use of a comprehensive fate and transport model USEtox. The characterization factors developed were later used to preform cradle-to-grave life cycle assessments of these ionic liquids. The life cycle assessments enabled us to see the differences between life cycle impacts of production steps and the hypothetical release of ionic liquids into the environment.

## 8.3 Future work on ionic liquid applications

A general and comprehensive CAILD model has now been developed and several case studies were solved to show the effectiveness of this approach. Below several suggestions that can advance the approach and expand the model furthermore are proposed.

Once more, we strongly emphasize that future research should focus on data collection as well as experimental measurements of physical and thermodynamic properties of ionic liquids. This will enable us to expand the available group contribution models to cover more diverse set of cations and anions including those which are less common or have not been studied yet. Further we also suggest development of new group contribution models of physical properties for which no model currently exists.

We also propose that process simulation models should be developed to measure the actual energy required to produce ionic liquids at a larger scale. This data would allow more accurate LCA analysis of ionic liquids enabling us to make more informed decisions about their environmental impacts before investing on a particular ionic liquid.

In addition, the studies on toxicity and corrosivity of ionic liquids are still limited. Therefore, we strongly propose that future research should focus on measuring and predicting these two properties of ionic liquids since it might be very costly to use an ionic liquid with a high degree of corrosivity or toxicity. This information will help in making an informed selection of an optimal ionic liquid for a particular application and it is necessary that before any investment on the production of ionic liquids in large scales can be made.

The use of the CAILD model, is not limited to the case studies we introduced in this dissertation. In this section of dissertation we propose several more cases for which a CAILD model can be developed accordingly to help in finding ionic liquids.

***Absorption chillers***

One of the potential where ionic liquids applications can be utilized is in absorption chiller, where the absorbent acts as a compressor. The cooling agent which is normally water would be absorbed into the absorbent agent, normally a molten salt such $Li^{+}Br^{-}$, and will get condensed. In the next step, water in the vapor phase will be separated from the absorbent in the generator by heating the solution. Furthermore, the water vapor gets cooled using the cooling water and then the liquefied vapor will be again evaporated to produce the chilling effect. Ionic liquids have shown the capability of replacing molten salts in closed-cycle absorption chiller, since they usually have an acceptable range of melting point and high decomposition temperature. The ionic liquid for this process needs to be optimized to have an acceptable crystallization temperature and high affinity towards water. An absorption chiller which performs using the optimal IL, designed through the CAILD model, would have a Coefficient of Performance (COP) comparable to that of a conventional absorption chiller without having their drawbacks such as easy crystallization.[298]

***Electrolytes***

One of the other applications of ionic liquids is as electrolyte in an electrolytic capacitor or in a battery. The high electrical conductivity of certain ionic liquids along with their tunable viscosity make them interesting for this application. An optimal ionic liquid with high ionic conductivity and electrochemical window which also has acceptable values of melting point and viscosity and is not corrosive and/or explosive can be designed through CAILD for electrochemical applications.[299]

***Lubricants***

One of the applications of ionic liquids which has attracted a lot of attention is their use as a lubricant agent. A low viscosity ionic liquid with low corrosivity can be used to reduce the friction between the metallic compounds in a particular system. A CAILD model can be developed to design an ionic liquid with good thermal properties, low viscosity and low melting point, and low corrosivity to be exploited as a lubricant of a system. Such anionic liquid would flow throughout the system to cool down the moving objects by absorbing the heat being produced.[300]

### *Surfactants*

A surfactant is a compound that has the capability of lowering the surface tension between two liquid phases or between a liquid and a solid. If further research in the future allows us to have precise group contribution models to estimate surface tension of ionic liquids as well as the accurate correlative models to predict the interfacial tension between an ionic liquid and other chemical compounds, a CAILD framework can be developed to design surfactants. The model would be able to design an optimal ionic liquid with an objective of lowering the surface tension between the two compounds of interest while having acceptable values for melting point, viscosity, toxicity and other necessary thermal properties.[301]

### *Cellulose dissolution*

Several studies have shown that certain type of ionic liquids can act as a good solvent for cellulose extraction process. Cellulose, the most abundant natural polymer on earth, can be extracted from plant residues, wood pulp or cotton by means of a solvent which can dissolve the cellulose out of the primary cell walls of green plants. The extracted cellulose can be mainly used to produce paperboard and paper. The advances on the field of ionic liquids have led to room temperature ionic liquids with relatively low viscosity that can dissolve cellulose. A CAILD model can be used to design an optimal ionic liquid with highest solvency power towards cellulose while also having a relatively low value of melting point and viscosity. In order to do that, group contribution models, such as UNIFAC, capable of predicting the solubility of a polymer (cellulose) in a solvent, should be advanced to include. The developed model would allow us to predict the activity coefficients of the cellulose in any ionic liquid of interest which can be further translated to the solubility of cellulose. This would help us find an ionic liquid, as a solvent for this process, with highest solvency power possible.[302]

# REFERENCES

(1) Rogers, R. D., & Seddon, K. R. (2003). Ionic liquids--solvents of the future?. *Science*, *302*(5646), 792-793.

(2) Armand, M., Endres, F., MacFarlane, D. R., Ohno, H., & Scrosati, B. (2009). Ionic-liquid materials for the electrochemical challenges of the future. *Nature materials*, *8*(8), 621-629.

(3) Walden, P. (1914). Molecular weights and electrical conductivity of several fused salts. *Bull. Acad. Imper. Sci.(St. Petersburg), 8*, 405-422.

(4) Welton, T. Room-temperature ionic liquids. Solvents for synthesis and catalysis. Chemical reviews 1999, 99, 2071-2084.

(5) Plechkova, N. V.; Seddon, K. R. Applications of ionic liquids in the chemical industry. Chemical Society Reviews 2008, 37, 123-150.

(6) Gani, R., Nielsen, B., & Fredenslund, A. (1991). A group contribution approach to computer-aided molecular design. *AIChE Journal*, *37*(9), 1318-1332.

(7) Achenie, L., Venkatasubramanian, V., & Gani, R. (Eds.). (2002). *Computer aided molecular design: theory and practice* (Vol. 12). Elsevier.

(8) Harper, P. M., Gani, R., Kolar, P., & Ishikawa, T. (1999). Computer-aided molecular design with combined molecular modeling and group contribution. *Fluid Phase Equilibria*, *158*, 337-347.

(9) Maranas, C. D. (1996). Optimal computer-aided molecular design: A polymer design case study. *Industrial & engineering chemistry research*, *35*(10), 3403-3414.

(10) Karunanithi, A. T., Achenie, L. E., & Gani, R. (2005). A new decomposition-based computer-aided molecular/mixture design methodology for the design of optimal solvents and solvent mixtures. *Industrial & engineering chemistry research*, *44*(13), 4785-4797.

(11) Karunanithi, A. T., Achenie, L. E., & Gani, R. (2006). A computer-aided molecular design framework for crystallization solvent design. *Chemical Engineering Science*, *61*(4), 1247-1260.

(12) Pretel, E. J., López, P. A., Bottini, S. B., & Brignole, E. A. (1994). Computer-aided molecular design of solvents for separation processes. *AIChE Journal*,*40*(8), 1349-1360.

(13) Karunanithi, A. T., & Mehrkesh, A. (2013). Computer-aided design of tailor-made ionic liquids. *AIChE Journal*, *59*(12), 4627-4640.

(14) Chong, F. K., Foo, D. C., Eljack, F. T., Atilhan, M., & Chemmangattuvalappil, N. G. (2014). Ionic liquid design for enhanced carbon dioxide capture by computer-aided molecular design approach. *Clean Technologies and Environmental Policy*, 1-12.

(15) Hada, S., Herring, R. H., Davis, S. E., & Eden, M. R. (2015). Multivariate characterization, modeling, and design of ionic liquid molecules. *Computers & Chemical Engineering*.

(16) Harini, M., Jain, S., Adhikari, J., Noronha, S. B., & Rani, K. Y. (2015). Design of an ionic liquid as a solvent for the extraction of a pharmaceutical intermediate. *Separation and Purification Technology*.

(17) Gardas, R. L., & Coutinho, J. A. (2008). Extension of the Ye and Shreeve group contribution method for density estimation of ionic liquids in a wide range of temperatures and pressures. *Fluid Phase Equilibria*, *263*(1), 26-32.

(18) Gardas, R. L., & Coutinho, J. A. (2008). A group contribution method for viscosity estimation of ionic liquids. *Fluid Phase Equilibria*, *266*(1), 195-201.

(19) Luis, P., Ortiz, I., Aldaco, R., & Irabien, A. (2007). A novel group contribution method in the development of a QSAR for predicting the toxicity (Vibrio fischeri EC 50) of ionic liquids. *Ecotoxicology and environmental safety*, *67*(3), 423-429.

(20) Carlisle, T. K., Bara, J. E., Gabriel, C. J., Noble, R. D., & Gin, D. L. (2008). Interpretation of CO2 solubility and selectivity in nitrile-functionalized room-temperature ionic liquids using a group contribution approach. *Industrial & Engineering Chemistry Research*, *47*(18), 7005-7012.

(21) Gardas, R. L., & Coutinho, J. A. (2008). A group contribution method for heat capacity estimation of ionic liquids. *Industrial & Engineering Chemistry Research*, *47*(15), 5751-5757.

(22) Lazzús, J. A. (2012). A group contribution method to predict the melting point of ionic liquids. *Fluid Phase Equilibria*, *313*, 1-6.

(23) Lazzús, J. A. (2012). A group contribution method to predict the thermal decomposition temperature of ionic liquids. *Journal of Molecular Liquids*, *168*, 87-93.

(24) Gharagheizi, F., Ilani-Kashkouli, P., Mohammadi, A. H., Ramjugernath, D., & Richon, D. (2012). Development of a group contribution method for determination of viscosity of ionic liquids at atmospheric pressure. *Chemical Engineering Science*, *80*, 326-333.

(25) Stoimenovski, J., MacFarlane, D. R., Bica, K., & Rogers, R. D. (2010). Crystalline vs. ionic liquid salt forms of active pharmaceutical ingredients: a position paper. *Pharmaceutical research*, *27*(4), 521-526.

(26) Lapkin, A. A., Plucinski, P. K., & Cutler, M. (2006). Comparative assessment of technologies for extraction of artemisinin. *Journal of natural products*, *69*(11), 1653-1664.

(27) Stoimenovski, J., MacFarlane, D. R., Bica, K., & Rogers, R. D. (2010). Crystalline vs. ionic liquid salt forms of active pharmaceutical ingredients: a position paper. *Pharmaceutical research*, *27*(4), 521-526.

(28) Karkamkar, A., Aardahl, C., & Autrey, T. (2007). Recent developments on hydrogen release from ammonia borane. *Material Matters*, *2*(2), 6-9.

(29) Rao, C. H., Venkatesan, K. A., Nagarajan, K., & Srinivasan, T. G. (2008). Dissolution of uranium oxides and electrochemical behavior of U (VI) in task specific ionic liquid. *Radiochimica acta*, *96*(7/2008), 403-409.

(30) Wu, B., Reddy, R., & Rogers, R. (2001). Novel ionic liquid thermal storage for solar thermal electric power systems. *Solar Engineering*, 445-452.

(31) Pham, T. P. T., Cho, C. W., & Yun, Y. S. (2010). Environmental fate and toxicity of ionic liquids: a review. *Water research*, *44*(2), 352-372.

(32) Pretti, C., Chiappe, C., Pieraccini, D., Gregori, M., Abramo, F., Monni, G., & Intorre, L. (2006). Acute toxicity of ionic liquids to the zebrafish (Danio rerio).*Green Chemistry*, *8*(3), 238-240.

(33) Zhao, D., Liao, Y., & Zhang, Z. (2007). Toxicity of ionic liquids. *Clean–soil, air, water*, *35*(1), 42-48.

(34) Ranke, J., Stolte, S., Störmann, R., Arning, J., & Jastorff, B. (2007). Design of sustainable chemical products the example of ionic liquids. *Chemical Reviews*,*107*(6), 2183-2206.

(35) Mehrkesh, A., & Karunanithi, A. T. (2014, July). New Perspective on Computer Aided Molecular Design: A Life Cycle Assessment Approach. In *Proceedings of the 8th International Conference on Foundations of Computer-Aided Process Design* (Vol. 34, p. 369). Elsevier.

(36) Li, X., Zhao, J., Li, Q., Wang, L., & Tsang, S. C. (2007). Ultrasonic chemical oxidative degradations of 1, 3-dialkylimidazolium ionic liquids and their mechanistic elucidations. *Dalton transactions*, (19), 1875-1880.

(37) Smiglak, M., Reichert, W. M., Holbrey, J. D., Wilkes, J. S., Sun, L., Thrasher, J. S., ... & Rogers, R. D. (2006). Combustible ionic liquids by design: is laboratory safety another ionic liquid myth?. *Chemical Communications*, (24), 2554-2556.

(38) Karunanithi, A. T., & Mehrkesh, A. (2013). Computer-aided design of tailor-made ionic liquids. *AIChE Journal*, *59*(12), 4627-4640.

(39) Cole, A. C., Jensen, J. L., Ntai, I., Tran, K. L. T., Weaver, K. J., Forbes, D. C., & Davis, J. H. (2002). Novel Brønsted acidic ionic liquids and their use as dual solvent-catalysts. *Journal of the American Chemical Society*, *124*(21), 5962-5963.

(40) Wilkes, J. S. (2004). Properties of ionic liquid solvents for catalysis. *Journal of Molecular Catalysis A: Chemical*, *214*(1), 11-17.

(41) Preiss, U., Bulut, S., & Krossing, I. (2010). In silico prediction of the melting points of ionic liquids from thermodynamic considerations: a case study on 67 salts with a melting point range of 337° C. *The Journal of Physical Chemistry B*, *114*(34), 11133-11140.

(42) Fredlake, C. P., Crosthwaite, J. M., Hert, D. G., Aki, S. N., & Brennecke, J. F. (2004). Thermophysical properties of imidazolium-based ionic liquids. *Journal of Chemical & Engineering Data*, *49*(4), 954-964.

(43) Hu, H. C., Soriano, A. N., Leron, R. B., & Li, M. H. (2011). Molar heat capacity of four aqueous ionic liquid mixtures. *Thermochimica Acta*, *519*(1), 44-49.

(44) Vila, J., Fernández-Castro, B., Rilo, E., Carrete, J., Domínguez-Pérez, M., Rodríguez, J. R.,... & Cabeza, O. (2012). Liquid–solid–liquid phase transition hysteresis loops in the ionic conductivity of ten imidazolium-based ionic liquids.*Fluid Phase Equilibria*, *320*, 1-10.

(45) Bradley, J-C., (2011), Alfa Aesar melting point data now openly available.

(46) Hughes, T. J., Syed, T., Graham, B. F., Marsh, K. N., & May, E. F. (2011). Heat capacities and low temperature thermal transitions of 1-hexyl and 1-octyl-3-methylimidazolium bis (trifluoromethylsulfonyl) amide. *Journal of Chemical & Engineering Data*, *56*(5), 2153-2159.

(47) Lourenço, C., Melo, C. I., Bogel-Łukasik, R., & Bogel-Łukasik, E. (2012). Solubility advantage of pyrazine-2-carboxamide: Application of alternative solvents on the way to the future pharmaceutical development. *Journal of Chemical & Engineering Data*, *57*(5), 1525-1533.

(48) Bradaric, C. J., Downard, A., Kennedy, C., Robertson, A. J., & Zhou, Y. (2003). Industrial preparation of phosphonium ionic liquids. *Green Chemistry*, *5*(2), 143-152.

(49) Coker, T. G., Ambrose, J., & Janz, G. J. (1970). Fusion properties of some ionic quaternary ammonium compounds. *Journal of the American Chemical Society*, *92*(18), 5293-5297.

(50) ChemFiles, Enabling Technologies-Ionic Liquids, Vol.5 No.6. [www.sigmaaldrich.com/content/dam/sigma-aldrich/docs/Aldrich/Brochure/alchemfile_v5_n6.pdf](www.sigmaaldrich.com/content/dam/sigma-aldrich/docs/Aldrich/Brochure/alchemfile_v5_n6.pdf)

(51) Li, H., Zhao, G., Liu, F., & Zhang, S. (2013). Physicochemical Characterization of MFm–-Based Ammonium Ionic Liquids. *Journal of Chemical & Engineering Data*, *58*(6), 1505-1515.

(52) Verevkin, S. P., Zaitsau, D. H., Emel'yanenko, V. N., Ralys, R. V., Yermalayeu, A. V., & Schick, C. (2013). Does alkyl chain length really matter? Structure–property relationships in thermochemistry of ionic liquids.*Thermochimica Acta*, *562*, 84-95.

(53) Nishida, T., Tashiro, Y., & Yamamoto, M. (2003). Physical and electrochemical properties of 1-alkyl-3-methylimidazolium tetrafluoroborate for electrolyte. *Journal of Fluorine Chemistry*, *120*(2), 135-141.

(54) Fletcher, S. I., Sillars, F. B., Hudson, N. E., & Hall, P. J. (2009). Physical properties of selected ionic liquids for use as electrolytes and other industrial applications. *Journal of Chemical & Engineering Data*, *55*(2), 778-782.

(55) Wachter, P., Schweiger, H. G., Wudy, F., & Gores, H. J. (2008). Efficient determination of crystallisation and melting points at low cooling and heating rates with novel computer controlled equipment. *The Journal of Chemical Thermodynamics*, *40*(10), 1542-1547.

(56) Zhang, Z. H., Tan, Z. C., Li, Y. S., & Sun, L. X. (2006). Thermodynamic investigation of room temperature ionic liquid. *Journal of thermal analysis and calorimetry*, *85*(3), 551-557.

(57) MacFarlane, D. R., Meakin, P., Amini, N., & Forsyth, M. (2001). Structural studies of ambient temperature plastic crystal ion conductors. *Journal of physics: condensed matter*, *13*(36), 8257.

(58) Lourenço, C., Melo, C. I., Bogel-Łukasik, R., & Bogel-Łukasik, E. (2012). Solubility advantage of pyrazine-2-carboxamide: Application of alternative solvents on the way to the future pharmaceutical development. *Journal of Chemical & Engineering Data*, *57*(5), 1525-1533.

(59) Liu, Q. S., Yang, M., Li, P. P., Sun, S. S., Welz-Biermann, U., Tan, Z. C., & Zhang, Q. G. (2011). Physicochemical properties of ionic liquids [C3py][NTf2] and [C6py][NTf2]. *Journal of Chemical & Engineering Data*, *56*(11), 4094-4101.

(60) Carvalho, P. J., Regueira, T., Santos, L. M., Fernandez, J., & Coutinho, J. A. (2009). Effect of Water on the Viscosities and Densities of 1-Butyl-3-methylimidazolium Dicyanamide and 1-Butyl-3-methylimidazolium Tricyanomethane at Atmospheric Pressure†. *Journal of Chemical & Engineering Data*, *55*(2), 645-652.

(61) Harris, K. R., Kanakubo, M., & Woolf, L. A. (2007). Temperature and pressure dependence of the viscosity of the ionic liquid 1-butyl-3-methylimidazolium tetrafluoroborate: viscosity and density relationships in ionic liquids. *Journal of Chemical & Engineering Data*, *52*(6), 2425-2430.

(62) Seddon, K. R., Stark, A., & Torres, M. J. (2000). Influence of chloride, water, and organic solvents on the physical properties of ionic liquids. *Pure and Applied Chemistry*, *72*(12), 2275-2287.

(63) Baker, S. N., Baker, G. A., Kane, M. A., & Bright, F. V. (2001). The cybotactic region surrounding fluorescent probes dissolved in 1-butyl-3-methylimidazolium hexafluorophosphate: effects of temperature and added carbon dioxide. *The Journal of Physical Chemistry B*, *105*(39), 9663-9668.

(64) Tokuda, H., Tsuzuki, S., Susan, M. A. B. H., Hayamizu, K., & Watanabe, M. (2006). How ionic are room-temperature ionic liquids? An indicator of the physicochemical properties. *The Journal of Physical Chemistry B*, *110*(39), 19593-19600.

(65) Jacquemin, J., Husson, P., Padua, A. A., & Majer, V. (2006). Density and viscosity of several pure and water-saturated ionic liquids. *Green Chemistry*,*8*(2), 172-180.

(66) Almeida, H. F., Passos, H., Lopes-da-Silva, J. A., Fernandes, A. M., Freire, M. G., & Coutinho, J. A. (2012). Thermophysical properties of five acetate-based ionic liquids. *Journal of Chemical & Engineering Data*, *57*(11), 3005-3013.

(67) Ghatee, M. H., Zare, M., Moosavi, F., & Zolghadr, A. R. (2010). Temperature-dependent density and viscosity of the ionic liquids 1-alkyl-3-methylimidazolium iodides: experiment and molecular dynamics simulation. *Journal of Chemical & Engineering Data*, *55*(9), 3084-3088.

(68) Freire, M. G., Teles, A. R. R., Rocha, M. A., Schröder, B., Neves, C. M., Carvalho, P. J., ... & Coutinho, J. A. (2011). Thermophysical characterization of ionic liquids able to dissolve biomass. *Journal of Chemical & Engineering Data*,*56*(12), 4813-4822.

(69) Schreiner, C., Zugmann, S., Hartl, R., & Gores, H. J. (2009). Fractional walden rule for ionic liquids: examples from recent measurements and a critique of the so-called ideal KCl line for the Walden Plot†. *Journal of Chemical & Engineering Data*, *55*(5), 1784-1788.

(70) Deive, F. J., Rivas, M. A., & Rodríguez, A. (2011). Thermophysical properties of two ionic liquids based on benzyl imidazolium cation. *The Journal of Chemical Thermodynamics*, *43*(3), 487-491.

(71) Restolho, J., Serro, A. P., Mata, J. L., & Saramago, B. (2009). Viscosity and surface tension of 1-ethanol-3-methylimidazolium tetrafluoroborate and 1-methyl-3-octylimidazolium tetrafluoroborate over a wide temperature range.*Journal of Chemical & Engineering Data*, *54*(3), 950-955.

(72) Neves, C. M., Carvalho, P. J., Freire, M. G., & Coutinho, J. A. (2011). Thermophysical properties of pure and water-saturated tetradecyltrihexylphosphonium-based ionic liquids. *The Journal of Chemical Thermodynamics*, *43*(6), 948-957.

(73) Crosthwaite, J. M., Muldoon, M. J., Dixon, J. K., Anderson, J. L., & Brennecke, J. F. (2005). Phase transition and decomposition temperatures, heat capacities and viscosities of pyridinium ionic liquids. *The Journal of Chemical Thermodynamics*, *37*(6), 559-568.

(74) Sanchez, L. G., Espel, J. R., Onink, F., Meindersma, G. W., & Haan, A. B. D. (2009). Density, viscosity, and surface tension of synthesis grade imidazolium, pyridinium, and pyrrolidinium based room temperature ionic liquids. *Journal of Chemical & Engineering Data*, *54*(10), 2803-2812.

(75) Gaciño, F. M., Regueira, T., Lugo, L., Comuñas, M. J., & Fernández, J. (2011). Influence of molecular structure on densities and viscosities of several ionic liquids. *Journal of Chemical & Engineering Data*, *56*(12), 4984-4999.

(76) Zhang, S., Sun, N., He, X., Lu, X., & Zhang, X. (2006). Physical properties of ionic liquids: database and evaluation. *Journal of physical and chemical reference data*, *35*(4), 1475-1517.

(77) Ionic Liquid Database-(IL Thermo). NIST Standard reference database # 147 (http://ILThermo.boulder.nist.gov/ILThermo/mainmenu.uix). Last accessed data of “05/10/2013”.

(78) Holbrey, J. D., & Seddon, K. R. (1999). Ionic liquids. *Clean Products and Processes*, *1*(4), 223-236.

(79) Rogers, R. D., & Seddon, K. R. (2003). Ionic liquids--solvents of the future?.*Science*, *302*(5646), 792-793.

(80) Ohno H. Electrochemical Aspects of Ionic Liquids. (2005) Hoboken: John Wiley & Sons Inc.

(81) Wood, N., & Stephens, G. (2010). Accelerating the discovery of biocompatible ionic liquids. *Phys. Chem. Chem. Phys.*, *12*(8), 1670-1674.

(82) Yan, T., Burnham, C. J., Del Pópolo, M. G., & Voth, G. A. (2004). Molecular dynamics simulation of ionic liquids: The effect of electronic polarizability. *The Journal of Physical Chemistry B*, *108*(32), 11877-11881.

(83) Cadena, C., & Maginn, E. J. (2006). Molecular simulation study of some thermophysical and transport properties of triazolium-based ionic liquids. *The Journal of Physical Chemistry B*, *110*(36), 18026-18039.

(84) Kroon, M. C., Buijs, W., Peters, C. J., & Witkamp, G. J. (2007). Quantum chemical aided prediction of the thermal decomposition mechanisms and temperatures of ionic liquids. *Thermochimica Acta*, *465*(1), 40-47.

(85) Vatamanu, J., Borodin, O., & Smith, G. D. (2010). Molecular insights into the potential and temperature dependences of the differential capacitance of a room-temperature ionic liquid at graphite electrodes. *Journal of the American Chemical Society*, *132*(42), 14825-14833.

(86) Achenie, L., Venkatasubramanian, V., & Gani, R. (Eds.). (2002). *Computer aided molecular design: theory and practice* (Vol. 12). Elsevier.

(87) Gani, R., & Brignole, E. A. (1983). Molecular design of solvents for liquid extraction based on UNIFAC. *Fluid Phase Equilibria*, *13*, 331-340.

(88) Buxton, A., Livingston, A. G., & Pistikopoulos, E. N. (1999). Optimal design of solvent blends for environmental impact minimization. *AIChE Journal*, *45*(4), 817-843.

(89) Sinha, M., Achenie, L. E., & Gani, R. (2003). Blanket wash solvent blend design using interval analysis. *Industrial & engineering chemistry research*,*42*(3), 516-527.

(90) Chavali, S., Lin, B., Miller, D. C., & Camarda, K. V. (2004). Environmentally-benign transition metal catalyst design using optimization techniques. *Computers & chemical engineering*, *28*(5), 605-611.

(91) Karunanithi, A. T., Achenie, L. E., & Gani, R. (2006). A computer-aided molecular design framework for crystallization solvent design. *Chemical Engineering Science*, *61*(4), 1247-1260.

(92) Chemmangattuvalappil, N. G., Eljack, F. T., Solvason, C. C., & Eden, M. R. (2009). A novel algorithm for molecular synthesis using enhanced property operators. *Computers & Chemical Engineering*, *33*(3), 636-643.

(93) Harper, P. M., Gani, R., Kolar, P., & Ishikawa, T. (1999). Computer-aided molecular design with combined molecular modeling and group contribution. *Fluid Phase Equilibria*, *158*, 337-347.

(94) Samudra, A. P., & Sahinidis, N. V. (2013). Optimization-based framework for computer-aided molecular design. *AIChE Journal*, *59*(10), 3686-3701.

(95) Sheldon, T. J., Folic, M., & Adjiman, C. S. (2006). Solvent design using a quantum mechanical continuum solvation model. *Industrial & engineering chemistry research*, *45*(3), 1128-1140.

(96) Odele, O., & Macchietto, S. (1993). Computer aided molecular design: a novel method for optimal solvent selection. *Fluid Phase Equilibria*, *82*, 47-54.

(97) Gardas, R. L., & Coutinho, J. A. (2008). A group contribution method for viscosity estimation of ionic liquids. *Fluid Phase Equilibria*, *266*(1), 195-201.

(98) Valderrama, J. O., Reátegui, A., & Rojas, R. E. (2009). Density of ionic liquids using group contribution and artificial neural networks. *Industrial & Engineering Chemistry Research*, *48*(6), 3254-3259.

(99) Qiao, Y., Ma, Y., Huo, Y., Ma, P., & Xia, S. (2010). A group contribution method to estimate the densities of ionic liquids. *The Journal of Chemical Thermodynamics*, *42*(7), 852-855.

(100) Ye, C., & Shreeve, J. N. M. (2007). Rapid and accurate estimation of densities of room-temperature ionic liquids and salts. *The Journal of Physical Chemistry A*, *111*(8), 1456-1461.

(101) Gardas, R. L., & Coutinho, J. A. (2008). Extension of the Ye and Shreeve group contribution method for density estimation of ionic liquids in a wide range of temperatures and pressures. *Fluid Phase Equilibria*, *263*(1), 26-32.

(102) Huo, Y., Xia, S., Zhang, Y., & Ma, P. (2009). Group contribution method for predicting melting points of imidazolium and benzimidazolium ionic liquids.*Industrial & engineering chemistry research*, *48*(4), 2212-2217.

(103) Gardas, R. L., & Coutinho, J. A. (2009). Group contribution methods for the prediction of thermophysical and transport properties of ionic liquids. *AIChE Journal*, *55*(5), 1274-1290.

(104) Gardas, R. L., & Coutinho, J. A. (2008). A group contribution method for heat capacity estimation of ionic liquids. *Industrial & Engineering Chemistry Research*, *47*(15), 5751-5757.

(105) Roughton, B. C., Christian, B., White, J., Camarda, K. V., & Gani, R. (2012). Simultaneous design of ionic liquid entrainers and energy efficient azeotropic separation processes. *Computers & Chemical Engineering*, *42*, 248-262.

(106) Luis, P., Ortiz, I., Aldaco, R., & Irabien, A. (2007). A novel group contribution method in the development of a QSAR for predicting the toxicity (Vibrio fischeri EC 50) of ionic liquids. *Ecotoxicology and environmental safety*, *67*(3), 423-429.

(107) Fredenslund, A., Jones, R. L., & Prausnitz, J. M. (1975). Group-contribution estimation of activity coefficients in nonideal liquid mixtures. *AIChE Journal*,*21*(6), 1086-1099.

(108) Wang, J., Sun, W., Li, C., & Wang, Z. (2008). Correlation of infinite dilution activity coefficient of solute in ionic liquid using UNIFAC model. *Fluid Phase Equilibria*, *264*(1), 235-241.

(109) Lei, Z., Zhang, J., Li, Q., & Chen, B. (2009). UNIFAC model for ionic liquids.*Industrial & Engineering Chemistry Research*, *48*(5), 2697-2704.

(110) Nebig, S., & Gmehling, J. (2011). Prediction of phase equilibria and excess properties for systems with ionic liquids using modified UNIFAC: Typical results and present status of the modified UNIFAC matrix for ionic liquids. *Fluid Phase Equilibria*, *302*(1), 220-225.

(111) Sinha, M., Achenie, L. E., & Ostrovsky, G. M. (1999). Environmentally benign solvent design by global optimization. *Computers & Chemical Engineering*,*23*(10), 1381-1394.

(112) Sahinidis, N. V., & Tawarmalani, M. (2000). Applications of global optimization to process and molecular design. *Computers & Chemical Engineering*, *24*(9), 2157-2169.

(113) Marcoulaki, E. C., & Kokossis, A. C. (2000). On the development of novel chemicals using a systematic synthesis approach. Part I. Optimisation framework. *Chemical Engineering Science*, *55*(13), 2529-2546.

(114) Xu, W., & Diwekar, U. M. (2005). Improved genetic algorithms for deterministic optimization and optimization under uncertainty. Part II. Solvent selection under uncertainty. *Industrial & engineering chemistry research*, *44*(18), 7138-7146.

(115) Karunanithi, A. T., Achenie, L. E., & Gani, R. (2005). A new decomposition-based computer-aided molecular/mixture design methodology for the design of optimal solvents and solvent mixtures. *Industrial & engineering chemistry research*, *44*(13), 4785-4797.

(116) Achenie, L. E. K., Gani, R., & Venkatasubramanian, V. (2002). Genetic algorithms based CAMD. *Computer Aided Molecular Design: Theory and Practice*, *12*, 95.

(117) Poling, B. E., Prausnitz, J. M., & O'connell, J. P. (2001). *The properties of gases and liquids* (Vol. 5). New York: McGraw-Hill.

(118) Aguirre, C. L., Cisternas, L. A., & Valderrama, J. O. (2012). Melting-point estimation of ionic liquids by a group contribution method. *International Journal of Thermophysics*, *33*(1), 34-46.

(119) Widegren, J. A., Saurer, E. M., Marsh, K. N., & Magee, J. W. (2005). Electrolytic conductivity of four imidazolium-based room-temperature ionic liquids and the effect of a water impurity. *The Journal of Chemical Thermodynamics*, *37*(6), 569-575.

(120) Nishida, T., Tashiro, Y., & Yamamoto, M. (2003). Physical and electrochemical properties of 1-alkyl-3-methylimidazolium tetrafluoroborate for electrolyte.*Journal of Fluorine Chemistry*, *120*(2), 135-141.

(121) Tokuda, H., Tsuzuki, S., Susan, M. A. B. H., Hayamizu, K., & Watanabe, M. (2006). How ionic are room-temperature ionic liquids? An indicator of the physicochemical properties. *The Journal of Physical Chemistry B*, *110*(39), 19593-19600.

(122) Tomida, D., Kenmochi, S., Tsukada, T., Qiao, K., & Yokoyama, C. (2007). Thermal conductivities of [bmim][PF6],[hmim][PF6], and [omim][PF6] from 294 to 335 K at pressures up to 20 MPa. *International Journal of Thermophysics*,*28*(4), 1147-1160.

(123) Van Valkenburg, M. E., Vaughn, R. L., Williams, M., & Wilkes, J. S. (2005). Thermochemistry of ionic liquid heat-transfer fluids. *Thermochimica Acta*,*425*(1), 181-188.

(124) Ge, R., Hardacre, C., Nancarrow, P., & Rooney, D. W. (2007). Thermal conductivities of ionic liquids over the temperature range from 293 K to 353 K.*Journal of Chemical & Engineering Data*, *52*(5), 1819-1823.

(125) Meindersma, G. W., Podt, A. J., & de Haan, A. B. (2005). Selection of ionic liquids for the extraction of aromatic hydrocarbons from aromatic/aliphatic mixtures. *Fuel Processing Technology*, *87*(1), 59-70.

(126) Gmehling, J. G., Anderson, T. F., & Prausnitz, J. M. (1978). Solid-liquid equilibria using UNIFAC. *Industrial & Engineering Chemistry Fundamentals*,*17*(4), 269-273.

(127) Anastas, P. T., & Warner, J. C. (2000). *Green chemistry: theory and practice*. Oxford university press.

(128) Jessop, P. G. (2011). Searching for green solvents. *Green Chemistry*, *13*(6), 1391-1398.

(129) MacFarlane, D. R., Tachikawa, N., Forsyth, M., Pringle, J. M., Howlett, P. C., Elliott, C. D., Davis, J. H., Watanabe, M., Simon, P., & Angell, C. A. (2014). Energy applications of ionic liquids. "*Energy & Environmental Science*, *7*, 232-250.

(130) Moens, L., Blake, D. M., Rudnicki, D. L., & Hale, M. J. (2003). Advanced thermal storage fluids for solar parabolic trough systems. *Journal of solar energy engineering*, *125*(1), 112-116.

(131) Armand, M., Endres, F., MacFarlane, D. R., Ohno, H., & Scrosati, B. (2009). Ionic-liquid materials for the electrochemical challenges of the future. *Nature materials*, *8*(8), 621-629.

(132) Paulechka, Y. U. (2010). Heat capacity of room-temperature ionic liquids: a critical review. *Journal of Physical and Chemical Reference Data*, *39*(3), 033108-1-033108-23.

(133) Gani, R., & Brignole, E. A. (1983). Molecular design of solvents for liquid extraction based on UNIFAC. *Fluid Phase Equilibria*, *13*, 331-340.

(134) Camarda, K. V., & Maranas, C. D. (1999). Optimization in polymer design using connectivity indices. *Industrial & Engineering Chemistry Research*, *38*(5), 1884-1892.

(135) Achenie, L., Venkatasubramanian, V., & Gani, R. (Eds.). (2003). *Computer aided molecular design: theory and practice*, *12*. Elsevier.

(136) Karunanithi, A. T., Achenie, L. E., & Gani, R. (2006). A computer-aided molecular design framework for crystallization solvent design. *Chemical Engineering Science*, *61*(4), 1247-1260.

(137) Folić, M., Adjiman, C. S., & Pistikopoulos, E. N. (2007). Design of solvents for optimal reaction rate constants. *AIChE journal*, *53*(5), 1240-1256.

(138) Samudra, A. P., & Sahinidis, N. V. (2013). Optimization-based framework for computer-aided molecular design. *AIChE Journal*, *59*(10), 3686-3701.

(139) Stavrou, M., Lampe, M., Bardow, A., & Gross, J. (2014). Continuous molecular targeting–computer-aided molecular design (CoMT–CAMD) for simultaneous process and solvent design for CO2 capture. *Industrial & Engineering Chemistry Research*, *53*(46), 18029-18041.

(140) McLeese, S. E., Eslick, J. C., Hoffmann, N. J., Scurto, A. M., & Camarda, K. V. (2010). Design of ionic liquids via computational molecular design.*Computers & chemical engineering*, *34*(9), 1476-1480.

(141) Roughton, B. C., Christian, B., White, J., Camarda, K. V., & Gani, R. (2012). Simultaneous design of ionic liquid entrainers and energy efficient azeotropic separation processes. *Computers & Chemical Engineering*, *42*, 248-262.

(142) Karunanithi, A. T., & Mehrkesh, A. (2013). Computer-aided design of tailor-made ionic liquids. *AIChE Journal*, *59*(12), 4627-4640.

(143) Chong, F. K., Foo, D. C., Eljack, F. T., Atilhan, M., & Chemmangattuvalappil, N. G. (2014). Ionic liquid design for enhanced carbon dioxide capture by computer-aided molecular design approach. *Clean Technologies and Environmental Policy*, 1-12.

(144) Hada, S., Herring, R. H., Davis, S. E., & Eden, M. R. (2015). Multivariate characterization, modeling, and design of ionic liquid molecules. *Computers & Chemical Engineering*.

(145) Valderrama, J. O., & Zarricueta, K. (2009). A simple and generalized model for predicting the density of ionic liquids. *Fluid Phase Equilibria*, *275*(2), 145-151.

(146) Valderrama, J. O., & Rojas, R. E. (2009). Critical properties of ionic liquids. Revisited. *Industrial & Engineering Chemistry Research*, *48*(14), 6890-6900.

(147) Valderrama, J. O., Martinez, G., & Rojas, R. E. (2011). Predictive model for the heat capacity of ionic liquids using the mass connectivity index. *Thermochimica Acta*, *513*(1), 83-87.

(148) a)Lazzús, J. A. (2012). A group contribution method to predict the thermal decomposition temperature of ionic liquids. *Journal of Molecular Liquids*, *168*, 87-93. b)Lazzús, J. A. (2012). A group contribution method to predict the melting point of ionic liquids. *Fluid Phase Equilibria*, *313*, 1-6.

(149) Harper, P. M., & Gani, R. (2000). A multi-step and multi-level approach for computer aided molecular design. *Computers & Chemical Engineering*, *24*(2), 677-683.

(150) Karunanithi, A. T., Achenie, L. E., & Gani, R. (2005). A new decomposition-based computer-aided molecular/mixture design methodology for the design of optimal solvents and solvent mixtures. *Industrial & engineering chemistry research*, *44*(13), 4785-4797.

(151) Pistikopoulos, E. N., & Stefanis, S. K. (1998). Optimal solvent design for environmental impact minimization. *Computers & Chemical Engineering*, *22*(6), 717-733.

(152) Venkatasubramanian, V., Chan, K., & Caruthers, J. M. (1994). Computer-aided molecular design using genetic algorithms. *Computers & Chemical Engineering*, *18*(9), 833-844.

(153) Marcoulaki, E. C., & Kokossis, A. C. (1999). Stochastic screening and scoping of separation sequences using detailed simulation models. *Computers & Chemical Engineering*, *23*, S97-S100.

(154) Gebreslassie, B. H., & Diwekar, U. M. (2015). Efficient ant colony optimization for computer aided molecular design: Case study solvent selection problem.*Computers & Chemical Engineering*, *78*, 1-9.

(155) Eljack, F. T., Abdelhady, A. F., Eden, M. R., Gabriel, F. B., Qin, X., & El-Halwagi, M. M. (2005). Targeting optimum resource allocation using reverse problem formulations and property clustering techniques. *Computers & chemical engineering*, *29*(11), 2304-2317.

(156) Chemmangattuvalappil, N. G., Solvason, C. C., Bommareddy, S., & Eden, M. R. (2010). Reverse problem formulation approach to molecular design using property operators based on signature descriptors. *Computers & Chemical Engineering*, *34*(12), 2062-2071.

(157) Patkar PR, Venkatasubramanian V. Genetic Algorithms Based CAMD. In: Achenie LEK, Gani R, Venkatasubramanian V, (2003). Editor. Computer Aided Molecular Design: Theory and Practice. Amsterdam: Elsevier Science.

(158) Anastas, P., & Eghbali, N. (2010). Green chemistry: principles and practice.*Chemical Society Reviews*, *39*(1), 301-312.

(159) Gupta, G. D., Misra, A., & Agarwal, D. K. (1991). Inhalation toxicity of furfural vapours: an assessment of biochemical response in rat lungs. *Journal of Applied Toxicology*, *11*(5), 343-347.

(160) Arts, J. H., Muijser, H., Appel, M. J., Kuper, C. F., Bessems, J. G., & Woutersen, R. A. (2004). Subacute (28-day) toxicity of furfural in Fischer 344 rats: a comparison of the oral and inhalation route. *Food and chemical toxicology*, *42*(9), 1389-1399.

(161) Gordon, C. J., Dyer, R. S., Long, M. D., & Fehlner, K. S. (1985). Effect of sulfolane on behavioral and autonomic thermoregulation in the rat. *Journal of Toxicology and Environmental Health, Part A Current Issues*, *16*(3-4), 461-468.

(162) Tilstam, U. (2012). Sulfolane: A versatile dipolar aprotic solvent. *Organic Process Research & Development*, *16*(7), 1273-1278.

(163) García, S., Larriba, M., García, J., Torrecilla, J. S., & Rodríguez, F. (2012). Separation of toluene from n-heptane by liquid–liquid extraction using binary mixtures of [bpy][BF 4] and [4bmpy][Tf 2 N] ionic liquids as solvent. *The Journal of Chemical Thermodynamics*, *53*, 119-124.

(164) Meindersma, G. W., & de Haan, A. B. (2008). Conceptual process design for aromatic/aliphatic separation with ionic liquids. *chemical engineering research and design*, *86*(7), 745-752.

(165) Ravilla, U. K., & Banerjee, T. (2012). Liquid liquid equilibria of imidazolium based ionic liquid+ pyridine+ hydrocarbon at 298.15 K: Experiments and correlations. *Fluid Phase Equilibria*, *324*, 17-27.

(166) Karunanithi, A. T., & Mehrkesh, A. (2013). Computer-aided design of tailor-made ionic liquids. *AIChE Journal*, *59*(12), 4627-4640.

(167) Zhang, S., Sun, N., He, X., Lu, X., & Zhang, X. (2006). Physical properties of ionic liquids: database and evaluation. *Journal of physical and chemical reference data*, *35*(4), 1475-1517.

(168) Ionic Liquid Database-(IL Thermo). NIST standard reference database #147. Available at:http://ILThermo.boulder.nist.gov/ILThermo/mainmenu.uix. Last accessed data: 05/10/2013.

(169) Holbrey, J. D., & Seddon, K. R. (1999). Ionic liquids. *Clean Products and Processes*, *1*(4), 223-236.

(170) Rogers, R. D., & Seddon, K. R. (2003). Ionic liquids--solvents of the future?.*Science*, *302*(5646), 792-793.

(171) Wood, N., & Stephens, G. (2010). Accelerating the discovery of biocompatible ionic liquids. *Phys. Chem. Chem. Phys.*, *12*(8), 1670-1674.

(172) Achenie LEK, Gani R, Venkatasubramanian V. (2003) Computer Aided Molecular Design: Theory and Practice. Amsterdam: Elsevier Science.

(173) Gani, R., & Brignole, E. A. (1983). Molecular design of solvents for liquid extraction based on UNIFAC. *Fluid Phase Equilibria*, *13*, 331-340.

(174) Buxton, A., Livingston, A. G., & Pistikopoulos, E. N. (1999). Optimal design of solvent blends for environmental impact minimization. *AIChE Journal*, *45*(4), 817-843.

(175) Sinha, M., Achenie, L. E., & Gani, R. (2003). Blanket wash solvent blend design using interval analysis. *Industrial & engineering chemistry research*,*42*(3), 516-527.

(176) Chavali, S., Lin, B., Miller, D. C., & Camarda, K. V. (2004). Environmentally-benign transition metal catalyst design using optimization techniques.*Computers & chemical engineering*, *28*(5), 605-611.

(177) Karunanithi, A. T., Achenie, L. E., & Gani, R. (2006). A computer-aided molecular design framework for crystallization solvent design. *Chemical Engineering Science*, *61*(4), 1247-1260.

(178) Chemmangattuvalappil, N. G., Eljack, F. T., Solvason, C. C., & Eden, M. R. (2009). A novel algorithm for molecular synthesis using enhanced property operators. *Computers & Chemical Engineering*, *33*(3), 636-643.

(179) Patkar PR, Venkatasubramanian V. (2003) Genetic Algorithms Based CAMD. In: Achenie LEK, Gani R, Venkatasubramanian V, editor. Computer Aided Molecular Design: Theory and Practice. Amsterdam: Elsevier Science.

(180) Fredenslund, A., Jones, R. L., & Prausnitz, J. M. (1975). Group‐contribution estimation of activity coefficients in nonideal liquid mixtures. *AIChE Journal*,*21*(6), 1086-1099.

(181) Wang, J., Sun, W., Li, C., & Wang, Z. (2008). Correlation of infinite dilution activity coefficient of solute in ionic liquid using UNIFAC model. *Fluid Phase Equilibria*, *264*(1), 235-241.

(182) Lei, Z., Zhang, J., Li, Q., & Chen, B. (2009). UNIFAC model for ionic liquids.*Industrial & Engineering Chemistry Research*, *48*(5), 2697-2704.

(183) Nebig, S., & Gmehling, J. (2011). Prediction of phase equilibria and excess properties for systems with ionic liquids using modified UNIFAC: Typical results and present status of the modified UNIFAC matrix for ionic liquids. *Fluid Phase Equilibria*, *302*(1), 220-225.

(184) Roughton, B. C., Christian, B., White, J., Camarda, K. V., & Gani, R. (2012). Simultaneous design of ionic liquid entrainers and energy efficient azeotropic separation processes. *Computers & Chemical Engineering*, *42*, 248-262.

(185) Dreiseitlová, J., Řehák, K., & Vreekamp, R. (2010). Mutual solubility of pyridinium-based tetrafluoroborates and toluene. *Journal of Chemical & Engineering Data*, *55*(9), 3051-3054.

(186) Corderí, S., González, E. J., Calvar, N., & Domínguez, Á. (2012). Application of [HMim][NTf 2],[HMim][TfO] and [BMim][TfO] ionic liquids on the extraction of toluene from alkanes: Effect of the anion and the alkyl chain length of the cation on the LLE. *The Journal of Chemical Thermodynamics*, *53*, 60-66.

(187) González, E. J., Requejo, P. F., Domínguez, Á., & Macedo, E. A. (2013). Phase equilibria of binary mixtures (ionic liquid+ aromatic hydrocarbon): Effect of the structure of the components on the solubility. *Fluid Phase Equilibria*,*360*, 416-422.

(188) García, S., Larriba, M., García, J., Torrecilla, J. S., & Rodríguez, F. (2010). Liquid−liquid extraction of toluene from heptane using 1-alkyl-3-methylimidazolium bis (trifluoromethylsulfonyl) imide ionic liquids. *Journal of Chemical & Engineering Data*, *56*(1), 113-118.

(189) Rogošić, M., Sander, A., & Pantaler, M. (2014). Application of 1-pentyl-3-methylimidazolium bis (trifluoromethylsulfonyl) imide for desulfurization, denitrification and dearomatization of FCC gasoline. *The Journal of Chemical Thermodynamics*, *76*, 1-15.

(190) Meindersma, G. W., Podt, A. J., & de Haan, A. B. (2006). Ternary liquid–liquid equilibria for mixtures of toluene+n-heptane+ an ionic liquid. *Fluid phase equilibria*, *247*(1), 158-168.

(191) González, B., Corderí, S., & Santamaría, A. G. (2013). Application of 1-alkyl-3-methylpyridinium bis (trifluoromethylsulfonyl) imide ionic liquids for the ethanol removal from its mixtures with alkanes. *The Journal of Chemical Thermodynamics*, *60*, 9-14.

(192) Naser, S. F., & Fournier, R. L. (1991). A system for the design of an optimum liquid-liquid extractant molecule. *Computers & chemical engineering*, *15*(6), 397-414.

(193) Mehrkesh, A.; Karunaithi, A.T. (2015) A Quantum Chemistry-based Correlative Approach for Prediction of the Melting point and Viscosity of Ionic Liquids, ready for submission.

(194) Metz, B., Davidson, O., de Coninck, H., Loos, M., & Meyer, L. (2005). Carbon dioxide capture and storage.

(195) Yu, C. H., Huang, C. H., & Tan, C. S. (2012). A review of CO2 capture by absorption and adsorption. *Aerosol and Air Quality Research*, *12*(5), 745-769.

(196) Rochelle, G. T. (2009). Amine scrubbing for CO2 capture. *Science*, *325*(5948), 1652-1654.

(197) Rao, A. B., & Rubin, E. S. (2002). A technical, economic, and environmental assessment of amine-based CO2 capture technology for power plant greenhouse gas control. *Environmental science & technology*, *36*(20), 4467-4475.

(198) Zhang, X., Zhang, X., Dong, H., Zhao, Z., Zhang, S., & Huang, Y. (2012). Carbon capture with ionic liquids: overview and progress. *Energy & Environmental Science*, *5*(5), 6668-6681.

(199) Anastas, P., & Eghbali, N. (2010). Green chemistry: principles and practice.*Chemical Society Reviews*, *39*(1), 301-312.

(200) Karunanithi, A. T., & Mehrkesh, A. (2013). Computer-aided design of tailor-made ionic liquids. *AIChE Journal*, *59*(12), 4627-4640.

(201) Zhang, S., Sun, N., He, X., Lu, X., & Zhang, X. (2006). Physical properties of ionic liquids: database and evaluation. *Journal of physical and chemical reference data*, *35*(4), 1475-1517.

(202) Ionic Liquid Database-(IL Thermo). NIST standard reference database #147. Available at: http://ILThermo.boulder.nist.gov/ILThermo/mainmenu.uix. Last accessed data: 05/10/2013.

(203) Holbrey, J. D., & Seddon, K. R. (1999). Ionic liquids. *Clean Products and Processes*, *1*(4), 223-236.

(204) Rogers, R. D., & Seddon, K. R. (2003). Ionic liquids--solvents of the future?.*Science*, *302*(5646), 792-793.

(205) Achenie LEK, Gani R, Venkatasubramanian V. (2003) Computer Aided Molecular Design: Theory and Practice. Amsterdam: Elsevier Science.

(206) Gani, R., & Brignole, E. A. (1983). Molecular design of solvents for liquid extraction based on UNIFAC. *Fluid Phase Equilibria*, *13*, 331-340.

(207) Buxton, A., Livingston, A. G., & Pistikopoulos, E. N. (1999). Optimal design of solvent blends for environmental impact minimization. *AIChE Journal*, *45*(4), 817-843.

(208) Sinha, M., Achenie, L. E., & Gani, R. (2003). Blanket wash solvent blend design using interval analysis. *Industrial & engineering chemistry research*,*42*(3), 516-527.

(209) Chavali, S., Lin, B., Miller, D. C., & Camarda, K. V. (2004). Environmentally-benign transition metal catalyst design using optimization techniques. *Computers & chemical engineering*, *28*(5), 605-611.

(210) Karunanithi, A. T., Achenie, L. E., & Gani, R. (2006). A computer-aided molecular design framework for crystallization solvent design. *Chemical Engineering Science*, *61*(4), 1247-1260.

(211) Chemmangattuvalappil, N. G., Eljack, F. T., Solvason, C. C., & Eden, M. R. (2009). A novel algorithm for molecular synthesis using enhanced property operators. *Computers & Chemical Engineering*, *33*(3), 636-643.

(212) Patkar PR, Venkatasubramanian V. (2003) Genetic Algorithms Based CAMD. In: Achenie LEK, Gani R, Venkatasubramanian V, editor. Computer Aided Molecular Design: Theory and Practice. Amsterdam: Elsevier Science.

(213) Klamt, A., Eckert, F., & Arlt, W. (2010). COSMO-RS: an alternative to simulation for calculating thermodynamic properties of liquid mixtures. *Annual review of chemical and biomolecular engineering*, *1*, 101-122.

(214) Eckert F, Klamt A. (2013) "COSMOtherm Version C3.0, Release 13.01. COSMOlogic GmbH & Co. KG, Leverkusen, Germany" Available at www.cosmologic.de.

(215) Palomar, J., Gonzalez-Miquel, M., Polo, A., & Rodriguez, F. (2011). Understanding the physical absorption of CO2 in ionic liquids using the COSMO-RS method. *Industrial & Engineering Chemistry Research*, *50*(6), 3452-3463.

(216) Anthony, J. L., Anderson, J. L., Maginn, E. J., & Brennecke, J. F. (2005). Anion effects on gas solubility in ionic liquids. *The Journal of Physical Chemistry B*, *109*(13), 6366-6374.

(217) Afzal, W., Liu, X., & Prausnitz, J. M. (2013). Solubilities of some gases in four immidazolium-based ionic liquids. *The Journal of Chemical Thermodynamics,63*, 88-94.

(218) Deng, Y., Husson, P., Delort, A. M., Besse-Hoggan, P., Sancelme, M., & Costa Gomes, M. F. (2011). Influence of an oxygen functionalization on the physicochemical properties of ionic liquids: density, viscosity, and carbon dioxide solubility as a function of temperature. *Journal of Chemical & Engineering Data*, *56*(11), 4194-4202.

(219) Jacquemin, J., Husson, P., Majer, V., & Gomes, M. F. C. (2007). Influence of the cation on the solubility of CO2 and H2 in ionic liquids based on the bis (trifluoromethylsulfonyl) imide anion. *Journal of solution chemistry*, *36*(8), 967-979.

(220) Anthony, J. L., Anderson, J. L., Maginn, E. J., & Brennecke, J. F. (2005). Anion effects on gas solubility in ionic liquids. *The Journal of Physical Chemistry B*, *109*(13), 6366-6374.

(221) Cadena, C., Anthony, J. L., Shah, J. K., Morrow, T. I., Brennecke, J. F., & Maginn, E. J. (2004). Why is CO2 so soluble in imidazolium-based ionic liquids?. *Journal of the American Chemical Society*, *126*(16), 5300-5308.

(222) Mehrkesh, A., & Karunanithi, A. T. (2015). A Quantum Chemistry-based Correlative Approach for Prediction of the Melting point and Viscosity of Ionic Liquids, ready for submission.

(223) Rogers, R. D., & Seddon, K. R. (2003). Ionic liquids--solvents of the future?. *Science*, *302*(5646), 792-793.

(224) Zhang, Y., Bakshi, B. R., & Demessie, E. S. (2008). Life cycle assessment of an ionic liquid versus molecular solvents and their applications. *Environmental science & technology*, *42*(5), 1724-1730.

(225) Righi, S., Morfino, A., Galletti, P., Samorì, C., Tugnoli, A., & Stramigioli, C. (2011). Comparative cradle-to-gate life cycle assessments of cellulose dissolution with 1-butyl-3-methylimidazolium chloride and N-methyl-morpholine-N-oxide. *Green Chemistry*, *13*(2), 367-375.

(226) Imperato, G., Koenig, B., & Chiappe, C. (2007). Ionic green solvents from renewable resources. *European journal of organic chemistry*, *2007*(7), 1049-1058.

(227) Pham, T. P. T., Cho, C. W., & Yun, Y. S. (2010). Environmental fate and toxicity of ionic liquids: a review. *Water research*, *44*(2), 352-372.

(228) Ventura, S. P., Marques, C. S., Rosatella, A. A., Afonso, C. A., Gonçalves, F., & Coutinho, J. A. (2012). Toxicity assessment of various ionic liquid families towards Vibrio fischeri marine bacteria. *Ecotoxicology and environmental safety*, *76*, 162-168.

(229) Bernot, R. J., Brueseke, M. A., Evans-White, M. A., & Lamberti, G. A. (2005). Acute and chronic toxicity of imidazolium-based ionic liquids on Daphnia magna. *Environmental Toxicology and Chemistry*, *24*(1), 87-92.

(230) Petkovic, M., Seddon, K. R., Rebelo, L. P. N., & Pereira, C. S. (2011). Ionic liquids: a pathway to environmental acceptability. *Chemical Society Reviews*,*40*(3), 1383-1403.

(231) Wells, A. S., & Coombe, V. T. (2006). On the freshwater ecotoxicity and biodegradation properties of some common ionic liquids. *Organic Process Research & Development*, *10*(4), 794-798.

(232) Matzke, M., Thiele, K., Müller, A., & Filser, J. (2009). Sorption and desorption of imidazolium based ionic liquids in different soil types. *Chemosphere*, *74*(4), 568-574.

(233) Mrozik, W., Jungnickel, C., Ciborowski, T., Pitner, W. R., Kumirska, J., Kaczyński, Z., & Stepnowski, P. (2009). Predicting mobility of alkylimidazolium ionic liquids in soils. *Journal of Soils and Sediments*, *9*(3), 237-245.

(234) Hu, S., Wang, A., Löwe, H., Li, X., Wang, Y., Li, C., & Yang, D. (2010). Kinetic study of ionic liquid synthesis in a microchannel reactor. *Chemical Engineering Journal*, *162*(1), 350-354.

(235) Wu, M., Wang, M., Liu, J., & Huo, H. (2007). *Life-cycle assessment of corn-based butanol as a potential transportation fuel* (No. ANL/ESD/07-10). Argonne National Laboratory (ANL).

(236) National Renewable Energy Laboratory (NREL), US LCI Database Project. (2010). U.S. Department of Energy: Golden, CO; www.nrel.gov/lci/about.html

(237) Chris Goemans, Athena Institute 2010 NREL US LCI Database – N.A. Electricity Generation by Fuel Type Update & Template Methodology.

(238) Rosenbaum, R. K., Bachmann, T. M., Gold, L. S., Huijbregts, M. A., Jolliet, O., Juraske, R., ... & Hauschild, M. Z. (2008). USEtox—the UNEP-SETAC toxicity model: recommended characterisation factors for human toxicity and freshwater ecotoxicity in life cycle impact assessment. *The International Journal of Life Cycle Assessment*, *13*(7), 532-546.

(239) Garcia, M. T., Gathergood, N., & Scammells, P. J. (2005). Biodegradable ionic liquids Part II. Effect of the anion and toxicology. *Green Chemistry*, *7*(1), 9-14.

(240) Ranke, J., Mölter, K., Stock, F., Bottin-Weber, U., Poczobutt, J., Hoffmann, J., ... & Jastorff, B. (2004). Biological effects of imidazolium ionic liquids with varying chain lengths in acute Vibrio fischeri and WST-1 cell viability assays.*Ecotoxicology and environmental safety*, *58*(3), 396-404.

(241) Docherty, K. M., & Kulpa Jr, C. F. (2005). Toxicity and antimicrobial activity of imidazolium and pyridinium ionic liquids. *Green Chemistry*, *7*(4), 185-189.

(242) Cho, C. W., Pham, T. P. T., Jeon, Y. C., Vijayaraghavan, K., Choe, W. S., & Yun, Y. S. (2007). Toxicity of imidazolium salt with anion bromide to a phytoplankton Selenastrum capricornutum: Effect of alkyl-chain length.*Chemosphere*, *69*(6), 1003-1007.

(243) Pham, T. P. T., Cho, C. W., Min, J., & Yun, Y. S. (2008). Alkyl-chain length effects of imidazolium and pyridinium ionic liquids on photosynthetic response of Pseudokirchneriella subcapitata. *Journal of bioscience and bioengineering*,*105*(4), 425-428.

(244) Bernot, R. J., Kennedy, E. E., & Lamberti, G. A. (2005). Effects of ionic liquids on the survival, movement, and feeding behavior of the freshwater snail, Physa acuta. *Environmental Toxicology and Chemistry*, *24*(7), 1759-1765.

(245) Ropel, L., Belvèze, L. S., Aki, S. N., Stadtherr, M. A., & Brennecke, J. F. (2005). Octanol–water partition coefficients of imidazolium-based ionic liquids.*Green Chemistry*, *7*(2), 83-90.

(246) Mrozik, W., Nichthauser, J., & Stepnowski, P. (2008). Prediction of the adsorption coefficients for imidazolium ionic liquids in soils using cyanopropyl stationary phase. *Pol. J. Environ. Stud*, *17*, 383-388.

(247) Centre for Environmental Research and Sustainable Technology (UFT), http://www.il-eco.uft.uni-bremen.de

(248) Eckelman, M. J., Mauter, M. S., Isaacs, J. A., & Elimelech, M. (2012). New perspectives on nanomaterial aquatic ecotoxicity: production impacts exceed direct exposure impacts for carbon nanotoubes. *Environmental science & technology*, *46*(5), 2902-2910.

(249) Neves, C. M., Freire, M. G., & Coutinho, J. A. (2012). Improved recovery of ionic liquids from contaminated aqueous streams using aluminium-based salts.*RSC Advances*, *2*(29), 10882-10890.

(250) Moser, P., Schmidt, S., & Stahl, K. (2011). Investigation of trace elements in the inlet and outlet streams of a MEA-based post-combustion capture process results from the test programme at the Niederaussem pilot plant. *Energy Procedia*, *4*, 473-479.

(251) Siedlecka, E. M., & Stepnowski, P. (2009). The effect of alkyl chain length on the degradation of alkylimidazolium-and pyridinium-type ionic liquids in a Fenton-like system. *Environmental Science and Pollution Research*, *16*(4), 453-458.

(252) Awad, W. H., Gilman, J. W., Nyden, M., Harris, R. H., Sutto, T. E., Callahan, J., ... & Fox, D. M. (2004). Thermal degradation studies of alkyl-imidazolium salts and their application in nanocomposites. *Thermochimica Acta*, *409*(1), 3-11.

(253) Itakura, T., Hirata, K., Aoki, M., Sasai, R., Yoshida, H., & Itoh, H. (2009). Decomposition and removal of ionic liquid in aqueous solution by hydrothermal and photocatalytic treatment. *Environmental Chemistry Letters*, *7*(4), 343-345.

(254) Anthony, J. L., Maginn, E. J., & Brennecke, J. F. (2001). Solution thermodynamics of imidazolium-based ionic liquids and water. *The Journal of Physical Chemistry B*, *105*(44), 10942-10949.

(255) Gathergood, N., Garcia, M. T., & Scammells, P. J. (2004). Biodegradable ionic liquids: Part I. Concept, preliminary targets and evaluation. *Green Chemistry*,*6*(3), 166-175.

(256) Anastas, P. T., & Warner, J. C. (2000). *Green chemistry: theory and practice*. Oxford university press.

(257) Pretti, C., Chiappe, C., Pieraccini, D., Gregori, M., Abramo, F., Monni, G., & Intorre, L. (2006). Acute toxicity of ionic liquids to the zebrafish (Danio rerio).*Green Chemistry*, *8*(3), 238-240.

(258) Weidema, B. & Roland, H. (2012). “ecoinvent data v2.2”

(259) Consultant, PRé. "SimaPro 7.3.3 LCA software", (2012). Amersfoort, Netherlands.

(260) Mehrkesh, A., & Karunanithi, A. T. (2013). Energetic ionic materials: How green are they? A comparative life cycle assessment study. *ACS Sustainable Chemistry & Engineering*, *1*(4), 448-455.

(261) Farahipour, R., & Karunanithi, A. T. (2014). Life Cycle Environmental Implications of CO2 Capture and Sequestration with Ionic Liquid 1-Butyl-3-methylimidazolium Acetate. *ACS Sustainable Chemistry & Engineering*, *2*(11), 2495-2500.

(262) Darke, G., Hawkins, T., Brand, A., Mckay, M., & Ismail, I. (2003). *Energetic, Low Melting Salts of Simple Heterocycles* (No. AFRL-PR-ED-TP-2003-007). AIR FORCE RESEARCH LAB EDWARDS AFB CA SPACE AND MISSILE PROPULSION DIV.

(263) Singh, R. P., Verma, R. D., Meshri, D. T., & Shreeve, J. N. M. (2006). Energetic Nitrogen-Rich Salts and Ionic Liquids. *Angewandte Chemie International Edition*, *45*(22), 3584-3601.

(264) Gao, H., Joo, Y. H., Twamley, B., Zhou, Z., & Shreeve, J. N. M. (2009). Hypergolic Ionic Liquids with the 2, 2-Dialkyltriazanium Cation. *Angewandte Chemie*, *121*(15), 2830-2833.

(265) Schneider, S., Hawkins, T., Rosander, M., Vaghjiani, G., Chambreau, S., & Drake, G. (2008). Ionic liquids as hypergolic fuels. *Energy & Fuels*, *22*(4), 2871-2872.

(266) Dontsova, K. M., Yost, S. L., Šimunek, J., Pennington, J. C., & Williford, C. W. (2006). Dissolution and transport of TNT, RDX, and Composition B in saturated soil columns. *Journal of Environmental Quality, 35*(6), 2043-2054.

(267) Talmage, S. S., Opresko, D. M., Maxwell, C. J., Welsh, C. J., Cretella, F. M., Reno, P. H., & Daniel, F. B. (1999). Nitroaromatic munition compounds: environmental effects and screening values. In *Reviews of environmental contamination and toxicology* (pp. 1-156). Springer New York.

(268) Alavi, G., Chung, M., Lichwa, J., D’Alessio, M., & Ray, C. (2011). The fate and transport of RDX, HMX, TNT and DNT in the volcanic soils of Hawaii: A laboratory and modeling study. *Journal of hazardous materials*, *185*(2), 1600-1604.

(269) Clausen, J., Robb, J., Curry, D., & Korte, N. (2004). A case study of contaminants on military ranges: Camp Edwards, Massachusetts, USA. *Environmental Pollution*, *129*(1), 13-21.

(270) Nipper, M., Carr, R. S., Biedenbach, J. M., Hooten, R. L., & Miller, K. (2002). Toxicological and chemical assessment of ordnance compounds in marine sediments and porewaters. *Marine pollution bulletin*, *44*(8), 789-806.

(271) Pimienta, I. S. O.; Elzey, S.; Boatz, J. A.; Gordon, M. S. Pentazole-Based Energetic Ionic Liquids: A Computational Study. *J. Phys. Chem. A* 2007, *111*, 691-703.

(272) Schmidt, M. W., Gordon, M. S., & Boatz, J. A. (2005). Triazolium-based energetic ionic liquids. *The Journal of Physical Chemistry A*, *109*(32), 7285-7295.

(273) Xue, H. O. N. G., & Shreeve, J. M. (2005). Energetic Ionic Liquids from Azido Derivatives of 1, 2, 4-Triazole. *Advanced Materials*, *17*(17), 2142-2146.

(274) Drake, G., Kaplan, G., Hall, L., Hawkins, T., & Larue, J. (2007). A new family of energetic ionic liquids 1-amino-3-alkyl-1, 2, 3-triazolium nitrates. *Journal of Chemical Crystallography*, *37*(1), 15-23.

(275) Drake, G. W., Hawkins, T. W., Boatz, J., Hall, L., & Vij, A. (2005). Experimental and Theoretical Study of 1, 5-Diamino-4-H-1, 2, 3, 4-Tetrazolium Perchlorate. *Propellants, Explosives, Pyrotechnics*, *30*(2), 156-163.

(276) Zorn, D. D., Boatz, J. A., & Gordon, M. S. (2006). Electronic structure studies of tetrazolium-based ionic liquids. *The Journal of Physical Chemistry B*, *110*(23), 11110-11119.

(277) Gutowski, K. E., Holbrey, J. D., Rogers, R. D., & Dixon, D. A. (2005). Prediction of the formation and stabilities of energetic salts and ionic liquids based on ab initio electronic structure calculations. *The Journal of Physical Chemistry B*, *109*(49), 23196-23208.

(278) Klapötke, T. M., Mayer, P., Miró Sabaté, C., Welch, J. M., & Wiegand, N. (2008). Simple, nitrogen-rich, energetic salts of 5-nitrotetrazole. *Inorganic chemistry*, *47*(13), 6014-6027.

(279) Darwich, C., Klapötke, T. M., Welch, J. M., & Suceska, M. (2007). Synthesis and Characterization of 3, 4, 5-Triamino-1, 2, 4-Triazolium 5-Nitrotetrazolate. *Propellants, Explosives, Pyrotechnics*, *32*(3), 235-243.

(280) Darwich, C., Klapötke, T. M., & Sabaté, C. M. (2008). 1, 2, 4-Triazolium-Cation-Based Energetic Salts. *Chemistry-A European Journal*, *14*(19), 5756-5771.

(281) Anastas, P. T., & Warner, J. C. (2000). *Green chemistry: theory and practice*. Oxford university press.

(282) Zhang, Y., Bakshi, B. R., & Demessie, E. S. (2008). Life cycle assessment of an ionic liquid versus molecular solvents and their applications. *Environmental science & technology, 42*(5), 1724-1730.

(283) Righi, S., Morfino, A., Galletti, P., Samorì, C., Tugnoli, A., & Stramigioli, C. (2011). Comparative cradle-to-gate life cycle assessments of cellulose dissolution with 1-butyl-3-methylimidazolium chloride and N-methyl-morpholine-N-oxide. *Green Chemistry*, *13*(2), 367-375.

(284) Ott, D., Kralisch, D., & Stark, A. (2010). Perspectives of Ionic Liquids as Environmentally Benign Substitutes for Molecular Solvents. *Handbook of Green Chemistry*.

(285) IPPC BAT Reference Document Large Volume Solid Inorganic Chemicals Family Process BREF for Soda Ash; (2004) European Soda Ash Producers Association; EUROPEAN CHEMICAL INDUSTRY COUNCIL: Brussels, Belgium.

(286) Atkins, P. (1997). *Physical Chemistry* 6$^{th}$ edition; W.H. Freeman and Company: New York, U.S.

(287) Greenlee, K. W., Henne, A. L., & Fernelius, W. C. (1946). Sodium amide. *Inorganic Syntheses, Volume 2*, 128-135.

(288) Jobelius, H. H., & Scharff, H. D. (2000). Hydrazoic acid and azides. *Ullmann's Encyclopedia of Industrial Chemistry*.

(289) Author, A. (1961). Proceedings of the Chemical Society. October 1961. *Proceedings of the Chemical Society*, *1961*(October), 357-396.

(290) Hudak, C. E., & Parsons, J. B. (1977). Industrial process profiles for environmental use: chapter 12: the explosives industry. In *Industrial process profiles for environmental use: chapter 12: the explosives industry*. NTIS.

(291) National Renewable Energy Laboratory (NREL), (2010). US LCI Database Project; US Department of Energy: Golden; www.nrel.gov/lci/about.html (accessed 02/19/2013)

(292) IEA Statistics and Balances; (2008). International Energy Agency; www.iea.org/stats/index.asp (accessed 02/19/2013)

(293) Bare, J. (2002). Tool for the Reduction and Assessment of Chemical and Other Environmental Impacts (TRACI); U.S. EPA: Cincinnati; www.epa.gov/nrmrl/std/traci/traci.html (accessed 02/19/2013)

(294) Eco-indicator 99 Manual for Designers: A damage oriented method for Life Cycle Assessment, Ministry of Housing, Spatial Planning and the environment; PRé Consultants B.V.: Amersfoort, Netherlands, 2000; www.pre-sustainability.com/download/manuals/EI99_Manual.pdf (accessed 02/19/2013)

(295) Anastas, P. T., & Warner, J. C. (2000). Green chemistry: theory and practice. Oxford university press.

(296) Jessop, P. G. (2011). Searching for green solvents. Green Chemistry, 13(6), 1391-1398.

(297) Gabriel, S., & Weiner, J. (1888). Ueber einige abkömmlinge des propylamins.Berichte der deutschen chemischen Gesellschaft, 21(2), 2669-2679.

(298) Zhang, X., & Hu, D. (2011). Performance simulation of the absorption chiller using water and ionic liquid 1-ethyl-3-methylimidazolium dimethylphosphate as the working pair. Applied Thermal Engineering, 31(16), 3316-3321.

(299) Galiński, M., Lewandowski, A., & Stępniak, I. (2006). Ionic liquids as electrolytes. Electrochimica Acta, 51(26), 5567-5580.

(300) Ye, C., Liu, W., Chen, Y., & Yu, L. (2001). Room-temperature ionic liquids: a novel versatile lubricant. Chemical Communications, (21), 2244-2245.

(301) Merrigan, T. L., Bates, E. D., Dorman, S. C., & Davis Jr, J. H. (2000). New fluorous ionic liquids function as surfactants in conventional room-temperature ionic liquids. Chemical Communications, (20), 2051-2052.

(302) Kosan, B., Michels, C., & Meister, F. (2008). Dissolution and forming of cellulose with ionic liquids. Cellulose, 15(1), 59-66.

## Appendix A: A comprehensive list of IL structural groups used in CAILD

| Cation | Structure | Valence | Anion | Structure | Groups | Valence |
|---|---|---|---|---|---|---|
| Imidazolium | 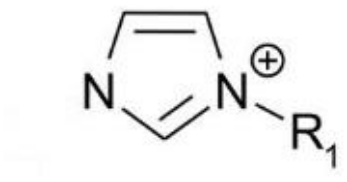<br>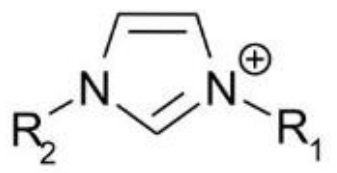<br>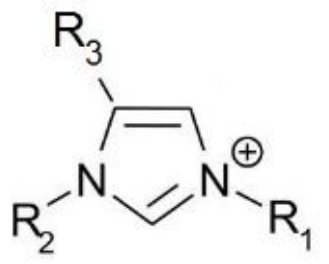<br>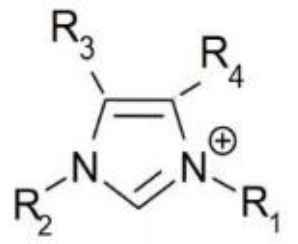<br>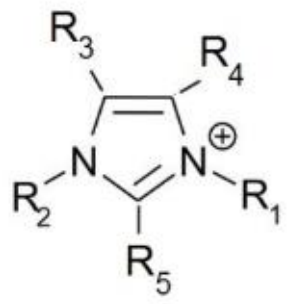 | 1<br>2<br>3<br>4<br>5 | $BF_4^-$ | 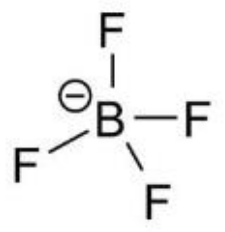 | $CH_3$ | 1 |
| Pyridinium | 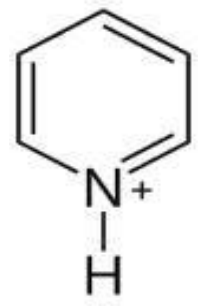 | Values 1,2,3,4,5,6 | $PF_6^-$ | 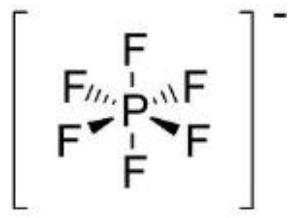 | $CH_2$ | 2 |
| Pyrrolidinium | 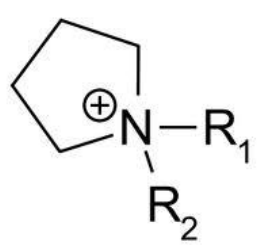 | Values 1,2,3,4,5,6 | $Tf_2N^-$ | 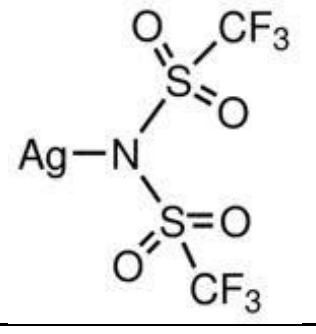 | CH | 3 |
| Ammonium | 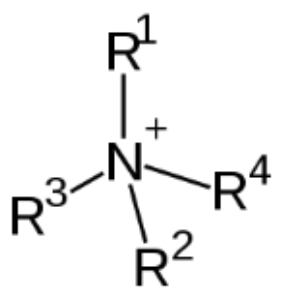 | Values 1,2,3,4 | Halogen | $Cl^-, Br^-$, .... | C= | 2 |

| Cation | Structure | Valence | Anion | Structure | Groups | Valence |
|---|---|---|---|---|---|---|
| Phosphonium | 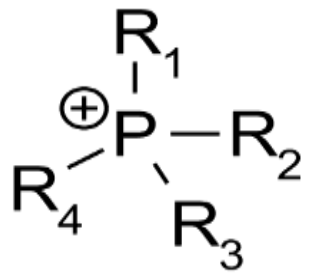 | Values 1,2,3,4 | Mesylate | 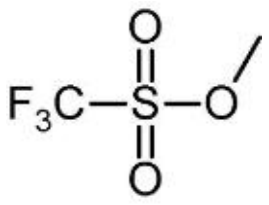 | C≡ | 1 |
| Piperidinium | 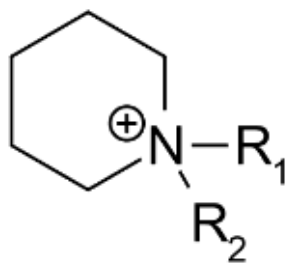 | Values 1,2,3,4,5,6,7 | Propionate | 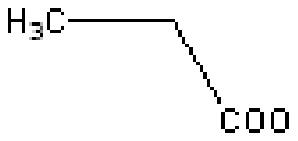 | OH | 1 |
| Triazolium | 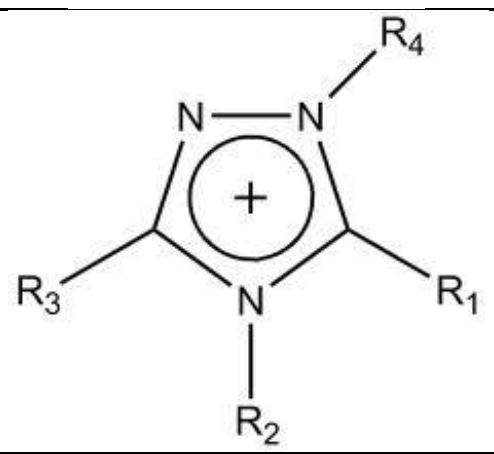 | Values 1,2,3,4,5 | Benzoate | 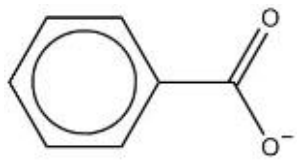 | ACH | 2 |
| Thiazolium | 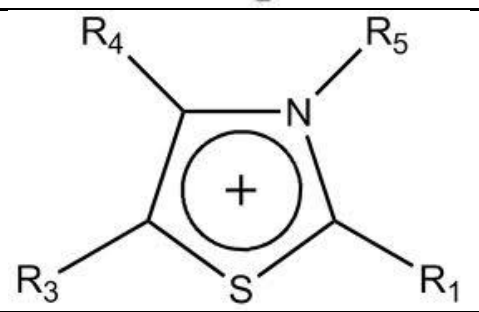 | Values 1,2,3,4 | Acetate | 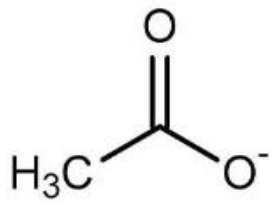 | AC | 3 |
| Pyrazinium | 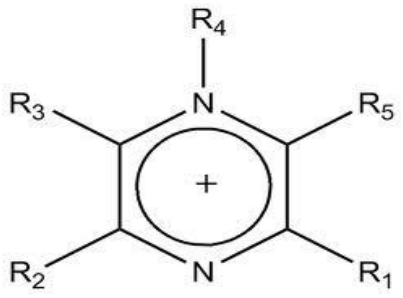 | Values 1,2,3,4,5 | Trifluoroacetate | 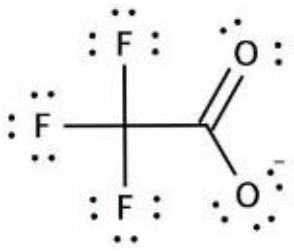 | $CH_3C=O$ | 1 |
| Sulfonium | 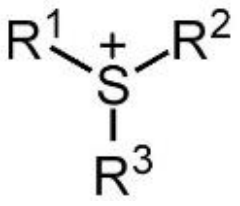 | Values 1,2,3 | Phosphate | 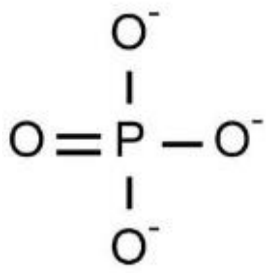 | $CH_2C=O$ | 2 |
| | | | Toluenesulfonate | 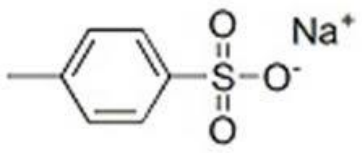 | HCO | 1 |
| | | | Hydrosulfate | 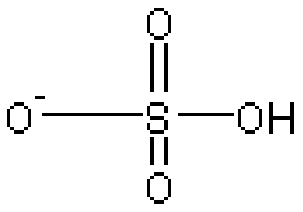 | $CH_3COO$ | 1 |
| | | | Azide | 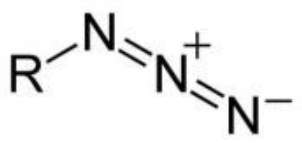 | $CH_2COO$ | 2 |

| Cation | Structure | Valence | Anion | Structure | Groups | Valence |
|---|---|---|---|---|---|---|
| | | | Perchlorate | 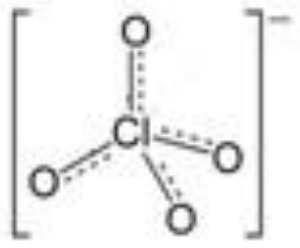 | HCOO | 1 |
| | | | | | $CH_3O$ | 1 |
| | | | | | $CH_2O$ | 2 |
| | | | | | CHO | 3 |
| | | | | | $CH_2NH_2$ | 1 |
| | | | | | $CHNH_2$ | 2 |
| | | | | | COOH | 1 |
| | | | | | COO | 1 |
| | | | | | $CH_3S$ | 1 |
| | | | | | $CH_2S$ | 2 |

## Appendix B: IL structural Groups used in CAILD for Aro/Ali separation

| Cation | Structure | Valences | Anion | Structure | Groups | Valence |
|---|---|---|---|---|---|---|
| Imidazolium | 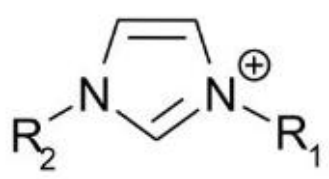 | 1,2 | $BF_4^-$ | 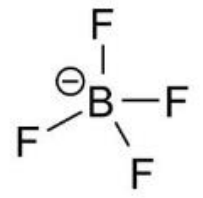 | $CH_2$-N | 2 |
| Pyridinium | 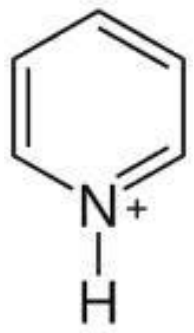 | 1,2 | $PF_6^-$ | 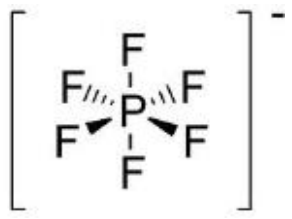 | $CH_3$-N | 1 |
| Pyrrolidinium | 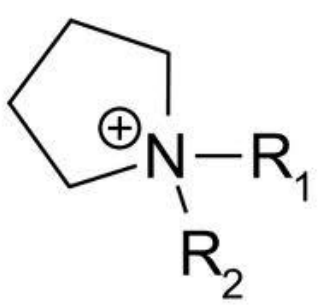 | 1,2 | $Tf_2N^-$ | 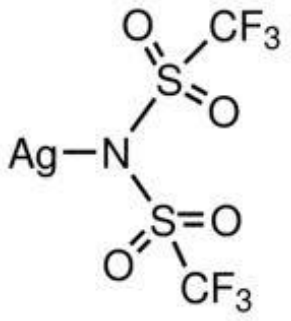 | $CH_2$-C | 2 |
| Ammonium | 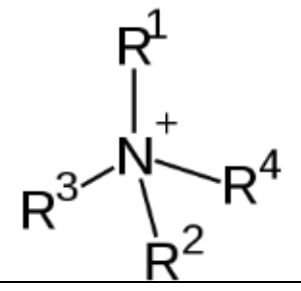 | 1,2,3,4 | Acetate | 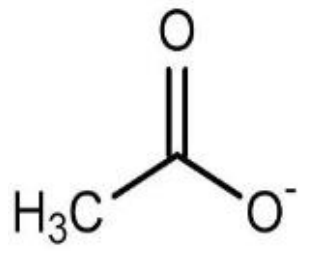 | $CH_3$-C | 1 |
| Phosphonium | 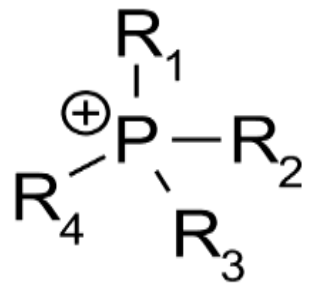 | 1,2,3,4 | Dicyanamide | 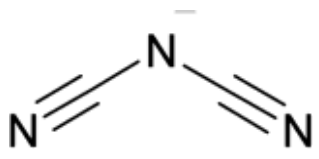 | (-$CH_2OH$) | 1 |
| Piperidinium | 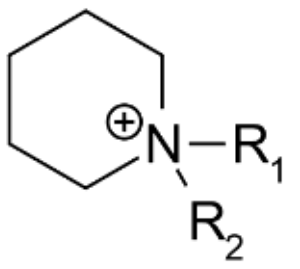 | 1,2 | Methyl sulfate | 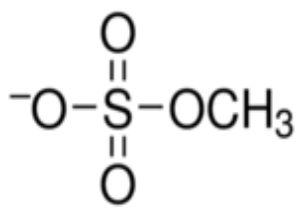 | (-$OCH_2$-) | 2 |
| Sulfonium | 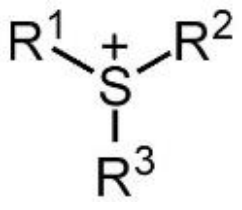 | 1,2,3 | Ethyl sulfate | 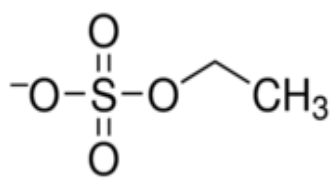 | (-$OCH_3$) | 1 |
| | | | $AlCl_4^-$ | 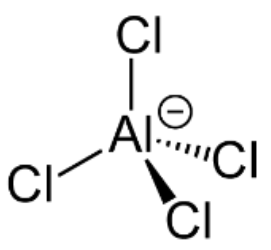 | Benzyl | 1 |
| | | | Trifluoromethanesulfonate | 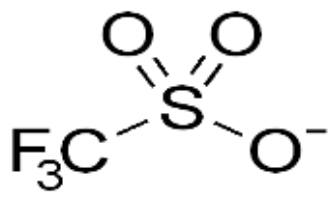 | | |

## Appendix C: UNIFAC parameters

**Table C-1:** UNIFAC Group parameters

| Groups | $R_k$ | $Q_K$ |
|---|---|---|
| $CH_2$_N | 13.8418 | 11.1728 |
| $CH_3$_N | 0.6141 | 0.1000 |
| $CH_2$_C | 3.2961 | 2.5364 |
| $CH_3$_C | 9.9824 | 9.4787 |
| -Benzyl | 40.5232 | 34.8127 |
| Benzene | 15.5701 | 14.0744 |
| Toluene | 22.8394 | 19.8815 |
| (-$OCH_2$) | 14.3225 | 12.0009 |
| (-$OCH_3$) | 21.9956 | 18.8325 |
| $CH_2OH$ | 7.3858 | 3.9473 |
| Imidazolium (Im) | 37.6794 | 25.3520 |
| Pyridinium (Py) | 27.7218 | 15.7289 |
| Acetate | 48.2941 | 58.7731 |
| $BF_4$ | 13.9150 | 15.1420 |
| dicyanamide | 14.7336 | 17.4006 |
| Ethyl sulfate | 15.8121 | 27.1192 |
| Methyl sulfate | 20.7785 | 28.2590 |
| $PF_6$ | 13.1277 | 9.7641 |
| Tetrachloroaluminate | 11.1443 | 17.7546 |
| $Tf_2N$ | 7.0149 | 20.7695 |
| Trifluoromethane-sulfonate | 10.6059 | 17.9915 |

## **Table C-2:** UNIFAC binary interaction parameters

| | $CH_2$_N | $CH_3$_N | $CH_2$_C | $CH_3$_C | Benzyl | **Benzene** | Toluene | (-$OCH_2$-) | (-$OCH_3$) | $CH_2OH$ | Im | Py | acetate | $BF_4$ | Dicyanamide | Ethyl sulfate | Methyl sulfate | $PF_6$ | Tetrachloroaluminate | $Tf_2n$ | Trifluoromethane-sulfonate |
|---|---|---|---|---|---|---|---|---|---|---|---|---|---|---|---|---|---|---|---|---|---|
| $CH_2$_N | 0.00000 | -1.16579 | 25.7773 | -6.45722 | 15.1489 | 1.48005 | 18.8038 | 5.88507 | -1.10437 | 2.02013 | -16.9582 | -5.79161 | -17.7573 | -15.99783 | -4.96941 | 0.05499 | -10.93569 | -0.80336 | 13.15483 | 15.05248 | 11.37369 |
| $CH_3$_N | -1.68084 | 0.00000 | 4.85109 | 5.06299 | 0.59709 | 2.01512 | 6.62040 | 5.40423 | 0.82860 | -1.20805 | -6.06809 | 0.37559 | 3.80043 | -1.34011 | -0.36836 | -0.18676 | -1.47925 | -0.05868 | 1.19747 | 1.23485 | -2.19861 |
| $CH_2$_C | 25.28551 | 5.99498 | 0.00000 | -47.09916 | 10.5870 | 15.0526 | -11.7414 | -19.6074 | -10.7467 | 7.95056 | 62.0555 | 75.0840 | -48.2112 | 15.77255 | 4.94174 | -21.71303 | -21.17017 | 48.3189 | -3.46539 | 16.69048 | 12.74337 |
| $CH_3$_C | -0.70579 | 5.06994 | 52.6398 | 0.00000 | 4.03457 | 4.29872 | -12.8932 | 40.9540 | -6.30428 | 2.09730 | 38.6775 | 49.7486 | 15.9366 | 29.41371 | 18.41411 | 11.70081 | 14.01750 | 10.8254 | -28.51723 | 21.49728 | 0.41066 |
| Benzyl | 14.00093 | 0.17840 | -11.2833 | 2.66468 | 0.00000 | 4.15252 | 2.02645 | 0.91927 | 0.35096 | 0.82274 | -4.84100 | 12.3605 | -17.1604 | -10.00296 | -1.72898 | -5.80848 | -13.08336 | 8.87472 | 5.45786 | 9.95203 | 8.21795 |
| Benzene | 2.30388 | 2.90971 | 18.3311 | 7.72913 | 1.14559 | 0.00000 | 0.89930 | -32.7881 | 0.29892 | 1.14945 | 2.54493 | 0.95936 | -0.39166 | 43.52084 | 30.26628 | 8.45436 | 10.15082 | 34.8469 | -17.31842 | 12.61899 | 17.02574 |
| Toluene | 15.64443 | 7.70630 | -18.0393 | -14.48392 | 2.79985 | 0.96391 | 0.00000 | 22.5853 | -0.84185 | 32.5409 | 37.5683 | 29.3082 | 3.13926 | 44.78354 | 33.04389 | 11.60019 | 14.54582 | 45.5756 | -7.33549 | -2.29163 | 21.66514 |
| (-$OCH_2$-) | 5.79003 | 4.96635 | -24.8178 | 44.89187 | 0.13243 | -33.7713 | 26.2788 | 0.00000 | 0.09726 | 0.96165 | -12.7951 | 21.0561 | -16.3011 | -8.05758 | -2.03792 | 1.09650 | -8.06744 | -0.36425 | 12.14903 | 13.34696 | 12.60302 |
| (-$OCH_3$) | -1.56356 | 0.38124 | -3.14236 | -6.94162 | 0.50097 | 0.83967 | -3.03527 | 1.00263 | 0.00000 | 0.73816 | -1.73568 | 3.08941 | 14.7912 | 10.15447 | 8.17613 | 13.91870 | 14.50267 | -22.7935 | -25.38931 | -3.77102 | 1.47467 |
| $CH_2OH$ | 1.95315 | -1.57426 | 7.94395 | 2.33672 | 0.13845 | 0.40771 | 34.9896 | 0.91483 | 0.36147 | 0.00000 | -3.45168 | -22.8505 | 5.23318 | -1.11358 | -2.11460 | -5.48680 | -3.60385 | 0.10905 | 2.39802 | 4.74375 | -1.98723 |
| Imidazolium | -17.16286 | -5.09030 | 57.6604 | 32.85272 | -5.33940 | 0.98912 | 25.7026 | -10.3675 | 0.24214 | -3.61899 | 0.00000 | 0.00000 | 0.00000 | 0.00000 | 0.00000 | 0.00000 | 0.00000 | 0.00000 | 0.00000 | 0.00000 | 0.00000 |
| Pyridinium | -5.53185 | 0.05547 | 84.3573 | 56.34316 | 16.6069 | 0.42367 | 31.6206 | 13.7473 | 1.17319 | -22.0478 | 0.00000 | 0.00000 | 0.00000 | 0.00000 | 0.00000 | 0.00000 | 0.00000 | 0.00000 | 0.00000 | 0.00000 | 0.00000 |
| acetate | -17.91113 | 2.94794 | -44.0661 | 11.43326 | -16.0744 | -1.77493 | 2.49335 | -16.8226 | 14.2412 | 3.60559 | 0.00000 | 0.00000 | 0.00000 | 0.00000 | 0.00000 | 0.00000 | 0.00000 | 0.00000 | 0.00000 | 0.00000 | 0.00000 |
| $BF_4$ | -17.19043 | -0.84555 | 11.9109 | 36.46595 | -9.14039 | 46.2113 | 45.1081 | -8.24678 | 7.28194 | -0.64549 | 0.00000 | 0.00000 | 0.00000 | 0.00000 | 0.00000 | 0.00000 | 0.00000 | 0.00000 | 0.00000 | 0.00000 | 0.00000 |
| Dicyanamide | -6.80295 | -0.42248 | -0.04199 | 21.01523 | -3.65490 | 30.1238 | 34.4092 | -3.34343 | 8.44273 | -2.60012 | 0.00000 | 0.00000 | 0.00000 | 0.00000 | 0.00000 | 0.00000 | 0.00000 | 0.00000 | 0.00000 | 0.00000 | 0.00000 |
| Ethyl sulfate | 1.69681 | -0.93907 | -27.5389 | 12.53314 | -1.11396 | 7.76601 | 15.5473 | 0.95280 | 13.2171 | -5.30778 | 0.00000 | 0.00000 | 0.00000 | 0.00000 | 0.00000 | 0.00000 | 0.00000 | 0.00000 | 0.00000 | 0.00000 | 0.00000 |
| Methyl sulfate | -11.52280 | -1.69572 | -22.9545 | 16.14358 | -12.3407 | 9.29195 | 15.0417 | -6.17558 | 14.0437 | -3.33744 | 0.00000 | 0.00000 | 0.00000 | 0.00000 | 0.00000 | 0.00000 | 0.00000 | 0.00000 | 0.00000 | 0.00000 | 0.00000 |
| $PF_6$ | -0.92455 | 0.49328 | 59.8144 | 9.51712 | 10.4494 | 35.0327 | 47.0096 | -0.17842 | -24.6162 | -0.43686 | 0.00000 | 0.00000 | 0.00000 | 0.00000 | 0.00000 | 0.00000 | 0.00000 | 0.00000 | 0.00000 | 0.00000 | 0.00000 |
| Tetrachloroaluminate | 5.26286 | 1.53392 | 1.44656 | -29.38851 | 6.91276 | -17.5325 | -6.62032 | 11.7053 | -14.6398 | 2.21476 | 0.00000 | 0.00000 | 0.00000 | 0.00000 | 0.00000 | 0.00000 | 0.00000 | 0.00000 | 0.00000 | 0.00000 | 0.00000 |
| $Tf_2n$ | 16.02941 | 0.32853 | -11.5435 | -23.38245 | 8.33547 | -12.7067 | -4.49287 | 10.7186 | -2.74782 | 5.19038 | 0.00000 | 0.00000 | 0.00000 | 0.00000 | 0.00000 | 0.00000 | 0.00000 | 0.00000 | 0.00000 | 0.00000 | 0.00000 |
| Trifluoromethane-sulfonate | 11.36804 | -2.21452 | 14.3549 | -3.59546 | 6.18483 | 14.2896 | 24.2871 | 10.9620 | 1.77800 | -2.26958 | 0.00000 | 0.00000 | 0.00000 | 0.00000 | 0.00000 | 0.00000 | 0.00000 | 0.00000 | 0.00000 | 0.00000 | 0.00000 |

## Appendix D: IL structural Groups used in CAILD for $CO_2$ capture

**Table D-1:** Structural groups used in CAILD to design an optimal IL for CO2 capture process

| Cation | Structure | Valences | Anion | Structure | Groups | Valence |
|---|---|---|---|---|---|---|
| Imidazolium | 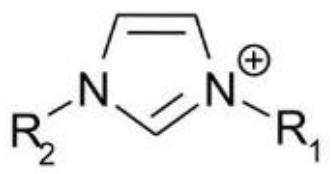 | 1,2 | $BF_4^-$ | 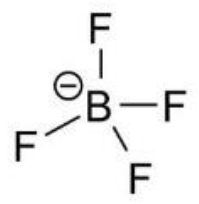 | $CH_2$ | 2 |
| Pyridinium | 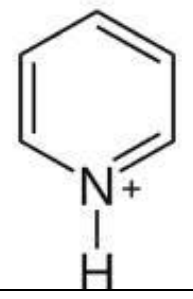 | 1,2 | $PF_6^-$ | 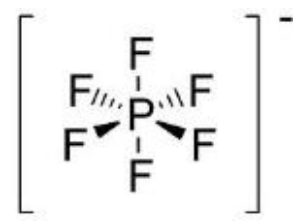 | $CH_3$ | 1 |
| Pyrrolidinium | 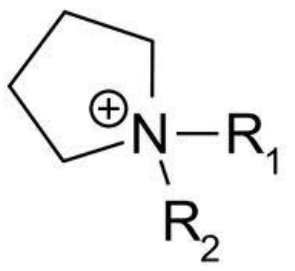 | 1,2 | $Tf_2N^-$ | 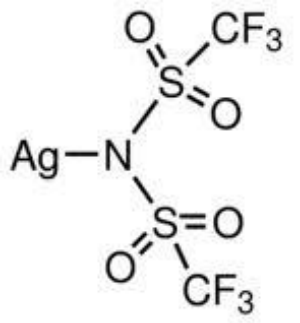 | (-OH) | 1 |
| Ammonium | 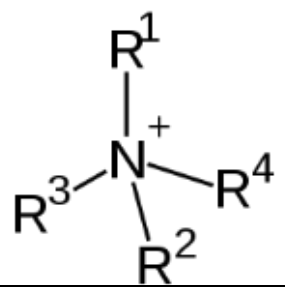 | 1,2,3,4 | Acetate | 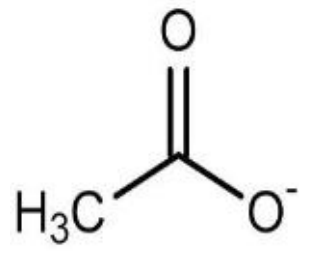 | (-O-) | 2 |
| Phosphonium | 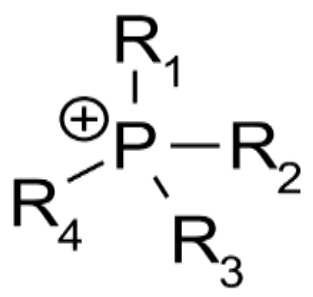 | 1,2,3,4 | Dicyanamide | 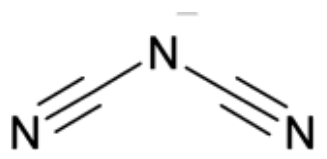 | Phenyl | 1 |
| Piperidinium | 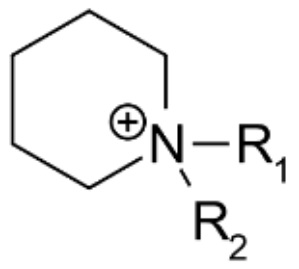 | 1,2 | Methyl sulfate | 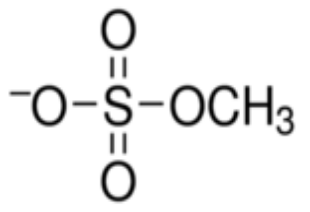 | Benzyl | 1 |
| Sulfonium | 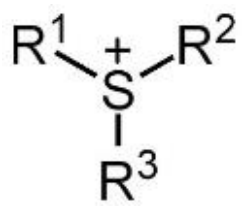 | 1,2,3 | Ethyl sulfate | 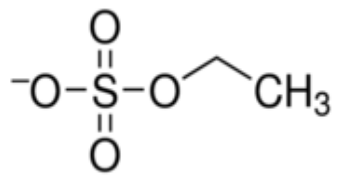 | | |
| | | | $AlCl_4^-$ | 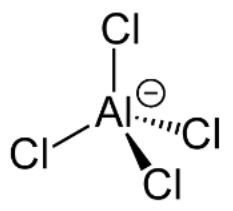 | | |
| | | | Trifluoromethanesulfonate | 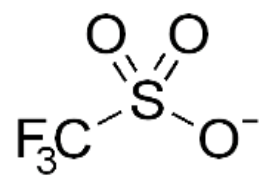 | | |

**Appendix E: Life Cycle Inventory of Ionic Liquids Production**

The life cycle inventory data of the production of Ionic Liquids (ILs) were derived from a combination of mass and energy balances from literature, theoretical calculations, and chemical process simulation. The methodology adopted to derive this inventory is presented below. Detailed material and energy flows for $[Bmim]^+[Br]^-$ is presented in Figure E-3, while a consolidated inventory of materials and energy requirements for the production of the other four ILs are presented in Table S1. The data sources for material and energy inputs are presented in Table S2.

**$[Bmim]^+[Br]^-$:** Chemical reaction and separation steps involved in the final synthesis of $[Bmim]^+[Br]^-$ using the reactants bromobutane and 1-methyl-imidazolium was modeled utilizing Aspen Plus. A schematic diagram of the process is shown in Figure E-2. Kinetics data needed to model the chemical reaction was derived from Shaozheng *et al.*[234] This experimental study used a micro-channel reaction system consisting of a micro-mixer and a tubular reactor to investigate the kinetics of the butylation of 1-methyl-imidazolium (MIM) towards the synthesis of the ionic liquid, $[Bmim]^+[Br]^-$, as shown in eqn. (E-1).

$$BrBu + MIM \rightarrow [Bmim]^+[Br]^- \qquad (E\text{-}1)$$

Our simulation included a micro-channel reactor with three tubes with an inner diameter of 1.8 mm and length of 1.13 m as specified in the experimental work by Shaozheng *et al.*[234] The residence time was 25 minutes. The molar flow rates of the reactants 1-bromobutane and 1-methylimidazole were set at 0.1 mol/hr and 0.08 mol/hr respectively. The reactants entered the reactor at 80℃ and 1 bar. The butylation process is an endothermic reaction and the reactor was simulated as an adiabatic reactor. The simulation results showed that the output flows were at about 33 ℃ and a conversion of 56% was achieved. The products were separated in a flash drum operating at 262℃ and 0.5 bar, where the unreacted 1-bromobutane and 1-methylimidazole were removed from the main product. The bottom product of the flash drum was enriched with ionic liquid having a purity of about 98%. The thermal energy needed for the separation step (flash drum) and for pre-heating the reactants was assumed to be provided by natural gas combusted in an industrial boiler. Electricity from grid was used for compression of gases, pumping the liquids, and cooling. The energy and materials required to produce 1 kg of $[Bmim]^{+}[Br]^{-}$ (based on the simulation results) are listed in the last block of Figure E-3. Inventory for the production of the reactant methyl imidazolium (MIM) was not available in any standard LCI database and therefore mass and energy needed was obtained from CPS results presented in Righi *et al.*[225] Inventory for the reactant bromobutane was also not available in any LCI database. Therefore we utilized the CPS mass and energy balances of chlorobutane provided in Righi *et al.*[225] to adjust the data for bromobutane (BuBr) as follows. Bromobutane is produced from hydrogen bromide and butanol through the reaction shown in eqn. (E-2).

$$HBr + C_4H_9OH \rightarrow C_4H_9Br + H_2O \qquad \text{(E-2)}$$

This reaction has the same stoichiometry and is similar to chlorobutane production except that hydrogen bromide is used as the reactant instead of hydrogen chloride. Inventory data for HBr was also not available in LCI databases. Hence, life cycle inventory of HCl (hydrogen chloride) was used since both reactions have the same stoichiometry and uses same process of production. The thermal and electrical energy needed to produce 1-chlorobutane reported in Righi *et al.*[225] was adjusted to 1-bromobutane based on its molecular weight and physical properties: heat of reaction and heat of vaporization. Similarly, the byproduct chlorobenzene in the LCI of HCl was replaced with equivalent amount of bromobenzene (same molar amount). Note that >95% of ecotoxicity impacts of the HCl production process is due to chlorobenzene release and substitution of it with equivalent amount of bromobenzene instead is a critical and crucial adjustment. Inventory data related to n-butanol was derived from Ecoinvent while data for other precursor materials (formaldehyde, Hydrobromic acid, ethylene glycol, ammonia, methanol, N & P fertilizers, and lime) were gathered from Ecoinvent or USLCI database (See Table S2). Whenever we had to use Ecoinvent data for chemicals derived from European databases we made a critical change of adjusting the energy mix to U.S. data (we refer to this as Ecoinvent adjusted to US). The energy and material data for the production of ionic liquid $[Bmim]^+[Br]^-$, are listed in Figure E-3.

**$[Bmim]^+[Cl]^-$:** Life cycle inventory for $[Bmim]^+[Cl]^-$ production was derived from CPS mass and energy balances collated from Righi *et al.*[225] The suggested industrial process for the production of $[Bmim]^+[Cl]^-$ was a three-step batch process.[225] Note that their inventory data was based on European energy mix while we applied US energy mix to their energy data to generate our inventory.

**$[Bmim]^+[BF_4]^-$:** The ionic liquid $[Bmim]^+[BF_4]^-$ can be produced through an anion exchange reaction as shown below

$$[Bmim]^+[Cl]^- + NaBF_4^- \longrightarrow [Bmim]^+[BF_4]^- + NaCl \quad \text{(E-3)}$$

To build the LCI for $[Bmim]^+[BF_4]^-$ production, inventory data of $NaBF_4$ from Ecoinvent database was combined with LCI data of $[Bmim]^+[Cl]^-$ discussed before. A stochometric calcualtion based on molar mass of reactants and products was used to estimate the amount of material needed to produce 1 kg of $[Bmim]^+[BF_4]^-$. As for the ion exchange reaction step, eqn. (E-3), the amount of thermal and electrical energy required was assumed to be similar to that of the final reaction step of $[Bmim]^+[Cl]^-$ production.

**$[Bmim]^+[PF_6]^-$:** The same approach was utilized to derive the LCI for production of $[Bmim]^+[PF_6]^-$. This ionic liquid was assumed to be produced through the below anion exchange reaction:

$$[Bmim]^+[Cl]^- + HPF_6 \longrightarrow [Bmim]^+[PF_6]^- + HCl \quad \text{(E-4)}$$

The inventory data for $HPF_6$ was not available in standard LCI databases and hence were derived based on theoretical estimations (using heat of reactions and heat of vaporization for reaction and separation steps) of mass and energy balances required to produce $HPF_6$ through the reaction shown below:

$$1\ H_3PO_4\ (aq) + 6\ HF\ (aq) \longrightarrow 1\ HPF_6(aq) + 4\ H_2O(l) \qquad \text{(E-5)}$$

For the upstream steps, the theoretical values of energy requirements were translated into industrisal scale data using conversion factors which were specifically developed for this purpose. These factors were derived by comparing the indutrial scale energy consumption of producing several common chemicals with their corresponding theoritcal energy requirements. In the final step of reaction tree (ion exchange step), eqn. E-4, the thermal and electrical energy required, were assumed to be similar to the final step of $[Bmim]^+[Cl]^-$ production.

**$[BPy]^+[Cl]^-$:** The ionic liquid $[BPy]^+[Cl]^-$ was assumed to be produced thorugh reaction shown in eqn. E-6.

$$BrBu + C_5H_5N\ (Pyridine) + C_4H_8O_2\ (Ethylacetate) \longrightarrow [BPy]^+[Br]^- \qquad \text{(E-6)}$$

Life cycle inventory of the precursor materials of the ionic liquid, pyridine and ethylacetate, were collected from Ecoinvent while the inventory for 1-bromobutane was assembled as discussed in the section of $[Bmim]^+[Br]^-$ production. The theoretical energy needed for the final step, eqn. E-6, was calculated based on heat of reaction and heat of vaporization as explained in the prevouis sections. The input material required to produduce 1 kg of the ionic liquid was calculated from the stoichiometry values of the reactants and products. The actual energy needed to calculate the life cycle inventory of this ionic liquid was estimated by applying the aformentioned conversion factors.

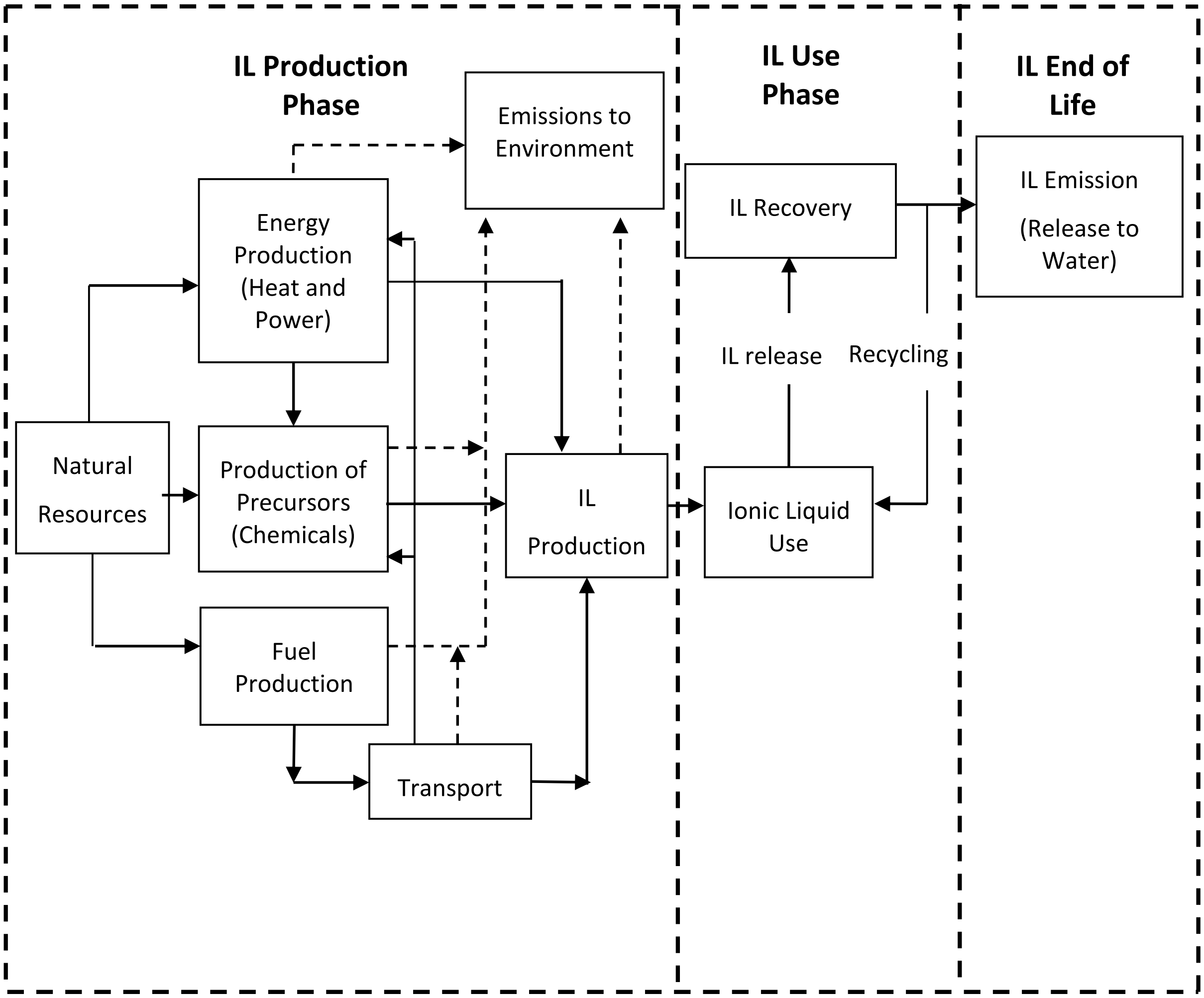

**Figure E-1:** System boundaries of the Cradle-to-Grave Life Cycle Assessment

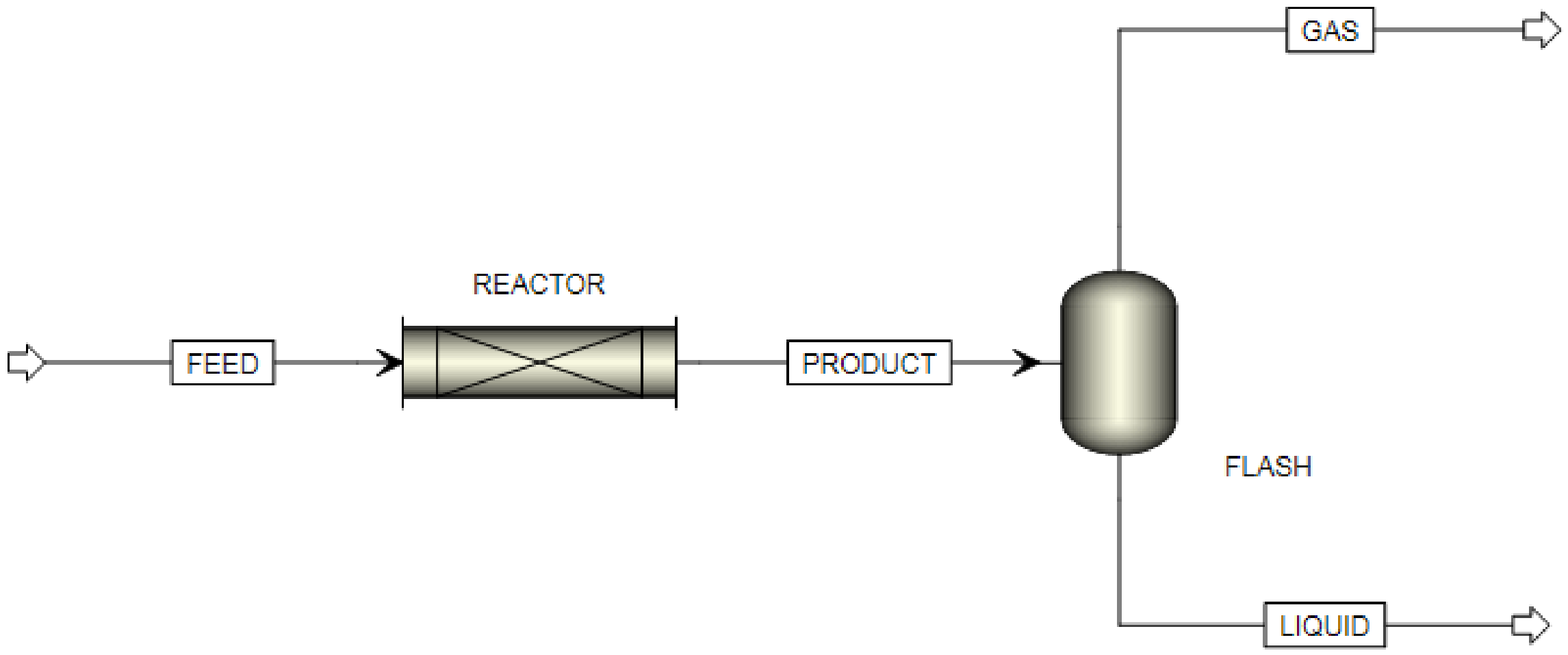

**Figure E-2:** A schematic of chemical process simulation for Production of $[Bmim]^{+}[Br]^{-}$

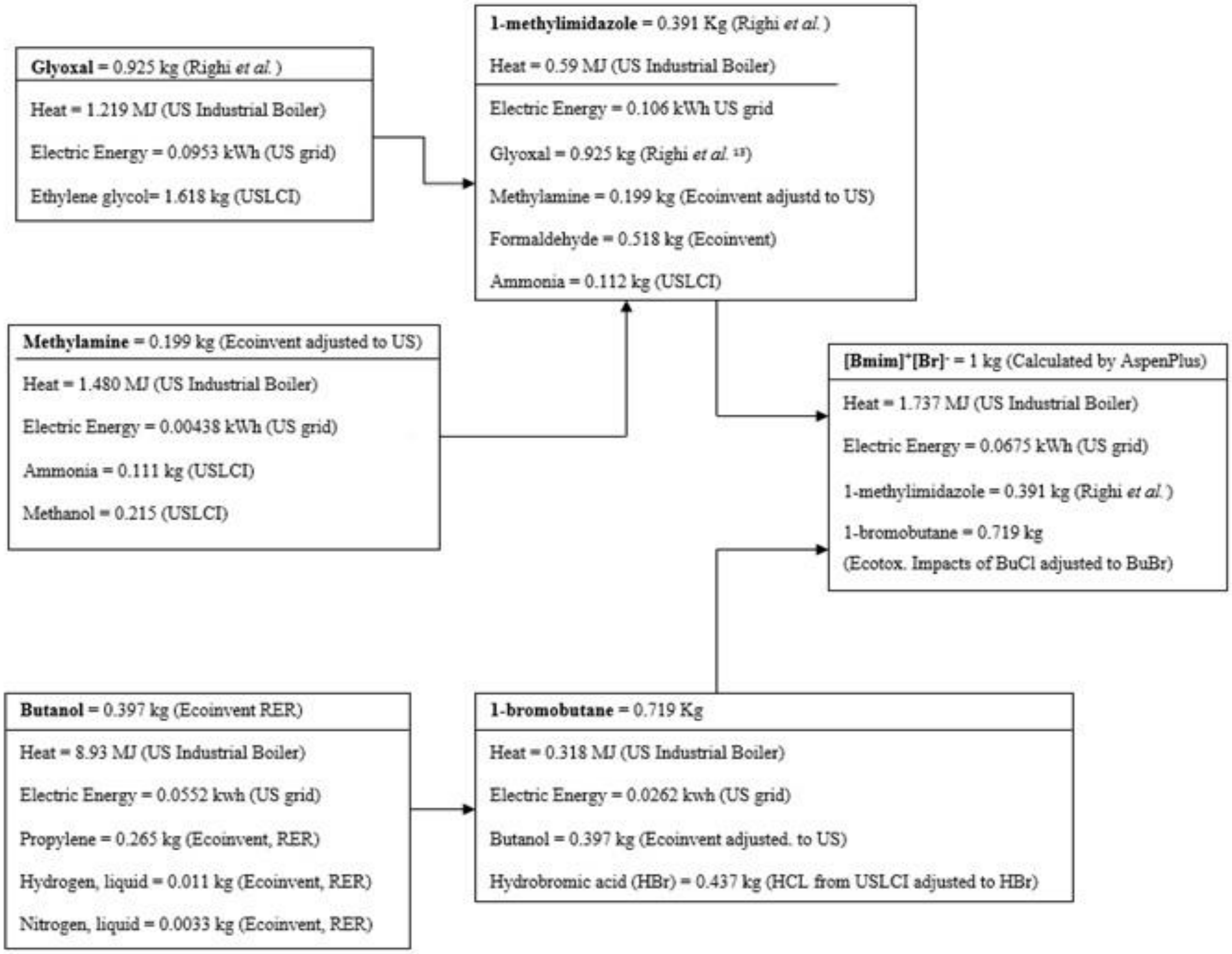

**Figure E-3:** Energy and material inventory for the production of $[Bmim]^+[Br]^-$. All values are adjusted to 1 kg of final product ($[Bmim]^+[Br]^-$)

**Table E-1:** Consolidated energy and material requirements for production of ILs

| Material (1 kg) | Inputs | | |
|---|---|---|---|
| | **Precursors** | **Thermal Energy (MJ)** | **Electricity (Kwh)** |
| $[Bmim]^+[Cl]^-$ | 1-methylimidazole (0.49 kg) [Righi *et al.*][224] | 1.5 | 0.058 |
| | 1-chlorobutane (0.61 kg) [Righi *et al.*][224] | | |
| | Ethyl acetate ( 0.04 kg) [USLCI] | | |
| $[Bmim]^+[BF_4]^-$ | Sodium tetrafluoroborate [$NaBF_4$] (0.48 kg) | 1.49 | 0.062 |
| | $[Bmim]^+[Cl]^-$ (0.77 kg) (Righi *et al.*) [224] | | |
| $[Bmim]^+[PF_6]^-$ | $HPF_6$ (0.51 kg) [This study] | 1.48 | 0.061 |
| | $[Bmim]^+[Cl]^-$ (0.61 kg) [Righi *et al.*] [224] | | |
| $[BPy]^+[Cl]^-$ | 1-chlorobutane (0.63 kg) [Righi *et al.*] [224] | 2.1 | 0.06 |
| | Pyridine (0.36 kg) [Ecoinvent adjusted to the US] | | |
| $HPF_6$ | Phosphoric acid (0.67 kg) [Ecoinvent, RER] | 1.17 | 0.316 |
| | Hydrogen fluoride (0.82 kg) [Ecoinvent, RER] | | |

**Table E-2:** Data sources

| Inventory | Reference |
|---|---|
| Glyoxal | Righi *et al.*[224] |
| 1-methylimidazole | Righi *et al.*[224] |
| Methylamine | Ecoinvent adjusted to US |
| Butanol | Ecoinvent adjusted. to US |
| Hydrobromic acid (HBr) | HCL from USLCI adjusted to HBr |
| Electric Energy | US grid, 2010 |
| Heat | US Industrial Boiler |
| 1-bromobutane | BuCl data adjusted to BuBr |
| Propylene | Ecoinvent, RER adjusted to the US |
| Hydrogen, liquid | Ecoinvent, RER adjusted to the US |
| Nitrogen, liquid | Ecoinvent, RER adjusted to the US |
| Ammonia | USLCI |
| Methanol | USLCI |
| Formaldehyde | Ecoinvent adjusted to the US |
| Ethylene glycol | USLCI |

| Inventory | Reference |
|---|---|
| $HPF_6$ | Calculated and adjusted to the US |
| $NaBF_4^-$ | Ecoinvent adjusted to the US |
| pyridine | Ecoinvent adjusted to the US |
| ethylacetate | Ecoinvent adjusted to the US |

**Table E-3:** Freshwater ecotoxicty Characterization Factors (CFs) of ionic liquids and conventional chemicals

| IL | CF (CTUe/kg) |
|---|---|
| $[Bmim]^+[Br]^-$ | 624.375 |
| $[Bmim]^+[Cl]^-$ | 747.448 |
| $[Bmim]^+[BF4]^-$ | 823.422 |
| $[Bmim]^+[PF6]^-$ | 927.05 |
| $[BPy]^+[Cl]^-$ | 1767.97 |
| Formaldehyde | 297.42 |
| Toluene | 55.91 |
| 2,3,4,6-tetrachlorophenol | 24927.8 |
| Furfural | 386.82 |
| Benzyl chloride | 818.11 |
| Benzylamine | 182.04 |
| Imidazole | 185.92 |
| 2-aminopyridine | 684.75 |
| Pyrene | 885597.45 |
| 2,6-diphenylpyridine | 71589.60 |
| 2,4-D Butyl ester | 16378.85 |
| Butanone | 12515.26 |
| 2,4,6-Trinitrotoluene | 9399.86 |

**Table E-4:** Ecotoxicity impacts related to different HCL production processes

| Process | Ecotoxicity Impacts ($CTU_e$) |
|---|---|
| Benzene Chlorination / EU | 8.22 |
| Reaction of propylene with chlorine (36% in $H_2O$) | 0.027 |
| Manheim process / EU | 0.353 |
| Reaction of $H_2$ with $Cl_2$ (chlorine) /EU | 0.553 |